\documentclass[%
reprint,
superscriptaddress,
showpacs,preprintnumbers,
 amsmath,amssymb,
 aps,
prd,
]{revtex4-1}

\usepackage{diagbox}

\usepackage[utf8]{inputenc}
\usepackage{graphicx}% Include figure files
\usepackage{dcolumn}% Align table columns on decimal point
\usepackage{bm}% bold math
\usepackage{bbold}
\usepackage{amssymb,amsmath}
\usepackage{color}
\definecolor{refcol}{RGB}{0,0,205}
\usepackage[colorlinks,linkcolor=refcol,citecolor=refcol,urlcolor=refcol]{hyperref}% add hypertext capabilities
\usepackage{color}
\usepackage{amsfonts}
\usepackage{subfigure}
\usepackage{array}

\newcommand{\Tr}{\ensuremath{\operatorname{Tr}}}
\newcommand{\tr}{\ensuremath{\operatorname{tr}}}

\newcolumntype{L}{>{\centering\arraybackslash}m{3cm}}

\definecolor{black}{rgb}{0,0,0}
\definecolor{blue}{rgb}{0,0,1}

\definecolor{green}{rgb}{0,1,0}

\definecolor{red}{rgb}{1,0,0}

\definecolor{gray}{rgb}{.5,.5,.5}

\definecolor{darkgreen}{rgb}{.0,.5,.0}

\def\Fig#1{Figure~\ref{#1}}

\def\Eq#1{Equation (\ref{#1})}
\def\eq#1{(\ref{#1})}
\def\eqref#1{(\ref{#1})}
\def\fig#1{Figure~\ref{#1}}
\def\Tab#1{Table~\ref{#1}}
\def\tab#1{Table~\ref{#1}}

\def\sec#1{Section~\ref{#1}}
\def\app#1{Appendix~\ref{#1}}

\def\lA0{{\langle A_0 \rangle}}
\def\bA0{{\bar{A}_0}}

\def\pslash{p\llap{/}}

\def\0#1#2{\frac{#1}{#2}}

\usepackage{xifthen}
\newcommand{\momarg}[1]{\ifthenelse{\isempty{#1}}{}{\left( #1 \right)}}

\definecolor{fabcol}{rgb}{0.56,0.00,1.00}

\graphicspath{{./figures/}{./}}

\begin{document}

\preprint{}

\title{The QCD phase structure at finite temperature and density} 

\author{Wei-jie Fu}
%\email[E-mail: ]{wjfu@dlut.edu.cn}
\affiliation{School of Physics, Dalian University of Technology, Dalian, 116024,
  P.R. China}

\author{Jan M. Pawlowski}
%\email[E-mail: ]{j.pawlowski@thphys.uni-heidelberg.de} 
\affiliation{Institut
 f\"{u}r Theoretische Physik, Universit\"{a}t Heidelberg,
 Philosophenweg 16, 69120 Heidelberg, Germany}
\affiliation{ExtreMe Matter Institute EMMI, GSI, Planckstr. 1, 64291
  Darmstadt, Germany}

 \author{Fabian Rennecke}
%\email[E-mail: ]{frennecke@bnl.gov}
\affiliation{Physics Department, Brookhaven National Laboratory,
  Upton, NY 11973, USA} 


\begin{abstract}
  We discuss the phase structure of QCD for $N_f=2$ and $N_f=2+1$
  dynamical quark flavours at finite temperature and baryon chemical
  potential. It emerges dynamically from the underlying fundamental
  interactions between quarks and gluons in our work. To this end,
  starting from the perturbative high-energy regime, we systematically
  integrate-out quantum fluctuations towards low energies by using the
  functional renormalisation group. By dynamically hadronising the
  dominant interaction channels responsible for the formation of light
  mesons and quark condensates, we are able to extract the phase
  diagram for \mbox{$\mu_B/T \lesssim 6$}.  We find a critical endpoint
    at
    \mbox{$(T_\text{\tiny CEP},{\mu_B}_{\text{\tiny CEP}})=(107,
      635)\,\text{MeV}$}. The curvature of the phase boundary at small
    chemical potential is $\kappa=0.0142(2)$, computed from the
    renormalised light chiral condensate $\Delta_{l,R}$.  Furthermore,
    we find indications for an inhomogeneous regime in the vicinity
    and above the chiral transition for $\mu_B\gtrsim 417$\,
    MeV. Where applicable, our results are in very good agreement with
    the most recent lattice results. We also compare to results from
    other functional methods and phenomenological freeze-out
    data. This indicates that a consistent picture of the phase
    structure at finite baryon chemical potential is beginning to
    emerge. The systematic uncertainty of our results grows large in
    the density regime around the critical endpoint and we discuss
    necessary improvements of our current approximation towards a
    quantitatively precise determination of QCD phase diagram.
\end{abstract}

                             % Classification Scheme.
\maketitle

%\tableofcontents 

\section{Introduction}\label{sec:intr}

The detailed understanding of the QCD phase structure at finite
temperature and density is not only essential for our understanding of
the formation of matter, but also for the interpretation and
prediction of the wealth of data collected at running and planned
heavy-ion experiments. For overviews on both, experimental
measurements and theoretical studies, see e.g.\ the reviews
\cite{Luo:2017faz,Adamczyk:2017iwn,Andronic:2017pug,Stephanov:2007fk,%
  Andersen:2014xxa,Shuryak:2014zxa,Fischer:2018sdj,Yin:2018ejt} and
references therein. While our experimental and theoretical
understanding of the phase structure at small densities has advanced
rapidly in the past decade, at large densities theoretical and
experimental investigations have been hampered by several
intricacies. They range, e.g., from the sign problem of lattice gauge
theory \cite{deForcrand:2010ys} to the influence of finite detector
efficiency on signatures of the phase transition \cite{Bzdak:2012ab}.

This leaves us with many highly relevant open questions regarding the
phase diagram, in particular the existence and location of a critical
end point (CEP) and the phase structure at small temperatures and
large densities. The relevance of a CEP derives from the fact that the
phase transition is of second order at this point. The resulting
critical long-range correlations can potentially be observed, e.g., in
particle number correlations measured in heavy-ion experiments, see
e.g.\ \cite{Rennecke:2019lus}.  Within the Beam Energy Scan (BES)
Program at RHIC, significant measurements have been performed in this
direction \cite{Adamczyk:2013dal, Adamczyk:2014fia, Luo:2015ewa,
  Adamczyk:2017iwn, Adamczyk:2017wsl}. This will be extended in BES
phase II. Experimental studies of the QCD phase structure are also
planned or run at other facilities with different collision energies
and luminosities, such as CBM at FAIR \cite{Friman:2011zz} and HADES
at GSI \cite{Agakishiev:2009am} in Germany, NA61/SHINE at CERN
\cite{Abgrall:2014xwa}, the NICA/MPD in Russia \cite{Sorin:2011zz},
J-PARC-HI in Japan \cite{Sakaguchi:2017ggo}, and HIAF in China
\cite{Yang:2013yeb}, see also \cite{Dainese:2019xrz, Alemany:2019vsk,
  Adamova:2019vkf}.

Theoretical investigations of the QCD phase structure have been
performed with first principle approaches to QCD, such as functional
approaches and lattice simulations, and with low energy effective
theories. In the past decade functional approaches like the functional
renormalisation group (fRG), see e.g.\ \cite{Braun:2008pi,
  Braun:2009gm, Mitter:2014wpa, Braun:2014ata, Rennecke:2015eba,
  Fu:2016tey, Cyrol:2016tym, Cyrol:2017ewj, Cyrol:2017qkl,
  Fu:2018qsk}, and Dyson-Schwinger equations (DSE), see e.g.\
\cite{Qin:2010nq, Fischer:2011mz, Fischer:2014ata, Shi:2014zpa,
  Gao:2016qkh, Fischer:2018sdj} have made significant progress in the
description of the QCD phase structure, for lattice simulations, see
e.g.\ \cite{Bazavov:2012vg, Borsanyi:2013hza, Borsanyi:2014ewa,
  Bonati:2015bha, Bellwied:2015rza, Bazavov:2017dus, Bazavov:2017tot,
  Bonati:2018nut, Borsanyi:2018grb, Bazavov:2018mes, Guenther:2018flo,
  Ding:2019prx}.

In the present work we evaluate the phase structure of $N_f=2$ and
$N_f=2+1$ flavour QCD within the fRG approach as introduced in
\cite{Wetterich:1992yh}, see also \cite{Ellwanger:1993mw,
  Morris:1993qb}. For QCD-related reviews see e.g.\
\cite{Berges:2000ew, Pawlowski:2005xe, Schaefer:2006sr, Gies:2006wv,
  Rosten:2010vm, Braun:2011pp, Pawlowski:2014aha}.  We built upon the
fRG-results for Yang-Mills theory in the vacuum, \cite{Cyrol:2016tym},
and at finite temperature, \cite{Cyrol:2017qkl}, as well as vacuum QCD
results in the quenched approximation, \cite{Mitter:2014wpa}, and in
full unquenched QCD, \cite{Braun:2014ata, Rennecke:2015eba,
  Cyrol:2017ewj}. This work is extended to unquenched QCD at finite
temperature and density, which gives us access to the chiral and
confinement-deconfinement phase structure of QCD in terms of QCD
correlation functions.

%%%%%%%%%%%%%%%%%%%%%%%%%%%%% 
\begin{figure}[t]
\includegraphics[width=1\columnwidth]{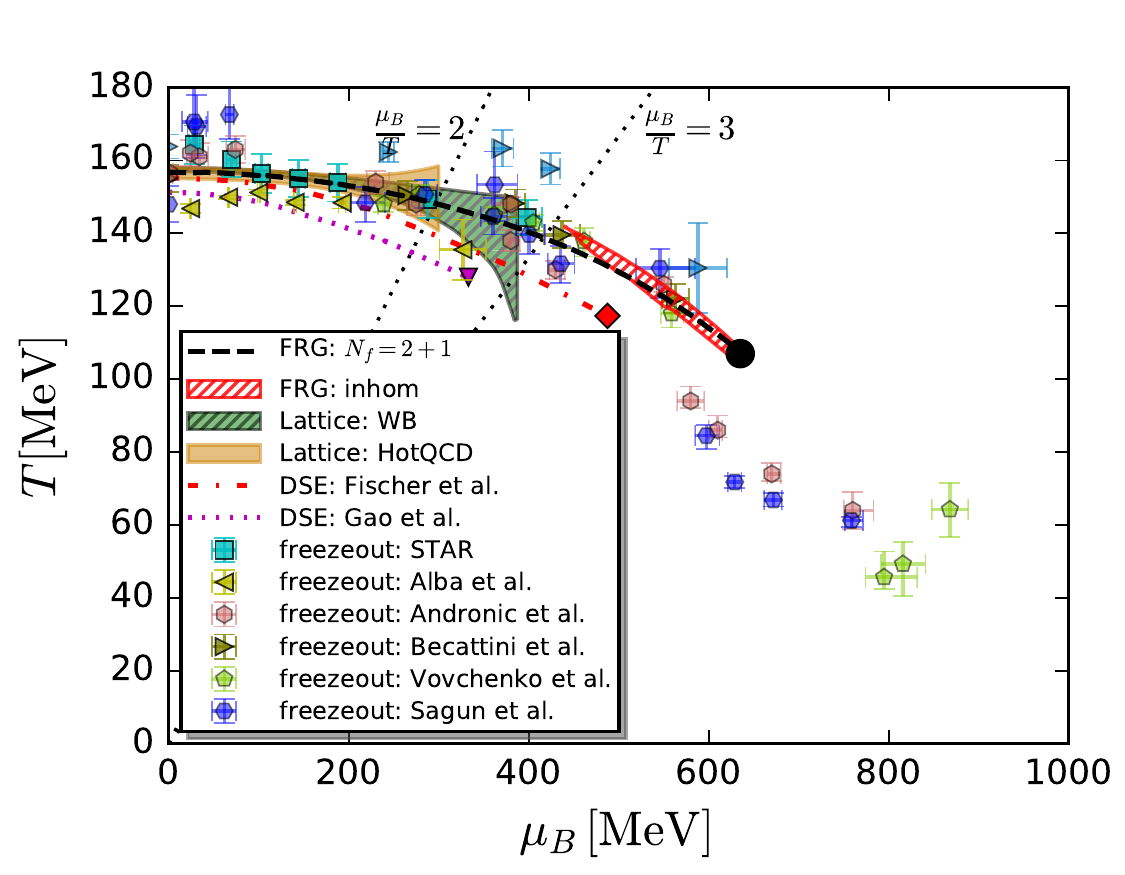}
\caption{Phase diagram for $N_f=2+1$ flavour QCD in comparison to
  other theoretical approaches and phenomenological freeze-out
  data. Our result for the chiral crossover is depicted by the black
  dashed line. The crossover temperature has been determined through
  the peak position of the thermal susceptibility of the renormalised
  light chiral condensate, $\partial_T \Delta_{l,R}$, at fixed baryon
  chemical potential $\mu_B$. For more details see
  \sec{sec:Conf+Chiral}, and in particular \Fig{fig:DeltalR}. We show
  dotted black lines for $\mu_B/T=2,3$ to indicate the reliability
  bounds for the
  lattice and functional methods.\\
  The phase boundary globally agrees well with recent lattice results.
  In particular the curvature of the phase boundary for small chemical
  potential, \mbox{$\kappa=0.0142(2)$}, is consistent with recent
  lattice results, $\kappa=0.0149(21)$ in \cite{Bellwied:2015rza},
  $\kappa=0.0144(26)$ in \cite{Bonati:2018nut}, and $\kappa=0.015(4)$
  in \cite{Bazavov:2018mes}, for an overview see
  \cite{DElia:2018fjp}. We find a critical end point at
  $(T_\text{\tiny CEP},{\mu_B}_{\text{\tiny CEP}})=(107,
  635)\,\text{MeV}$. Indications for an inhomogeneous regime close to
  the chiral phase transition for $\mu_B\gtrsim 420$\, MeV are
  depicted by the hatched red area. For quantitative statements in
  this area the current approximation has to be upgraded
  systematically. Accordingly the hatched red area also serves as a
  reliability bound for the current approximation. For more details
  see \sec{sec:Inhom} and
  Figure~\ref{fig:PhasediagramInhom}.\\[0.5ex]
  \textit{Other theoretical results}: lattice QCD based on an analytic
  continuation from the imaginary chemical potential
  \cite{Bellwied:2015rza} (WB), lattice QCD based on a Taylor
  expansion in chemical potential \cite{Bazavov:2018mes} (HotQCD), DSE
  approach with backcoupled quarks and a dressed vertex
  \cite{Fischer:2014ata} (Fischer {\it et al.}), and DSE calculations
  with a gluon model
  \cite{Gao:2015kea} (Gao {\it et al.}).\\[0.5ex]
  \textit{Freeze-out data}: \cite{Adamczyk:2017iwn} (STAR),
  \cite{Alba:2014eba} (Alba {\it et al.}), \cite{Andronic:2017pug}
  (Andronic {\it et al.}), \cite{Becattini:2016xct} (Becattini {\it et
    al.}), \cite{Vovchenko:2015idt} (Vovchenko {\it et al.}), and
  \cite{Sagun:2017eye} (Sagun {\it et al.}). Note that freeze-out data
  from Becattini {\it et al}.\ with (light blue) and without (dark
  green) afterburner-corrections are shown in two different
  colors.}\label{fig:phasediagramII}
\end{figure}
%%%%%%%%%%%%%%%%%%%%%%%%%%%%%

Our study cumulates in a prediction for the QCD phase diagram for
$\mu_B/T \lesssim 6$. For $N_f=2+1$ it is presented in
\Fig{fig:phasediagramII} on the next page together with a survey of
other theoretical predictions as well as a compilation of freeze-out
points from different incarnations of the hadron resonance gas. We
define the chiral phase boundary through the renormalised light chiral
condensate $\Delta_{l,R}$. At small and intermediate baryon chemical
potential the transition is a crossover. At $\mu_B = 0$ the
pseudocritical temperature is $T_c = 156$\,MeV. The curvature of the
chiral phase boundary at small chemical potential is
\mbox{$\kappa=0.0142(2)$}. With increasing $\mu_B$, the crossover
becomes sharper and we find a critical endpoint at
\begin{align}
(T_\text{\tiny CEP},{\mu_B}_{\text{\tiny CEP}})=(107, 635)\,\text{MeV}\,.
\end{align}
Our results for the chiral phase boundary are depicted by the black
dashed line in \Fig{fig:phasediagramII}.

In addition to a CEP, we also find indications for an inhomogeneous
regime for $\mu_B \gtrsim 420$\,MeV in the vicinity and above the
chiral phase boundary. It is given by the region in the phase diagram
where mesonic dispersion relation develop a minimum at nonvanishing
spatial momentum, for more details we refer to \sec{sec:Inhom}. This
indicates a potential instability towards the formation of an
inhomogeneous quark condensate. The region where this regime has
significant overlap with the homogeneous chiral condensate is shown by
the red hatched area in \Fig{fig:phasediagramII}. Within
this area, a competition between homogeneous and inhomogeneous quark
condensation has to be taken into account. Hence, this already
suggests that the systematic error of the present approximation grows
large for $\mu_B/T \gtrsim 3$.

In \Fig{fig:phasediagramII} also we compare our results to recent
predictions of lattice gauge theory for the phase structure at small
$\mu_B/T$ from the Wuppertal-Budapest Collaboration
\cite{Bellwied:2015rza} (WB) and the HotQCD Collaboration
\cite{Bazavov:2018mes} (HotQCD). Our result for the pseudocritical
temperature and the curvature of the phase boundary agree very well
with the lattice. We also show predictions of the DSE approach from
different groups, \cite{Fischer:2014ata} (Fischer {\it et al.}) and
\cite{Gao:2015kea} (Gao {\it et al.}). Finally we included the
freeze-out data from \cite{Adamczyk:2017iwn} (STAR),
\cite{Alba:2014eba} (Alba {\it et al.}), \cite{Andronic:2017pug}
(Andronic {\it et al.}), \cite{Becattini:2016xct} (Becattini {\it et
  al.}), and \cite{Vovchenko:2015idt} (Vovchenko {\it et al.}). The
freeze-out points are surprisingly close to our result for the chiral
phase boundary, even at larger $\mu_B$.  All in all, we see that a
consistent picture of the QCD phase boundary at finite density starts
to emerge form a culmination of results from different sources.

In order to discuss the implication for CEP searches, it is
instructive to convert $\mu_B$ to the center-of-mass beam energy per
nucleon, $\sqrt{s}$. Assuming the connection between these quantities
is captured by the statistical hadronisation scenario, one finds to a
very good approximation for central collisions the relation
$\sqrt{s} = (a/\mu_B-1)/b$ with $a = 1307.5\,\text{MeV}$ and
$b = 0.288\,\text{GeV}^{-1}$ \cite{Andronic:2017pug}. This yields for
our prediction of the location of the CEP the beam energy
\begin{align}
  \sqrt{s_\text{\tiny CEP}} \approx 3.7\, \text{GeV}\,.
\end{align}
This is clearly below the smallest beam energy of current BES
measurements of $\sqrt{s} = 7.7\,\text{GeV}$, but well within reach of
future experiments such as FAIR's SIS100 \cite{Friman:2011zz}, NICA
MPD \cite{Sorin:2011zz}, J-PARC HI \cite{Sakaguchi:2017ggo}, and
STAR's Fixed-Target (FXT) program \cite{Meehan:2016qon}, see also
\cite{Dainese:2019xrz, Alemany:2019vsk, Adamova:2019vkf}. Our results
therefore provide a strong motivation for CEP searches at these future
experiments.  Furthermore, the inhomogeneous regime appears to be also
within reach of heavy-ion collisions at small beam-energies.  Hence,
looking for experimental signatures of this regime might be a
worthwhile endeavour.

This work is organised as follows. In \sec{sec:FunRG-QCD} we
introduce the functional renormalisation group approach to QCD. In
\sec{sec:TruncG}, \ref{sec:correlation} we discuss in detail the
underlying systematic truncation scheme, and specify the flows for
correlation functions including the propagators, vertices, and the
effective potential. In \sec{sec:num} we present the numerical
results and discuss them in detail. In \sec{sec:SysFun} we
analyse the systematic error in the current approximation, also in
relation to that of other functional approaches. In \sec{sec:sum}
we close with a short summary. Many technical details of our
calculations are deferred to the appendices.

%%%%%%%%%%%%%%%%%%%%%%%%%%%%%%%%%%%%%%%%%%%%%%%%%%%%%%%%%%%%%
%%%%%%%%%%%%%%%%%%%%%%%%%%%%%%%%%%%%%%%%%%%%%%%%%%%%%%%%%%%%%

\section{Functional renormalisation group approach to QCD}
\label{sec:FunRG-QCD}

Here we discuss the functional renormalisation group approach to
QCD. Our study is based on and extends the work on QCD in
\cite{Braun:2007bx, Braun:2008pi, Braun:2009gm, Mitter:2014wpa,
  Braun:2014ata, Rennecke:2015eba, Cyrol:2016tym, Cyrol:2017ewj,
  Cyrol:2017qkl} and also draws from results on the QCD phase
structure within low energy effective theories (LEFTs),
\cite{Pawlowski:2014zaa, Fu:2015naa, Fu:2015amv, Fu:2016tey,
  Rennecke:2016tkm, Braun:2017srn, Fu:2018qsk, Fu:2018swz,
  Braun:2018bik}. For a selection of fRG studies within LEFTs see
\cite{Schaefer:2004en,Schaefer:2006ds, Herbst:2010rf, Skokov:2010wb,
  Skokov:2010uh, Braun:2011iz, Strodthoff:2011tz, Fukushima:2012xw,
  Aoki:2012mj, Kamikado:2012cp,Jiang:2012wm,Haas:2013qwp,
  Herbst:2013ufa, Herbst:2013ail, Tripolt:2013zfa, Grahl:2013pba,
  Mitter:2013fxa, Herbst:2013ufa, Morita:2013tu, Pawlowski:2014zaa,
  Helmboldt:2014iya,Khan:2015puu, Wang:2015bky,
  Mueller:2015fka,Eser:2015pka, Weyrich:2015hha, Fu:2015naa,
  Aoki:2015mqa, Fejos:2015xca, Jiang:2015xqz, Fu:2015amv, Fu:2016tey,
  Rennecke:2016tkm, Jung:2016yxl, Fejos:2016hbp, Almasi:2016zqf,
  Posfay:2016ygf, Yokota:2016tip, Springer:2016cji,Resch:2017vjs,
  Tripolt:2017zgc, Braun:2017srn, Fejos:2017kpq, Yokota:2017uzu,
  Almasi:2017bhq, Zhang:2017icm, Aoki:2017rjl, Braun:2018bik,
  Fejos:2018dyy, Braun:2018svj, Fu:2018qsk, Fu:2018swz, Sun:2018ozp,
  Wen:2018nkn, Yin:2019ebz, Leonhardt:2019fua,Li:2019nzj}. For
QCD-related reviews on the fRG approach see
\cite{Berges:2000ew, Pawlowski:2005xe, Schaefer:2006sr, Gies:2006wv,
  Rosten:2010vm, Braun:2011pp, Pawlowski:2014aha}.

%%%%%%%%%%%%%%%%%%%%%%%%%%%%%%%%%%%%%%%%
\subsection{QCD with dynamical hadronisation}
\label{sec:FlowQCD}

The fRG approach is based on an infrared regularisation for momentum
modes $p^2 \lesssim k^2$ of the theory at hand, where $k$ is the
infrared (IR) cutoff scale. This is achieved by adding a
momentum-dependent mass term to the action which vanishes for momenta
$p^2\gtrsim k^2$. Accordingly, the ultraviolet quantum fluctuations of
the theory with momenta $p^2 \gtrsim k^2$ are untouched. In turn, in
the presence of the infrared cutoff, quantum fluctuations with momenta
$p^2\lesssim k^2$ carry a mass proportional to $k$ and are therefore
suppressed.  The presence of this regulator leads to a scale dependent
effective action $\Gamma_k$, which includes quantum fluctuations of
momentum modes $p^2 \gtrsim k^2$.

Thus, in the ultraviolet (UV) at large cutoff scales,
$k^2/\Lambda_\textrm{QCD}^2\gg 1$, the effective action $\Gamma_k$
tends towards the bare QCD action, which is well under control with
perturbation theory.  This can be used as the starting point of a
renormalisation group (RG) evolution of $\Gamma_k$, where the IR
cutoff scale $k$ plays the r$\hat{\textrm{o}}$le of an RG scale. By
evolving $\Gamma_k$ from the UV to the IR, quantum fluctuations at a
given momentum scale $p^2 \propto k^2$ are included successively, and
the full quantum theory of QCD is resolved as $k \rightarrow
0$. $\Gamma_{k=0}$ is the full quantum effective action of QCD. Its
flow $k \partial_k\Gamma_k$, is given by the Wetterich equation
\cite{Wetterich:1992yh}, see also \cite{Ellwanger:1993mw,
  Morris:1993qb}. It is evident from the discussion above that the fRG
is a practical realisation of the Wilson RG.

At low energies or momenta the dynamical degrees of freedom of QCD are
hadrons rather than quarks and gluons. The offshell dynamics relevant
for the flow equation is then dominated by the lightest hadrons. At
small and intermediate densities, these are the pseudo-Goldstone
bosons of spontaneous chiral symmetry breaking, in particular the
pseudoscalar pions $\bm\pi$, and the scalar resonance $f_0(500)$, or
$\sigma$. Formulated in terms of the fundamental degrees of freedom,
this requires taking care of the scalar-pseudoscalar four-quark
  interaction channels where these mesons emerge as resonances, as
well as their scatterings. This is done conveniently by introducing a
composite field
\begin{align}\label{eq:phi}
  \phi\propto \bar q (T_f^0 ,
  i \gamma_5 T_f^a) q\,, 
\end{align}
where $q$ is a Dirac spinor with $N_f$ flavours and $N_c$ colours.
$T_f^a\,, a \!=\! 1,\dots , N_f^2 -1 $ are the generators of the
$SU(N_f)$ flavour group and $T^0_f = \mathbb{1}/\sqrt{2 N_f}$. For
example, in the two-flavour case with up and down quarks, $q=(u,d)$ we
have the tensor structure $( i \gamma_5 \bm{\sigma}/2)$, where
$\bm\sigma =(\sigma_1,\sigma_2,\sigma_3)$ are the Pauli matrices. This
part of $\phi$ then corresponds to the three pions. The full field
$\phi$ reflects the underlying $SU(N_f)$ flavour symmetry for $N_f=2$,
which translates into a $O(4)$ symmetry for $\phi$ in case of isospin
symmetric matter. In the physically more relevant case of $N_f=2+1$
with the light $l=(u,d)$ quarks, the heavy $s$ quark and $q=(l,s)$ we
simply embed \eq{eq:phi} accordingly. This is discussed later, see
\eq{eq:SU3embed}.

The systematic introduction of the composite field $\phi$ in the fRG
approach is done via dynamical hadronisation, see
Refs.~\cite{Gies:2001nw, Gies:2002hq, Pawlowski:2005xe,
  Floerchinger:2009uf}.  The present formulation follows
\cite{Pawlowski:2005xe, Braun:2014ata, Mitter:2014wpa, Cyrol:2017ewj}:
we introduce a scale-dependent composite field $\hat\phi_k(q,\bar q)$
and $\phi=\langle \hat\phi_k\rangle$ through a source
$\int_x\! J_\phi \,\hat\phi_k$ and a respective regulator term. This
field carries scalar and pseudoscalar quantum numbers. The effective
action is then given by a modified Legendre transformation of the
Schwinger functional $W_k[J]$ w.r.t.\ the current
\begin{align}\label{eq:J}
  J=(J_A, J_c,J_{\bar c}, J_q,J_{\bar q},J_\phi)\,,
\end{align}
including that of the
composite field $\hat\phi_k$. This leads us to 
\begin{align}\label{eq:GDynHad}
  \Gamma_k[\Phi]=  \int_x J \cdot \Phi - W_k[J]  -
  \Delta S_k[\Phi]
  \,,  
\end{align}
with the currents $J[\Phi]$ defined by the equations of motion (EoM),
\begin{align}\label{eq:EoMJ}
  J= \frac{\delta\left(\Gamma_k[\Phi]+\Delta
  S_k[\Phi]\right)}{\delta\Phi}\,,  
\end{align}
and the cutoff term
\begin{align}\label{eq:DeltaS}
  \Delta S_k[\Phi]=\frac12 \int_p \Phi(-p) \cdot R_k(p)\cdot
  \Phi(p)\,.
\end{align}
The cutoff term implements the IR regularisation though a momentum
dependent mass-like term, as discussed above.  The components of the
superfield $\Phi$ in \eq{eq:GDynHad}, \eq{eq:DeltaS} are gluons,
ghosts, quarks and the scalar-pseudoscalar mesonic field $\phi$,
\begin{align}\label{eq:PhiQCD}
  \Phi=(\Phi_{f}, \phi)\,,\qquad  \Phi_{f}=(A,c,\bar c,q,\bar q)\,,\quad
  \phi=(\sigma, \bm\pi)\,, 
\end{align}
and $q=l$ with $l=(u,d)$ for $N_f=2$, and $q=(l,s)$ for
$N_f=2+1$. Note also that the cutoff term includes an infrared
regularisation with $R_\phi$ for the composite field. This leads us to
the regulator matrix
\begin{align}
  R_k= \begin{pmatrix}
    R_{A} & 0 & 0& 0 & 0& 0\\[1ex]
    0 & 0 & -R_c & 0 & 0& 0\\[1ex]
    0 & R_c  &  0  & 0 & 0& 0\\[1ex]
      0 & 0 & 0  & 0 & -R_q& 0\\[1ex]
      0 & 0 & 0 & R_q & 0 & 0\\[1ex]
 0 & 0 & 0 & 0 & 0      & R_\phi
\end{pmatrix}\,.
\label{eq:QCD-RegulatorMatrix}\end{align}
The regulator $R_k(p)$ specifies the momentum dependence of the
mass-like cutoff. It is chosen such that it suppresses quantum
fluctuations with momenta smaller than the cutoff scale,
$p^2\lesssim k^2$, while it leaves the UV physics with momenta
$p^2\gtrsim k^2$ unaffected. Furthermore, to recover the full quantum
effective action at $k = 0$, we demand
$R_{k\rightarrow 0} \rightarrow 0$.

This setup is closely related to a two-particle irreducible (2PI) or
rather two-particle point irreducible (2PPI) formulation. If also
considering the density channel $\bar q\gamma_0 q$ it resembles
density functional theory, for a more detailed discussion of these
relations for generic composite fields see \cite{Pawlowski:2005xe}. 

The scale evolution of $\hat\phi_k$, or rather its expectation value
$\langle \partial_t \hat\phi_k \rangle $, can be chosen freely. Its
choice corresponds to a reparameterisation of the theory. We emphasise
that the mean field $\phi =\langle \hat\phi_k\rangle$ in the effective
action is $k$-independent. Note also that on the EoS of the auxiliary
field the effective action reduces to the standard effective action of
QCD in terms of the fundamental fields,
$\Gamma_{\textrm{QCD}}[\Phi_\textrm{f}]$: The EoM of the composite
field including the cutoff term entail a vanishing current,
\begin{align}\label{eq:EoMphi}
  J_\phi=  \left. \frac{\delta\Gamma_{k=0}[\Phi]}{
  \delta\phi} \right|_{\phi_\textrm{EoM}[\Phi_f]}=0\,.
\end{align}
Since the composite field is introduced only trough the source
$J_\phi$ and its regulator $R_\phi$ in the first place, the EoM
\eq{eq:EoMphi} removes $\hat\phi_k$ from the path integral at
vanishing cutoff, reducing it to the standard gauge fixed path
integral of QCD. We are led to
\begin{align}\label{eq:GQCD=GDynHad}
  \Gamma_\textrm{QCD}[A,c,\bar c,q,\bar q]=
  \Gamma_{k=0}[\Phi_{f},\phi_\textrm{EoM}[\Phi_f]]\,.
\end{align}
At finite cutoff scale the composite field can also be eliminated.
There, vanishing of the current $J_\phi$ is obtained for
$\delta(\Gamma_k+\Delta S_k)/ \delta\phi =0$, and the infrared
regularised path integral at $J_\phi=0$ still depends on the cutoff
term of the composite field. This amounts to inserting a UV-irrelevant
four-quark interaction in the classical QCD action. This procedure
does not spoil the renormalisability as a pointlike NJL-type
interaction does, but solely provides an IR regularisation of the
respective resonant four-quark interaction.

In summary this setup encodes the full information of the QCD
correlation functions but also allows for a simple access to bound
state information such as the Bethe-Salpeter wave functions, see
\cite{Mitter:2014wpa, Braun:2014ata, Cyrol:2017qkl, Alkofer:2018guy}.
Note also that in general the QCD correlation functions now involve
derivatives of the composite field. As an important example we
consider general four-quark vertices. They are given by functional
derivatives of the QCD effective action $\Gamma_{\textrm{QCD}}$ on the
EoM $\Phi_{f,\textrm{EoM}}=0$ at $T$. At finite temperature
$\Phi_{f,\textrm{EoM}}$ contains a nonvanishing temporal gluon
background field, $A_{0,\textrm{EoM}}$, which carries the information
of confinement, see \cite{Braun:2007bx, Braun:2010cy, Fister:2013bh,
  Herbst:2015ona}.

If the composite field $\phi$ is simply
proportional to a quark bilinear as indicated in \eq{eq:phi}, the
four-quark derivatives of \eq{eq:GQCD=GDynHad} lead us to
  \begin{align}\nonumber 
   & \hspace{-.6cm}\Gamma^{(4)}_{\textrm{QCD}, q_1 \bar q_1 q_2\bar q_2}[0] =
    \Gamma^{(4)}_{0, q_1 \bar q_1 q_2\bar q_2}[0,
    \phi]\\[1ex]\nonumber 
    &+
  \int_x \frac{\delta^2\phi(x)}{\delta q_1 \delta \bar q_1 }
      \Gamma^{(3)}_{0, q_2 \bar q_2\phi}[0,\phi]+
      \textrm{permut.}\\[1ex]
    & +
    \int_{x,y} \frac{\delta^2\phi(x)}{\delta q_1 \delta \bar q_1 }
    \frac{\delta^2\phi(y)}{\delta q_2 \delta \bar q_2 }
    \Gamma^{(2)}_{0, \phi\phi}[0,\phi]+\textrm{permut.}\,.
 \label{eq:4q} \end{align}
In \eq{eq:4q} it is understood that $\phi=\phi_\textrm{EoM}[\Phi_f]$,
and we have restricted ourselves to the $T=0$ case with 
$\Phi_{f,\textrm{EoM}}=0$. In \eq{eq:4q} we have also introduced our
notation for $n$-point correlation functions or vertices,
\begin{align}\label{eq:Gn}
  \qquad \Gamma^{(n)}_{\Phi_{i_1}\cdots \Phi_{i_n}}[\Phi]=
  \0{\delta^n\Gamma[\Phi]}{\delta\Phi_{i_1}
  \cdots\delta\Phi_{i_n}}\,. 
\end{align}
If we choose the composite field such that it completely absorbs a
given momentum channel in the four-quark scattering, the first term on
the right hand side in \eq{eq:4q} vanishes and we are left with
exchange terms of the composite field. For example, for the
pseudoscalar channel the second and third line in \eq{eq:4q} comprise
terms with a pion propagator $(1/\Gamma^{(2)})_{\pi\pi}$ with two
Bethe-Salpeter wave functions $\Gamma^{(3)}_{q\bar q\pi}$ attached.

Note also that nontrivial terms such as in the second and third line
of \eq{eq:4q} only occur in correlation functions of the fundamental
fields $\Phi_f$ for more than one quark-antiquark pair. For example,
for the quark two-point function or inverse propagator we find
schematically
\begin{align}
  \Gamma^{(2)}_{\textrm{QCD}, q_1 \bar q_1}=
  \Gamma^{(2)}_{0, q_1 \bar q_1}+\int \frac{\delta^2\phi}{\delta q_1
  \delta \bar q_1 }
  \Gamma^{(1)}_{0, \phi}
  =
  \Gamma^{(2)}_{0, q_1 \bar q_1}\,,
  \label{eq:qbarq} \end{align}
with $\phi=\phi_\textrm{EoM}[\Phi_f]$. Then the term proportional to
$\Gamma^{(1)}_{0,\phi}$ vanishes as the latter is the EoM for $\phi$.
Finally we use $\Phi_f=0$ as in \eq{eq:4q}.

For the dynamical hadronisation in the present work we use the option
of absorbing the dominant four-quark interaction channel
completely. This is achieved by choosing the flow of $\hat\phi_k$ such
that the scalar-pseudoscalar channel in the effective four-quark
interaction of the light quarks $u,d$ is eliminated. Formally, this
can be viewed as a successive bosonisation in terms of a
Hubbard-Stratonovich transformation of this interaction channel at
each RG scale $k$. For more details see \sec{sec:DynHad}. This puts
forwards the general hadronisation relation
\begin{align}\label{eq:HadABC}
  \langle   \partial_t \hat \phi_k\rangle =&\dot{A}_k \,\bar{q}\tau q
                                             +\dot{B}_k
                                             \,\phi +\dot{C}_k \,\hat e_\sigma\,,  
\end{align}
with $\tau=(T_2^0, i \gamma_5 \bm{T}_2)$ and $T_f^a$ defined in
\eq{eq:phi}. The vector $\hat e_\sigma$ points in the
$\sigma$-direction, with the convenient normalisation
$\hat e_\sigma=1/\sqrt{2 N_f}$ matching that of $\tau$. Since we only
consider the dynamical hadronisation of the $\bm\pi$--$\sigma$
channel of the light quarks, we embed the $SU(2)$ scalar-pseudoscalar
field $(\sigma,\bm\pi)$ into $SU(3)$. This implies in
\eq{eq:HadABC},
\begin{align}\label{eq:SU3embed}
  \tau^{a}_{N_f=2+1} = \begin{pmatrix}\tau^a
    & 0 \\[1ex] 0 & 0\end{pmatrix}\,,
\end{align}
in a slight abuse of notation. The hadronisation functions
$\dot{A}_k$, $\dot{B}_k$ and $\dot{C}_k$ can be chosen freely. The
hadronisation function $\dot{A}_k$ controls the overlap of $\phi$ with
the scalar-pseudoscalar channel, while $\dot{B}_k$ simply changes the
wave function renormalisation $Z_\phi$ of the composite field. The
latter function can e.g.\ be used to conveniently choose
$Z_\phi=1$. As it does not change the parameterisation of the theory
we discard it in the following, using $\dot{B}_k=0$.

Finally, the hadronisation function $\dot{C}_k$ introduces a shift in
the scalar field $\sigma$. In general, this shift can be used to
entirely absorb the quark mass of the respective quark flavour,
leading to chiral quarks. However, this comes at the expense of
diffusing the symmetry properties of the effective action. For
example, if we start with a $\mathbf{Z}_2$-symmetric effective
potential $V(\sigma) = V(-\sigma)$ necessary (but not sufficient) for
chiral symmetry, a shift in $\sigma$ introduces odd powers in the
$\sigma$ field in the effective potential.  Of course, chiral symmetry
is not lost, but the respective symmetry transformation is not simply
$\sigma\to -\sigma$ anymore. A prominent example is the $\sigma$-mass
term $1/2 m_\sigma^2 \sigma^2$. A shift in $\sigma$ with
$\sigma \to \sigma - c_\sigma/m_\sigma^2 $ leads to a linear term in
the effective action,
\begin{align}\label{eq:LinTerm}
  \Gamma_k[\Phi] \propto -\int_x c_\sigma\, \sigma\,, 
\end{align}
while still keeping the original $\sigma$-mass term. In the presence
of higher powers of $\sigma$, more $\sigma$-odd terms are generated as
well as changing the coefficients of the $\sigma$-even
terms. Consequently we expect that $\dot{C}_k$ cannot be chosen freely
if we restrict ourselves to dynamical hadronisation setups that keep
chiral symmetry apparent. Indeed this constraint leads us to
$\dot{C}_k\equiv 0$, see \app{app:DynHad-c}.

The linear term \eq{eq:LinTerm} plays a special r$\hat{\textrm{o}}$le
in the effective action. To see this more clearly, we solve the
Legendre transform \eq{eq:GDynHad} for the Schwinger functional and
use the EoM for the current leading to $J[\Phi]$ defined in
\eq{eq:EoMJ}. With these definitions \eq{eq:LinTerm} implies that
$J[\Phi]$ contains the term $- c_\sigma$. Then, the Schwinger
functional reads,
\begin{align}\label{eq:noc_sigma}
  W_k[J ] = \int_ x J\cdot \Phi - \Gamma_k[\Phi]
  - \Delta S_k[\Phi]\,, 
\end{align}
where $J=J[\Phi]$. Evidently the $c_\sigma$-dependences in the first
two terms on the right hand side of \eq{eq:noc_sigma} cancel each
other. We conclude already from here that the flow equation of the
effective action with dynamical hadronisation should be explicitly
$c_\sigma$-independent. Moreover, necessarily the left hand side of
\eq{eq:noc_sigma} is also $c_\sigma$-independent.  Consequently, the
$c_\sigma$-dependence of the current in the Schwinger functional is
canceled by that of $W_k[0]$. This leaves us with
\begin{align}\label{eq:Wwoc_sigma}
  W_k[J] = \widehat{W}_k[\hat J] \quad\textrm{with}\quad
  \widehat W_{k}=\left.W_{k}\right|_{c_\sigma=0}  \,,
\end{align}
and $J=J[\Phi]$ satisfies the EoM \eq{eq:EoMJ}. The current is shifted
in the $J_\sigma$-direction,
$\hat J[\Phi]=J[\phi]+c_\sigma \delta_{\Phi \sigma}$, in
components: 
\begin{align}\label{eq:Jwoc_sigma}
 \hat J=(J_A, J_c,J_{\bar c}, J_q,J_{\bar q}, J_\sigma
+c_\sigma, J_\pi)\,. 
\end{align}
The shifted current $\hat J[\Phi]$ in \eq{eq:Jwoc_sigma} does not
depend on $c_\sigma$, that is $\partial_{c_\sigma} \hat J\equiv
0$. With these relation we finally arrive at a convenient form of the
effective action,
\begin{align}\label{eq:GDynHadfin}
 \Gamma_k[\Phi]=  \int_x J \cdot \Phi - \widehat W_{k}[\hat J]  -
  \Delta S_k[\Phi]
  \,,  
\end{align}
We would like to elucidate the above relations within a simple example
which is relevant in the present context: Consider a composite
$\sigma$-field that is just proportional to the scalar quark bilinear,
\begin{align} \label{eq:hatsigma-Example}
  \hat\sigma_k =- D_k \,\hat{\bar q} \hat q\,,
\end{align}
with a cutoff dependent prefactor $D_k$. Note that the hats indicate
that this relation holds on the level of the fluctuation fields in the
path integral. We also emphasise that our example
\eq{eq:hatsigma-Example} has an overlap with the first term in the
dynamical hadronisation flow \eq{eq:HadABC} with
$\dot{A}_k=-\partial_t D_k$, but is not identical:
$\langle \hat{\bar q} \hat q\rangle = \bar q q + G_{\bar q
  q}[\Phi]$. Importantly, the linear term $-\int_x c_\sigma \sigma$ in
the effective action \eq{eq:LinTerm} is in one to one correspondence
to the quark (current) mass term in the classical action,
$S_\textrm{QCD} \propto m^0_q \int \hat{\bar q} \hat{q}$. Obviously,
the latter term can be absorbed into a shift of the source term for
$\hat\sigma$ with
\begin{align}\label{eq:absorbtion}
  - m^0_q \int \hat{\bar q} \hat{q}  + \int_x J_\sigma \hat\sigma_k =
  \int_x \hat J_\sigma \hat\sigma_k\,, 
\end{align}
where 
\begin{align}\label{eq:c_sigmaExample}
  \hat J_\sigma = J_\sigma + c_\sigma\,, \quad \textrm{with}
  \quad c_\sigma
  =\frac{m^0_q}{D_k}\,.
\end{align}
We conclude that the Schwinger functional with a source for the
composite field $\hat \sigma_k$ in \eq{eq:hatsigma-Example} and the
classical QCD action with a quark current mass term,
$S_\textrm{QCD} \propto m^0_q \int \hat{\bar q}\hat{ q}$, is that in
the chiral limit with $m^0_q=0$ and a shifted current for the
composite field. However, $m^0_q=0$ simply is $c_\sigma=0$, see
\eq{eq:c_sigmaExample}, and we arrive at
\begin{align}\label{eq:WhatExample}
  W_k[J_\sigma] = \left. W_k[\hat J_\sigma]\right|_{m_q=0}=
  \widehat
  W_k[\hat J_\sigma]\,.
\end{align}
Note that \eq{eq:WhatExample} entails that the full dynamics of QCD in
the presence of explicit chiral symmetry breaking is that of the
theory with a composite condensate field $\hat\sigma$ with full chiral
symmetry: The explicit chiral symmetry breaking is completely
  absorbed in a shift of the external current $J\to \hat J$. 

%%%%%%%%%%%%%%%%%%%%%%%%%%%%% 
\begin{figure}[t]
\includegraphics[width=0.98\columnwidth]{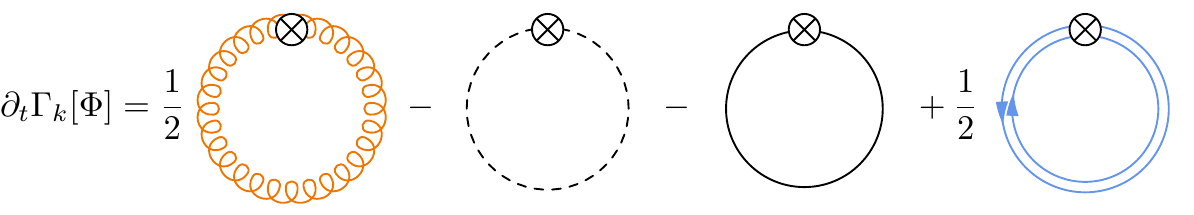}
\caption{Flow of the effective action of QCD. The first three
  diagrams arise from the gluon, ghost, and the quark degrees of
  freedom respectively. The last diagram is that of the mesonic
  contribution.  The double line with the up-down arrows indicates the
  nature of the mesons as quark-antiquark composites. The crossed
  circles indicate the regulator insertion in the flow
  equation. }\label{fig:fleq}
\end{figure}
%%%%%%%%%%%%%%%%%%%%%%%%%%%%%
%
In summary this leads us to the flow equation for QCD with dynamical
hadronisation, \cite{Pawlowski:2005xe, Rennecke:2015},
\begin{align} \label{eq:FlowQCD}
\begin{split}
&\partial_{t}\Gamma_{k}[\Phi]+
  \int \langle   \partial_t \hat \phi_{k,i}\rangle\, \left( \frac{\delta
    \Gamma_{k}[\Phi]}{\delta
                          \phi_i}+ c_\sigma \delta_{i\, \sigma}\right) \\[1ex]
  &\quad= \frac12 \Tr\! \big(G_k[\Phi]\,\partial_t R_k\big)+
     \frac{1}{2}\Tr\! \bigg(G_{\phi\Phi_j}[\Phi] \frac{\delta
     \langle \partial_t \hat \phi_k\rangle }{\delta \Phi_j} \,R_\phi\bigg)\,,
\end{split}
\end{align}
with the RG time $t=\ln (k/\Lambda)$, and
\begin{align}\label{eq:Gk}
  G_k[\Phi]= \frac{1}{\Gamma^{(2)}_k[\Phi]+R_k}\,,\qquad
  G_{\Phi_i \Phi_j}[\Phi]=
  \left( G_k[\Phi]\right)_{\Phi_i \Phi_j}\,. 
\end{align}
The shift in the dynamical hadronisation term in the first line in
\eq{eq:FlowQCD} subtracts the explicit chiral symmetry breaking term
in $\Gamma^{(1)}_\sigma$. The $t$-derivative $\partial_t \Gamma_k$ in
\eq{eq:FlowQCD} is taken at fixed $c_\sigma$ as the $\dot{c}_\sigma$
terms from $\partial_t \Gamma_k$ cancel with that from
$\partial_t \int J\cdot\Phi \propto \partial_t \int (\partial_t
J_\textrm{sym})\cdot\Phi - \int \dot{c}_\sigma\sigma$. The
$ (\partial_t J_\textrm{sym})$-terms are cancel by those from the
Schwinger functional in \eq{eq:GDynHadfin}. We emphasise that
\eq{eq:FlowQCD} is explicitly $c_\sigma$-independent, and
  is a novel representation of dynamical hadronisation.

Note also that \eq{eq:FlowQCD} with \eq{eq:HadABC} entails that we do
not have to specify $\hat \phi_k$ but only the expectation value of
its flow, $\langle \partial_t \hat\phi_k\rangle $, as done in
\eq{eq:HadABC}. The second term in the first line of \eq{eq:FlowQCD}
accounts for the cutoff dependence of the reparameterisation that
originates in the cutoff dependence of $\hat\phi_k$. It can be
understood as a generalised anomalous dimension of $\phi$. Indeed, for
\begin{align}\label{eq:GenAn}
  \langle\partial_t \hat\phi\rangle = -\frac12  \eta_\phi \phi\,, 
\end{align}
this term carries nothing but the anomalous rescaling of the composite
field $\phi$. However, we emphasise again that $\partial_t$ is a
derivative at fixed mean superfield $\Phi$, and in particular at fixed
$\phi$. The second term in the second line also accounts for the
cutoff dependence of the reparameterisation via
$\partial_t \hat \phi_k$.

The first term in the second line of \eq{eq:FlowQCD} is the standard
functional flow. The trace sums over momenta, internal indices and
species of fields including the composite field. More explicitly it
reads
\begin{align}\label{eq:FullProp}
\begin{split}
  \frac12 \Tr G_k\,\partial_t R_k=&\,
  \frac12 \Tr G_{AA} \,\partial_t R_A -\Tr\, G_{c\bar c} \,\partial_t R_c\\[1ex]
  &-
     \Tr G_{q\bar q}\,\partial_t R_q+ \frac{1}{2}\Tr
                                 G_{\phi\phi}\, \partial_{t} R_{\phi} \,, 
\end{split}
\end{align}
where all propagators are $\Phi$-dependent.  We show the flow equation
\eq{eq:FlowQCD} with $\partial_t \hat \phi=0$ diagrammatically in
\Fig{fig:fleq} for QCD. The first gluon and ghost diagrams constitute
the glue contributions, and the last two arise from the matter sector,
which are the quark and meson degrees of freedom,
respectively. Note that the scalar and pseudoscalar mesonic
  loops are present here because we introduced them explicitly with
  dynamical hadronisation. This is nothing but a convenient
  parametrisation of resonant interaction channels. We emphasise that
  the absense of explicit loops for other hadrons, e.g.\ other mesons
  and baryons, does not entail that their dynamics is not included.

%%%%%%%%%%%%%%%%%%%%%%%%%%%%%%%%%%%%%%%%%
\section{Truncation scheme}\label{sec:TruncG}
In this section we discuss in detail the expansion scheme for the
effective action and the flow equations used in the present work. This
includes in particular a discussion of the approximations used, and
their quantitative or qualitative validity.

\subsection{Vertex expansion}
In QCD, \eq{eq:FlowQCD} can only be solved within given truncations to
the effective action. The systematic expansion scheme behind such
approximations is the vertex expansion. It is an expansion of the
effective action in powers of the field, the expansion coefficients
being the $n$-point functions $\Gamma^{(n)}(p_1,...,p_n)$. Typically,
this expansion shows a rather good convergence if no divergent
exchange processes or couplings are present. In Landau gauge QCD the
strong couplings related to gluon exchange grow towards the infrared
but tend to zero for momenta or cutoff scales below approximately one
GeV, reflecting the QCD mass gap, see \cite{Mitter:2014wpa,
  Braun:2014ata, Rennecke:2015eba, Rennecke:2015, Cyrol:2017ewj}. This
behaviour supports the apparent convergence of the vertex
expansion. Another source for divergent exchange processes are
resonant interactions that potentially spoil or at least slow down the
convergence of the vertex expansion. In particular for this reason
dynamical hadronisation is crucial as it controls the resonant
interaction channels and allows to take into account multi-scatterings
of resonances in a technically accessible way via the respective
effective potential of the composite fields.

\subsection{Expansion about vacuum QCD \& finite temperature
  Yang-Mills theory}\label{sec:ExpVac+FinT}
  
We can rely on quantitative results for vacuum Yang-Mills theory,
\cite{Cyrol:2016tym} and $N_f=2$ flavour QCD,
\cite{Mitter:2014wpa,Cyrol:2017ewj}, and finite temperature Yang-Mills
theory, \cite{Cyrol:2017qkl}, as input as well as benchmark tests for
our current computations. State-of-the-art truncations which involve a
large set of correlation functions have been used in these works.

Furthermore, it is well-known that mild momentum-dependences of vertex
functions and propagators are well captured by scale-dependent
dressing functions; for investigations in QCD and low energy effective
models see e.g.\ \cite{ Helmboldt:2014iya, Pawlowski:2014zaa,
  Braun:2014ata, Rennecke:2015, Rennecke:2016tkm}. This suggests an
approximation scheme in which only the dominant nontrivial momentum
dependences are taken into account explicitly, while the rest of the
dependences is approximated by scale-dependent dressings.

This approach has been successfully applied to vacuum QCD in
\cite{Braun:2014ata, Rennecke:2015eba, Rennecke:2015} and for QCD at
finite temperature and imaginary chemical potential in
\cite{Braun:2009gm}. In these works QCD flows have been expanded about
the full gluon and ghost two-point functions of Yang-Mills theory. All
other correlation functions considered there have been approximated
with scale-dependent dressings. Schematically we decompose the
two-point functions as
\mbox{$\Gamma^{(2)} \simeq Z \, S^{(2)}_{\textrm{kin}}+$
  (\textit{additional tensor structures})}. For example, for the
quark, $S^{(2)}_{\textrm{kin}}$ carries the Dirac term, while the mass
term is buried in the additional tensor structures. For the gluon this
entails for the two-point function is
$\Gamma^{(2)}_{A A}(p) \simeq Z_A(p^2) \, S^{(2)}_{AA}(p)$, but we
have to consider different Lorentz tensors. The nontrivial momentum-
and scale-dependence of the kinetic term of a given field is fully
captured by the anomalous dimensions,
\begin{align}\label{eq:eta}
  \eta_{\Phi_i}= -\frac{\partial_t Z_{\Phi_i}}{Z_{\Phi_i}} \,.
\end{align}
Note that \eq{eq:eta} only schematically provides the anomalous dimension, we
have neither specified the projection procedure nor the momentum
evaluation. We further exemplify the setup with the relevant case of
the gluon two-point function. There, one also utilises the observation
that the anomalous dimension $\eta_A$ of the gluon can be
parameterised in terms of the running coupling
$\alpha^{\textrm{YM}}_k$ in Yang-Mills theory,
\begin{align}\label{eq:eta(alpha)}
  \eta_{A,k}^\textrm{YM}= \eta_A^{\textrm{glue}}\left(\alpha^{\textrm{YM}}_k,
  (\bar m^{\textrm{YM}}_{A,k})^2\right)\,,
\end{align}
with the Yang-Mills running coupling $\alpha^{\textrm{YM}}_k$ and the
transversal gluon mass parameter,
\begin{align}\label{eq:mA}
m_{A,k}^2= \left. \frac{1}{3( N_c^2 -1)}\frac{1}{k^2}
  \Pi_{\mu\nu}^{\perp}(p){\Gamma^{(2)}_{AA}}^{aa}_{\mu\nu}(p) \right|_{p=0}\,,
\end{align}
where $\Pi_{\mu\nu}^{\perp}$ is the transversal projection operator,
see \eq{eq:Piperp}. The renormalised gluon mass parameter $\bar m_A^2$
differs from $m_A^2$ in \eq{eq:mA} by an appropriate renormalisation.
This renormalisation has to be applied to all coupling parameters and
fields and is discussed in detail in the next
\sec{sec:TruncAct}. In \eq{eq:eta(alpha)} the Yang-Mills gluon
mass parameter $(\bar m^{\textrm{YM}}_{A,k})^2$ has been
used. $\eta_A^{\textrm{glue}}$ stands for all diagrams involving
gluons and ghosts that contribute to the gluon anomalous dimension.
This parameterisation of the glue part of the anomalous dimension
allows for a simple representation of the anomalous dimension of full
QCD,
\begin{align}\label{eq:etaQCD}
  \eta_{A,k}= \eta_A^{\textrm{glue}}(\alpha_k, \bar m_{A,k}^2)
  +\eta^\textrm{quark}_{A,k}\,, 
\end{align}
with the full QCD coupling $\alpha_k$ and renormalised mass parameter
$\bar m_{A,k}^2= m_{A,k}^2/Z_A$, respectively.  Importantly, this
setup also takes care of the backreaction of the quark fluctuations to
the pure glue and ghost diagrams, for a detailed description see
\cite{Braun:2014ata, Rennecke:2015}. A benchmark test concerns the
full gluon propagator in unquenched two-flavour QCD, which has been
shown to be in quantitative agreement with the lattice results in
\cite{Rennecke:2015}, more details can be found in \app{app:YMtoQCD}.

A simpler approximation is given by the `direct sum'
\begin{align}\label{eq:etaQCD-DS}
  \eta_{A,k}\approx  \eta_{A,k}^{\textrm{YM}}
  + \eta^\textrm{quark}_{A,k}\,.  
\end{align}
With \eq{eq:etaQCD-DS} the backreaction of the quark loop on the pure
glue couplings and propagators in $\eta_A^{\textrm{glue}}$ is
neglected.  While such an approximation also lacks full RG invariance,
it works well for initial scales of $\Lambda\approx 10 -20 $\, GeV.
Alternatively a readjustment of the scales can be done on top of the
simple approximation \eq{eq:etaQCD-DS}. A respective benchmark test
for the full gluon propagator in unquenched two-flavour QCD has been
done in \cite{Rennecke:2015}, for more details see
\app{app:YMtoQCD}. The direct-sum approximation has been used very
successfully in many functional applications to the phase structure of
QCD, notably in fRG application to QCD, \cite{Braun:2009gm}, as well
as the DSE applications \cite{Fischer:2011mz, Fischer:2012vc,
  Fischer:2013eca, Fischer:2014ata, Fischer:2014vxa, Eichmann:2015kfa,
  Mitter:2017iye, Fischer:2018sdj, Isserstedt:2019pgx,
  Gunkel:2019xnh}.  We also emphasise that the approximation may get
even better for large chemical potential, as the latter lifts the
fermi sea and successively more quark fluctuations are buried in
it. This effect reduces the impact of the quark loop on the purely
gluonic correlation functions.  However, this reasoning does not apply
straightforwardly to the quark-gluon vertex, and has to be taken with
a grain of salt beyond the onset of baryon density. Still, in summary
the approximation is well-founded and well-tested.

Here we apply this promising approximation scheme to the unquenched
QCD gluon propagator at vanishing temperature as well as the glue
contribution at finite temperature. Let us describe this procedure
with an expansion of the finite temperature and density theory about
the vacuum. Schematically this is described by the following
separation of the flow for a given correlation function at finite $T$
and $\mu$,
\begin{align}\label{eq:DeltaGQCD}
  \partial_t \Gamma^{(n)}_{T,\mu}(p) = \partial_t
  \Gamma^{(n)}_{0,0}(p) +   \partial_t
  \Delta \Gamma^{(n)}_{T,\mu}(p) \,,
\end{align}
where we suppressed the subscript $k$, and $\Delta \Gamma^{(n)}$ is
defined by,
\begin{align}
  \label{eq:DeltaG}
  \Delta \Gamma^{(n)}_{T,\mu}(p) =  \Gamma^{(n)}_{T,\mu}(p) -
  \Gamma^{(n)}_{0,0}(p)  \,.
\end{align}
Its flow is given by the difference of the flow
diagrams for the correlation function at finite $T,\mu$ and the
vacuum,
\begin{align}\label{eq:FlowDeltaG}
  \partial_t  \Delta \Gamma^{(n)}_{T,\mu}(p) = \textrm{Flow}^{(n)}_{T,\mu}(p) -
  \textrm{Flow}^{(n)}_{0,0}(p) \,,
\end{align}
with
\begin{align}\label{eq:Flown}
  \textrm{Flow}^{(n)}_{\Phi_{i_1}\cdots\Phi_{i_n}}= \frac{\delta^n}{\delta \Phi_{i_1}\cdots
  \Phi_{i_n}}
  \left[\frac12 \Tr\, G_k[\Phi]\,\partial_t R_k\right] \,. 
\end{align}
While the first term on the right hand side of \eq{eq:FlowDeltaG}
depends on $ \Delta \Gamma^{(n)}_{T,\mu}(p) $ and
$\Gamma^{(n)}_{0,0}(p) $, the second term is a function of only the
latter vacuum correlation functions. Accordingly, the flows
\eq{eq:FlowDeltaG} are closed equations for the set
$\{\Delta\Gamma^{(n)}\}$ of thermal and density corrections of general
correlation functions $\Gamma^{(n)}_{T,\mu}$ with the given input of
the set of vacuum correlation functions $\{\Gamma^{(n)}_{0,0}\}$.

In summary this allows for the use of scale-dependent dressings for
$\Delta \Gamma^{(n)}$ with a mild momentum dependence at finite
temperature and density, while still maintaining the quantitative
nature of the approximation: Quantitative results for vacuum QCD and
finite temperature Yang-Mills theory \cite{Mitter:2014wpa,
  Cyrol:2016tym, Cyrol:2017ewj, Cyrol:2017qkl} are used for the
nontrivial momentum dependences of this part of the correlation
functions. Note that this setup also allows for the use of
quantitative results from other approaches such as lattice simulations
and DSE computations.

In the present work we apply this approach to the gluon two-point
function $\Gamma^{(2)}_{AA}$. Its density corrections and thermal
quark fluctuations are computed with the input of the quantitatively
reliable vacuum QCD results from \cite{Cyrol:2017ewj} and the finite
temperature Yang-Mills results in \cite{Cyrol:2017qkl}.  For the ghost
two-point function $\Gamma^{(2)}_{c \bar c}$ we only use input from
vacuum QCD from \cite{Cyrol:2017ewj}.  The details can be found in
\sec{sec:GlueSector}.

\subsection{Truncation for the effective action}\label{sec:TruncAct}

Having captured the nontrivial momentum-dependence of the gluon
propagation, we adopt the following truncation for the Euclidean
scale-dependent effective action $\Gamma_k$ for both, $N_f=2$ and
$N_f=2+1$ flavours, 
\begin{align}
  \Gamma_{k}=&\int_{x} \bigg\{\frac{1}{4}F^a_{\mu\nu}F^a_{\mu\nu}+
               Z_{c} \,(\partial_{\mu}\bar{c}^a)D_{\mu} ^{ab}c^b+
               \frac{1}{2\xi}(\partial_{\mu} A^a_{\mu})^2\nonumber\\[1ex]
             &+ \frac12 \int_p A^a_\mu(-p) \left(
               \Gamma^{(2)}_{AA}{}^{ab}_{\mu\nu}(p) -Z_A \Pi_{\mu\nu}^{\perp}
               \delta^{ab}p^2
               \right) A^b_\nu(p)
               \nonumber\\[1ex]
             &+\bar{q}\,\bigl[Z_{q}\, (\gamma_{\mu} D_{\mu}-\gamma_0 \hat \mu)+
               m_s(\sigma_s) \bigr]\, q\,\nonumber  \\[1ex]
             &  -{\lambda_{q}}\,
               \Big[(\bar{q}\,\tau^0 q)^2 + (\bar{q}\,
               \bm{\tau} q)^2\Big]+h\,\bar{q}\,
               \left(\tau^0\sigma+\bm\tau
               \cdot\bm{\pi}\right)\,q\nonumber \\[1ex]
             &  +\frac{1}{2}Z_{\phi}\,(
               \partial_{\mu}\phi)^2 
  +V_{k}(\rho,A_0)-
               c_\sigma\,\sigma-\frac{1}{\sqrt{2}}\, c_{\sigma_s}\, \sigma_s\bigg\}\,,
               \label{eq:action}
\end{align}
with $\int_{x}=\int_0^{1/T}d x_0 \int d^3 x$ and
$\tau=1/2 (\mathbb{1}, i \gamma_5 \bm \sigma)$ being defined in
\eq{eq:phi}. Note that all couplings depend on the RG scale $k$,
  which in most cases is omitted for the sake of clarity. The
effective potential
\begin{align}\label{eq:EffPot}
  V_k(\rho,A_0) = V_{\textrm{glue},k}(A_0) + V_{\textrm{mat},k}(\rho,A_0) \,,
\end{align}
carries a dependence on the mesonic field
$\rho= (\sigma^2 +\vec \pi^2)/2$, and the temporal gauge field $A_0$,
which is related to the Polyakov loop $L[A_0]$. The expectation values
of these fields are approximate order parameters for the chiral phase
transition ($\rho)$ and the confinement-deconfinement phase transition
($A_0$ or $L[A_0]$), for more details see \sec{sec:TPot+Order}. In
\eq{eq:EffPot} we separated the contribution from the gluonic loop in
\eq{eq:FlowQCD}, see \fig{fig:fleq}, that is $V_{\textrm{glue},k}$ and
that of the quark and meson loops, $V_{\textrm{mat},k}$. We also
neglected the subleading $\rho$-dependences of the gluon loop,
$V_{\textrm{glue},k}(A_0,\rho)\to V_{\textrm{glue},k}(A_0) $.

The gluon two-point function $\Gamma^{(2)}_{AA}$ has been already
discussed in the previous \sec{sec:ExpVac+FinT}. The subtraction with
$Z_A p^2$ in the second line removes the $A^2$-contribution coming
from the $F_{\mu\nu}^2$-term for avoiding double counting. In the
present fRG approach the theory is fully determined by the strong
running coupling $\alpha_{s,\Lambda}$ at the initial cutoff scale
$k=\Lambda$ in the UV, the light current quark mass $m_l^0$ for
$N_f=2$ and additionally the strange current quark $m_s^0$ for
$N_f=2+1$ with the choice
\begin{align}\label{eq:Zin1}
  Z_{\Phi,k=\Lambda} = 1\,.
\end{align}
With dynamical hadronisation the current quark masses are encoded in
$c_\sigma$ and $c_{\sigma_s}$. Their values in the physical case are
fixed by the pion mass and the kaon mass. Alternatively, the pion and
kaon decay constants can be used to pin down these parameters. This is
explained below. The values for an initial cutoff $\Lambda=20\,$GeV
can be found in Table~\ref{tab:InitialValues} at the end of this
Section.

The field strength tensor $F^a_{\mu\nu}$ and the covariant derivatives
in the ghost and quark kinetic terms also carry a cutoff dependence,
and are defined and discussed in detail in \sec{sec:StrCoup},
\eq{eq:Fmunu} and \eq{eq:CovDer} respectively. The $Z_{\Phi_i}$'s are
the cutoff dependent wave function renormalisations of the respective
component fields $\Phi_i$ of the superfield defined in
\eq{eq:PhiQCD}. Note that the effective action is renormalisation
group invariant (but $k$-dependent) w.r.t. the underlying
renormalisation group scale $\mu$ of the theory at $k=0$, for more
details see \cite{Pawlowski:2005xe}. This suggests the introduction of
\textit{renormalised} fields
\begin{align}\label{eq:RenPhi}
  \bar \Phi = Z_{\Phi}^{1/2} \Phi\,,
\end{align}
as well as \textit{renormalised} coupling parameters, for example
\begin{align}\label{eq:RenMass-Yuk}
  \bar m_\Phi = \frac{m_\Phi}{Z^{1/2}_\Phi}\,, \quad \quad \bar h=
  \frac{h}{Z_\phi^{1/2} Z_q}\,.
\end{align}
Since we perform our computations in Euclidean spacetime, the
renormalised masses in \eq{eq:RenMass-Yuk} are \textit{curvature}
masses in contradistinction to the physical pole masses
$m_{\Phi,\textrm{pol}}$. Within the fRG setup this difference has been
discussed in detail in \cite{Helmboldt:2014iya}. For example, the
constituent quark masses are curvature masses. In turn, the pion pole
mass $m_{\pi,\textrm{pol}}$ is used for fixing the value of explicit
chiral symmetry breaking, see Table~\ref{tab:InitialValues}. For more
details on the definition of the mesonic pole masses see
\sec{sec:MatterProps}.

Similar relations hold for the other coupling parameters and are
discussed later. Note that the epithet \textit{renormalised} is
related to the $k$-dependence. Note also that the renormalised fields
carry classical dispersions if we use fully momentum-dependent
$Z_\Phi$'s in \eq{eq:RenPhi}. Accordingly in pole renormalisation
schemes the respective masses are pole masses. Moreover, observables
are provided in terms of the renormalised fields and couplings, and we
shall discuss physics in terms of these objects.

The quark chemical potential matrix $\hat \mu$ considered here is
understood to couple to all flavours equally,
$\hat\mu = \textrm{diag}(\mu_q, \mu_q, \mu_q)$. Thus, it is directly
related to the baryon chemical potential $\mu_B = 3 \mu_q$.  The
constituent quark mass $\bar m_s$ also has to be read as a diagonal
flavour matrix $\textrm{diag}(0,0,\bar m_s)$.

In the present work we do not consider dynamical hadronisation of the
full $N_f=3$ flavour multiplet
$T^a_{N_f}(\sigma_a + i\,\gamma_5 \pi_a)$, with the $U(N_f)$
generators $T^a_{N_f}$, $a=0,..., N_f^2-1$. In the singlet-octet basis
they are given by $T^0_{N_f}=\mathbb{1}_{N_f\times N_f}/\sqrt{2 N_f}$
and the $SU(N_f)$ generators $T^i_{N_f}$ for $i=1,...,
N_f^2-1$. Instead, we only consider part of the embedded $N_f=2$
multiplet: the scalar $\sigma$ as well as the pseudoscalar pions. The
four-quark interaction that gives rise to the corresponding resonances
is given by $\lambda_{q}$. The other mesons of the scalar and
pseudoscalar nonets are too heavy for playing a significant
r$\hat{\textrm{o}}$le in the offshell dynamics. Of the strange part of
the multiplet we only consider the scalar $\sigma_s$. The two scalars
$\sigma,\sigma_s$ are related to the ones in the singlet-octet basis
via
\begin{align}\label{eq:SigmaSigmas}
  \begin{pmatrix} \sigma \\ \sigma_s \end{pmatrix}  =
  \frac{1}{\sqrt{3}} \begin{pmatrix} \sqrt{2} &  1
      \\ 1 & -\sqrt{2} \end{pmatrix}
  \begin{pmatrix} \sigma_0 \\ \sigma_8 \end{pmatrix} \,,
\end{align}
for more details see e.g.\ \cite{Schaefer:2008hk, Herbst:2013ufa,
  Mitter:2013fxa, Rennecke:2016tkm, Fu:2018qsk, Wen:2018nkn}. Both,
light and strange quark constituent masses are given in terms of
condensates,
\begin{align}\label{eq:ml+s_const}
  \bar m_l =\frac12 \bar h \,\bar \sigma\,,\qquad  \qquad  \bar m_s =
  \frac{1}{\sqrt{2}} \bar h\, \bar \sigma_s\,,   
\end{align}
with the renormalised Yukawa coupling and the renormalised
$\sigma,\sigma_s$-expectation value $\bar \sigma= Z_\phi^{1/2} \sigma$
and $\bar \sigma_s= Z_{\phi}^{1/2} \sigma_s$ respectively, see
\eq{eq:RenPhi} and \eq{eq:RenMass-Yuk}. Their values are determined
via the explicit breaking terms $c_\sigma\,\sigma$ and
$1/\sqrt{2}\, c_{\sigma_s}\,\sigma_s$, as well as dynamical chiral
symmetry breaking. Both phenomena lead to nonvanishing
$\sigma_\textrm{EoM}$ and $\sigma_{s,\textrm{EoM}}$ and hence to
nonvanishing renormalised expectation values
$(\bar \sigma,\bar \sigma_s)_\textrm{EoM}$. The explicit breaking
coefficient $c_\sigma$ in the light sector is fixed with the physical
pion mass. Alternatively one could use the ratio of the pion decay
constant $f_{\pi}$ with that in the chiral limit, $f_{\pi,\chi}$ for
fixing $c_\sigma$, that is $f_{\pi}/f_{\pi,\chi}\approx 93/88$,
\cite{Tanabashi:2018oca}. Here, we choose the pion mass instead of
this ratio as it is more easily accessible. Furthermore, small errors
in the determination of the pion mass only propagate to small errors
in other observables. In turn, the explicit breaking coefficient
$c_{\sigma_s}$ may be either determined by e.g.\ the kaon mass or the
ratio $f_K/f_\pi \approx 111/93$, \cite{Tanabashi:2018oca}. We
  refer to \app{app:Scales} for a more detailed discussion of the
  scale setting.

  The relative size of the explicit breaking coefficients,
  $c_{\sigma_s}/c_\sigma$ may also be determined by their relation to
  the current quark masses $m_l^0$ and $m_s^0$. For large momentum
  scales of the order of the electro-weak scale $\sim 90$\,GeV, that
  is in the absence of any chiral dynamics, the constituent quark
  masses \eq{eq:ml+s_const} reduce to the current quark masses. Note
  also that the renormalised mesonic quantities tend towards bare ones
  for large momentum or cutoff scales as $Z_\phi\to 1$. Moreover,
  $Z_q\to 1$ due to the Landau gauge. Accordingly,
  $\bar \sigma, \bar\sigma_s \to \sigma, \sigma_s$ and
  $\bar h, \bar m_q\to h, \bar m_q$. This limit entails,
\begin{align}\label{eq:currentmasses} 
  \sigma_\textrm{EoM}=\frac{ 2 m_l^0}{h} =
  \frac{c_\sigma}{m^2_\phi}\,,\quad 
   \sigma_{s,\textrm{EoM}}=\frac{\sqrt{2   }m_s^0}{h} =
  \frac{c_{\sigma_s}}{\sqrt{2} m^2_\phi}
  \,, 
\end{align}
where $m_\phi^2$ is the unique mesonic mass function for large
momentum scales. \Eq{eq:currentmasses} entails that the ratio of the
current quark masses agrees with that of the explicit breaking
parameters,
\begin{align}\label{eq:ctom}
  \frac{c_{\sigma_s}}{c_\sigma}= \frac{m^0_s}{m_l^0} \approx
  27.46(15)(41) \,,  
\end{align}
in $N_f=2+1$ flavour computations. The estimate comes from lattice
computations, see \cite{Aoki:2013ldr}. \Eq{eq:currentmasses} and
\eq{eq:ctom} also enable us to relate the chiral condensates to
$c_\sigma$- and $c_{\sigma_s}$-derivatives, for more details see
\app{app:chiralcond}.

Finally, one may also adjust the constituent strange quark mass
$\bar m_s$ or the difference to the constituent light quark mass
$\bar m_l$,
\begin{align}\label{eq:Deltabarmsl}
  \Delta\bar m_{sl}=\bar m_s-\bar m_l\,,
\end{align}
on the basis of quantitative functional or lattice results in the
Landau gauge. 

In the present work we compute observables in the light
quark sector. We use that \textit{offshell} flavour-mixing terms
are small, as they always involve the propagator of a heavy mesonic
state. This is in stark contradistinction to \textit{onshell}
flavour-mixing terms, which are e.g.\ maximal in the pseudoscalar
sector due to the axial anomaly. Accordingly, light quark and gluon
correlation functions are sensitive to strange quark fluctuations only
via the gluon propagator or rather the gluon dressing. The latter
carries the momentum and RG running of the strange-quark--induced
vacuum polarisation, and it is well-known from respective quantitative
$N_f=2$ flavour computations that the gluon propagator is almost
insensitive to changes of the quark mass. This has been studied in
\cite{Cyrol:2017qkl}, where the pion mass (and hence the light quark
mass) has been changed from very light quarks to heavy ones in the
range $60\, \textrm{MeV}\lesssim m_\pi \lesssim 300\,
\textrm{MeV}$. This change had no effect on the gluon propagator
within the systematic error bars of the result. This analysis carries
over readily to the present $N_f=2+1$ flavour computation. Indeed, the
influence of the strange quark mass is even smaller due to the smaller
relative importance of the strange quark and its more effective
decoupling in the infrared due to the larger explicit chiral symmetry
breaking.

Consequently we use a simple approximation to the strange sector. The
chiral offshell dynamics are dominated by the pions, and we approximate
\begin{align}
\label{eq:sigmas}
  \bar \sigma_{s,\textrm{EoM}} =&\,  \frac{1}{\sqrt{2} }
                                  \bar \sigma_{\textrm{EoM}} +
                                  \frac{\sqrt{2}}{\bar h} \Delta \bar m_{sl}\,, 
\end{align}
With \eq{eq:sigmas} the chiral dynamics in the strange quark sector
are the same as in the light quark sector.

All these determinations have to be taken with a grain of salt due to
the rough nature of our approximation of the strange quark
sector. While this approximation has to be improved for an access to
observables with strangeness, the observables considered here depend
only very mildly on the difference between the constituent light and
strange quark masses, \eq{eq:Deltabarmsl},
\begin{align}\label{eq:barms-range}
  100\,\textrm{MeV} \lesssim \Delta\bar m_{sl}\lesssim 200\, \textrm{MeV}\,.
\end{align}
For the determination of $\Delta\bar m_{sl}$ in the present work we
use the ratio of the decay constants, which relates the current
strange quark mass directly to observables. In the mean field
approximation in low energy effective theories we typically have
$f_\pi\approx \bar\sigma_\textrm{EoM}$ and
$f_K\approx 1/2 \bar \sigma_\textrm{EoM}
+1/\sqrt{2}\,\bar\sigma_{s,\textrm{EoM}}$. We emphasise that these
relations do not hold true in QCD as the determination of the decay
constant requires the full momentum dependence of the quark mass
functions; for a detailed discussion of the pion decay constant see
\app{app:Scales}.  However, the cutoff scale (and momentum) dependence
of both the light and the strange quark masses are very similar (after
being rescaled by their value at vanishing momentum). Hence we
conclude that the mean field relations should hold even
quantitatively for the ratios of the decay constants. Using
\eq{eq:sigmas} in the mean field approximation for the ratio of kaon
decay constant and pion decay constant, this is well adjusted with
$\Delta \bar m_{sl}=120$\,MeV,
\begin{align}\label{eq:fKfpi}
  \frac{f_K}{f_\pi} \approx
  1+ \frac12 \frac{\Delta \bar m_{sl}}{\bar m_l}
  \quad \stackrel{\Delta
  \bar m_{sl}=120\,\textrm{MeV}}
  {\longrightarrow}
  \quad 
  \frac{f_K}{f_\pi } \approx 1.17\,.
\end{align}
\Eq{eq:fKfpi} is in good agreement with the actual value
$\frac{f_K}{f_\pi } \approx 1.19$, see \cite{Tanabashi:2018oca}. More
accurately we may derive the mass difference with a comparison to
$N_f=2+1$ lattice QCD (with isospin symmetry). The ratio of the decay
constant is determined with $f_K/f_\pi = 1.194(5)$,
\cite{Aoki:2013ldr}, and provides $\Delta\bar m_{sl}\approx 135$\,MeV.

For our choice $\Delta \bar m_{sl}=120$\,MeV, the constituent strange
quark mass is found to be $\bar m_s = 467$\,MeV, in good agreement
with quark model values. Note however, that the Landau gauge
constituent strange quark mass is considerably higher,
$\bar m_s \gtrsim 500$\,MeV, see e.g.\ \cite{Fischer:2018sdj}, which
would amount to a current strange quark mass of
$\Delta \bar m_{sl} \gtrsim 150$\,MeV in the present approach. This is
further discussed in \app{app:chiralcond}. We emphasise again that a
variation of $\Delta \bar m_{sl}$ within this range does not influence
our results for light quark observables.
 
Next we discuss the treatment of the four-quark sector in
\eq{eq:action}: We only consider the four-quark interaction
$\lambda_q$ of the $N_f = 2$ scalar-pseudoscalar multiplet as well as
the corresponding Yukawa interaction $h$ and the mesonic composite
fields. In turn, the Dirac term depends on all quark fields, so either
$u,d$ in the $N_f=2$ flavour case or, $u,d,s$ in the $N_f=2+1$ flavour
case. Accordingly, the mesonic field $\phi=(\sigma,\bm{\pi})$ in
\eq{eq:action} is in the $O(4)$-representation with $\rho=\phi^2/2$
for both, $N_f=2$ and $N_f=2+1$.  The linear term $-c_\sigma\,\sigma$
breaks the two flavour chiral symmetry explicitly, and leads to the
current quark masses for $u,d$ quarks. We assume light isospin
symmetry here, so they have identical masses. They are absorbed in a
shift of $\sigma$, leading to current quark mass terms from the Yukawa
term.  In the present setup the $c_\sigma$-term is generated from the
$u,d$-quark mass term via dynamical hadronisation with an appropriate
choice of $\dot{C}_k$ in \eq{eq:HadABC}, for more details see
\sec{sec:DynHad}. It is now apparent that the shifted
$\sigma$-current coupled to the dynamical hadronisation flow
$\dot{\phi}_k$ is chirally symmetric, as it does not depend on
$c_\sigma$,
\begin{align}\label{eq:GammaSigmaSym}
  \partial_{c_\sigma}\left[ \frac{\delta \Gamma_k}{
  \delta \sigma }+ c_\sigma\right]\equiv 0\,.
\end{align}
In turn, we do not consider offshell fluctuations from the four-quark
interactions with strangeness: In \eq{eq:action} neither four-quark
terms nor mesonic fields with strangeness are included.  This
approximation is based on the observation that at $T,\mu=0$ even the
two-flavour scalar-pseudoscalar terms produce negligible contributions
for cutoff scales $k\gtrsim 500\,$MeV above the onset of chiral
symmetry breaking, \cite{Mitter:2014wpa, Cyrol:2017qkl}. For cutoff
scales in the vicinity of the onset of chiral symmetry breaking and
below, $k\lesssim 500\,$MeV the other two-flavour channels and even
more so the $s$-quark channels are not dynamical anymore due to their
large mass scales. Indeed, sizable contributions at $T,\mu=0$ are only
triggered by the pion channel. This is in line and supports chiral
perturbation theory.

\begin{table}[t]
  \begin{center}
  \begin{tabular}{|c ||c|| c |}
    \hline & & \\[-1ex]
    Observables & Value & Parameter in $\Gamma_\Lambda$\\[1ex]
    \hline& & \\[-2ex]
    $m_{\pi,\textrm{pol}}$ [MeV]      & 137& \begin{tabular}{rcl}  2: $c_\sigma$&=&
                                                                                    3.6\,$\text{GeV}^3$ \\[1ex]
                                               2+1:  $ c_\sigma$&=& 3.6\,$\text{GeV}^3 $\end{tabular}      \\ [1ex]
    \hline & & \\[-1ex] 
    $f_K/f_\pi $   & 1.17    & 
                                2+1:    $\Delta \bar m_{sl}$= 120\, MeV  \\[1ex]
    \hline & & \\[-2ex]
    $\alpha_s$   &   \begin{tabular}{l} RG-consistency: \\
                       \sec{sec:RGconsistency}\end{tabular}    &  \begin{tabular}{rcl} 2:
                                                                            $\alpha_{s,\Lambda}$ &=& 0.21\\[1ex]
                                                                            2+1:  $  \alpha_{s,\Lambda}$ &=&0.235\end{tabular}
    \\[1ex]
    \hline& &\\[-2ex]
    \hline & & \\[-2ex]
    $\bar m_l$ [MeV]      &  \begin{tabular}{rl} 2:
                               & 367 \\[1ex]
                               2+1:  & 347 \end{tabular} & \begin{tabular}{rcl}  2: $a$=0.008\,\quad $b$=2\,\textrm{GeV} \\[1ex]
                                                             2+1:  $a$=0.034\,\quad $b$=2\,\textrm{GeV} \end{tabular} 
    \\[1ex]  
    \hline& &\\[-2ex]
    \hline & & \\[-2ex]
    $f_\pi$ [MeV]      &  \begin{tabular}{rl} 2:
                            & 96.0 \\[1ex]
                            2+1: & 93.0 \end{tabular} & \noindent\rule{1cm}{0.4pt}
    \\[1ex]
    \hline& &\\[-2ex] 
    $\bar m_s$ [MeV]      &  467  &  \noindent\rule{1cm}{0.4pt}
    \\[1ex]  
    \hline& &\\[-2ex] 
    $\bar m_\sigma$ [MeV]      &  \begin{tabular}{rl} 2:
                                    & 531\\[1ex]
                                    2+1:  & 510\end{tabular} & 
                                                          \noindent\rule{1cm}{0.4pt} \\[1ex]  
    \hline
  \end{tabular}
  \caption{\emph{Upper part}: observables and related fundamental QCD
    couplings at the initial cutoff scale $\Lambda=20$\,GeV: the
    strong coupling $\alpha_{s,\Lambda}$, the pion pole mass
    $m_{\pi,\textrm{pol}}$ via $c_\sigma$, and the ratio $f_K/f_\pi$
    via the constituent quark mass difference
    $\Delta \bar m_{sl}=\bar m_s-\bar m_l$ for the present $N_f=2$
    and $N_f=2+1$ flavour computations.\\[1ex]
    \emph{Middle part}: IR enhancement of quark-gluon coupling
    $\alpha_{\bar q A q}\to (1+a)\,\alpha_{\bar q A q}$ below
    $k\approx b$. The value of $a$ is adjusted with the constituent
    light quark mass $\bar m_l$, for more details see
    \app{app:IR-enhancement}. This phenomenological IR-enhancement
    effectively accounts for the effect of nonclassical tensor
    structures in the quark-gluon vertex which are missing in the
    present approximation. If taking the full quark-gluon vertex into
    account, this is not necessary and $\bar m_l$ is a prediction, see
    \cite{Mitter:2014wpa,Cyrol:2017ewj}.\\[1ex]
    \emph{Lower part}: predictions in vacuum QCD: the strange
    constituent quark masses $\bar m_s$ and the $\sigma$-mass
    $\bar m_\sigma$. Also $f_\pi$ is a prediction as we have fixed the
    pion pole mass $m_{\pi,\textrm{pol}}$ instead of $f_\pi$. Fixing
    the latter relative to the pion decay constant $f_{\pi,\chi}$ in
    the chiral limit (in the present work $f_{\pi,\chi} = 88.6$\,MeV
    for $N_f=2+1$) would have been the more physical but less
    accessible choice, for a detailed discussion see
    \app{app:Scales}.}
  \label{tab:InitialValues}
  \end{center}\vspace{-1cm}
\end{table}
Note however, that e.g.\ in the vicinity of the phase transition,
kaons and the eta meson may play a r$\hat{\textrm{o}}$le. Neglecting
this is part of our current systematic error. We also emphasise that
at large densities we expect relevant offshell contributions from
diquark and/or density channels.  The importance of the additional
$N_f=2$ flavour channels has been investigated thoroughly in effective
theories in \cite{Braun:2017srn, Braun:2018bik, Zhang:2017icm},
leading to a semiquantitative agreement of both approximations up to
large densities after an appropriate rescaling including the critical
region found in the present work for $N_f=2$ flavour QCD, see
\sec{sec:PhaseStructure} and \fig{fig:phasediagram}. This estimate is
fully confirmed in a QCD study with the fRG in
\cite{Braun:2019a}. Note that the observed dominance of the
scalar-pseudoscalar channel for $N_f=2 $ flavours in
\cite{Braun:2017srn, Braun:2018bik, Braun:2019a} translates to
$\mu_B/T\lesssim 7$ in the present $N_f=2+1$ flavour study, and
includes the respective critical region, see \fig{fig:phasediagramII}.

Accordingly, the $N_f =2$ and $N_f =2+1$
theories differ by the Dirac term in the effective action, and hence
by the respective additional $s$-quark loops. This is important for
purely gluonic correlation functions and amounts to a relative change
in the physics scale $\Lambda_\textrm{QCD}$ as well as the respective
$\beta$-functions. This is very similar to respective DSE
computations, a difference being the backreaction onto the purely
gluonic diagrams which is partly taken into account in the present
work.

In summary the couplings $\alpha_{s,\Lambda}$, $c_\sigma$ and
$\Delta \bar m_{sl}$ (or the ratio of the current quark masses
$m_s^0/m_l^0=c_{\sigma_s}/c_\sigma$, see \eq{eq:ctom}) with
$Z_{\Phi,\Lambda}=1$ in the initial action $\Gamma_\Lambda$ at the
initial UV scale $\Lambda=20$\,GeV are fixed by the fundamental
parameters of QCD; for the strong coupling and the current quark
masses, as well as phenomenological infrared enhancement parameters
$a,b$ for the quark-gluon coupling, see
Table~\ref{tab:InitialValues}. The latter phenomenological parameters,
$(a,b)$, effectively account for the infrared effects of the missing
nonclassical tensor structures of the quark-gluon vertex, see
\app{app:IR-enhancement}.

% 
%%%%%%%%%%%%%%%%%%%%%%%%%%%%%
\begin{figure}[t]
\includegraphics[width=0.98\columnwidth]{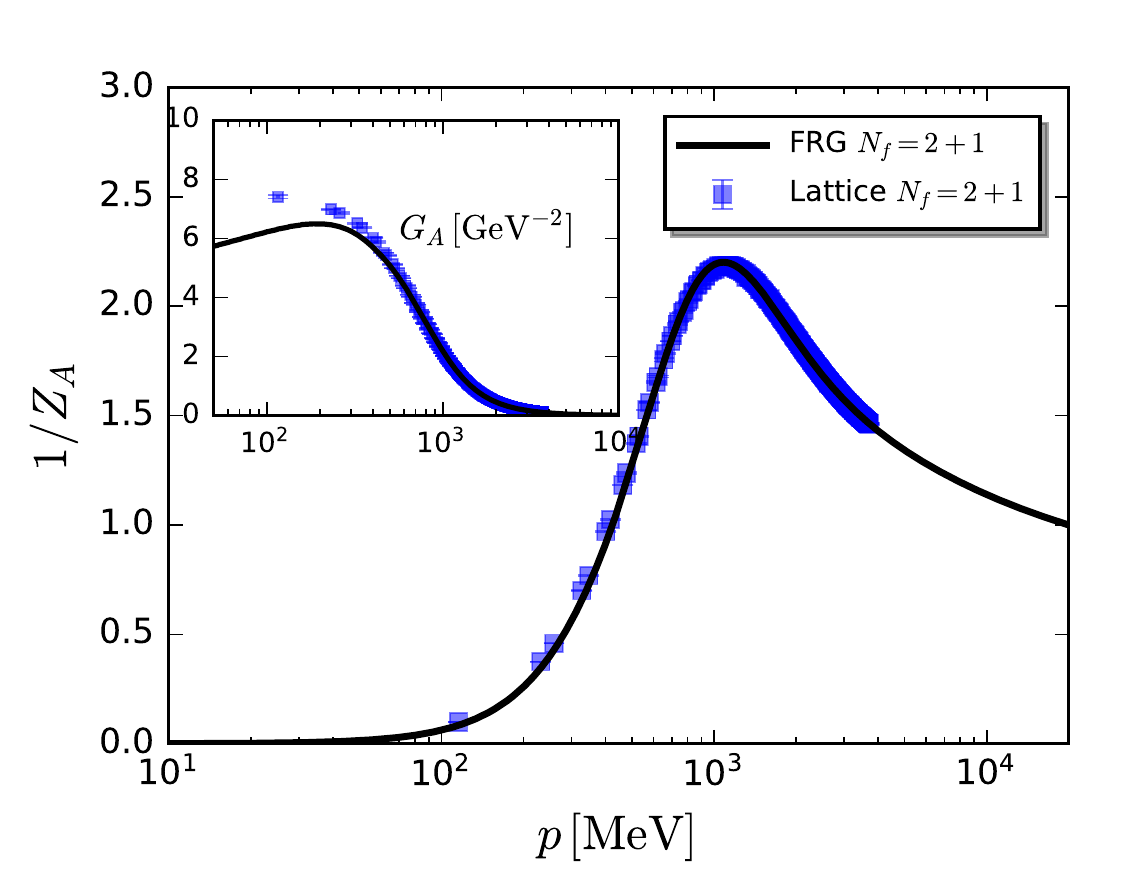}
\caption{$N_f=2+1$ gluon dressing function $1/Z_{A}(p^2)$ and gluon
  propagator $G_A=1/[Z_{A}(p^2)\, p^2]$ (inset) in the vacuum as
  functions of momenta in comparison to lattice results based on the
  $N_f=2+1$ configurations (domain wall fermions) of the RBC/UKQCD
  collaboration, e.g.\ \cite{Blum:2014tka}. The lattice results are
  continuum extrapolated from data with
  $\beta=2.25,\, a^{-1} = 2.359(7)\,\textrm{GeV},\, m_\pi=139.2(5)$\,MeV
  and
  $\beta=2.13,\, a^{-1} = 1.730(4)\,\textrm{GeV},\, m_\pi=139.2(4)$\,MeV,
  see \cite{Zafeiropoulos:2019flq,Boucaud:2018xup,BDRSZ:2019}. The
  momentum dependence of the fRG results is given by $p^2=k^2$, in
  accordance with the same identification of the $N_f=2$ flavour input
  data; for more details see
  \sec{sec:GlueSector}.}\label{fig:inZA_lattice}
\end{figure}
%%%%%%%%%%%%%%%%%%%%%%%%%%%%%
%
Strictly speaking, absolute momentum scales should be measured in the
pion decay constant in the chiral limit. In the present approximation
it is computed as
\begin{align}\label{eq:fpichiral}
  f_{\pi,\chi}=88.6\, \textrm{MeV} \quad \textrm{for}\quad  N_f=2+1\,,
\end{align}
see \eq{eq:fpichi}. The details of the scale setting procedure are
discussed in \app{app:Scales}. Apart from the quantitative precision
of the predictions for observables in Table~\ref{tab:InitialValues} in
particular for $N_f=2+1$, the $N_f=2+1$ gluon dressing is a prediction
within the current setup. It agrees quantitatively with the respective
lattice results, see \fig{fig:inZA_lattice}. Note that the gluon
propagator shows the expected deviations in the deep infrared: the
$N_f=2$ flavour propagator in \cite{Cyrol:2017ewj} is of the scaling
type, related to a different infrared closure of the Landau gauge as
that implicitly defined by the lattice gauge fixing. This has been
discussed at length in the literature, see \cite{Cyrol:2017ewj} and
references therein. Importantly, as part of the gauge fixing, it has
no impact on observables.

\subsection{Strong couplings $\alpha_s$ from gluon, ghost-gluon and
  quark-gluon vertices}\label{sec:StrCoup}
The quark-gluon and purely gluonic correlation functions in the
current approximation give rise to different `avatars' of the strong
couplings $\alpha_{\Phi_1\cdots \Phi_n}$ with
$\Phi_i=A_\mu, c,\bar c, q,\bar q$ and $n=3,4$. For details see \cite{
  Mitter:2014wpa,Cyrol:2016tym,Cyrol:2017ewj,Cyrol:2017qkl}, for
similar considerations in quantum gravity see
\cite{Eichhorn:2018akn,Eichhorn:2018ydy}. They are defined by the
respective vertex dressings $\lambda_{\Phi_1\cdots\Phi_n}$, divided by
the wave function renormalisations $\prod Z_{\Phi_i}^{1/2}$ of the
attached legs evaluated at vanishing momentum, see e.g.\
\cite{Cyrol:2017ewj}. Here we consider the couplings of the purely
gluonic sector,
\begin{align}\label{eq:alphasGluon}
  \alpha_{A^3} = \frac{1}{4\pi} \frac{\lambda_{A^3}^2}{Z_A^{3}}\,,
  \quad \alpha_{A^4} =
  \frac{1}{4\pi}  \frac{\lambda_{A^4}}{Z_A^{2}}\,,\quad
  \alpha_{\bar c Ac} = \frac{1}{4\pi} \frac{
  \lambda_{A c \bar c}^2}{Z_A\,Z_c^2 }\,,
\end{align}
and that of the matter sector,
$\alpha_{\bar q A q}=(\alpha_{ \bar lAl } ,\alpha_{ \bar s Ac})$ with
two entries for the light quark-gluon and strange quark-gluon
couplings respectively,
\begin{align}\label{eq:alphasMatter}
  \alpha^{\ }_{\bar l A l} = \frac{1}{4\pi}
  \frac{\lambda_{ A l \bar l }^2}{Z_A\,Z_l^2 }\,, \qquad
  \alpha_{\bar s A s} = \frac{1}{4\pi}
  \frac{\lambda_{A s \bar s }^2}{Z_A\,Z_s^2 }\,.
\end{align}
Note that the strong couplings \eq{eq:alphasGluon},
\eq{eq:alphasMatter} are natural definitions of gluon exchange
couplings. For example, the quark-gluon couplings involve two
quark-gluon vertex functions and one gluon dressing $1/Z_A$, related
to the one-gluon scattering of two quark currents, see
\Fig{fig:flow4q}.

For perturbative and semiperturbative scales $k\gtrsim 1- 3\,$GeV
these couplings are related by modified Slavnov-Taylor identities
(mSTIs). This property also holds true in the quantitative
approximation
\cite{Mitter:2014wpa,Cyrol:2016tym,Cyrol:2017ewj,Cyrol:2017qkl} for
the purely gluonic couplings, for more details see in particular 
\cite{Cyrol:2017ewj}. We emphasise that in the present setup gauge
invariance is encoded in these mSTIs. They arise from BRST-variations
of the gauge fixed action in the presence of the regulator terms, see
e.g.\ \cite{Ellwanger:1994iz,Bonini:1994kp,DAttanasio:1996tzp,%
  Becchi:1996an,Reuter:1997gx,Freire:2000bq,%
  Igarashi:2001mf,Pawlowski:2003sk,Pawlowski:2005xe,%
  Wetterich:2017aoy,Igarashi:2019gkm}. For smaller momenta and cutoff
scales $k\lesssim 1-3\,$GeV they start to differ for two reasons:
first of all the gluon mass scale starts to influence the
running. More importantly the mSTIs only relate the longitudinal
couplings, while the flow depends on the transversal couplings. It can
be shown that some of the transversal $\alpha$'s cannot be identified
with their longitudinal counterparts as otherwise the gluon mass scale
would be absent, see
\cite{Cyrol:2016tym,PhysRevD.26.1453,Aguilar:2011ux,%
  Aguilar:2011xe,Aguilar:2019jsj}.  However, in the absence of the
gluon mass scale in the propagator the theory lacks confinement,
\cite{Braun:2007bx,Braun:2010cy,Fister:2013bh}.

Here we compute the flows of the three-gluon coupling, $\alpha_{A^3}$,
and the quark-gluon couplings $\alpha_{\bar q A q}$. From the vacuum
results for the couplings in \cite{Braun:2014ata, Cyrol:2017ewj} we
infer that the four-gluon coupling runs closely to $\alpha_{A^3}$, and
the ghost-gluon coupling runs closely to $\alpha_{\bar q A q}$ in the
relevant momentum regime $k\gtrsim 1$\, GeV. In the deep infrared all
these couplings. However, except of the ghost-gluon coupling,
  all the avatars of the strong coupling decay for these scales, see
  \Fig{fig:alphas} in \sec{sec:RGconsistency}, and
  \cite{Braun:2014ata, Cyrol:2017ewj}. The ghost-gluon coupling only
  enters in diagrams that are suppressed except for the deep infrared
  with $k\lesssim 100$\,MeV, a more detailed discussion of this fact
  is provided in \sec{sec:gluoniccouplings}.  In summary the differences
  lead to negligible effects and we use
\begin{align}\label{eq:IDalphaGlue}
  \alpha_{A^4} = \alpha_{A^3}=\alpha_{\textrm{glue}}\,,
  \qquad  \alpha_{\bar c Ac} \simeq\alpha_{
 \bar q Aq} \,.
\end{align}
For gluons and quarks this leads us to covariant derivatives in the
fundamental and adjoint representation of $SU(N_c)$ respectively,
\begin{align}\nonumber 
  D_{\mu}&=\partial_{\mu}-i Z_{A}^{1/2} g_{\bar q A q}
           A^a_{\mu} t^a\,,\\[1ex]
  D^{ab}_{\mu}&=\partial_{\mu} \delta^{ab}-Z_{A}^{1/2}
                g_{\textrm{\tiny{glue}}} f^{abc} A^c_{\mu}\,,
                \label{eq:CovDer}
\end{align}
where
 \begin{align}
   g_{\textrm{\tiny{glue}}} =\sqrt{4\pi{\alpha}_{\textrm{\tiny{glue}}}}\,,\qquad
   g_{\bar q  A q } =\sqrt{4\pi{\alpha}_{\bar q A q }} \,.
\end{align}
The covariant derivative in the ghost terms is an adjoint one but
carries $g_{\bar q A q }$.  In \eq{eq:CovDer}, $f^{abc}$ are the
structure constants for the $SU(N_c)$ group with
\begin{align}\label{eq:group}
  [t^a,t^b]=i f^{abc} t^c\,,\qquad \qquad 
  \Tr\, t^a t^b=\frac{1}{2}\delta^{ab}\,.
\end{align}
The terms in the first two lines on the r.h.s.\ of \eq{eq:action} are
built from the operators in the classical QCD action. The gluonic
field strength tensor reads
\begin{align}\label{eq:Fmunu}
   F^a_{\mu\nu}&=Z_{A}^{1/2}\left(\partial_{\mu}A^a_{\nu}
                 -\partial_{\nu}A^a_{\mu}+Z_{A}^{1/2}g_{\textrm{\tiny{glue}}}
                 f^{abc}A^b_{\mu}A^c_{\nu}\right)\,,
\end{align}
and  the Landau gauge $\xi=0$  is chosen in this work.

A similar truncation has been found to be successful in describing the
transition from the quark-gluon regime at high energy to the hadronic
one at low energy in the vacuum \cite{Braun:2014ata}. The truncation
\eq{eq:action} for the effective action takes into account all order
scatterings of the resonant scalar and pseudoscalar channels of the
four-fermi interaction through the scale-dependent effective potential
$V_k(\rho,L,\bar L)$. It is complementary to taking into account a
Fierz-complete four-fermi basis as done in \cite{Mitter:2014wpa,
  Cyrol:2017ewj, Braun:2017srn, Braun:2018bik, Leonhardt:2019fua}. A
respective two flavour study with a Fierz complete basis and all order
interactions of resonant channels is work in progress,
\cite{fQCD:2019a}.

%%%%%%%%%%%%%%%%%%%%%%%%%%%%%%%%%%%%%%%%%%%%%%%%%%%%%%%%%%%%%

\subsection{Dynamical hadronisation in 
  the $\sigma-\pi$ channel}\label{sec:DynHad}

As discussed in \sec{sec:FunRG-QCD}, we take into account the resonant
interactions in the $\sigma-\bm{\pi}$ channel with the help of
dynamical hadronisation. We use \eq{eq:HadABC} with ${\dot B}_k=0$,
${\dot C}_k=0$,
\begin{align}\label{eq:HadAB0}
  \langle   \partial_t \hat \phi_k\rangle =&\dot{A}_k \,\bar{q}\tau q\,, 
\end{align}
in the flow equation \eq{eq:FlowQCD}. Our strategy is to choose the
coefficients in \eq{eq:HadAB0} such that the running of the
four-quark interaction $\lambda_q$ in \eq{eq:action} is exactly
cancelled. This amounts to a scale dependent bosonisation of this
channel. We emphasise that this only transfers the information carried
by four-quark interaction from the quark-gluon sector to the meson
sector. It is a practical way to enter the symmetry broken phase in
the presence of resonant channels and still retain the full
information of the underlying dynamics of the fundamental degrees of
freedom. In case of the scalar-pseudoscalar channel considered here, a
resonance at vanishing momentum indicates the formation of a quark
condensate $\langle \bar q q \rangle$ and therefore chiral symmetry
breaking.

%
%%%%%%%%%%%%%%%%%%%%%%%%%%%%%
\begin{figure}[t]
\includegraphics[width=0.98\columnwidth]{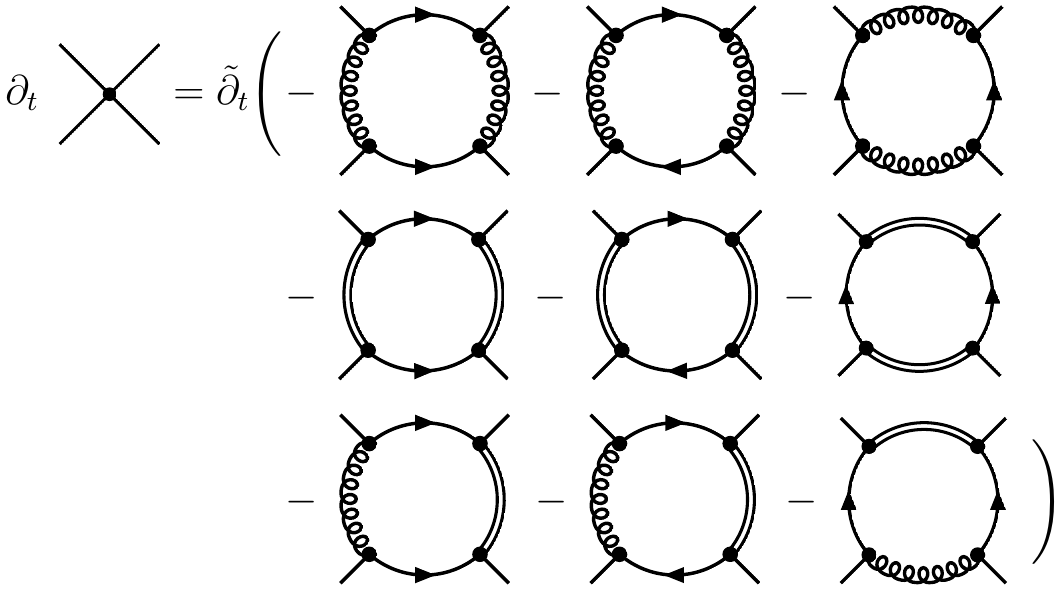}
\caption{Flow equation for the four-quark coupling in the present
  approximation. Further diagrams with the nonresonant part of the
  scalar-pseudoscalar channel and other four-quark tensor structures,
  as well as that with higher order couplings are dropped. The black
  dots denote the full vertices and the various lines represent for
  the full propagators. The derivative
  $\tilde \partial_t=\left. \partial_t\right|_{\{\Gamma^{(n)}_k \}} $
  only hits the $k$-dependence of the regulator, e.g.\
  $\tilde\partial_t G_k=-G_k \dot{R}_k G_k$. The first line with the
  quark-gluon diagrams generates the four-quark interaction through
  gluon-exchange at large scales. The second line takes into account
  the self-interaction of the resonant four-quark interaction in terms
  of meson exchange diagrams. The last line comprises the mixed
  diagrams with meson and gluon exchange. Gluon-exchange dominates the
  flow for large and intermediate energy scales, while the
  meson-exchange diagrams dominate at low energy, see
  \app{app:dtlambda}. Accordingly, in the present work we keep the
  quark-gluon and quark-meson diagrams and drop the mixed
  ones.}\label{fig:flow4q}
\end{figure}
%%%%%%%%%%%%%%%%%%%%%%%%%%%%%
%
To this end, we project the flow equation on the flow of the
scalar-pseudoscalar four-quark vertex, $\lambda_q$ in \eq{eq:action}
at vanishing momentum. We define the respective dimensionless,
renormalised four-quark coupling, Yukawa coupling and hadronisation
function as
\begin{align}\label{eq:RenCoup}
  \bar\lambda_q= \frac{\lambda_q\,k^2}{Z_q^2}\,,\qquad \bar h=
  \frac{h}{Z_\phi^{1/2} Z_q}\,,\qquad
  \dot{\bar A}=  \frac{Z_\phi^{1/2}}{Z_q} \dot {A}\,k^2\,, 
\end{align}
where we suppressed the subscript $k$, and the renormalised Yukawa
coupling has already been introduced in \eq{eq:RenMass-Yuk}. With
these definitions the flow of the renormalised four-quark function
reads
\begin{align}
  \partial_{t}\bar{\lambda}_{q} - 2\left( 1+\eta_q\right)
  \bar{\lambda}_{q}
  -\bar{h} \,\dot{\bar A}
  =  \overline{\textrm{Flow}}^{(4)}_{(\bar q\tau q)(\bar q\tau q)}\,,
 \label{eq:flow4q}
\end{align}
where 
\begin{align}\label{eq:barFlown} 
  \overline{\textrm{Flow}}^{(n)}_{\Phi_{i_1}\cdots\Phi_{i_n}} =
  \frac{1 }{\prod_{j=1}^n Z^{1/2}_{{\Phi_{i_j}}}\,
  k^{d_{\lambda_{{\Phi_{i_1}\cdots\Phi_{i_n}}}}}}
  \textrm{Flow}^{(4)}_{\Phi_{i_1}\cdots\Phi_{i_n}}\,, 
\end{align}
are the dimensionless, renormalised flow diagrams.

The projection on the scalar-pseudoscalar four-quark channel in
\eq{eq:barFlown} is done along the lines of \cite{Braun:2011pp} with
an expansion of the flow in terms of quark-bilinears. This is
indicated by the subscript ${}_{(\bar q\tau q)(\bar q\tau q)}$.
$\textrm{Flow}^{(n)}$ is defined in \eq{eq:Flown}, and
$d_{\lambda_{{\Phi_{i_1}\cdots\Phi_{i_n}}}}$ is the canonical momentum
dimension of the vertex dressing
$\lambda_{\Phi_{i_1}\cdots\Phi_{i_n}}$.  Now we resort to fully
hadronised flows in the $\sigma$--$\bm\pi$ channel by demanding
\begin{align}\label{eq:FullDynHad}
  \bar\lambda_q \equiv 0\,,\qquad \forall k\,.
\end{align}
With this, all diagrams with four-quark vertices $\bar \lambda_q$ are
absent and the flow is depicted by \Fig{fig:flow4q}. The diagrams in
the first line on the r.h.s.\ in \Fig{fig:flow4q} arise from the
exchange of gluons and those in the second line from mesons. The
respective expressions are given in \app{app:dtlambda},
\eq{eq:dtlambdaA} and \eq{eq:dtlambdaAphi}. The mixed diagrams with
gluon and meson exchange, shown in the third line in \Fig{fig:flow4q},
are neglected here, since the dynamics of gluons and mesons dominate
in approximately disjoint regions of the RG scale and the mixed
diagrams are comparatively small in either regime, cf.\
\app{app:dtlambda}.

Inserting \eq{eq:FullDynHad} into \eq{eq:flow4q} leaves us with an
equation for the hadronisation function,
\begin{align}
  \dot{\bar A}&=-\frac{1}{\bar{h}}\,\overline{
                \textrm{Flow}}^{(4)}_{(\bar q\tau q)(\bar q\tau q)}\,.
 \label{}
\end{align}
This choice entails that the quantum fluctuations contributing to the
quark scattering in the $\sigma$--$\bm\pi$ channel are transferred
completely to effective hadronic degrees of freedom, the $\sigma$ and
$\bm\pi$ fields, their propagation, self-scattering and scattering
with quark-antiquark pairs.

The original four-quark interaction is described with instantaneous
meson propagators and a Yukawa interaction of the $\sigma,\bm\pi$ with
a quark-antiquark pair. Multi-scatterings of the resonant channels are
encoded in the effective potential $V_k$. The flow of the respective
couplings and propagators is discussed in \sec{sec:correlation}.

%%%%%%%%%%%%%%%%%%%%%%
\subsection{Thermodynamic potential \& order
  parameters}\label{sec:TPot+Order}

The equilibrium thermodynamic potential density $\Omega$ is given by
\begin{align}
  \Omega[\Phi;T,\mu]&=\Big(\frac{T}{\cal V}\Big)
                      \Gamma_{k=0}[\Phi]\,,\label{eq:thermpo}
\end{align}
where $\cal V$ is the spatial volume. $\Omega[\Phi;T,\mu]$ is nothing
but $V_{k=0}(\rho, L,\bar L)$ in \eq{eq:EffPot}. It depends on the
given background $\Phi$, the temperature $T$ and the quark chemical
potential $\mu$. In the present case we only consider homogeneous
background fields and the infinite volume limit. \Eq{eq:thermpo} can
be obtained from the evolution of the effective action $\Gamma_{k}$
with \eq{eq:FlowQCD}. To simplify the calculations, we split
$\Gamma_{k}$ into two parts,
\begin{align}
  \Gamma_k[\Phi]=\Gamma_{\text{\tiny{glue}},k}[\Phi]+
  \Gamma_{\text{\tiny{matt}},k}[\Phi]\,,\label{eq:Gasplit}
\end{align}
where the glue sector corresponds to the first two loops in
\Fig{fig:fleq}, and the matter sector to the latter two. The
thermodynamic potentials related to the two parts are dealt with
differently in this work; that is, we do not evolve the flow of
$\Gamma_{\text{\tiny{glue}},k}$, but instead replace it with the
QCD-enhanced glue potential \cite{Haas:2013qwp,Herbst:2013ufa}, to
wit,
 \begin{align}
   \Big(\frac{T}{\cal V}\Big)\Gamma_{\text{\tiny{glue}},k=0}[\Phi]  = 
   V_{\text{\tiny{glue}}}(L, \bar L)\,. \label{eq:Vglue}
\end{align}
In \eq{eq:Vglue} we have introduced the traced Polyakov loops
$L,\bar L$ with
\begin{align}
  L(\bm{x})=\frac{1}{N_c} \left\langle \Tr\, {\cal P}(\bm x)\right\rangle
  \,,\quad  \bar L (\bm{x})=\frac{1}{N_c} \langle
  \Tr\,{\cal P}^{\dagger}(\bm x)\rangle \,,\label{eq:Lloop}
\end{align}
and 
\begin{align}
  {\cal P}(\bm x)= \mathcal{P}\exp\Big(ig\int_0^{\beta}
  d\tau \hat A_0(\bm{x},\tau)\Big)\,. \label{eq:Ploop}
\end{align}
Here, $\mathcal{P}$ on the r.h.s.\ standing for the path ordering.  In
\eq{eq:Ploop} the gauge field is the fluctuation field and
$A_0=\langle \hat A_0 \rangle$ is the temporal mean gauge field in
$\Phi$.  Accordingly, the expectation values \eq{eq:Lloop} are
nontrivial functions of the mean field $ A_0$ and
\begin{align}\label{eq:diffL}
  L[A_0] \neq \frac{1}{N_c}  \Tr\, {\cal P}[A_0]\,.
\end{align}
This has been discussed at length in \cite{Braun:2007bx,
  Marhauser:2008fz, Braun:2010cy, Fister:2013bh, Herbst:2015ona}, for
related work see e.g.\ \cite{Fukushima:2017csk, Fukushima:2012qa,
  Reinosa:2014ooa, Reinosa:2014zta, Maelger:2017amh, Maelger:2018vow}.

However, in \cite{Herbst:2015ona} it
has been shown that the difference in \eq{eq:diffL} can largely be
attributed to a temperature-dependent rescaling.

In the present work this difference is ignored, it will be discussed
elsewhere. We approximate
$ L[A_0] \approx \frac{1}{N_c} \Tr\, {\cal P}[A_0]$ which allows us to
utilise pure glue lattice results for the expectation value $L$ and
the correlations of the Polyakov loop for the construction of the
Polyakov loop potential. The glue potential $V_{\text{\tiny{glue}}}$
employed in this work is that computed in \cite{Lo:2013hla}, where the
quadratic fluctuations of the Polyakov loop,
$\langle {\cal P}[A_0](\bm x) {\cal P}[A_0](\bm y)\rangle$, are taken
into account. We specify the glue potential in
\app{app:gluepot}. Finally, one obtains the thermodynamic potential
density as follows
\begin{align}\label{eq:fullThermodyn}
  \Omega[\Phi;T,\mu]&=V_{\text{\tiny{glue}}}(L, \bar
                      L)+V_\text{mat}(\rho,L,\bar L)-c_\sigma\sigma\,. 
\end{align}
The potential \eq{eq:fullThermodyn} allows us to access both, the
confinement-deconfinement phase transition or crossover via the
Polyakov loop expectation value $L$,
\begin{align}\label{eq:EoMPolloop}
  \left.  \0{\partial\Omega[\Phi;T,\mu]}{\partial L}
  \right|_{\Phi=\Phi_\textrm{EoM}} =
  \left.  \0{\partial \Omega[\Phi;T,\mu]}{\partial
  \bar L}\right|_{\Phi=\Phi_\textrm{EoM}} =0\,, 
\end{align}
where $\Phi_\textrm{EoM}$ now includes
$L_\textrm{EoM}, \bar L_\textrm{EoM} $ instead of
$A_{0,\textrm{EoM}} $ in a slight abuse of notation.

In turn, explicit and spontaneous chiral symmetry breaking is encoded
in the expectation value
$\langle \sigma\rangle := \sigma_\textrm{EoM}$ of the $\sigma$-field,
defined with the respective EoM,
\begin{align}\label{eq:EoMsigma}
  \left.  \0{\partial V_{k=0}(\rho,L,\bar L)}{\partial \sigma}
  \right|_{\Phi_\textrm{EoM}} = c_\sigma\,. 
\end{align}
The expectation value $\sigma_\textrm{EoM}$ is an order parameter for
chiral symmetry breaking similar to the magnetisation in the Ising
model. It is closely related to the chiral condensate of the light
quarks, $\langle \bar u u +\bar d d\rangle/2$, which is a function of
the constituent quark mass of the light flavours,
\begin{align}\label{eq:mqlight}
 \bar m_l= \frac12\, \bar h \bar \sigma_\textrm{EoM}\,.
\end{align}
The potential \eq{eq:fullThermodyn} also gives us access to the
thermodynamics of the system which will be discussed elsewhere.

%%%%%%%%%%%%%%%%%%%%%%%%%%%%%%%%%%%%%%%%%%%%%%%%%%%%%%%%%%%%%
%%%%%%%%%%%%%%%%%%%%%%%%%%%%%%%%%%%%%%%%%%%%%%%%%%%%%%%%%%%%%

\section{Correlation functions}
\label{sec:correlation}

In this section we discuss the correlation functions presented in
\eq{eq:action}, and derive their flow equations. This includes the
propagators of all fields and the respective anomalous dimensions
(\sec{sec:propa}), the strong couplings related to pure glue,
ghost-gluon and quark-gluon vertices (Sections
\ref{sec:gluoniccouplings}, \ref{sec:QuarkGluon}), the four-quark
scattering and the Yukawa coupling between pions, $\sigma$ and the
quarks due to dynamical hadronisation (\sec{sec:Yukawa}), and the flow
of the effective potential (\sec{sec:EffPot}).

\subsection{Propagators and anomalous dimensions}
\label{sec:propa}

\begin{figure}[t]
\includegraphics[width=0.98\columnwidth]{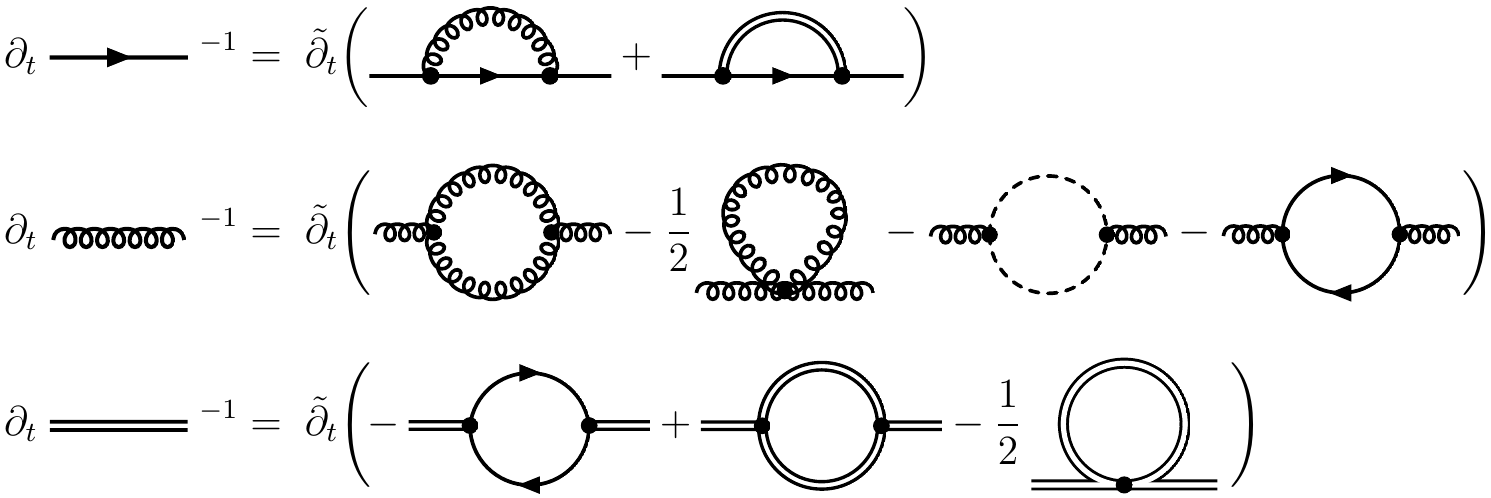}
\caption{Flow equations for the inverse propagators of quarks, gluons,
  mesons respectively, in the present approximation; in particular the
  tadpole contribution from two-quark--two-meson as well as the
  remnant four-quark scatterings are missing, as is the ghost-gluon
  tadpole. The derivative $\tilde \partial_t$ is defined in the
  caption of \Fig{fig:flow4q}. The left hand sides stand for, in a
  slight abuse of notation,
  $\partial_t \Gamma_k^{(2)}= \partial_t \left( G_k^{-1} -R_k\right)$,
  where we have dropped the $-R_k$ in the
  depiction.}\label{fig:propas}
\end{figure}

The flow equation for the (inverse) propagators,
$\partial_t \Gamma^{(2)}_k(p)$ is obtained by taking the second
derivative w.r.t.\ the fields of the Wetterich equation
\eq{eq:FlowQCD}.  They are depicted in \Fig{fig:propas} for the quark,
gluon, and meson fields, respectively. We are led to the anomalous
dimensions $\eta_{\Phi_i}$ with $\Phi_i=(A,q,\bar q,\phi)$. While the
$\eta_A$ comprises the full flow of the inverse gluon propagator, the
mass terms of quarks and mesons are field dependent. They are captured
by the Yukawa term and the second derivative of the effective
potential respectively.

\subsubsection{Mesons and quarks}\label{sec:MatterProps}

The full quark and meson two-point functions (at vanishing pion fields
$\bm\pi=0$ and vanishing quark chemical potential $\mu_q=0$) are given
by
\begin{align}\nonumber 
  \Gamma^{(2)}_{q\bar q}(p) = &\,Z_{q}(p)\,\left[ i \pslash +
                                M_q(p)\right]\,, \\[1ex]
  \Gamma^{(2)}_{\phi\phi}(p) =&\, Z_{\phi}(p)\,p^2 + m_\phi^2\,, 
  \label{eq:G2qphi}
\end{align}
where both the wave function renormalisations and the mass terms also
depend on the chosen mesonic field $\phi$. The notation in
\eq{eq:G2qphi} is that in \cite{Mitter:2014wpa,Cyrol:2017ewj}, where
the full momentum dependence of the meson and quark two-point function
of two-flavour QCD in the vacuum has been computed. The running quark
mass parameter in the present work is given by
$\bar m_{l,k} = M_{l,k}(0)$, and this relation holds on a quantitative
level, see the Figures~\ref{fig:Mq-ml2} and \ref{fig:Mq-ml2+1} in
\app{app:Scales}. Evidently, the fully momentum-dependent wave
function renormalisations $Z_q(p)$ and mass functions $M_q(p)$ of the
quarks are uniquely defined as the prefactor of the Dirac tensor
structure and the scalar tensor structure respectively,
\begin{align}\nonumber 
  Z_{q_i}(p) =&\,  \frac{1}{4 N_c } \frac{1}{p^2 }\tr\pslash \,
               \Gamma^{(2)}_{q_i\bar q_i}(p)  \,,\\[1ex]
  M_{q_i}(p) = &\, \frac{1}{4 N_c} \frac{1}{Z_{q_i}(p)}
                 \tr\Gamma^{(2)}_{q_i\bar q_i}(p)  \,,
  \label{eq:Z_qM_q}
\end{align}
where $\tr$ stands for the Dirac and colour trace, and we do not sum
over the flavour index $i=1,...,3$. The different tensor structures
for the kinetic term and the mass term provide us also with unique
projections of the flow equation on the flows $\partial_t Z_q(p)$ and
$\partial_t M_q(p)$ respectively. The latter flow is related to the
flow of the Yukawa coupling and is discussed in
Chapter~\ref{sec:Yukawa}. The flow of the wave function
renormalisation is encoded in the anomalous dimension \eq{eq:eta} of
the quarks. In the present work we do not introduce a thermal
splitting of the wave function renormalisation in parts longitudinal
and transversal to the heat bath, but use the transversal part
throughout. This leads us to
\begin{align}
  \eta_{q_i}(p_0,\bm{p})=\frac{1}{4 N_c}\frac{1}{
  Z_{q,k}(p_0,\bm{p})}\mathrm{Re}
  \left[\frac{i}{\bm{p}^2}\mathrm{tr}
  \, \bm{\gamma}\cdot\bm{p}\,\partial_t
  \Gamma^{(2)}_{\bar q_i q_i,k}(p)\right]\,,
  \label{eq:etapsiGen}
\end{align}
%

%
%%%%%%%%%%%%%%%%%%%%%%%%%%%%%
%%%%%%%%%%%%%%%%%%%%%%%%%%%%%
%

\noindent
where no sum over the flavours $i$ is taken.  It is well-known that
the anomalous dimension of the quark carries a mild momentum
dependence, for results in the present fRG setting as well as
respective DSE and lattice results see
\cite{Mitter:2014wpa,Cyrol:2017ewj,Williams:2015cvx,Fischer:2018sdj}. In
particular, from \cite{Mitter:2014wpa,Cyrol:2017ewj} we infer a very
small momentum dependence of $\eta_q$ in the regime $p^2\leq
k^2$. Hence, we use $\bm p=0$ which facilitates the explicit
expressions,
\begin{align}
  \eta_{q_i}(p_0)=\frac{1}{4\, Z_{q_i,k}(p_0)}
          \mathrm{Re}\left[i\frac{\partial}{\partial \bm{p}^2}\mathrm{tr}
            \, \bm{\gamma}\cdot\bm{p} \,\partial_t
  \Gamma^{(2)}_{\bar q_i q_i,k}(p)\right]_{\bm{p}=0}\,,    \label{eq:etapsi}
\end{align}
where $p_0$ is a small, but nonvanishing, frequency, for more details
see \app{app:etaq}. Its nontrivial choice is also related to fact
that the expression in the square bracket is complex for nonvanishing
chemical potential. This is well-understood and originates in the
Silver-Blaze property: At vanishing temperature correlation functions
involving a quark field $q(\omega)$ below density onset are real
functions of the complex frequency variable $\omega- i\,\mu_q$, for a
detailed discussion in the fRG and 2PI approaches see
\cite{Marko:2014hea,Khan:2015puu,Fu:2015amv}. Therefore we project the
flow on its real part, denoted by $\mathrm{Re}$. The trace in
\eq{eq:etapsi} projects on the Dirac tensor structure in the quark
two-point function, that is its kinetic part. The explicit expression
of $\eta_{q,k}$ is also given in \app{app:etaq}.

It follows from the parameterisation \eq{eq:G2qphi} that the mesonic
wave function renormalisations and masses are given by
\begin{align}\nonumber 
  Z_\phi(p) =&\, \frac{1}{p^2}\left[ \Gamma_{\phi\phi}^{(2)}(p^2) -
                 \Gamma_{\phi\phi}^{(2)}(0 ) \right]\,,\\[1ex]
  m_\phi^2 = &\,\Gamma_{\phi\phi}^{(2)}(0)\,.
\label{eq:ZphiMphi}
\end{align}
In contradistinction to the quark wave function renormalisations and
masses, that of the mesonic degrees of freedom do not allow for unique
definitions as they cannot be distinguished by their tensor
structures. Accordingly, one can easily shift momentum dependence from
the wave function renormalisation $Z_\phi$ to the mass $m_\phi^2$ in
\eq{eq:G2qphi}. Furthermore, for given
$\phi_0 = (\sigma_0 ,\bm \pi=0)$, the $\sigma$ and $\bm \pi$ parts of
$\Gamma^{(2)}_{\phi\phi}$ are different,
\begin{align}\nonumber 
  \Gamma^{(2)}_{\sigma\sigma}(p) =&\, Z_{\sigma,k}(p)\,p^2 + m_\sigma^2
                                    \,, \\[1ex]
  \Gamma^{(2)}_{\pi_i\pi_i}(p) =&\, Z_{\pi,k}(p)\,p^2 + m_\pi^2\,. 
                                 \label{eq:sigmapi2}
\end{align}
The $\sigma$ and pion masses are given by the curvature of the mesonic
potential $V_k$ in the respective field directions,
\begin{align}
  m_\pi^2 =  \partial_\rho V_k \,, &\quad & m_\sigma^2 =  m_\pi^2 +
                                         2 \rho\,\partial_\rho^2 V_k\,,
\label{eq:MesonicM}
\end{align}
and are hence called curvature masses, see e.g.\
\cite{Helmboldt:2014iya}. Evidently, the pion mass vanishes in the
chiral limit as it is simply given by the EoM for $c_\sigma = 0$. The
two masses differ by a higher $\rho$-derivative term which is only
present for $\rho_0\neq 0$, that is $\sigma_0\neq 0$ as
  $\bm\pi\equiv 0$. The wave function renormalisations follow
similarly,
\begin{align}
  Z_{\pi}(p) =Z_\phi(p) \,,& \quad& Z_{\sigma}(p) =  Z_\phi(p)
                                          +\rho\partial_\rho Z_\phi(p)\,.
\label{eq:MesonicZ}\end{align}
In the present approximation we do not consider the dependence of the
wave function renormalisations on the mesonic fields. These terms
amount to momentum-dependent mesonic self-scatterings that vanish at
vanishing momentum. These terms are suppressed at low energies which
is the only regime where the offshell mesonic fluctuations can play a
r$\hat{\textrm{o}}$le in the first place. This leaves us with a
uniform wave function renormalisation which we define via the pion
two-point function,
\begin{align}\label{eq:ZphiZpi}
  Z_\phi(p) = Z_\pi(p)\,.
\end{align}
A natural definition of $Z_\pi$ comes from the definition of the pion
pole mass
\begin{align}
  \bar\Gamma^{(2)}_{\pi_i\pi_i}(p_0 = i m_{\pi,\textrm{pol}},\,
  \bm{p}^2 = 0) = 0\,,
\end{align}
with
$\bar \Gamma^{(2)}_{\phi\phi}(p_0,\bm{p})=
\Gamma_{\phi\phi}^{(2)}(p_0,\bm{p})/Z_\phi(p_0,\bm{p})$, which implies
\begin{align}\label{eq:fullPoleMass}
  m_{\pi,\textrm{pol}}^2 = \frac{1}{Z_\pi(i m_{\pi,\textrm{pol}},0)} m_\pi^2\,,
\end{align}
see e.g.\ \cite{Helmboldt:2014iya}. In the present Euclidean setup we
are restricted to $p_0^2 \geq 0$. The approximation
$Z_\pi(i m_{\pi,\textrm{pol}},0) \approx Z_\pi(0,0) \equiv Z_\pi(0)$
is therefore the optimal choice for the wave function renormalisation
in \eq{eq:fullPoleMass}, as it is closest to the pion pole. Since this
pole, in turn, is close to the origin in the region where mesons are
dynamically relevant, this approximation is even quantitative if
$Z_\pi(p)$ is only mildly momentum dependent in the small region
$|p^2| \lesssim m_\pi^2$. Indeed, this has been demonstrated to be
true for cutoffs $\lesssim 1 $\,GeV within low energy effective
theories in \cite{Helmboldt:2014iya}. Hence, we arrive at
\begin{align}\label{eq:PoleMass} 
  m_{\pi,\textrm{pol}}^2 = \frac{1}{Z_\pi(0)} m_\pi^2\,,
\end{align}
which also implies that the pole mass and the (renormalised) curvature
mass of the pion agree. Therefore we consider explicitly the pion wave
function renormalisation, and with \eq{eq:ZphiZpi} we arrive at
\begin{align}\label{eq:Zphi}
  Z_\phi(p) = \frac{1}{3} \delta_{i j} \frac{1}{p^2}\left[
  \Gamma^{(2)}_{\pi_i\pi_j}(p)
  - \Gamma^{(2)}_{\pi_i\pi_j}(0)\right]\,. 
\end{align}
For momenta close to the pion pole this agrees well with the pole
renormalisation. In the present work we consider $p^2\geq 0$ and hence
the optimal choice is $\bm p^2 =0$ and $p_0 =0$, where the latter
limit is taken first.  This leads us to
\begin{align}\label{eq:NexttoPole}
  Z_\phi(0) = \frac{1}{3}\delta_{i j} \left[
  \frac{\partial\Gamma^{(2)}_{\pi_i\pi_j}}{\partial \bm{p}^2}\right](p=0)
  \,,
\end{align}
with the pion pole mass \eq{eq:PoleMass}. As for the quarks we have
neglected the thermal splitting of the wave function renormalisation
in parts longitudinal and transversal to the heat bath.  Instead we
use the transversal part throughout. The validity of this
approximation has been checked explicitly in e.g.\ \cite{Yin:2019ebz}.

In the present work we also approximate the full momentum dependence
of the quark and meson propagators in the diagrams by
cutoff-scale--dependent wave function renormalisations and masses. The
cutoff scale dependence carries the -averaged- momentum dependence at
$p^2 \approx k^2 $. Accordingly, this approximation provides even
semiquantitative results in the flow if the momentum dependence of
the propagators is small for momenta $p^2 \lesssim k^2$. Moreover, the
spatial momentum $\bm q$ loop integrals of flows of correlation
functions at vanishing external spatial momenta $\bm p_i=0$ are peaked
at $\bm q^2 \approx k^2$ for generic cutoffs, and $\bm q^2 = k^2$ for
the present cutoff choice. Both properties originate in the spatial
momentum measure
\begin{align}\label{eq:SpatialLoop}
  \int \frac{d^3 q}{(2 \pi)^3} \,{\cal I}_k(\bm{q}^2)=
  \frac{1}{2 \pi^2}\int_0^\infty d q\, q^2\, 
  {\cal I}_k(q^2)\,, 
\end{align}
with ${\cal I}_k({q}^2>k^2) =0$ for the present cutoff choice
\eq{eq:Regs}. We conclude that if approximating the full $Z_\phi(p)$
with \eq{eq:Zphi} at $p=k$, the flow diagrams with meson propagators
are well-approximated.  This leads us to the following approximation
of the full momentum dependence of the meson two point function in the
right hand side of the flow equations,
\begin{align}\nonumber 
  \Gamma^{(2)}_{\phi\phi} (p) =&\, \bar Z_{\phi,k} \,p^2 +
                                 m_\phi^2 \,,\\[1ex]
  \partial_t
  \bar Z_{\phi,k} =&\,\frac{\partial_t \Gamma^{(2)}_{\phi\phi} (0,k)-
                     \partial_t \Gamma^{(2)}_{\phi\phi} (0) }{k^2}\,,
\label{eq:barZphi}\end{align}
where $(0,k)$ stands for $(p_0=0,\bm{p}^2=k^2)$. In summary, best use
of these consideration is made if $\bar Z_\phi$ with \eq{eq:barZphi}
is used in the flow diagrams, while $Z_\phi(0)$ defined in
\eq{eq:NexttoPole} is used for determining the current quark masses
via the pole mass of the pion in the vacuum.

It is left to derive the flow equations for the wave function
renormalisations $\bar Z_\phi$ and $Z_\phi(0)$. Both can be read of
from the anomalous dimension \eq{eq:eta} computed from the
$k$-derivative of \eq{eq:Zphi}. We use that 
\begin{align}
  \eta_{\phi}(0,\bm{p})=
                           -\frac{\delta_{ij}}{3Z_{\phi}(0,\bm{p})}
                            \frac{\partial_t
                            \Gamma_{\pi_i\pi_j}^{(2)}(0,\bm{p})
  -\partial_t \Gamma_{\pi_i\pi_j}^{(2)}(0,\bm{p})}{\bm{p}^2}
  \,,
                           \label{eq:etaphipk}
\end{align}
lead to the flow of $\bar Z_\phi$ with
\begin{align}\label{eq:dtbarZphi}
  \frac{1}{\bar Z_\phi} \partial_t \bar Z_\phi:= - \eta_\phi(0,k)\,,
  \quad \textrm{with}\quad
  \bar Z_{\phi,\Lambda}=Z_{\phi,\Lambda}\equiv 1\,.
\end{align}
Note that \eq{eq:etaphipk} and \eq{eq:dtbarZphi} imply that
$\bar Z_\phi\neq Z_\phi(0,k)$. This can be seen from the
$t$-derivative of $Z_\phi(0,k)$, 
\begin{align}
  \partial_t \left[\log Z_\phi(0,k) \right]= - \eta_{\phi}(0,k) 
                           +\frac{k}{Z_{\phi}(0,k)} 
                            \frac{\partial 
                            Z_\phi}{\partial |\bm{p}|}(0,k)\,, 
                           \label{eq:etaphipkfull}
\end{align}
where the second term comes from the spatial momentum $|\bm{p}| =k$ in
$Z_\phi(0,k)$.  We emphasise again that the present approximation is
based on the mild momentum dependence of the mesonic wave function
renormalisation for $p^2 \lesssim k^2$ of both low energy effective
theories as well as full QCD. In the vacuum this has been shown in
\cite{Helmboldt:2014iya, Mitter:2014wpa, Cyrol:2017ewj}.  Note,
however, that at finite density a nontrivial momentum dependence of
the meson wave function renormalisation can be induced by a modulated
spatial structure, e.g.\ due to an inhomogeneous quark condensate. As
discussed below, we find indications for such a regime here.

For the determination of the pion mass in the vacuum we also require
the wave function renormalisation at vanishing momentum,
$Z_\phi(0)$. Its flow is read-off from \eq{eq:NexttoPole}, 
\begin{align}
  \eta_{\phi}(0)&=
                  -\frac{1}{3Z_{\phi}(0)}\delta_{ij}\left[
                  \frac{\partial}{\partial  \bm{p}^2}\partial_t
                  \Gamma^{(2)}_{\pi_i\pi_j}\right](p=0)\,.
                           \label{eq:etaphi}
\end{align}
The explicit expression for \eq{eq:etaphipk} and \eq{eq:etaphi} are
deferred to \app{app:etaphi}.  Note that $Z_\phi(0)$ and $Z_\phi(0,k)$
agree at $k=0$, while $\bar Z_\phi$ does not. This deviation can be
used for a systematic error estimate, since the wave function
renormalisations only enter via $\eta_\phi$. Accordingly, it is the
difference $\eta_\phi(0,k)+\partial_t \ln Z_\phi(0,k)$, that is the
last term on the right hand side of \eq{eq:etaphipkfull}, which is
missing in the flows. We have checked numerically that this difference
does not play any r$\hat{\textrm{o}}$le.

In the current approximation, the pion pole mass in the vacuum is
given by
\begin{align}
  m_{\pi,\text{pol}}=\frac{m_\pi}{\sqrt{Z_{\phi,k=0}(0)}}\,.
  \label{eq:mpionp0}
\end{align}
Similarly we approximate
\begin{align}
  m_{\sigma,\text{pol}}\approx
  \frac{m_\sigma}{\sqrt{Z_{\phi,k=0}(0)}}\,.
  \label{eq:msigmap0}
\end{align}
However, we emphasise that \eq{eq:msigmap0} carries a qualitatively
larger systematic error in comparison to \eq{eq:mpionp0}: Apart from
the fact that the physical meaning of the $\sigma$-resonance in the
scalar four-quark channel is still debated, \cite{Jaffe:1976ig,
  Jaffe:1976ih}, it requires a small momentum behaviour for a larger
momentum range $|p^2| \lesssim m_{\sigma,\text{pol}}^2$. Nonetheless,
in an extension of the current setup to Minkowski frequencies the
respective observable provides the position of the lowest resonance in
the scalar channel.

Within this setup we adjust the pion mass in the vacuum,
$m_{\pi,\textrm{pol}}=137$ MeV, by fixing $c_\sigma$. For the
systematic error estimate we also have run the vacuum flows with
$\eta_{\phi}(0)$ instead of $\eta_{\phi}(0,k)$.  Starting from the
same initial conditions we get a pion pole mass of 140 MeV, which
gives us a $\sim 2\%$ error. Moreover, the purely fermionic
observables, e.g.\ for $N_f=2+1$ the constituent light quark mass
$\bar m_l \!=\! 347$\,MeV ($\eta_{\phi}(0,k)$), 343\,MeV
($\eta_{\phi}(0)$) and the four fermion coupling
$\bar h_k^2/(2\bar m_{\pi,k}^2) \!=\! 2.73 \times 10^3\,
\text{GeV}^{-2}$ ($\eta_{\phi}(0,k)$),
$2.80 \times 10^3\, \text{GeV}^{-2}$ ($\eta_{\phi}(0)$) also give us a
systematic error estimate of $\lesssim 2$\,\%. Technically, this
stability of the fermionic observables comes about as $\eta_{\phi}$
plays a subleading r$\hat{\textrm{o}}$le for purely fermionic
observables. In summary both the actual small changes as well as the
formal argument for the fermionic observables support the present
approximation.

At finite chemical potential the situation changes significantly: The
mesonic dispersion develops a minimum for nonvanishing momentum.
This entails that $Z_\phi(0,k)$ and $Z_\phi(0)$ may and do differ
qualitatively. Moreover, such a behaviour indicates an inhomogeneous
regime, for more details see \sec{sec:Inhom} and
\fig{fig:PhasediagramInhom}. Still, we have checked that this does not
affect the purely fermionic observables.

\subsubsection{Gluons and ghosts}\label{sec:GlueSector}

It is left to specify the gluon and ghost anomalous dimensions. The
ghost propagator has been shown to be rather insensitive to quark
contributions for $N_f=2$ and $N_f=2+1$ flavours as well as finite
temperatures $T\lesssim 1$\,GeV, see e.g.\ \cite{Cyrol:2017qkl} and
references therein.  For this reason we use the ghost anomalous
dimension at vanishing temperature from \cite{Cyrol:2017ewj},
\begin{align}
  \eta_{c}=&-\left. \frac{\partial_t
             Z^{\text{QCD}}_{c, k=0}(p)}{
             Z^{\text{QCD}}_{c, k=0}(p)}\right|_{p=k}\,.\label{eq:etaCT}
\end{align}
Note that \eq{eq:etaCT} only enters explicitly the ghost triangle in
the vacuum flow of the three-gluon vertex depicted in
\fig{fig:3pointsAll}, see \sec{sec:gluoniccouplings}. Implicitly,
$\eta_c$ is also present via the input from
\cite{Cyrol:2017ewj,Cyrol:2017qkl} and in the temperature and density
fluctuations of purely gluonic two- and three-point functions.

For the gluon anomalous dimension we decompose the flow for the
inverse gluon propagator into the pure glue diagrams and the quark
loop, see \Fig{fig:propas}. As discussed in \sec{sec:ExpVac+FinT}, we
utilise quantitative results for vacuum two-flavour QCD,
\cite{Cyrol:2017ewj}, and finite temperature Yang-Mills theory,
\cite{Cyrol:2017qkl}. Then, the gluon anomalous dimension $\eta_{A,k}$
at finite temperature and density is decomposed into a sum of three
parts,
\begin{align}
  \eta_{A}=&\eta_{A,\textrm{vac}}^{\text{QCD}}+
             \Delta\eta_{A}^{\text{glue}}+
             \Delta\eta_{A}^{q}\,.\label{eq:etaAT}
\end{align}
The first term on the r.h.s.\ of the equation above accounts for the
vacuum contribution to the gluon anomalous dimension. In turn,
$\Delta\eta_A^A$ and $\Delta \eta^q_A$ account for the medium
contributions to the gluon anomalous dimension, from gluon loops and
quark loops respectively.

For $N_f=2$ we infer it directly from the corresponding gluon dressing
function $Z^\textrm{QCD}_{A,k=0}(p)$ from \cite{Cyrol:2017ewj} with
\begin{align}
  \left.\eta_{A,\textrm{vac}}^{\text{QCD}}\right|_{N_f=2}=
  &-\left. \frac{\partial_t
    Z^{\text{QCD}}_{A, k=0}(p)}{
    Z^{\text{QCD}}_{A, k=0}(p)}\right|_{p=k}\,.
                                                   \label{eq:etaAQCD2}
\end{align}
Full RG-invariance of the procedure is then obtained by rewriting the
anomalous dimension as a function of the running coupling
$\alpha_{s,k}$ as done in \cite{Braun:2014ata,Rennecke:2015}. This is
described further in \app{app:YMtoQCD}. In the present work we simply
adjust the input coupling $\alpha_{s,\Lambda}$ consistent with its
implicit value given in $\eta_A$: We choose $\alpha_{s,\Lambda}$ such
that the quark-gluon and purely gluonic couplings show the same
ultraviolet running. Their running is proportional to $1/2 \eta_A$ and
$3/2 \eta_A$ respectively, the rest of the $\beta$-functions is given
by diagrams proportional to $\alpha_{s,k}$. Only for the consistent
initial coupling both runnings can agree as they should. For more
details see \sec{sec:RGconsistency}.

We also emphasise that instead of the fRG data from
\cite{Cyrol:2017ewj} we also could have taken lattice data or other
data from other functional approaches such as the DSE. As mentioned
before, it is an important feature of the current setup that results
obtained within other approaches can be systematically included. This
allows for systematic improvements and hence enhances the reliability
of the current approach.

For $N_f=2+1$ we include the contribution of the $s$-quark via its
flow. Then we use
\begin{align}
  \eta_{A,\textrm{vac}}^{\text{QCD}}=
  &\left.
    \eta_{A,\textrm{vac}}^{\text{QCD}}\right|_{N_f=2}+
    \eta^s_{A,\textrm{vac}}\,,\label{eq:etaAQCD2+1}
\end{align}
with
\begin{align}
  \eta^s_{A,\textrm{vac}}=\, -\frac{1}{2 (N_c^2-1)}
  \,\partial_{p^2}\left[\Pi_{\mu\nu}^{\perp}\overline{{
  \textrm{Flow}}}^{(2)}_{AA}{\,}^{aa}_{\mu\nu}(p)
  \right]^{(s)}_{p=0}\,.\label{eq:etas}
\end{align}
The transversal projection operator $\Pi^\perp$ in \eq{eq:etas} is
defined in \eq{eq:Piperp}. The superscript $^{(s)}$ denotes the
strange quark contribution. As in the two-flavour case the initial
coupling $\alpha_{s,\Lambda}$ is adjusted by RG-consistency. For more
details see \sec{sec:RGconsistency}.

Now we proceed to the gluon anomalous dimension at finite temperature
and density. The difference to the vacuum anomalous dimension is
comprised in the second and third term on the right hand side of
\eq{eq:etaAT}.  Here, $\Delta\eta_{A,T}^{q}$ in \eq{eq:etaAT} denotes
the contribution of the quark loop at finite temperature and quark
chemical potential $\mu_q$. With \eq{eq:barFlown} it reads
\begin{align}
  \Delta\eta_{A}^{q}=
  &\, -\frac{1}{2 (N_c^2-1)}
    \frac{\Pi_{\mu\nu}^{\text{M}}(p)}{p^2}
    \,\Bigg(\left[\overline{{
    \textrm{Flow}}}^{(2)}_{AA}{\,}^{aa}_{\mu\nu}(p)\right]^{(q)}_{T,\mu} 
    \nonumber \\[1ex]
  &\,\hspace{1cm}-\left[\overline{{
    \textrm{Flow}}}^{(2)}_{AA}{\,}^{aa}_{\mu\nu}(p)\right]^{(q)}_{T,\mu=0} 
    \Bigg)\Bigg|_{\substack{p_0=0\\|\bm{p}|=k}}\,,\label{eq:etaAQL}
\end{align}
where the superscript $^{(q)}$ indicates the contribution of the
different quark flavours.  The vacuum contribution is subtracted in
\eq{eq:etaAQL}, and we have used the transverse magnetic tensor in the
projection,
\begin{align}
  \Pi_{\mu\nu}^{\text{M}}(p)=
  &(1-\delta_{\mu 0})(1-\delta_{\nu 0})
    \Big(\delta_{\mu\nu}-\frac{p_{\mu}p_{\nu}}{
    \bm{p}^2}\Big)\,.\label{eq:magproj}
\end{align}
Moreover, the contraction of Lorentz and group indices is implicitly
understood in \eq{eq:etaAQL}. The explicit expression for
$\Delta\eta_{A}^{q}$ can be found in \eq{eq:DeltaAqexpl}.

The thermal contribution of the glue sector is encoded in the second
term on the r.h.s.\ of \eq{eq:etaAT}, i.e.,
$\Delta\eta_{A}^{\text{glue}}$. It has been discussed in
\cite{Cyrol:2017qkl} that it is well-captured by a thermal screening
mass, its inclusion is described in detail in \app{app:gluon}. Note
that the present setup does not take into account the full
backcoupling of the thermal and density fluctuations of the quark in
the pure glue diagrams. A respective error estimate is discussed in
\app{app:YMtoQCD}. There it is shown that this approximation, the
`direct sum' of the contributions, already works semiquantitatively
for the much bigger backcoupling effects related to adding the quarks
to pure glue in the vacuum. In the present work we take the vacuum
backcoupling fully into account, and only apply the direct sum
approximation to the much smaller thermal and density
contributions. This suggests a quantitative nature of the current
setup.

%%%%%%%%%%%%%%%%%%%%%%%%%%%%%%%%%%%%%%%%%%%%%%%%%%%%%%%%%%%%%

\subsection{`Avatars' of the strong coupling $\alpha_s$}
\label{sec:vertices}

In the present truncation we consider different `avatars' of the
strong coupling, introduced in \sec{sec:StrCoup}, as well as the
Yukawa coupling between quarks and the scalar-pseudoscalar mesons,
introduced via dynamical hadronisation in \sec{sec:DynHad}.

\subsubsection{Quark-gluon couplings}\label{sec:QuarkGluon}

\begin{figure}[t]
\includegraphics[width=0.97\columnwidth]{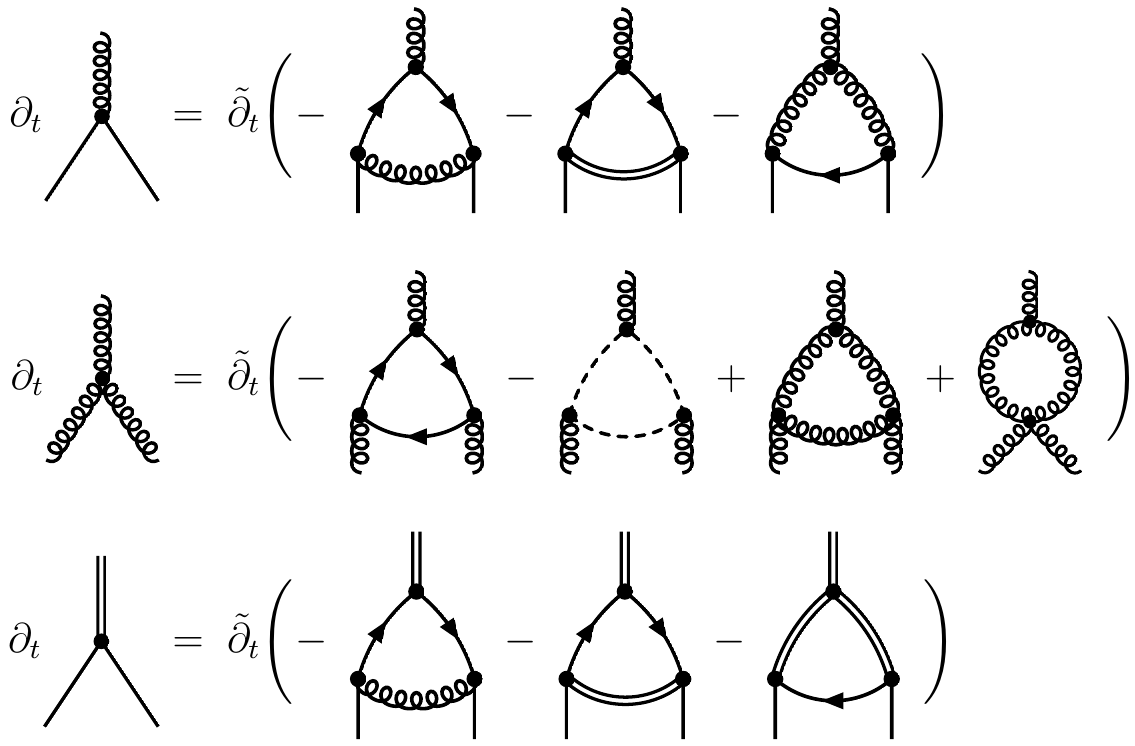}
\caption{Flow equations for the three-point functions computed here:
  quark-gluon, three-gluon and Yukawa couplings.  In the present
  approximation, the diagrams with higher order vertices and the
  remnant four quark vertices are dropped. $\tilde\partial_t$ is
  defined in the caption of \Fig{fig:flow4q}.}\label{fig:3pointsAll}
\end{figure}

The strong couplings and their running with the RG scale, especially
the quark-gluon couplings
$g_{\bar{q}Aq}=(\alpha_{\bar l A l},\alpha_{\bar s A s})$, play a
significant r$\hat{\textrm{o}}$le in dynamical chiral symmetry
breaking, see e.g., \cite{Gies:2002hq, Braun:2006jd, Braun:2014ata,
  Mitter:2014wpa} for more detailed discussions. As we discuss in
\ref{sec:StrCoup}, different strong couplings are identical in the
perturbative regime due to the mSTIs, however, they deviate from each
other in the nonperturbative or even semiperturbative regime
\cite{Cyrol:2017ewj}. In \Fig{fig:3pointsAll} we show the flow
equations for the quark-gluon and three-gluon couplings. The first and
third diagrams on the r.h.s.\ of the equation in the first line are
the usual QCD contributions, while the second one arises from the
mesonic fluctuations. Thus the relevant flow equation for the
quark-gluon coupling is given by the projection of the flow of
$\Gamma^{(3)}_{A q\bar q }$ on the classical tensor structure
$T^{{(1)}}_{A q\bar q }=S^{(3)}_{A q\bar q }/g$. This leads us to
\begin{align}
  \partial_t g_{\bar{q} A  q}=
  &\left( \frac12\eta_A+\eta_q\right) g_{\bar{q}Aq}
    \nonumber \\[1ex]
  &-\frac{1}{8(N_c^2-1)}\tr\,\left[\left[
    \overline{{\textrm{Flow}}}^{(3)}_{A q\bar{q} }\right]^{a}_{\mu} 
    \left[ T^{{(1)}}_{A q\bar q }\right]^a_{\mu}\right]\big(\{p\})\,,
 \label{eq:dtg}
\end{align}
where the trace sums over Dirac indices and the fundamental
representation of gauge group. The classical tensor structure is given
by
\begin{align}\label{eq:SqbarqA}
  \left[T^{{(1)}}_{A q\bar q }\right]^a_{\mu}
  = \left[\frac{1}{g}S^{(3)}_{A q\bar q }
  \right]^a_{\mu} = -i \gamma_\mu t^a\,,
\end{align}
where the coupling has been divided out. The contracted flow in
\eq{eq:dtg} is normalised with the trace of the classical tensor
structure squared, 
\begin{align}
  \tr \,\left[\left[ T^{(1)}_{A q\bar{q} }\right]^{a}_{\mu} 
  \left[ T^{(1)}_{A q\bar q }\right]^a_{\mu}\right]
  = - 8 (N_c^2 -1)\,.
\end{align}
In \eq{eq:dtg}, $\{p\}$ denotes the set of external momenta for the
three-point vertex. The momentum for the gluon is chosen to be
vanishing, and that for the quark as illustrated in \app{app:etaq}. We
divide the term in the second line of \eq{eq:dtg} into two parts,
i.e., $\partial_t g_{\bar{q}Aq} \big |_A$ and
$\partial_t g_{\bar{q}Aq}\big |_{\phi}$, which correspond to the
relevant contributions from the quark-gluon and quark-meson couplings,
respectively. Their expressions are given in \eq{eq:dtgA} and
\eq{eq:dtgphi}.

It is left to discuss the effect of the current approximation with
only the classical tensor structure. In quantitatively reliable
approximations to QCD the initial conditions at a perturbative initial
scale $\Lambda\gtrsim 10$\,GeV, as described in \app{app:initial},
lead to QCD physics at vanishing cutoff scale $k=0$. In particular,
this holds for the physics of spontaneous chiral symmetry breaking,
see \cite{Mitter:2014wpa,Cyrol:2017ewj} for quenched and unquenched
$N_f=2$ flavour results respectively. From these computations we also
know that not only the classical tensor structure of the quark-gluon
vertex carries the fluctuations important for chiral symmetry
breaking.  In turn this implies, that an approximation with only the
classical vector tensor structure may lack interaction strength. This
typically calls for a phenomenological infrared enhancement of the
quark-gluon coupling $\alpha_{\bar q A q}$ that compensates for the
missing tensor structures. This is common to all functional approaches
within approximations that lack full quark gluon vertices, for related
discussions in DSEs see \sec{sec:SysFun} and and e.g.\
\cite{Qin:2010nq,Fischer:2011mz,Fischer:2012vc,Fischer:2013eca,%
  Fischer:2014ata,Fischer:2014vxa,Williams:2015cvx,Eichmann:2015kfa,%
  Gao:2015kea,Gao:2016qkh}, for a recent review see
\cite{Fischer:2018sdj}.

%
%%%%%%%%%%%%%%%%%%%%%%%%%%%%%
%%%%%%%%%%%%%%%%%%%%%%%%%%%%%
%

In the present work we follow the approach proposed in
\cite{Braun:2014ata}, where the infrared enhancement is directly
introduced in the infrared part of the flow. More details can be found
in \app{app:IR-enhancement}. Here, we simply summarise the results:

In the present setup it turns out that the required infrared
enhancement is rather small, only for cutoff scales smaller than
2\,GeV the flow is enhanced with a factor 1.034, see \eq{eq:IR-fct}
and \eq{eq:IR-enhance} in \app{app:IR-enhancement}.  While being
small, its effect is significant as the constituent quark mass of the
light quarks would be $m_q \!=\! 217$\,MeV without the
enhancement. Thus, there is only a fine line between `too little' (or
no) and `too much' chiral symmetry breaking in QCD. This has already
been observed in \cite{Mitter:2014wpa,Cyrol:2017ewj}, where a
quantitative approximation to two-flavour QCD in the vacuum was
studied. See also \cite{Vujinovic:2018nko} for a related two-flavour
DSE study with $N_c=2$. This peculiar behaviour of first principle QCD
which may also has significant consequences for the interplay of
confinement and chiral symmetry breaking at finite temperature and
density, can be phrased as: \textit{QCD is living on the edge}. In
\cite{Mitter:2014wpa,Cyrol:2017ewj}, in particular all tensor
structures and momentum dependences of the quark-gluon vertex have
been included selfconsistently. There it has been observed that the
nonclassical tensor structures, in particular the chiral symmetry
breaking ones, contribute significantly to the size of dynamical
chiral symmetry breaking.

Still, we interpret the smallness of the enhancement as a sign of
robustness and quantitative reliability of the present setup: At
finite $\mu_B$ the coupling decreases and we are sensitive to the
functional form of how the infrared enhancement is switched of at
larger momentum scales. Note that the latter scale also includes
temperature and chemical potential. Accordingly, the smaller the
infrared enhancement is the less influence this can have. In turn, in
the presense of a significant infrared enhancement this calls for a
systematic study of its shape.

\subsubsection{Gluonic couplings}
\label{sec:gluoniccouplings}

The gluonic couplings includes the three- and four- gluon couplings
$g_{A^3}$, $g_{A^4}$, and the ghost-gluon coupling $g_{\bar{c} A
  c}$. In the vacuum we adopt the approximations: $g_{A^4}=g_{A^3}$
and $g_{\bar{c}A c }=g_{\bar{l}A l }$, see also
\eq{eq:IDalphaGlue}. This is consistent with the results obtained in
\cite{Cyrol:2017ewj,Braun:2014ata} for cutoff and momentum scales
$k\gtrsim 1$\,GeV. For smaller cutoff scales $k\lesssim 1$\, GeV the
purely gluonic couplings, $g_{A^3}$, $g_{A^4}$, decay rapidly and the
relative difference is subleading. In turn, the ghost-gluon coupling
$g_{\bar{c}A c }$ approaches a sizable value
($\alpha_{\bar{c}A c }(k=0) \approx 3$). However, in the present
setup, the ghost-gluon coupling only contributes explicitly to the
vacuum part of the flow of the three-gluon vertex in the ghost
triangle in \fig{fig:3pointsAll}. This is a subleading diagram except
in the deep infrared for $k\lesssim 100$\, MeV. In this regime it
leads to the zero-crossing of the three-gluon vertex and a small
negative IR value of the coupling. This infrared limit is not
reproduced in the present approximation, but this deep-IR property has
no effect on the present results.

This leaves us with the quark-gluon couplings $\alpha_{\bar q Aq}$
with the flows \eq{eq:dtg} at finite temperature and density, and the
three-gluon coupling $\alpha_{A^3}$. We strive for RG-consistency also
at finite temperature and density and identify the thermal
contributions of the quark-gluon and ghost-gluon couplings. For the
computations we use that of the quark-gluon coupling, \eq{eq:dtg},
which is the most relevant for the chiral dynamics of the theory. This
entails for the medium contribution
$\partial_t \Delta g_{A^3}= \partial_t g_{A^3}-\partial_t
g_{A^3,\textrm{vac}}$
\begin{align}
  \partial_t \Delta g_{A^3}=
  &
    \partial_t \Delta g_{\bar{q}Aq} \,.
 \label{eq:dtDeltagA3}
\end{align}
The vacuum part has been computed in \cite{Braun:2014ata} and we use
the respective flows
\begin{align}\label{eq:dtgA3}
  \nonumber
  \partial_t g_{A^3}
  &=\frac{2}{3} \eta_A\, g_{A^3} +
    \frac{i}{12 N_c (N_c^2-1)}\\[1ex]\nonumber
  &\quad \times \lim_{p\rightarrow 0}
    \bigg(\frac{1}{p^2}\,p_\rho f^{abc} \left[
    \overline{{\textrm{Flow}}}^{(3)}_{A^3} (p,-p)\right]^{abc}_{\mu\nu\rho}
    \delta_{\mu\nu}\bigg)\\[1ex]
  &\quad +\partial_t \Delta g_{A^3}\,. 
\end{align}
The momentum configuration of
$ \overline{{\textrm{Flow}}}^{(3)}_{A^3}(p,-p)$ is chosen such that
the three external gluons carry incoming momenta $p$, $-p$ and 0
respectively. The contraction with $\delta_{\mu\nu}\,p_\rho f^{abc} $
and the normalisation comes from the projection on the classical
tensor structure $T^{{(1)}}_{A^3}$,
\begin{align}\nonumber 
  &\,  \left[T^{{(1)}}_{A^3}\right]^{abc}_{\mu\nu\rho}(p_1,p_2,p_3)
    = \frac{1}{g} \left[S^{(3)}_{A^3}
    \right]^{abc}_{\mu\nu\rho}(p_1,p_2,p_3)\\[1ex] 
  = &   i\, f^{abc} \left[ (p_2 -p_1)_\rho \delta_{\mu\nu}+ (p_1 -p_3)_\nu
      \delta_{\mu\rho}+ (p_3 -p_2)_\mu \delta_{\nu\rho}\right]\,. 
    \label{eq:SA3}
\end{align}
As in \eq{eq:SqbarqA}, the strong coupling has been divided out.  For
the diagrammatic representation of the quark-, gluon-, ghost- triangle
diagrams and the tadpole diagram from the four-gluon coupling, see
\Fig{fig:3pointsAll}.

\subsubsection{RG-consistency of the 'avatars' of the strong
  coupling}\label{sec:RGconsistency}

The flows computed in the present work utilise input from the
quantitative $N_f=2$ flavour vacuum QCD \cite{Cyrol:2017ewj} and
finite temperature YM computations \cite{Cyrol:2017qkl}: Parts of the
gluon and ghost anomalous dimensions $\eta_{A}$ and $\eta_{c}$ are
taken from these works, see \sec{sec:GlueSector}. The input is given
in terms of the running scale $k$ in either $N_f=2$ flavour vacuum QCD
or finite temperature Yang-Mills theory. The current setup differs in
several aspects. First of all, we also consider $N_f=2+1$
flavours. Moreover, the regulators chosen in the present work differ
from that in \cite{Cyrol:2017ewj,Cyrol:2017qkl}. These differences are
fully accounted for if one resolves the anomalous dimension as a
function of $\alpha_{s}$ and further parameters, that is
$\eta_{A,k}=\eta_A(\alpha_{s,k})$. This has been put forward in
\cite{Braun:2014ata,Rennecke:2015}, see also \app{app:YMtoQCD}.

This scale matching entails in particular that the strong coupling at
the initial scale has to be chosen consistently with that implicit in
$\eta_A(\alpha_s)$. In the present work we guarantee the above
RG-consistency in the following sense: In the perturbative regime the
$\beta$-functions of the different `avatars' of the strong coupling,
i.e.\ the three-gluon coupling \eq{eq:alphasGluon} and the quark-gluon
coupling \eq{eq:alphasMatter} have to agree. Their flows are given by
\eq{eq:dtgA3} and \eq{eq:dtg} respectively. In the perturbative regime
we have $g_{A_3} = g_{\bar q A q} = g_s$, which entails schematically
\begin{align}
  \partial_t g_{A^3}-\partial_t g_{\bar q A q}= \eta_A(\alpha_s)\, g_s   +
  \textrm{diagrams}(\alpha_s) \stackrel{!}{=} 0\,, 
\label{eq:RG-consistency}\end{align}
where '$\textrm{diagrams}(\alpha_s)$' stands for the difference of the
diagrams for the three-gluon and quark-gluon vertices depicted in
\Fig{fig:3pointsAll}. The constraint \eq{eq:RG-consistency} is the
requirement of RG-consistency of the couplings. For a given initial
cutoff scale $k=\Lambda$ it defines the consistent initial condition
for the strong coupling. In particular, it adjusts for the relative
RG-scaling in $N_f=2$ and $N_f=2+1$ flavour QCD. The couplings
$\alpha_{\bar q Aq},\alpha_{\bar s As},\alpha_{A^3}$ determined by
this procedure agree well with each other for cutoff scales
$k\gtrsim 5$\, GeV, and hence are RG-consistent, see
\Fig{fig:alphas}. For smaller cutoff scales, $k\lesssim 5$\, GeV the
different couplings start to separate which is also seen in
\cite{Cyrol:2017ewj}. This is expected and necessary, as for smaller
scales we enter the nonperturbative regime which is influenced by the
confining gluon gap. Moreover, the initial coupling
$\alpha_{s,\Lambda}$ derived from \eq{eq:RG-consistency} agrees
quantitatively with that in \cite{Cyrol:2017ewj}. Both properties
constitute nontrivial reliability tests of the current
approach. Moreover, in \app{app:YMtoQCD} it is shown that the
procedure adapted here only leads to small quantitative effects even
if going from Yang-Mills theory to $N_f=2$ flavour QCD, which requires
a far larger scale correction. In summary the combined checks suggest
a small systematic error of the current procedure, the attraction of
which is its simplicity.

\subsection{Yukawa coupling}\label{sec:Yukawa}

In the full effective action without approximation the Yukawa coupling
in \eq{eq:action} is field-dependent, $h=h(\rho)$, and the
$\rho$-dependence takes into account higher scattering processes of
quarks with the scalar-pseudoscalar mesonic channel
\cite{Pawlowski:2014zaa}. This already entails that strictly speaking
we have to deal with two Yukawa couplings at the expansion point
\mbox{$\phi_0=(\sigma_0,\bm\pi=0)$},
\begin{align}\nonumber 
  h_\pi =
  &\, h(\rho_0) =\Gamma^{(3)}_{(\bar q \bm \tau q) \bm\pi}[
    \Phi_0]\,,\\[1ex] 
  h_\sigma =
  &\, h(\rho_0) + \rho\, h'(\rho_0)=
    \Gamma^{(3)}_{(\bar q \tau^0 q) \sigma}[\Phi_0]\,.
\label{eq:2Yuks}\end{align}
Both vertices are derived from $h(\rho)$ and are present in the flow
equations for correlation functions. In the present work we consider
the $\rho$-dependence of the Yukawa coupling as subleading and
identify $ h_\sigma = h_\pi=h$. The corresponding error is minimised
by identifying $h=h_\pi$ as we have three pions and only one radial
mode $\sigma$. Moreover, the pions are lighter except in the scaling
regime in the vicinity of the critical point. Hence already the
diagrams with one pion mode give bigger contributions than those with
the $\sigma$. In the critical region the $\sigma$, as the critical
mode, becomes massless. However, this regime is exceedingly small and
a detailed analysis of its features will be presented elsewhere.

This leaves us with the flow equation of $h=h_\pi$, or rather that of
$\bar h$,
\begin{align}
  \partial_t \bar{h}
  &=\left( \frac12 \eta_{\phi}+\eta_{q}\right)\,\bar{h}-
    \bar m^2_\pi\,\dot{\bar{A}}+
    \overline{\textrm{Flow}}^{(3)}_{  (\bar q \bm\tau q) \bm\pi}\,,
                      \label{eq:hpiflow}
\end{align}
where $\bar m^2_\pi(\bar \rho)= \bar V'(\bar \rho)/k^2$, and the
subscript ${}_{(\bar q \bm\tau q) \bm\pi}$ indicates the projection on
the pseudoscalar part of the Yukawa coupling. The diagrams
contributing to
$\overline{\textrm{Flow}}^{(3)}_{ (\bar q \bm\tau q) \bm\pi}$ are
shown in the third line of \Fig{fig:3pointsAll}. Note that, as
discussed in \sec{sec:DynHad}, the hadronisation function
$\dot{\bar{A}}$ explicitly enters the flow of the Yukawa
coupling. This highlights the fact that dynamical hadronisation, in
contrast to conventional bosonisation, stores the full dynamical
information of the four-quark correlation in the
constituent-quark--meson sector. The ratio $\bar h^2/(2\bar m_\phi^2)$
then is identical to $\bar\lambda_q$ (without dynamical hadronisation)
in the chirally symmetric phase, reflecting the smooth transition from
fundamental to composite degrees of freedom in QCD.

However, instead of using the flow equation for
$\Gamma^{(3)}_{q \bar q \pi_i}[\Phi_0]$, it is simpler to use the fact
that the scalar part of the quark two-point function is proportional
to $ Z_q \sigma_0 h(\rho_0)$. Note that this only holds true within a
dynamical hadronisation of the theory which keeps the formulation
maximally symmetric: The only term that breaks explicitly chiral
symmetry are the linear terms in $\sigma$ and $\sigma_s$.  As shown in
\app{app:DynHad-c} this implies ${\dot C}_k\equiv 0$ in
\eq{eq:HadABC}. Then, the flow of $\bar h$ can be deduced from that of
the quark two-point function, see \fig{fig:propas}, with
\begin{align}
  \partial_t \bar{h}
  &=\left( \frac12 \eta_{\phi}+\eta_{q}\right)\,\bar{h}-
    \bar m^2_\pi\,\dot{\bar{A}}+
    \frac{1}{\bar \sigma}\mathrm{Re} \,
    \overline{\textrm{Flow}}^{(2)}_{\bar q \tau^0 q}
    \,,\label{eq:hflow}
\end{align}
where the subscript ${}_{(\bar q \tau^0 q)}$ indicates the projection
on the scalar part $\bar q \tau^0 q= (\bar u u+\bar d d)/2$ of the
quark two-point function. More details can be found in
\app{app:DynHad-c}.

Both, \eq{eq:hpiflow} and \eq{eq:hflow} provide the flow of $\bar
h$. This is different from the flows of $h_\pi$ and $h_\sigma$, which
differ by a term proportional to $h'(\rho)$. In the present case we
use \eq{eq:hflow} for $\partial_t \bar h$, as it has the simpler
diagrammatic part.

\subsection{Effective potential}\label{sec:EffPot}

The effective potential $V_{k}(\rho,L,\bar L)$ in \eq{eq:action}
receives direct contributions from the quark and meson loop in
\Fig{fig:fleq}. Via the quark loop of the flow equation for the gluon
two-point function the gluon propagator also develops a
$\phi$-dependence. The latter $\phi$-dependence then also propagates
to the ghost loop. However, both dependences are negligible and are
dropped in the following. This leaves us with a flow for the effective
potential which only receives contributions from the quark and meson
loop in \eq{eq:FlowQCD}, see also \fig{fig:fleq}. Owing to the
coupling between the gluon background field $A_0$ and the quarks, the
quark loop also induced a dependence on the Polyakov loop to the
effective potential. The flow equation for $\partial_t V_k$,
\eq{eq:VflowApp}, is discussed in detail in \app{app:flowV}.

Here we only emphasise the important parts of \app{app:flowV}. In the
present work we resort to a Taylor expansion of the effective
potential. In general this can be done about a $k$-dependent expansion
point,
\begin{align}
  \bar V_{\textrm{mat},k}(\bar \rho,L,\bar L)
  &=\sum_{n=0}^{\infty}\frac{\bar
    \lambda_{n,k}}{n!}(\bar \rho-\bar \kappa_k)^n\,,
    \quad\,  \bar\rho = Z_\phi \rho\,,\label{eq:VbarSeries}
\end{align}
where $\bar\rho$ is the renormalised field \eq{eq:RenPhi}.  On the
right hand side of \eq{eq:VbarSeries} we have suppressed the
dependences on $L,\bar L$ of the expansion coefficients:
$\lambda_{n,k}=\lambda_{n,k}(L,\bar L)$ and
$\kappa_k=\kappa_k(L,\bar L)$.  Moreover, in \eq{eq:VbarSeries} we
have implicitly assumed its convergence.  It has been shown in
\cite{Pawlowski:2014zaa} that the most rapid convergence of this
expansion is achieved if the expansion point $\kappa$ is kept fixed,
for applications see also \cite{Braun:2014ata,Fu:2015naa,
  Rennecke:2016tkm}. Then the only $k$-dependence of the expansion
point in \eq{eq:VbarSeries} comes from the mesonic wave function
$Z_\phi$.

In the present work we resort to an expansion about the flowing
solution of the quantum equation of motion,
$\kappa_k = \bar\rho_{\textrm{EoM},k}$ with
\begin{align}
  \frac{\partial}{\partial \bar \rho}\Big(\bar V_{k}(\bar
  \rho,L,\bar L)-\bar c_k
  \bar \sigma \Big)\bigg
  \vert_{\bar\rho=\bar\rho_{\textrm{EoM},k}}
  =0\,.   \label{eq:EoMBarRho}
\end{align}
Using the flowing expansion point \eq{eq:EoMBarRho} facilitates the
access to the expectation value of the $\sigma$-field. The expansion
is described in more detail in \sec{app:flowV}.

The price we pay for this simple access to the EoM of the $\sigma$
field is that this expansion shows bad convergence properties for
large chemical potential. This is but one of the reasons why our
current analysis is limited to $\mu_B/T\lesssim 6$. For general
considerations concerning the stability of this expansion scheme see
\cite{Pawlowski:2014zaa,Grossi:2019urj}. Respective $k$-independent
expansion schemes, full effective potentials and a detailed stability
analysis will be considered in future work.

%
%%%%%%%%%%%%%%%%%%%%%%%%%%%%%
\begin{figure*}[t]
\includegraphics[width=0.98\textwidth]{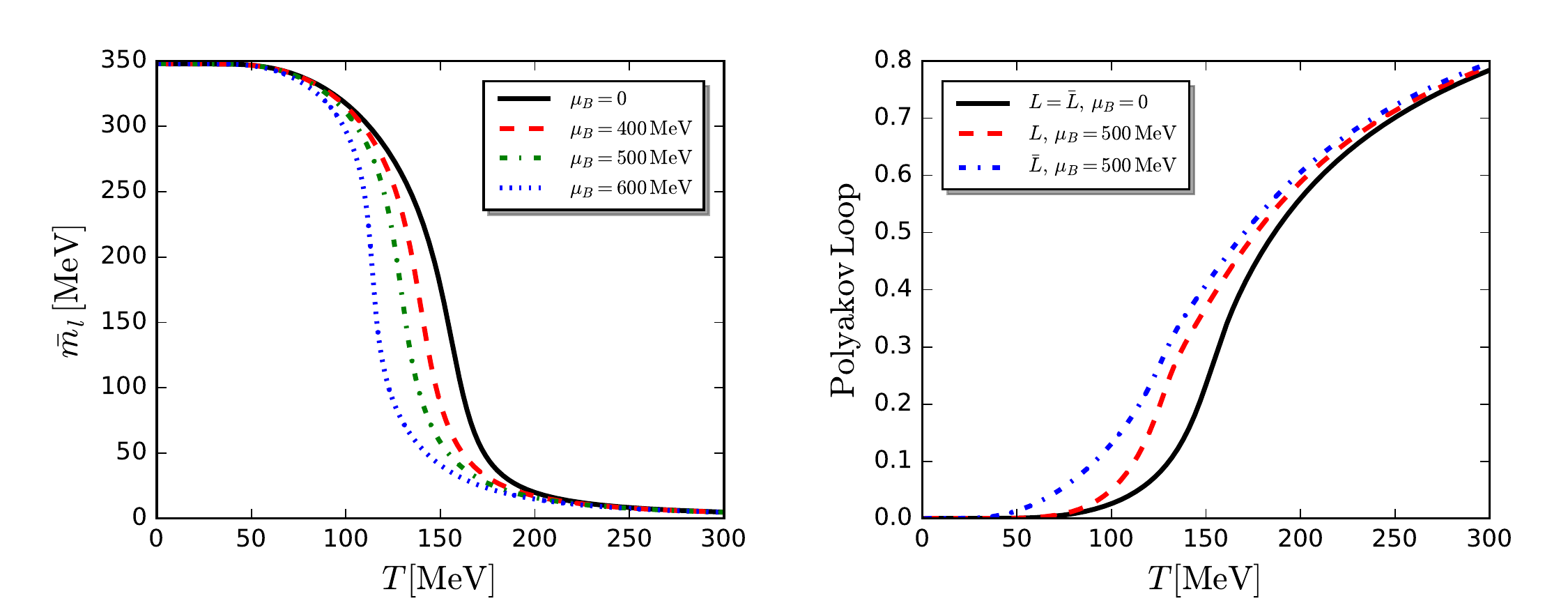}
\caption{Left panel: $N_f=2+1$ constituent mass $\bar m_l$ of light
  quarks as a function of temperature $T$ at various baryon chemical
  potentials
  $\mu_B$.\\[1ex]
  Right panel: Polyakov loops $L$, $\bar L$ as functions of
  temperature $T$ at different baryon chemical potentials
  $\mu_B$.}\label{fig:mf}
\end{figure*}
%%%%%%%%%%%%%%%%%%%%%%%%%%%%%
%

\section{Results and discussion}
\label{sec:num}

In this section we discuss a selection of numerical results obtained
in the present setup. We concentrate on $N_f=2+1$ flavour QCD, and use
the corresponding results for $N_f=2$ only for the comparison of the
$N_f=2$ and $N_f=2+1$ flavour QCD phase structures. Confinement and
chiral symmetry breaking is discussed in
\sec{sec:Conf+Chiral}. Indications for an inhomogeneous phase at large
chemical potential are shown in \sec{sec:Inhom}. The strong couplings
and propagators at finite $T$ and $\mu_B$ are discussed
\sec{sec:Props+Couplings}. In \sec{sec:PhaseStructure} we present
phase structure of $N_f=2$ and $N_f=2+1$ flavour QCD at finite
temperature and density.

In practice, we numerically solve the coupled set of flow equations
discussed in the previous section together with the initial conditions
discussed in \app{app:num}. As it has been discussed extensively in
the literature, e.g.\, \cite{Gies:2002hq, Braun:2005uj, Braun:2009ns,
  Braun:2010qs, Mitter:2014wpa,Braun:2014ata,Pawlowski:2014aha,
  Rennecke:2015eba,Cyrol:2017ewj}, the running couplings in the bound
state sector are governed by an IR attractive fixed point in the
chirally symmetric regime. As a consequence, if we initialise the flow
in the perturbative regime, the initial conditions are fixed by the
initial values of the strong coupling and the quark masses.  As long
as the initial meson masses are larger than the cutoff scale, our
results are independent of the choice for the initial values in the
meson sector, i.e.\ independent of any low-energy effective
parameters. The low-energy sector emerges dynamically in our case, and
the corresponding parameters cannot be tuned, but are predicted.

Note that the mutual coupling of the different $n$-point functions
through their respective flow equations leads to a resummation of
infinitely many diagrams upon integration of these equations. The
renormalisation and regularisation provided by the fRG gives us access
to the nonperturbative IR regime of QCD.

\subsection{Chiral and confinement-deconfinement order parameters \&
  signatures for the CEP}\label{sec:Conf+Chiral}

Observables and order parameters are derived from fields with
nonvanishing expectation values: The scalar condensates
$\bar\sigma_\textrm{EoM}$ and $\bar\sigma_{s,\textrm{EoM}}$ are
directly related to dynamical chiral symmetry breaking, and the
Polyakov loop and its conjugate, $L, \bar L$, are related to
confinement, as introduced in \sec{sec:TPot+Order}. Note however that
these order parameters are not observables themselves.

In particular, the mesonic fields $\phi=(\sigma,\bm\pi)$ are
introduced as low energy effective fields that carry the dynamics of
the scalar-pseudoscalar four-quark interaction. They carry the same
quantum numbers as the related mesons but their direct physics
interpretation in terms of onshell mesons has to be taken with a grain
of salt. Nonetheless, the \textit{renormalised} light chiral
condensate and the \textit{reduced} chiral condensate, $\Delta_{l,R}$
and $\Delta_{l,s}$ respectively, are given in terms of the
unrenormalised expectation values $\sigma_\textrm{EoM}$ and
$\sigma_{s,\textrm{EoM}}$, for more details see
\app{app:chiralcond}. In the present work we concentrate on the
renormalised light chiral condensate $\Delta_{l,R}$,
\begin{align}\label{eq:DeltalR}
  \Delta_{l,R}(T,\mu_q) = -\frac{c_\sigma}{{2 \cal N}_{R}}
  \left[ \sigma_\textrm{EoM}(T,\mu_q) -
  \sigma_\textrm{EoM}(0,0)\right]\,, 
\end{align}
with a normalisation $-c_\sigma/(2 {\cal N}_{R})$ which is fixed to
the lattice, see also \eq{eq:chiralcondren}. The reduced condensate is
also discussed in \app{app:chiralcond}, but carries a larger
systematic error in the present approximation.

In turn, the Polyakov loop is an order parameter for confinement in
pure Yang-Mills theory, but does not carry a direct physics
interpretation in QCD with dynamical quarks. Despite these
deficiencies both, $\bar \sigma_\textrm{EoM}$ and
$(L,\bar L)_\textrm{EoM}$ play a distinguished r$\hat{\textrm{o}}$le
for the low energy dynamics of QCD.
%
%%%%%%%%%%%%%%%%%%%%%%%%%%%%%
\begin{figure}[t]
\includegraphics[width=0.98\columnwidth]{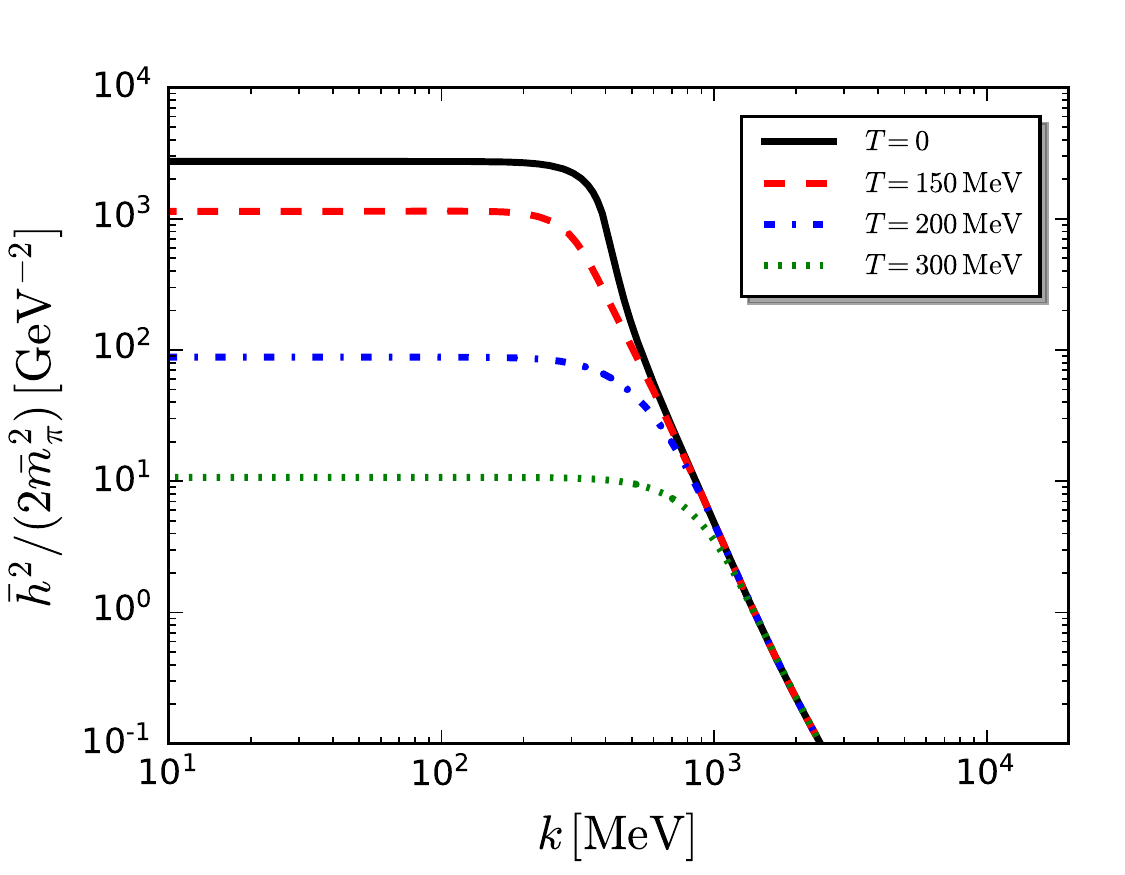}
\caption{ $N_f=2+1$ effective four-quark coupling
  $\bar h^2/(2\bar m_\pi^2)$ in the pseudoscalar channel as a
  function of the RG scale $k$ at different temperatures and
  $\mu_B=0$.}\label{fig:lambda}
\end{figure}
%%%%%%%%%%%%%%%%%%%%%%%%%%%%%
%

In \Fig{fig:mf} we show the constituent mass of light quarks and the
Polyakov loop as functions of the temperature for several values of
$\mu_B$. The size of the quark mass entails the strength of dynamical
chiral symmetry breaking triggered by the scalar-pseudoscalar
four-quark interaction. We see that both transitions are crossovers
for small baryon chemical potential as is well-known due to explicit
chiral and center symmetry breaking, see e.g.\
\cite{Aoki:2006we}. They are getting sharper and shift to smaller
temperatures with increasing $\mu_B$. As $L$ and $\bar L$ are related
to the free energy of a single quark and antiquark respectively, they
are different at finite $\mu_B$: Heuristically speaking, it is easier
for an antiquark to propagate in a system with an excess of quarks,
and hence $\bar L$ is larger than $L$ at finite, positive $\mu_B$.

\begin{figure}[b]
\includegraphics[width=0.98\columnwidth]{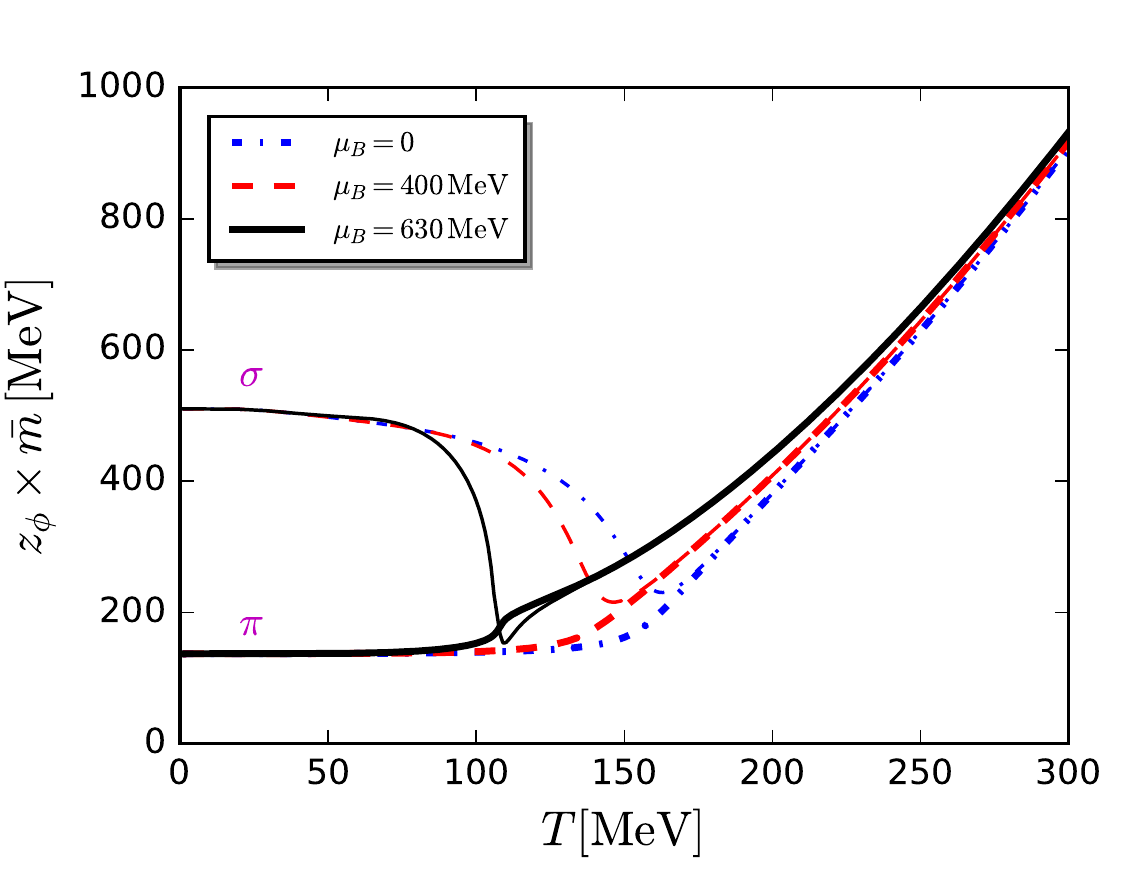}
\caption{$N_f=2+1$ meson masses $\bar m_{\pi}$, $\bar m_{\sigma}$ as
  function of the temperature at different baryon chemical potentials
  $\mu_B$. They are rescaled by the multiplicative constant
  $z_\phi \equiv (\bar Z_\phi/Z_\phi(0))^{1/2}_{T,\mu_B=0}$. As a
  result, $z_\phi\, \bar m$ coincides with the pole mass defined in
  \eq{eq:PoleMass} and \eq{eq:msigmap0} in the vacuum. The in-medium
  behavior is that of the curvature masses defined in
  \eq{eq:RenMass-Yuk}. For more details see \sec{sec:MatterProps}.
}\label{fig:mesonmass}
\end{figure}

As discussed in particular in \sec{sec:DynHad}, the four-quark
interaction plays a central r$\hat{\textrm{o}}$le for chiral symmetry
breaking. The physical picture is that, as the strong coupling
increases in strength with decreasing energy scale, the scattering of
quark-antiquark pairs, mediated by gluon exchange, becomes stronger
and eventually resonant. If this resonance occurs in the channel which
carries the quantum numbers of a scalar quark condensate,
$\langle \bar q q \rangle$, chiral symmetry is spontaneously
broken. Dynamical hadronisation allows us to accurately capture the
dynamics of the resonant quark-antiquark scattering channels in both
the symmetric and the spontaneously broken phase. As argued in
\sec{sec:Yukawa}, the resulting effective four-quark interaction of
the scalar-pseudoscalar channel is the given by the ratio
$\bar h^2/(2\bar m_\pi^2)$. In \Fig{fig:lambda} we show it as a
function of the RG scale $k$ for several values of the
temperature. Starting from large $k$, the coupling rises as expected
from the increasingly strong quark-antiquark scattering. At $T = 0$
and around $k \approx 400$\,MeV, the increasingly steep rise indicates
the resonance. The coupling turns into a pion-exchange interaction
below this scale through dynamical hadronisation. Since the pion mass
and the pion-quark Yukawa coupling are essentially constant in the
chirally broken phase (cf. Figures \ref{fig:mesonmass} and
\ref{fig:h}), this interaction becomes approximately constant as
well. The thermal suppression of the strong coupling discussed below
leads to a suppression of the four-quark interaction with increasing
temperature. As a consequence, the resonance disappears and chiral
symmetry is restored at sufficiently high temperature. This is
discussed in more detail in \cite{Braun:2006jd}.

%
%%%%%%%%%%%%%%%%%%%%%%%%%%%%%
\begin{figure}[t]
\includegraphics[width=0.98\columnwidth]{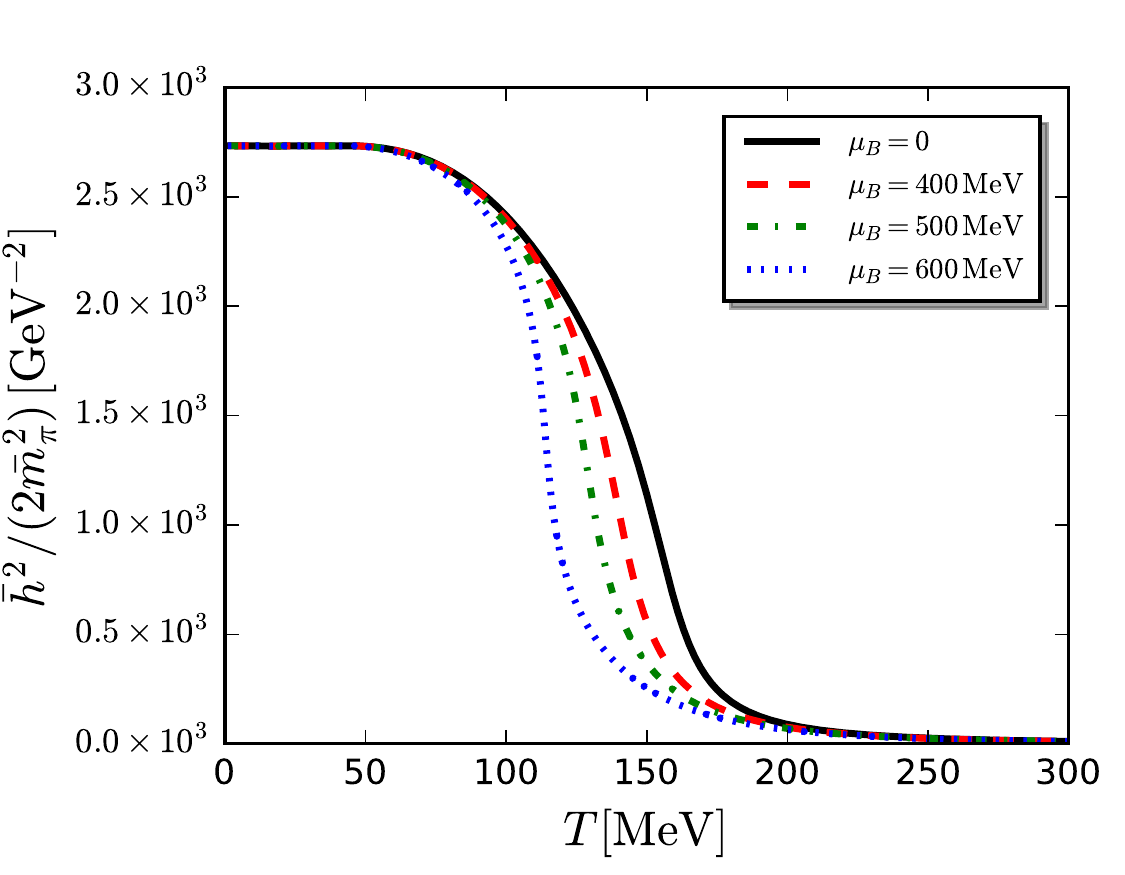}
\caption{$N_f=2+1$ effective four-quark coupling
  $\bar h^2/(2\bar m_\pi^2)$ at vanishing cutoff scale, $k=0$, in the
  pseudoscalar channel as a function of temperature $T$ at various
  baryon chemical potentials $\mu_B$.}\label{fig:lambdamuB}
\end{figure}
%%%%%%%%%%%%%%%%%%%%%%%%%%%%%
%
In \Fig{fig:lambdamuB} we show the effective four-quark interaction
$\bar h^2/(2\bar m_\pi^2)$ as a function of $T$ for $k=0$ for various
chemical potentials. The insensitivity to $\mu_B$ at vanishing
temperature is a consequence of the Silver Blaze property discussed in
\sec{sec:MatterProps}. For correlation functions below onset density,
it implies that their dependence on $\mu_B$ is given solely by a
$\mu_B$-dependent shift of the frequencies of external legs that carry
baryon number. This is discussed in detail in \cite{Khan:2015puu}. For
the present case, see \app{app:etaq}. For increasing temperature, the
four-quark interaction melts down earlier and more rapidly with
increasing $\mu_B$. Consequently, the chiral transitions occurs also
more rapidly and at lower temperatures.

%
%%%%%%%%%%%%%%%%%%%%%%%%%%%%%
\begin{figure*}[t]
\includegraphics[width=0.98\textwidth]{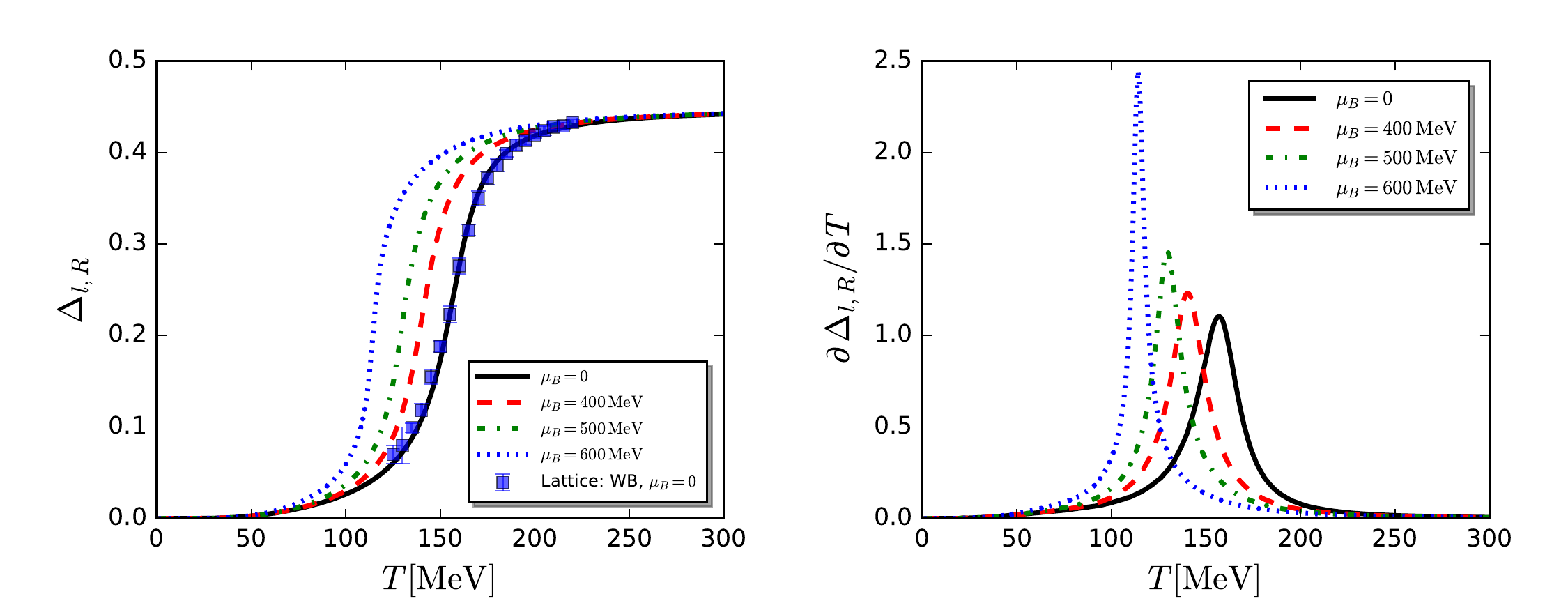}
\caption{Left panel: $N_f=2+1$ renormalised light chiral condensate,
  $\Delta_{l,R}$, see Eqs. (\ref{eq:chiralcondren}),
  (\ref{eq:RenCondSigma}), as a function of temperature $T$ at various
  baryon chemical potentials $\mu_B$. The normalisation constant in
  \eq{eq:chiralcondren}, \eq{eq:RenCondSigma} is chosen to match the
  scale in the lattice calculation
  \cite{Borsanyi:2010bp}.\\[1ex]
  Right panel: Thermal susceptibility of $N_f=2+1$ renormalised light
  chiral condensate, $\partial_t \Delta_{l,R}$, as a function of
  temperature $T$ at various baryon chemical potentials
  $\mu_B$.}\label{fig:DeltalR}
\end{figure*}
%%%%%%%%%%%%%%%%%%%%%%%%%%%%%
%

%
%%%%%%%%%%%%%%%%%%%%%%%%%%%%%
%%%%%%%%%%%%%%%%%%%%%%%%%%%%%
%

The order parameter for chiral symmetry breaking used in the present
work is the renormalised light chiral condensate $\Delta_{l,R}$, see
\app{app:chiralcond}, in particular \eq{eq:RenCondSigma} and
\Fig{fig:DeltalR_muB0}.  In \Fig{fig:DeltalR} (left panel) we show
$\Delta_{l,R}$ as a function of temperature at various baryon chemical
potentials. At $\mu_B=0$ we compare our result to the continuum
extrapolation of lattice results at the physical point from
\cite{Borsanyi:2010bp} and find very good agreement.  The
corresponding thermal susceptibility,
$\partial \Delta_{l,R}/\partial T$, is shown in the right plot of
\Fig{fig:DeltalR}. We use the peak position to define the
pseudocritical temperature.  We note that the pseudocritical
temperature $T_c=156$\,MeV at $\mu_B=0$ is in quantitative agreement
with the lattice prediction. Similarly, one can use the inflection
points of $\bar m_l$ and the Polyakov loop in \Fig{fig:mf} to extract
$T_c$, which yield 155 MeV and 153 MeV respectively, at vanishing
baryon chemical potential. While these are not unique definitions of
the crossover temperature, they match well with each other.  For a
more detailed comparison of chiral order parameters in the crossover
regime see e.g.\ \cite{Pawlowski:2014zaa}.

The dynamics of the system at different $T$ and $\mu_B$ are reflected
in the meson masses. They are shown in \Fig{fig:mesonmass}. The
splitting of the pion and $\sigma$ meson mass due to the chiral
condensate is clearly visible. Furthermore, since $\bar m_\sigma$ is
directly related to the inverse correlation length of the system, it
has a dip at the chiral crossover. The closer the system is to the
CEP, the lighter the sigma becomes at the transition until it
eventually dips below $\bar m_\pi$. This indicates that the system
enters the critical region, as visible in \Fig{fig:mesonmass} for
$\mu_B = 630$\,MeV. At the CEP, the sigma is exactly massless since it
is the critical mode of the chiral phase transition.

We also find that in the chirally symmetric phase $m_\pi$ and
$m_\sigma$ become degenerate and grow rapidly with the increase of
$T$. This entails that the mesonic degrees of freedom decouple quickly
when the temperature of the system is above $T_c$. In this regime the
dynamics are governed by the fundamental degrees of freedom of QCD.
This will be discussed in more detail below. We note that the
inclusion of only the resonant scalar-pseudoscalar interaction channel
in the two-flavour subspace of 2+1 flavours entails that $U(1)_A$ is
maximally broken. Hence, the mesons in the light sector that receive a
positive contribution to their mass from the axial anomaly,
$\eta^\prime$ and $\bm{a}_0$, are fully decoupled here. We defer a
more complete study of chiral symmetry restoration, including the
anomalous effects, to future work.

%
%%%%%%%%%%%%%%%%%%%%%%%%%%%%%
\begin{figure*}[t]
\includegraphics[width=0.98\textwidth]{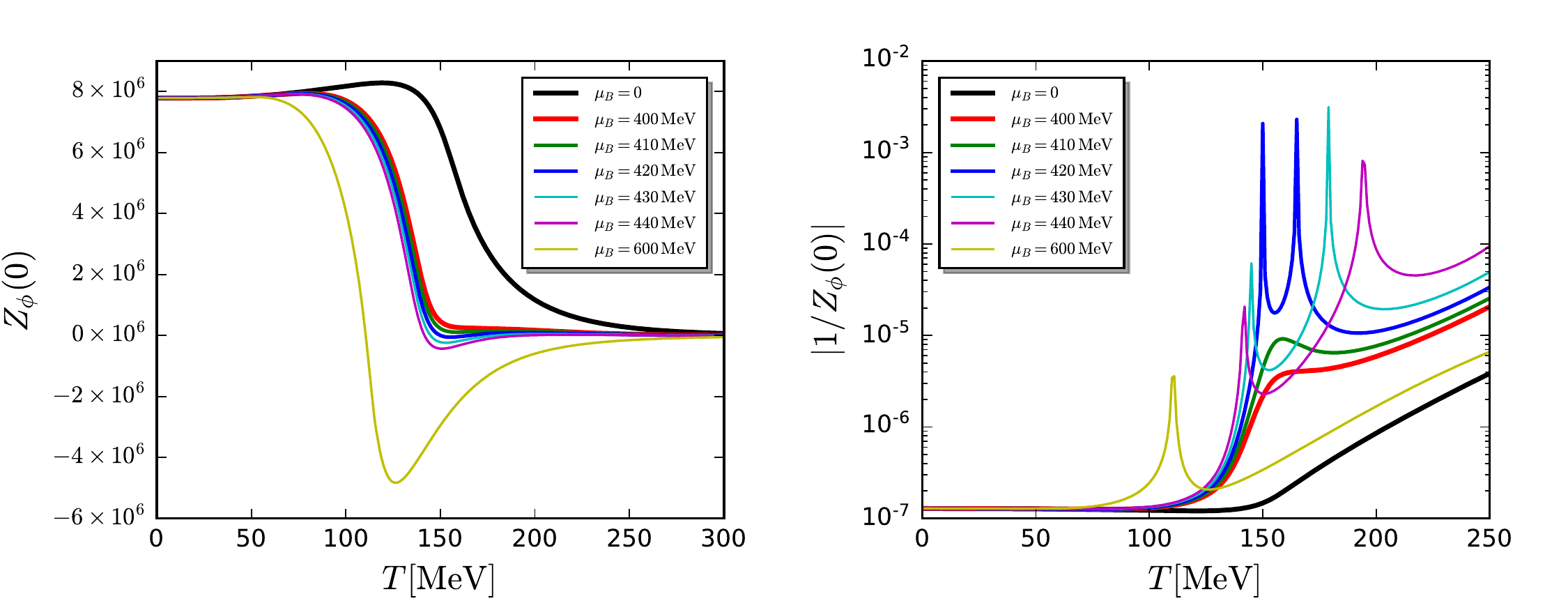}
\caption{Left panel: $Z_{\phi}(0)$ at vanishing cutoff scale as a
  function of the temperature $T$ for various baryon chemical
  potentials $\mu_B$. Negative $Z_{\phi}(0)$, see \eq{eq:NexttoPole},
  entails a negative spatial momentum slope of the dispersion
  $\Gamma^{(2)}_{\phi\phi}(p)$ at vanishing momentum. This
  indicates a minimum of the dispersion at a finite $\bm p^2 \neq 0$. \\[1ex]
  Right panel: $\big|1/Z_{\phi}(0)\big|$ for the data in the left
  panel. Between the spikes the wave function renormalisation is
  negative at $\bm p^2 = 0$.}\label{fig:Zphip0}
\end{figure*}
%%%%%%%%%%%%%%%%%%%%%%%%%%%%%
%

\subsection{Inhomogeneous regime at large chemical
  potential}\label{sec:Inhom}

At finite chemical potential the situation changes significantly: The
mesonic dispersion develops a minimum at nonvanishing momentum.  This
is signaled by $Z_\phi(0)$ at $k=0$ as defined in \eq{eq:ZphiZpi} and
\eq{eq:NexttoPole} turning negative. Indeed this happens for baryon
chemical potentials $\mu_B\gtrsim 420$\, MeV for a temperature regime
that widens with increasing chemical potential, see
\fig{fig:Zphip0}. The left panel shows $Z_\phi(0)$ as a function of
temperature for different baryon chemical potentials, while the right
panel shows $1/|Z_\phi(0)|$, which highlights the negative regime
bounded by the spikes. This property entails that the mesonic
two-point function develops a negative slope at vanishing momentum
implying a minimum at nonvanishing spatial momentum,
$\bm{p}^2_\textrm{min}\neq 0$. This signals a spatially modulated or
inhomogeneous regime. Vanishing of the wave function renormalisation,
as the coefficient of $\bm{p}^2$ in the meson propagator, can be
indicative of an instability towards the formation of an inhomogeneous
quark bound state, such as a chiral density wave. Whether or not
inhomogeneous condensation occurs cannot be answered on the basis of
only the meson wave function renormalisation. Such an analysis
requires a systematic study of the competing effects of potential
homogeneous and inhomogeneous resonances in quark-antiquark scattering
channels. This will be deferred to future work.  The relation between
a minimum of the dispersion relation of bound states and inhomogeneous
phases has been discussed in detail in the past decade within low
energy effective theories at and beyond mean field.  For reviews see
e.g.\ \cite{Buballa:2014tba,Anglani:2013gfu,Thies:2006ti,Schon:2000qy,
  Pisarski:2019cvo}, for applications with the functional RG see e.g.\
\cite{Carignano:2014jla,Braun:2015fva,Roscher:2015xha,Tripolt:2017zgc}.

This regime is indicated in the phase diagram in
\fig{fig:PhasediagramInhom}. The blue shaded area corresponds to the
region with $Z_\phi(0) < 0$. The hatched red area denotes where this
region overlaps with a sizable homogeneous chiral condensate. Note
that the chiral phase boundary lies within this regime for baryon
chemical potentials $\mu_B\gtrsim 500$\, MeV. Hence, in particular in
the hatched red region a competition between homogeneous and
inhomogeneous quark-antiquark pairing is expected. This competing
order effect can alter the phase structure. The hatched red region
should therefore also be interpreted as a systematic error on our
phase boundary.  Accordingly, the present approximation should be
upgraded by inhomogeneous four-quark interactions at larger
$\mu_B$. Most notably, the CEP lies in this regime.

The connection between the CEP and inhomogeneous phases has been
investigated within low energy effective theories, e.g.\
\cite{Carignano:2014jla}. It has been shown that the relative position
of the onset of the inhomogeneous regime and the chiral CEP is related
to the $\sigma$ mass gap, $m_\sigma$, and the constituent light quark
mass, $m_l$. For $m_l= m_\sigma/2$ the two onsets agree on the mean
field level. In this case the phases with restored chiral symmetry,
the homogeneous phase and the inhomogeneous phase meet at a singe
point, the Lifshitz point. For $m_l\geq m_\sigma/2$ the location of
the (potential) chiral CEP is at larger baryon chemical potential as
that for the onset of the inhomogeneous regime. In the present case we
have $m_l\geq m_\sigma/2$ with $m_l/m_\sigma=347/510\approx 0.68$ in
the vacuum. However, it has been argued in \cite{Pisarski:2018bct}
that strong IR fluctuations wipe out the Lifshitz point. This could
even lead to the destruction of the CEP.  We emphasise that a detailed
investigation of this regime goes beyond the scope of the present work
but will be discussed in a forthcoming publication. However, we
emphasise again that our observation of a minimum at nonzero spatial
momentum in the mesonic dispersion is a strong indication for very
interesting physics in this regime.

\subsection{Dynamics and sequential
  decoupling}\label{sec:Props+Couplings}

\begin{figure*}[t]
\includegraphics[width=0.98\textwidth]{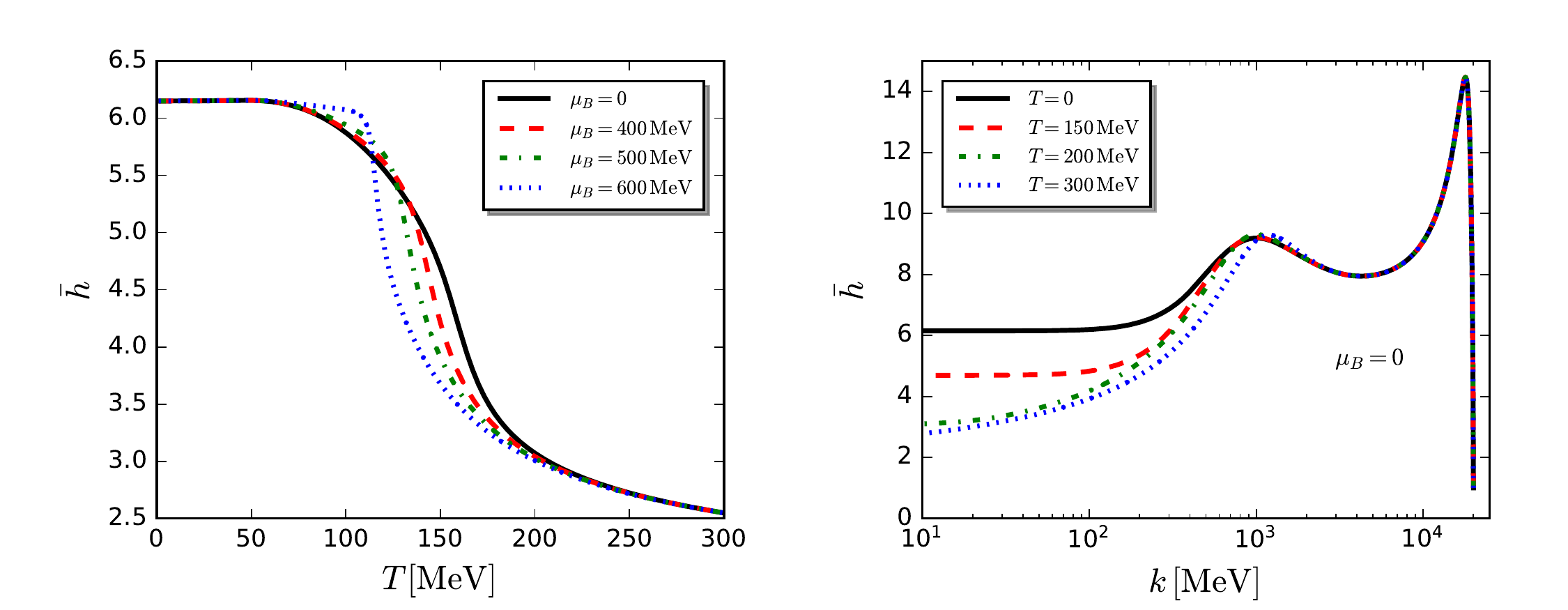}
\caption{Left panel: $N_f=2+1$ Yukawa coupling at vanishing cutoff
  scale, $k=0$, as a function of the temperature $T$ at different
  baryon
  chemical potentials $\mu_B$.\\[1ex]
  Right panel: $N_f=2+1$ Yukawa coupling as a function of the RG scale
  $k$ at different temperatures $T$ and vanishing baryon chemical
  potential $\mu_B=0$.}\label{fig:h}
\end{figure*}

The relevance of the different fields for the dynamics of the system
at different energy scales is encoded in the strength of the vertices
and the propagator dressings. For example, we have already seen, that
the effective four quark coupling $\bar h^2/(2 \bar m^2_\pi)$ decays
rapidly for large momentum or cutoff scales, see \fig{fig:lambda} and
\fig{fig:lambdamuB}. Accordingly, the quark-self coupling gets
irrelevant in the UV. Moreover, for increasing temperature and
chemical potential the effective four quark coupling decreases
further. In the present parameterisation with dynamical hadronisation
this decoupling is solely triggered by the increase of the pion mass
function $\bar m^2_\pi$, for the temperature dependence of the mesonic
mass functions see \fig{fig:mesonmass}. In turn, the Yukawa
interaction does not show a significant momentum scale and parameter
dependence, see \fig{fig:h} and discussion there. Similar observation
apply to the gluonic sector. All avatars of the strong coupling show a
rather strong scale dependence, as do the gluon and meson
dressings. This will be discussed in the following. In combination,
the different dynamics and scale dependence of the couplings, wave
function renormalisations and mass functions allow us to provide a
consistent picture of a sequential decoupling of gluon, quark and
mesonic degrees of freedom with decreasing momentum and cutoff
scales. This leads to the natural emergence of low energy effective
theories (LEFTs) of QCD for scales below $\sim 1$\, GeV and is
discussed at the end of the present Section.

  %
%%%%%%%%%%%%%%%%%%%%%%%%%%%%%
\begin{figure}[t]
\includegraphics[width=0.98\columnwidth]{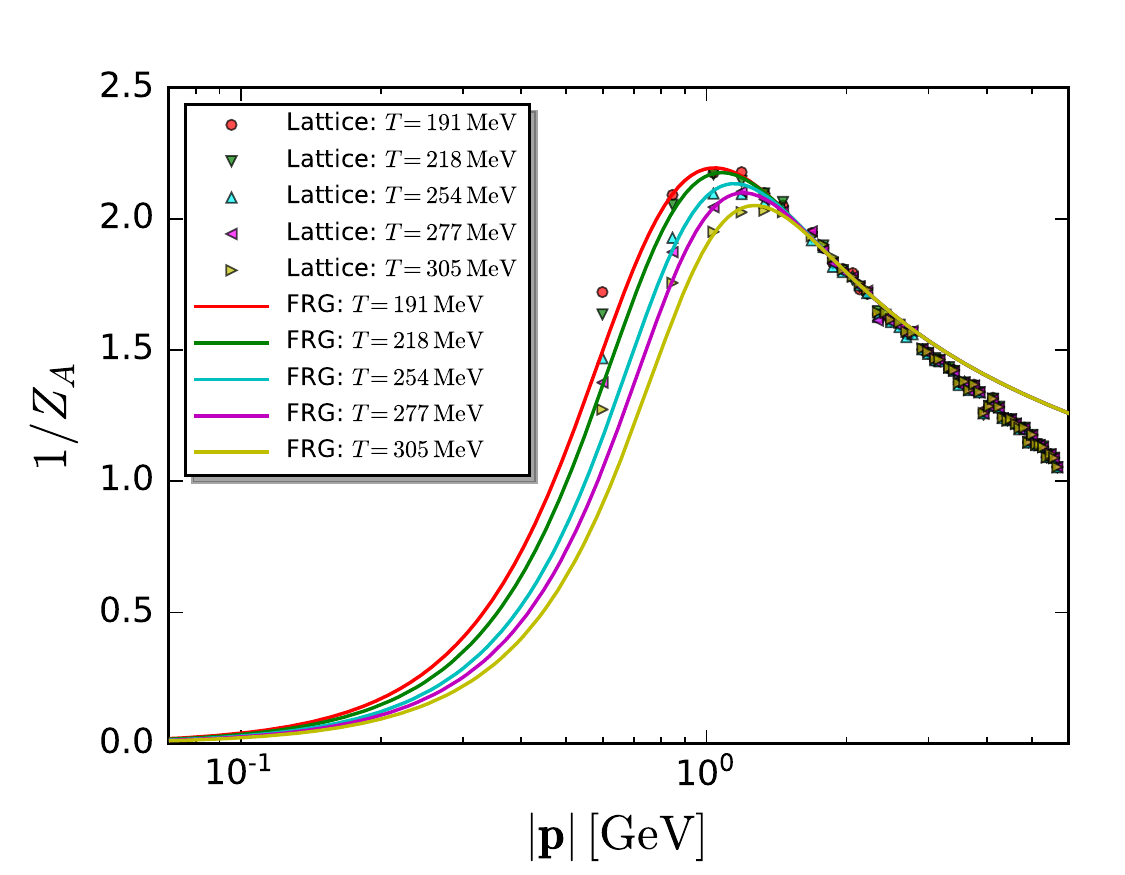}
\caption{$N_f \!=\! 2 \!+\! 1$ gluon dressing function $1/Z_{A}$ as a
  function of spatial momenta $|{\bm p}|$ at different temperatures in
  comparison to the $N_f \!=\! 2 \!+\! 1 \!+\! 1$ lattice data from
  \cite{Ilgenfritz:2017kkp}, based on the tmfT configurations
  \cite{Burger:2013hia}, for the tmfT simulation setup see also
  \cite{Baron:2011sf,Baron:2010bv}. The lattice results depicted here
  are obtained for $\beta=2.1$, $a=0.0646\,$fm and an approximate pion
  mass of $m_\pi = 369(15)$\,MeV leading to a pseudocritical
  temperature of $T_\textrm{deconf}=193(13)(2)$\,MeV, for more details
  see \cite{Ilgenfritz:2017kkp}.}\label{fig:inZAlat}
\end{figure}
%%%%%%%%%%%%%%%%%%%%%%%%%%%%%
%
We start our analysis of the dynamics and decoupling properties of the
system with the gluon dressing. Here we also provide a further
benchmark test of the present approach at vanishing baryon chemical
potential: In \Fig{fig:inZAlat} we compare our results at finite $T$
and $\mu_B = 0$ to lattice results on the magnetic gluon dressing
function from \cite{Ilgenfritz:2017kkp} based on the tmfT
configurations \cite{Burger:2013hia}; for this simulation setup see
also \cite{Baron:2011sf,Baron:2010bv}. Note that the lattice results
are obtained for $N_f \!=\! 2 \!+\! 1 \!+\! 1$ flavours, with
$\beta=2.1$, $a=0.0646\,$fm and an approximate pion mass of
$m_\pi = 369(15)$\,MeV leading to a critical temperature of
$T_\textrm{deconf}=193(13)(2)$\,MeV. Hence, agreement with our results
can only be expected at large temperatures and the corresponding
spatial momentum scales. In this regime, the vacuum constituent quark
masses are small against their thermal masses and the perturbative
running of $N_f=3$ vs $N_f=4$ is not yet dominant. Furthermore, the
effects of the additional quark flavour and the different masses are
subleading. In order to facilitate this comparison, we rescaled our
momentum scale by $|\bm{p}| \rightarrow 1.07 |\bm{p}| $ as to match
our results in the semiperturbative regime around $p \approx
2$\,GeV. This is necessary since, owing to the different mass scales,
also the momentum scales of our results and the lattice results are
inherently different. Indeed, we find good agreement for the
temperatures, 191, 218, 254, 277 and 305\,MeV, for
$0.8\,\textrm{GeV} \lesssim |\bm{p}| \lesssim 3$\!  GeV. For larger
momenta, our result show the correct leveling-off expected from the
logarithmic momentum dependence in the perturbative regime. For
smaller momenta, both the different mass scales and the different
infrared gauge fixing come into play. Note also that the lattice data
are subject to lattice artefacts at large momenta, as is the case in
the $N_f=2$ vacuum data, see \fig{fig:inZA_lattice2-2+1}.

%
%%%%%%%%%%%%%%%%%%%%%%%%%%%%%
\begin{figure*}[t]
\includegraphics[width=0.98\textwidth]{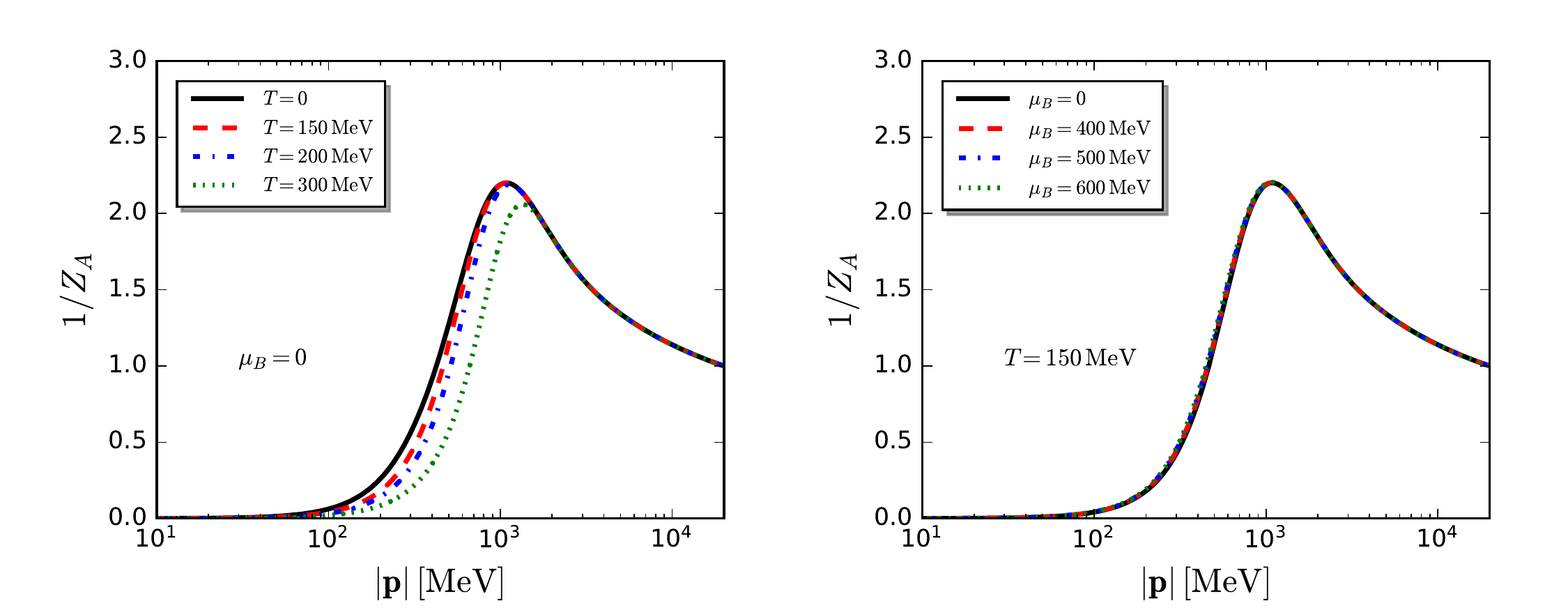}
\caption{$N_f=2+1$ gluon dressing function $1/Z_{A}$ at vanishing
  Matsubara frequency, $p_0=0$, as a function of spatial momenta
  $|{\bm p}|$ at different temperatures (left panel) and baryon
  chemical potentials (right panel). As for the $N_f=2$ flavour input
  data and the $N_f=2+1$ vacuum results, see \fig{fig:inZA_lattice},
  we use the identification $\bm {p}^2=k^2$.}\label{fig:inZA}
\end{figure*}
%%%%%%%%%%%%%%%%%%%%%%%%%%%%%
%
In \Fig{fig:inZA} we show the gluon dressing function
$1/Z_{A}(|\bm{p}|)=1/Z_{A,k=|\bm{p}|}$ at vanishing Matsubara
frequency $p_0=0$ as a function of spatial momenta for different
temperatures (left panel) and baryon chemical potentials (right
panel). While the gluon dressing shows a mild dependence on the
temperature, it is essentially independent on the baryon chemical
potential. Note that the respective dependences are more significant
in the gluon propagator $G_A= 1/(p^2 Z_A(p) )$ shown in \fig{fig:GA}
in \app{app:gluon}. However, we emphasise that the flows of the
couplings and observables computed in the present work depend on the
gluon dressing rather than on the gluon propagator.

%
%%%%%%%%%%%%%%%%%%%%%%%%%%%%%
\begin{figure*}[t]
\includegraphics[width=0.98\textwidth]{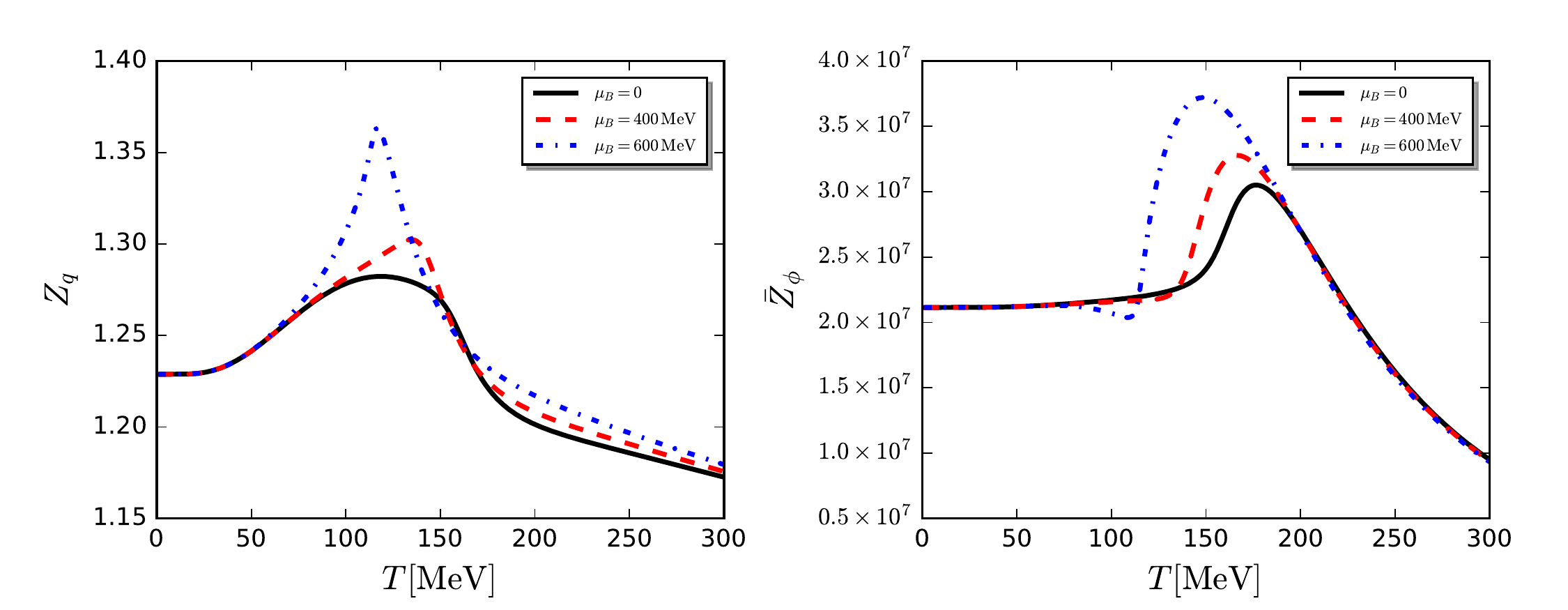}
\caption{$N_f=2+1$ quark (left panel, $Z_q$) and meson (right panel,
  $\bar Z_\phi$) wave function renormalisations at vanishing cutoff
  scale, $k=0$, as functions of temperature $T$ at different baryon
  chemical potentials $\mu_B$.}\label{fig:Zpsi}
\end{figure*}
%%%%%%%%%%%%%%%%%%%%%%%%%%%%%
%
The suppression of the thermal and chemical potential dependences
discussed above are linked to the mass gap of the gluon propagator in
the Landau gauge. This mass gap signals confinement, see
\cite{Braun:2007bx,Marhauser:2008fz,Braun:2010cy,%
  Fister:2013bh,Herbst:2015ona}, and manifests itself through the
nonmonotonicity of the curves depicted in \Fig{fig:inZA}. The
backbending of the gluon propagator at small momenta also signals
positivity violation \cite{Alkofer:2000wg}, and results in a negative
spectral function at low frequencies \cite{Cyrol:2018xeq}.  With
increasing temperature above $T_c$, the chromoelectric part of the
gluon is subject to increasingly strong Debye-screening. As discussed
in \sec{sec:GlueSector}, we define the gluon dressing via the
chromomagnetic contribution. Hence, the thermal suppression in the
left plot of \Fig{fig:inZA} is due to nonperturbative magnetic
screening. For more technical details on the in-medium gluon
propagator we refer to \sec{sec:GlueSector} and \app{app:gluon}.

Overall, the temperature dependence of the gluon dressing is a
subleading effect in the temperature region relevant for the phase
transition. Below the critical temperature it is mostly affected in
the infrared, where the mass gap suppresses gluon fluctuations anyway.
While thermal corrections are also present in the pure gauge theory,
density corrections are only triggered by the quark contributions
discussed in \sec{sec:GlueSector}. We find that the gluon dressing
function is almost insensitive to $\mu_B$. This is shown in the right
plot of \Fig{fig:inZA}, and also in \fig{fig:GA} for the propagators
itself. For temperatures in the vicinity of the crossover, we find a
slight enhancement of the gluon propagator with increasing $\mu_B$ for
momenta below $\sim 1$\,GeV. The reason is that the quark contribution
to the gluon propagator, i.e.\ the vacuum polarisation, is suppressed
at finite $\mu_B$, which can be read-off from the explicit expression
in \eq{eq:DeltaAqexpl}. This counteracts the color screening of
quarks.  These findings support the qualitative or even
semiquantitative approximations in functional applications (fRG and
DSE), where either both or part of the thermal and density dependence
of the gluon dressing is left out.

%
%%%%%%%%%%%%%%%%%%%%%%%%%%%%%
%%%%%%%%%%%%%%%%%%%%%%%%%%%%%
%

The quark and meson wave function renormalisations and their
dependence on $T$ and $\mu_B$ are shown in \Fig{fig:Zpsi}.  Owing to
our choice of regulators, shown in \eq{eq:Rq} and \eq{eq:Rphi},
$Z_{q,k}$ and $Z_{\phi,k}$ enter the system of flow equations only
indirectly through the corresponding anomalous dimensions,
\eq{eq:eta}. Consequently, the flows of the wave function
renormalisations do not have to be integrated and their initial values
are irrelevant. We have chosen $Z_{k=\Lambda}=1$ for convenience, see
\eq{eq:Zin1}. One observes that there is a bump in $Z_{q}$ as a
function of $T$ during the phase transition, and it evolves into a
sharp peak with the increase of the baryon chemical potential. This
indicates that constituent quark fluctuations are enhanced in the
vicinity of the phase transition at finite $\mu_B$. We also find that
the meson wave function renormalisation decreases after chiral
symmetry is restored. This effect drives the decoupling of mesons from
the physical spectrum above $T_c$, which is also visible in
\Fig{fig:mesonmass} and discussed in the previous section.

Interestingly, the renormalised Yukawa coupling, \eq{eq:RenMass-Yuk},
is relatively stable over all scales. Accordingly, the quark-meson
vertex running counterbalances the strong scale dependence of the
meson wave function renormalisation. This is explicitly seen in
\Fig{fig:h}: On the left side we show the temperature and chemical
potential dependence of the physical Yukawa coupling, $\bar
h_{k=0}$. On the right side we show its running with $k$ at different
temperatures. The Yukawa coupling is related to the expectation value
of the renormalised $\bar \sigma$-field and the constituent mass of
the light quarks via
$\bar h_{k=0}=2\bar m_q/\langle \bar \sigma \rangle$. In the symmetric
phase, the fast decrease of $\bar h$ with increasing $T$ triggers a
more rapid melting of $\bar m_q$ as compared to
$\langle \bar \sigma \rangle$.  It is evident from the running of the
Yukawa coupling in the right panel of \Fig{fig:h} that thermal
corrections only set in if the RG scale is smaller than the
temperature scale, $k\lesssim 2 \pi T$. For larger $k$ and the 3d
spatial regulators used here, see \app{app:threshold}, they are
exponentially suppressed with $\exp( -k/(2 \pi T))$ and there is only
the vacuum running. For a respective discussion of the thermal
properties including also other regulators we refer to
\cite{Fister:2015eca}. Furthermore, we have explicitly checked that
our results are independent of the initial value of the Yukawa
coupling: The Yukawa coupling is an irrelevant coupling that is
generated by the flow. The memory of its initial condition is
washed-out by an IR-attractive fixed point in the regime of small
strong coupling. This is in line with the discussion at the beginning
of this section.

This stability of the renormalised Yukawa coupling together with the
rapid rise of the meson mass parameters $\bar m_\sigma, \bar m_\pi$
entails the decoupling of the scalar-pseudoscalar mesonic channel of
the four-quark interaction in the chirally symmetric phase at large
scales, discussed in \sec{sec:Conf+Chiral}.

%
%%%%%%%%%%%%%%%%%%%%%%%%%%%%%
\begin{figure}[t]
\includegraphics[width=0.98\columnwidth]{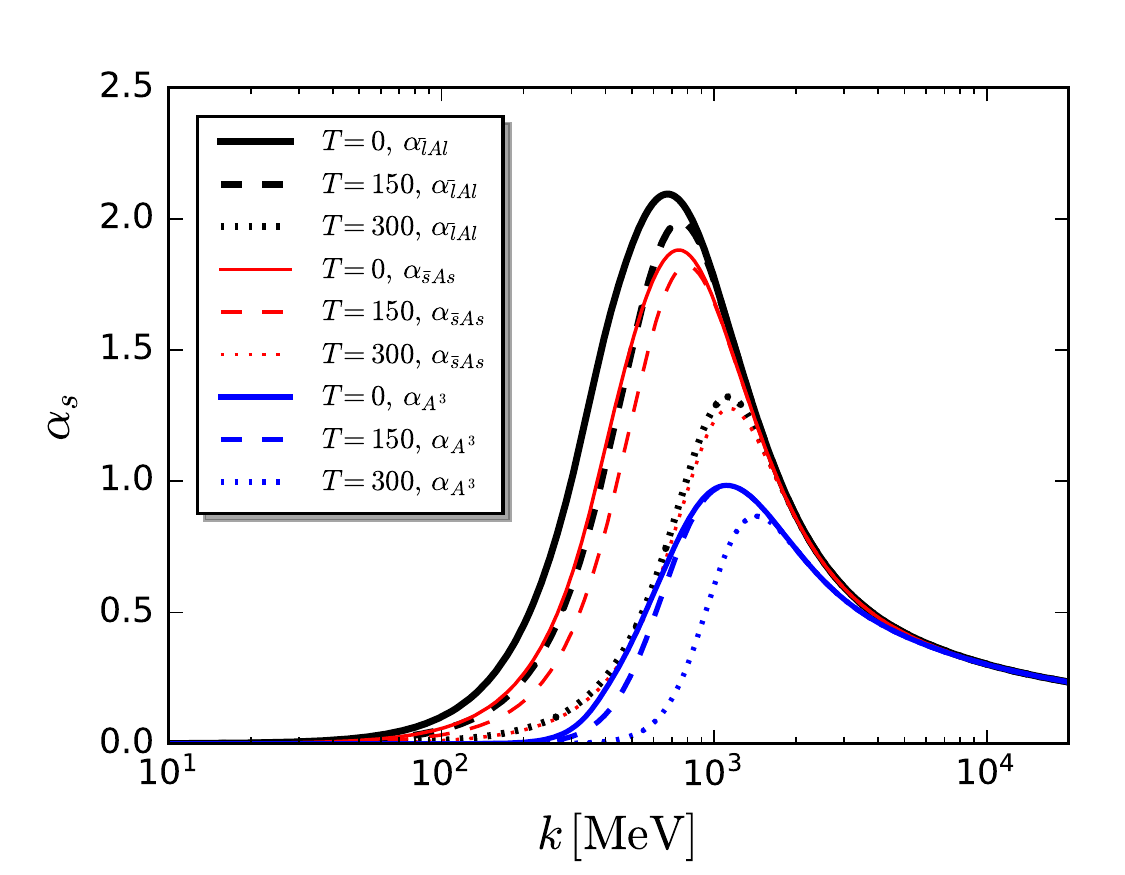}
\caption{$N_f=2+1$ quark-gluon coupling for light quarks
  ($\alpha_{\bar l A l}$) and strange quarks ($\alpha_{\bar s A s}$)
  quarks, and three-gluon coupling ($\alpha_{A^3}$) as functions of
  the RG scale $k$ for several values of the temperature $T$ and for
  vanishing baryon chemical potential, $\mu_B=0$.}\label{fig:alphas}
\end{figure}
%%%%%%%%%%%%%%%%%%%%%%%%%%%%%
%

For larger scales, $k\gtrsim 1$ GeV, the gluon exchange interactions
are dominating the system, while for smaller scales, $k\lesssim 1$
GeV, the gluons decouple. This is clearly seen in \Fig{fig:alphas},
where we show the running of the different strong couplings: The
quark-gluon and the three-gluon couplings.  We have distinguish
between the quark-gluon coupling for $u$ and $d$ quarks, denoted by
$\alpha_{\bar l A l}$, and for the $s$ quark, $\alpha_{\bar s A
  s}$. The quark-gluon couplings $\alpha_{\bar l A l}$ and
$\alpha_{\bar s A s}$ match for momentum scales $k$ above $\sim 1$
GeV. In turn, the strange quark coupling is slightly smaller at lower
scales. This is expected since quark number conservation of the strong
interactions implies that there is no flavour mixing in the
quark-gluon interactions. Hence, only strange quarks contribute to the
leading quantum corrections of $\alpha_{\bar s A s}$, resulting in a
suppression relative to the leading corrections to
$\alpha_{\bar l A l}$.

The quark-gluon couplings and the three-gluon coupling agree with each
other, and are also consistent with perturbation theory, for scales
above $\sim 5$\,GeV \cite{Braun:2014ata, Cyrol:2017ewj}. One also
finds that for scales $k\lesssim 5$\, GeV, the light flavour coupling
$\alpha_{\bar l A l}$ and the strange coupling $\alpha_{\bar s A s}$
grows bigger than $\alpha_{A^3}$, which is qualitatively consistent
with the calculation in \cite{Cyrol:2017ewj}. For scales
$k \lesssim 1-2$\,GeV, $\alpha_{\bar l A l}$, $\alpha_{\bar s A s}$
and $\alpha_{A^3}$ deviate from each other pronouncedly, and all
strong couplings are suppressed significantly. This behavior in the IR
region is due to the running of the gluon dressing function shown in
the left panel of \Fig{fig:inZA} and our definition of the strong
couplings in \eq{eq:alphasGluon} and \eq{eq:alphasMatter}. Since they
effectively describe gluon exchange, the suppression of glue dynamics
in the nonperturbative regime due to the dynamically generated gluon
mass gap is transferred to the RG invariant strong couplings.
Apparently, the three-gluon coupling is more suppressed than the
quark-gluon vertex in the IR, since more gluons are attached to the
former, which is verified in \Fig{fig:alphas}.  The dependence of the
strong couplings on the temperature is consistent with our expectation
inferred from the results of the gluon propagator. The strong
couplings decrease with the increase of the temperature, which
indicates the interaction between gluons and quarks gets weaker at
high temperature. We have also investigated the dependence of the
strong couplings on the baryon chemical potential, and find a small
$\mu_B$- dependence in the $\mu_B$ region of our interest.

We close this Section with a discussion of sequential decoupling and
the natural emergence of low energy effective theories (LEFT) in the
present fRG approach to QCD for low cutoff scales $k\lesssim 1$\,GeV:
The flows of matter vertices and propagators (quarks and mesons) are
driven by the tree-level four-point single field exchange couplings
with either quark or meson legs in the current approach.  We emphasise
that this is not due to the approximations used here but originates in
the one loop completeness of the flow equation for the effective
action, \eq{eq:FlowQCD}. Relevant examples are provided by the flows
of the quark-gluon and Yukawa couplings depicted in
\fig{fig:3pointsAll}, and the flow of the four quark coupling depicted
in \fig{fig:flow4q}. These flows contain all single field exchange
couplings considered in the present approach: The four-quark single
gluon exchange coupling and the four-quark single meson exchange
coupling.  Note that the potential third single field exchange
coupling, the two-quark--two-meson single quark exchange coupling
simply comes from a reordering of the building blocks in the
diagrams. It also originates in the fundamental building blocks in the
QCD matter sector considered here, the scalar-pseudoscalar channel of
the four-quark interaction. Accordingly, the two-quark--two-meson
single quark exchange coupling simply measures the strength of the
Yukawa coupling and has no physical interpretation. In particular, it
does not entail a possible decoupling.  This leaves us with
\begin{align}\label{eq:ExchangeCouplings}
  g^2_{\bar q A q}\,, \qquad \qquad \frac{\bar h^2 }{1
  +\tilde m_\phi^2}\,, 
\end{align}
with $\phi=\sigma,\bm \pi$ and the dimensionless mass functions
$\tilde m$ defined in \eq{eq:DimlessMass}. Note also that these
exchange couplings generate genuine four-field interactions in the
flow, such as the four-quark and four-meson vertices. Consequently,
the strength of these higher order couplings is tightly linked to that
of the single field exchange couplings, the analysis of which suffices
in the absense of additional resonant interaction channels besides the
scalar-pseudoscalar one.

\begin{figure}[t]
\includegraphics[width=0.98\columnwidth]{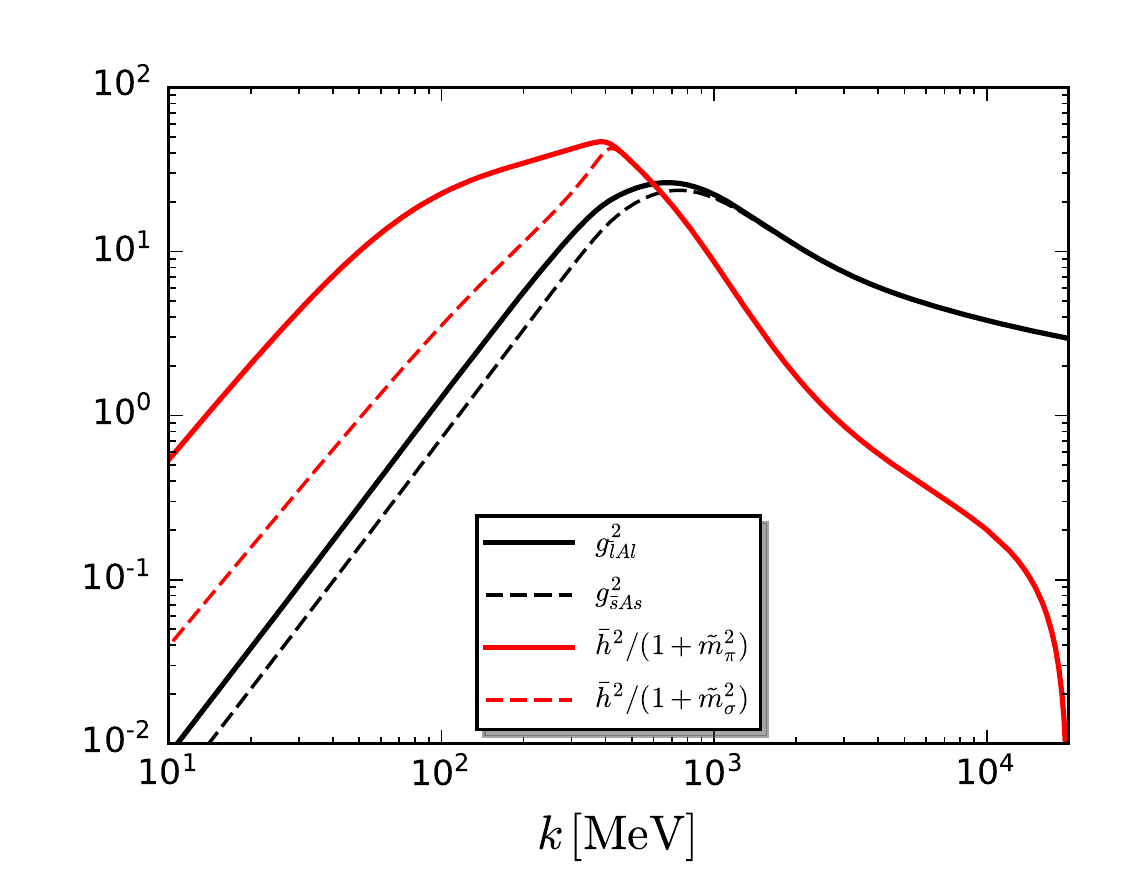}
\caption{Dimensionless four-quark single gluon exchange coupling
  ($g^2_{\bar l A l}$ and $g^2_{\bar s A s}$) and four-quark single
  meson exchange couplings ($\bar h^2 /(1+\tilde m_\pi^2)$ and
  $\bar h^2 /(1+\tilde m_\sigma^2)$) as functions of the cutoff scale
  $k$ in the vacuum. Gluons, quarks, and mesons decouple sequentially
  from the matter dynamics.  }\label{fig:ExchangeCouplings}
\end{figure}

In \fig{fig:ExchangeCouplings} we depicted the two fundamental single
field exchange couplings defined in \eq{eq:ExchangeCouplings}. The UV
dominance of the gluonic exchange coupling for cutoff scales
$k\gtrsim 1$\,GeV is clearly visible.  In turn, for cutoff scales
$k\lesssim 1$\,GeV the mesonic exchange coupling takes over gradually
with $g_{\bar l A l}^2 = \bar h^2/(1+m_\pi^2)$ at $k\approx 600$\,
MeV, while the gluonic coupling decays more quickly. Note that the
exchange couplings only provide the qualitative picture; the
respective flow diagram also have different combinatorial factors
from the different colour and flavour traces. For the four-quark flow
a details comparison is done in \app{app:dtlambda}, and is summarised
in \fig{fig:flow4fermi}.
  
\begin{figure}[t]
\includegraphics[width=0.98\columnwidth]{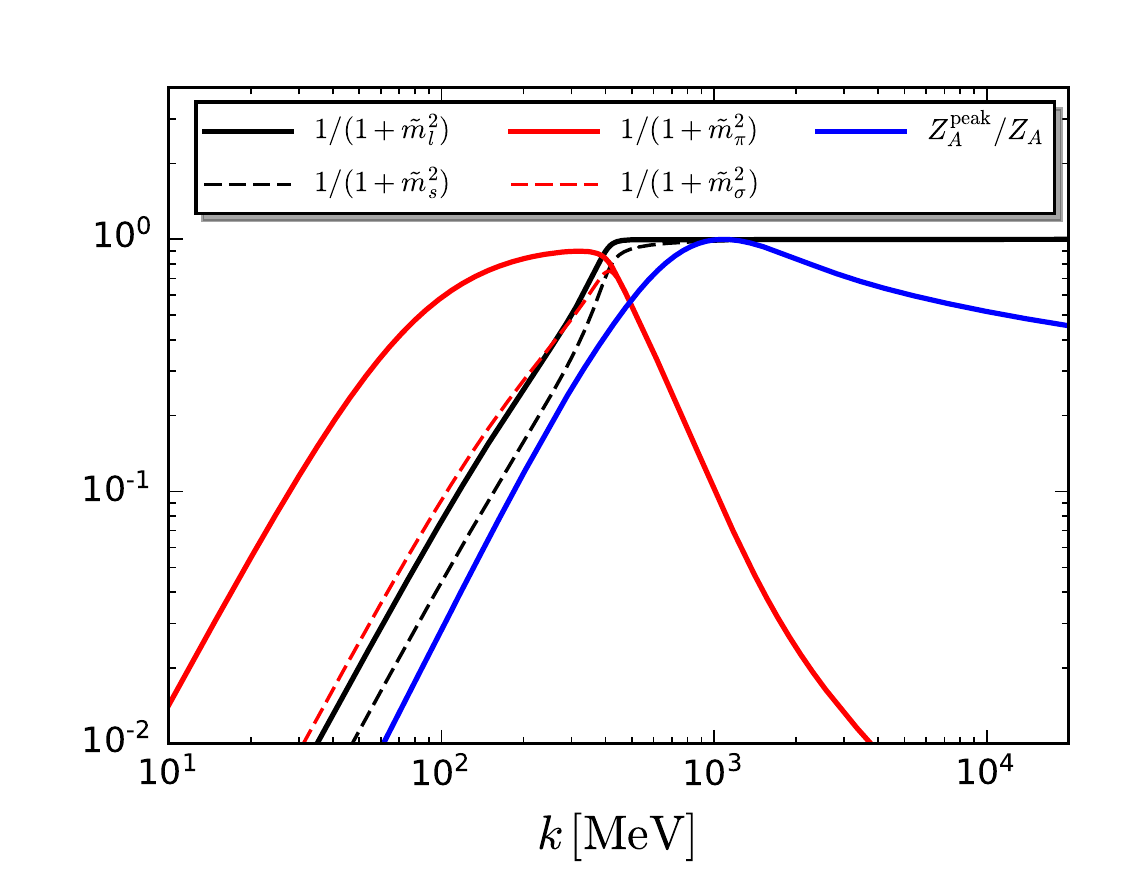}
\caption{Dimensionless propagator gapping $1/(1+\tilde m^2_{\Phi_i})$
  for $\Phi_i=l,s,\sigma, {\bm \pi}$. For comparison, we also show the
  gluon dressing function $Z_A^{\textrm{peak}}/Z_A$ with the solid
  blue line. Here, $1/Z_A$ is normalised by its peak position at
  $k_\textrm{peak}$, that is
  $Z_A^{\textrm{peak}}=Z_A(k_\textrm{peak})$: for cutoff scales
  $k\gtrsim k_\textrm{peak}$ the $k$-dependence of $Z_A$ is dominated
  by the running of the gluon wave function renormalisation, for
  $k\lesssim k_\textrm{peak}$ it shows the gluon gapping. In summary,
  the sequential decoupling of gluon, quark and meson dynamics towards
  the IR is evident here.}\label{fig:PropagatorGapping}
\end{figure}

In the matter-dominated regime the best indicator for the further
decoupling is given by the respective propagator gapping
\begin{align}\label{eq:PropagatorGapping}
    \frac{1}{1+\tilde m_q^2} \,,\qquad \frac{1}{1+\tilde m_\phi^2}\,,
\end{align}
with $\phi=\sigma,\bm \pi$, $q=l,s$ and the $\tilde m$ defined in
\eq{eq:DimlessMass}. These gapping functions are depicted in
\fig{fig:PropagatorGapping}, where also the full gluon dressing is
displayed for the sake of comparison. Note that the latter does not
only entail the gapping information for cutoff scales smaller than
that of the peak postion of the dressing function,
$k\lesssim k_\textrm{peak}$, but also the running of the gluon wave
function for $k\gtrsim k_\textrm{peak}$. Qualitatively, this carries
the same information as that depicted in \fig{fig:ExchangeCouplings}:
gluonic fluctuations decouple first. However,
\fig{fig:ExchangeCouplings} carries the full decoupling which also
requires the coupling strengths of quark-gluon vertex and Yukawa
vertex respectively.

%
%%%%%%%%%%%%%%%%%%%%%%%%%%%%%
%%%%%%%%%%%%%%%%%%%%%%%%%%%%%
%
In turn, after the decoupling of the gluonic sector the interactions
between quark and mesons only depend on the Yukawa coupling, and the
respective propagator gapping is sufficient for the decoupling
information: the propagator gappings of quarks and mesons clearly
entail that diagrams with quark propagators decouple first (first
strange, than the light quark diagrams). The $\sigma$-field decouples
at roughly the same scale as the light quarks and finally the
contributions of pion diagrams tend towards zero below the pion mass
scale.
      
In summary, the sequential decoupling is clearly seen. First the
gluonic dynamics decouples from the matter sector, then the quark and
$\sigma$ exchange diagrams decouple and finally the pion
decouples. For example, this already entails that in the vacuum chiral
perturbation theory emerges naturally in this setup, for
investigations of respective low energy parameters see e.g.\
\cite{Divotgey:2019xea,Eser:2019pvd}. Moreover, it allows us to also
investigate the validity bounds of chiral perturbation theory. Note
also that the ordering in the sequential decoupling is changes at
finite density. In particular the $\sigma$ mode decouples later in the
vicinity of the CEP, where, as the critical mode, gets massless.

%
%%%%%%%%%%%%%%%%%%%%%%%%%%%%%
%%%%%%%%%%%%%%%%%%%%%%%%%%%%%
%
This leads us to an intriguing and simple picture: Initiating the QCD
flow at a large perturbative cutoff scale
$k=\Lambda^{\textrm{UV}}_\textrm{QCD}$, here
$\Lambda^{\textrm{UV}}_\textrm{QCD}= 20$\,GeV, the flow towards
smaller scales leaves us with the dynamics of quarks and mesons for
$k\ll 1$\, GeV. In the present setup with dynamical hadronisation we
flow into a Polyakov loop enhanced quark meson model. In turn, without
dynamical hadronisation we arrive at a Polyakov loop enhanced NJL-type
model.  We also emphasise that while the offshell dynamics of gluons
decouples, the gluon sector still leaves its imprint in terms of the
quantum equations of motion for the gluonic background $A_0$ or
$L,\bar L$.

Hence, the underlying assumption of LEFTs, that high-energy degrees of
freedom are integrated out, can be made explicit with the present
setup.  Moreover, this can also provide valuable reliability checks
for LEFTs. In summary this provides an explanation and further
reliability. The sequential decoupling demonstrated here adds further
justification for the \textit{unreasonable effectiveness} of LEFTs in
QCD. For a selection of fRG work on QCD LEFTs see
\cite{Schaefer:2004en,Schaefer:2006ds, Herbst:2010rf, Skokov:2010wb,
  Skokov:2010uh, Braun:2011iz, Strodthoff:2011tz, Fukushima:2012xw,
  Aoki:2012mj, Kamikado:2012cp,Jiang:2012wm,Haas:2013qwp,
  Herbst:2013ufa, Herbst:2013ail, Tripolt:2013zfa, Grahl:2013pba,
  Mitter:2013fxa, Herbst:2013ufa, Morita:2013tu, Pawlowski:2014zaa,
  Helmboldt:2014iya, Braun:2014fua, Khan:2015puu, Wang:2015bky,
  Mueller:2015fka,Eser:2015pka, Weyrich:2015hha, Fu:2015naa,
  Aoki:2015mqa, Fejos:2015xca, Jiang:2015xqz, Fu:2015amv, Fu:2016tey,
  Rennecke:2016tkm, Jung:2016yxl, Fejos:2016hbp, Almasi:2016zqf,
  Posfay:2016ygf, Yokota:2016tip, Springer:2016cji,Resch:2017vjs,
  Tripolt:2017zgc, Braun:2017srn, Fejos:2017kpq, Yokota:2017uzu,
  Almasi:2017bhq, Zhang:2017icm, Aoki:2017rjl, Braun:2018bik,
  Fejos:2018dyy, Braun:2018svj, Fu:2018qsk, Fu:2018swz, Sun:2018ozp,
  Wen:2018nkn, Yin:2019ebz, Leonhardt:2019fua,Li:2019nzj}.

We emphasise that the emergence of QCD LEFTs in the infrared at scales
$k\lesssim 1$\,GeV within the current approach allows for their
\textit{systematic} improvement in terms of \textit{QCD-assisted}
LEFTs, and constitutes important progress in the setup and the
qualitative and quantitative reliability of low energy effective
theories.  Of course, QCD LEFTs in general rely on QCD
information. For example, the parameters of QCD LEFTs used for QCD at
finite temperature and density are typically fixed with vacuum
observables such as the pion decay constant and meson masses as well
as quark constituent masses. Further QCD input ranges from the
enhancement of the models with a Polyakov loop potential that carries
information about the confinement-deconfinement phase transition.

A step in the direction towards using dynamical or fluctuation
information of QCD is done with the implementation of the QCD-running
of external parameters such as the critical temperature parameter
$T_0$ in the Polyakov loop potential as suggested in
\cite{Schaefer:2007pw}. A first example for a \textit{QCD-assisted}
LEFTs is then provided by \cite{Haas:2013qwp,Herbst:2013ufa}, where
the phenomenological approach in \cite{Schaefer:2007pw} was confirmed
and improved by results for the Polyakov loop potential from
fRG-computations in Yang-Mills theory
\cite{Braun:2007bx,Fister:2013bh} and $N_f=2$ flavour QCD
\cite{Braun:2009gm}. Another example is provided by
\cite{Springer:2016cji}, where the UV initial conditions for the QCD
LEFT were computed from QCD flows. 

The results and in particular the flows of the current work can be
readily used as external input for QCD LEFTs, hence lifting them to
\textit{QCD-assisted} LEFTs. An interesting and relevant example under
progress is provided by fluctuation observables such as cumulants of
net-particle multiplicity distributions, e.g.\
\cite{Fu:2016tey,Fu:2018swz}. A qualitative improvement of the
computations there is given by utilising the quark-gluonic flows from
the present work as external input. This provides quantitative
reliability of the results at large density or small $\sqrt{s}$
relevant for the search of the critical end point.

\subsection{Phase Structure of $N_f=2$ and $N_f=2+1$ flavour
  QCD}\label{sec:PhaseStructure}

\begin{figure}[t]
  \includegraphics[width=.98\columnwidth]{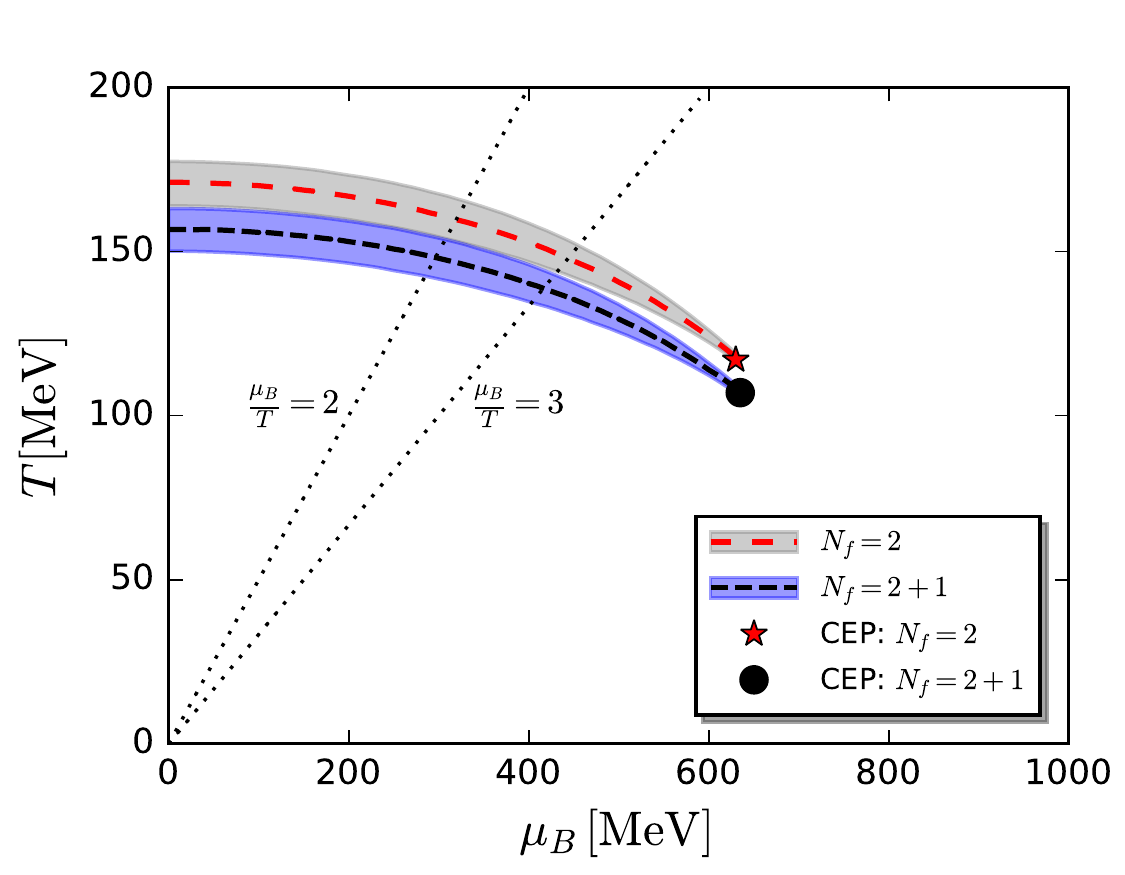}
  \caption{Phase diagram in the plane of the temperature and the
    baryon chemical potential. The gray and blue bands denote the
    crossover transitions for the $N_f=2$ and 2+1 flavour QCD,
    respectively; and the red star and black circle are their relevant
    CEP. The bands are determined through the 80$\%$ peak height of
    $ \partial \Delta_{l,R}/\partial T$ at fixed $\mu_B$. The black
    and red dashed lines depict the peak positions for the $N_f=2+1$
    and $N_f=2$, respectively. We also provide the two lines of
    $\mu_B/T=2,3$ related to reliability bounds for both lattice and
    functional methods.  }\label{fig:phasediagram}
\end{figure}

In \Fig{fig:phasediagram} we show our results on the phase structure
in $T$-$\mu_B$ plane for $N_f = 2$ and $N_f = 2+1$ flavours. In both
cases, as discussed in \sec{sec:Conf+Chiral}, we define the
pseudocritical temperature via the thermal susceptibility of the
renormalised light chiral condensate,
$\partial \Delta_{l,R}/\partial T$. At vanishing baryon chemical potential,
the respective chiral transition temperatures are 
\begin{align}\label{eq:Tcmu=0}
  T_{c,{\tiny{N_f=2}}}=171\,\textrm{MeV}\,,\qquad T_{c,{\tiny{N_f=2+1}}}
  =156\,\textrm{MeV}\,. 
\end{align}
The width of the transition is taken to be the full width at 80\% of
the maximum of this susceptibility.  For $N_f =2$, the crossover phase
boundary is given by the long-dashed red line and its width by the
gray band. The corresponding CEP is denoted by the red star. For
$N_f = 2+1$, the short-dashed black line marks the phase boundary and
the blue band its width. The black point shows our result for the CEP
of $N_f = 2+1$ QCD.

At small chemical potential lattice results provide a benchmark test
for the present fRG computations. This includes the curvature of the
phase boundary. It is a sensitive measure for the correct relative
strength of quantum, thermal and density fluctuations. The curvature
coefficient $\kappa$ of the phase boundary at vanishing baryon
chemical potential is the quadratic expansion coefficient of
$T_c(\mu_B)$ around $\mu_B=0$, i.e.,
\begin{align}
  \frac{T_c(\mu_B)}{T_c}
  &=1-\kappa \left(\frac{\mu_B}{T_c}\right)^2+
    \lambda \left(\frac{\mu_B}{T_c}\right)^4+\cdots\,,
 \label{eq:curv}
\end{align}
with $T_c=T_c(\mu_B=0)$. We refer, e.g., to \cite{Fischer:2018sdj} for
additional relevant discussions.  As for the transition temperature,
the curvature of the phase boundary of a crossover depends on its
definition \cite{Pawlowski:2014zaa}. By using the renormalised light
chiral condensate, we ensure comparability with the lattice results in
\cite{Bellwied:2015rza, Bazavov:2018mes}.  To extract the curvature,
we fit our numerical results for $T_c(\mu_B)/T_c$ by the polynomial
given in \eq{eq:curv} for orders 2 and 4 in $\mu_B/T_c$ within the
regions $\mu_B/T \leq 2$ and 3, and read-off the corresponding
quadratic coefficient $\kappa$. Our final result for the curvature
with its numerical error is then given by the mean and the standard
deviation of all these coefficients.
%
%%%%%%%%%%%%%%%%%%%%%%%%%%%%%
%%%%%%%%%%%%%%%%%%%%%%%%%%%%%
%

The curvature for the two-flavour pseudocritical line is determined to
be
\begin{align}
  \kappa_{_{\tiny{N_f=2}}}=0.0176(1)\,.
\end{align}
For the curvature of the 2+1 flavour phase boundary we find
\begin{align}
  \kappa_{_{\tiny{N_f=2+1}}}=0.0142(2)\,.
\end{align}
As has been emphasised before, our results are consistent with recent
$N_f = 2+1$ lattice results, e.g.\ $\kappa=0.0149(21)$ in
\cite{Bellwied:2015rza} and $\kappa=0.015(4)$ in
\cite{Bazavov:2018mes}. Note also that
$\kappa_{_{\tiny{N_f=2}}} \approx \kappa_{_{\tiny{N_f=2+1}}}
(T_{c,{\tiny{N_f=2}}}/T_{c,{\tiny{N_f=2+1}}})^2=0.171$. This entails
that the two dimensionless phase transition lines given by
$T_c(\mu)/T_c(0)$ in \eq{eq:curv} agree for small baryon chemical
potentials. This even holds for a large $\mu_B$-range as can be
already deduced from \Fig{fig:phasediagram}.

All results, including DSE, are summarised in \tab{tab:kappas}. For a
comparison to other results and potential implications for
experimental CEP searches, we also refer to the discussion in the
introduction, \sec{sec:intr}, and in particular
\Fig{fig:phasediagramII}.
\begin{table}[t]
  \begin{center}
  \begin{tabular}{|c || c | c |}
    \hline  & & \\[-2ex]
    \backslashbox{reference}{curvature $\kappa$}  & $N_f=2$
              & $N_f=2+1$  \\[1ex]
    \hline &  & \\[-1ex]
    fRG:  this work   & 0.0176(1) & 0.0142(2)  \\[1ex]
    \hline &  & \\[-1ex]
    lattice: \cite{Bellwied:2015rza}  & 
                                            \noindent\rule{1cm}{0.4pt}
              & 0.0149(21)  \\[1ex]
    \hline &  &  \\[-1ex]
    lattice:  \cite{Bonati:2018nut} 
            & \noindent\rule{1cm}{0.4pt} & 0.0144(26)
    \\[1ex]
    \hline &  &  \\[-1ex]
    lattice:  \cite{Bazavov:2018mes} 
            & \noindent\rule{1cm}{0.4pt} & 0.015(4)
    \\[1ex]\hline &  &  \\[-1ex]
    DSE:  \cite{Fischer:2014ata} 
            & \noindent\rule{1cm}{0.4pt} & 0.0238
    \\[1ex] 
    \hline && \\[-1ex]
    DSE:  \cite{Gao:2015kea} 
            & \noindent\rule{1cm}{0.4pt} & 0.038
    \\[1ex] \hline &  &  \\[-1ex]
    lattice:  \cite{Allton:2002zi} 
            & 0.0078(39) 
              & \noindent\rule{1cm}{0.4pt}   \\[1ex] \hline &  &  \\[-1ex]
    lattice:  \cite{deForcrand:2002hgr} 
            & 0.0056(6) 
              & \noindent\rule{1cm}{0.4pt}  \\[1ex]
    \hline 
  \end{tabular}
  \caption{Curvature coefficients $\kappa$, see \eq{eq:curv}: from the
    fRG: present work; Lattice collaborations: \cite{Bellwied:2015rza}
    (WB), \cite{Bonati:2018nut} (Bonati {\it et al.}),
    \cite{Bazavov:2018mes} (hotQCD), \cite{Allton:2002zi} (Allton {\it et al.}), 
    \cite{deForcrand:2002hgr} (Forcrand and Philipsen), Lattice overviews 
    \cite{Philipsen:2007rj,DElia:2018fjp}; DSE: \cite{Fischer:2014ata} (Fischer {\it et
      al.}), \cite{Gao:2015kea} (Gao {\it et al.}), DSE overview
    \cite{Fischer:2018sdj}.}
  \label{tab:kappas}
 \end{center}
\end{table}
At finite baryon chemical potential lattice simulations are hampered
by the sign problem, resulting in increasing statistical and
systematic errors with increasing $\mu_B$. The former are shown in
\Fig{fig:phasediagramII}, the latter are unknown. The respective
validity bounds are indicated by the dashed lines at $\mu_B/T=2$ and
$\mu_B/T=3$ in \Fig{fig:phasediagram}. For $\mu_B/T$ close to the
crossover line and larger than $\mu_B/T=3$ also the approximations
used in the present work start to lose reliability: In the presence of
the nontrivial mesonic dispersions, the phase structure can be
altered, cf.\ \sec{sec:Inhom}.  Accordingly, results in this regime,
including our result for the location of the CEP, have to be
interpreted with care.

We find CEPs for $N_f=2$ and $N_f=2+1$ flavour QCD at
\begin{align}\label{eq:CEP2}
  (T_{_{\tiny{\text{CEP}}}},{\mu_B}_{_{\tiny{\text{CEP}}}})_{_{\tiny{N_f=2}}}
  =(117, 630)\,\textrm{MeV}\,,
\end{align}
and
\begin{align}\label{eq:CEP2+1}
  (T_{_{\tiny{\text{CEP}}}},{\mu_B}_{_{\tiny{\text{CEP}}}})_{_{\tiny{N_f=2+1}}}
  =(107, 635)\,\textrm{MeV}\,,
\end{align}
respectively. This amounts to rather large values of
${\mu_B}_{_{\tiny{\text{CEP}}}}/T_{_{\tiny{\text{CEP}}}}$ of
\begin{align}\nonumber 
  N_f=2:
  &\   \frac{{\mu_B}_{_{\tiny{\text{CEP}}}}}{T_{_{\tiny{\text{CEP}}}}}
    = 5.38\,,\\[1ex]
  N_f=2+1:
  &\  \frac{{\mu_B}_{_{\tiny{\text{CEP}}}}}{T_{_{\tiny{\text{CEP}}}}}
    = 5.93\,. 
\end{align}
Our results for the CEP are extracted from the onset of critical
scaling on the chiral phase boundary. As discussed in \sec{sec:EffPot}
and \app{app:flowV}, we expand the effective potential about its
running minimum. The flow of the minimum is proportional to the
correlation length
$1/\partial_{\bar\sigma}^2 \bar V_k\big|_{\textrm{EoM}}$ (cf.\
\eq{eq:miniflow}), and hence becomes unstable in the critical
region. We use this instability to determine where the system enters
the critical region from below in $\mu_B$. To pin down the location of
the CEP, we exploit the well established fact that, owing to the very
small pion mass and strong mesonic fluctuations close to the CEP, the
critical region is at most a few MeV wide in $\mu_B$; see e.g.\
\cite{Herbst:2010rf}. We emphasise that the critical properties are
solely driven by the quark-meson fluctuations and hence QCD LEFT
investigations are fully applicable here. Accordingly, to very high
accuracy, the CEP is located at the point on the phase boundary where
the running minimum becomes unstable.

We emphasise that the reduced reliability of our results at large
$\mu_B$ is not of conceptual origin, as opposed to the sign problem on
the lattice. It is related to correlation functions that have been
neglected in our truncation of the effective action. Improvements
require the systematic inclusion of additional momentum-dependences of
correlation functions. Moreover, a Fierz complete basis of the
four-quark tensor structures should be taken into account, including
potentially dynamical hadronisation of additional resonant channels.
This is work in progress within the fQCD collaboration \cite{fQCD}.

A physically intriguing source for part of the systematic error in the
present work is related to the indications of an inhomogeneous regime
we discussed in \sec{sec:Inhom}. To highlight the potential effect on
the phase boundary, we show the region where the meson wave function
renormalisation at vanishing spatial momentum is negative,
$Z_\phi(0) < 0$, with the shaded blue area in
\Fig{fig:PhasediagramInhom}. Note that we only scanned the phase
diagram for $\mu_B < {\mu_B}_{_{\tiny{\text{CEP}}}}$. Hence, the
boundary of the shaded blue region at $\mu_B = 635$\,MeV is of no
physical significance. As discussed in \sec{sec:Inhom}, the
possibility of inhomogeneous condensation leads to a competition
between potential resonances in the homogeneous and inhomogeneous
quark-antiquark interaction channels. Regarding the phase structure
itself, this is most relevant when condensation occurs in the first
place. The region where the mesonic dispersion with a minimum at
nonvanishing spatial momentum overlaps with a sizable condensate,
defined in \sec{sec:Inhom} and the caption of
\Fig{fig:PhasediagramInhom}, is given by the hatched red area in this
figure. There, the phase structure could be altered significantly by
possible inhomogeneities. To put his into perspective, we also added a
selection of freeze-out points to \Fig{fig:PhasediagramInhom}. It is
very intriguing that the inhomogeneous regime could be probed in
heavy-ion collisions at small beam energies of roughly
$\sqrt{s} \lesssim 6$\,GeV.

%%%%%%%%%%%%%%%%%%%%%%%%%%%%%%%%%%%%%%%%%%%%%%%%%%%%%%%%%%%
\section{Critical end point and systematic error estimates in
  functional approaches}\label{sec:SysFun}

Here we combine the up-to-date CEP results of functional approaches
for a first estimate of the most likely region of the CEP. To that end
we discuss the respective approximations and the potential effect of
missing fluctuations. A full discussion of the respective systematic
errors is deferred to future work as it goes far beyond the scope of
the present work.

The results for the phase structure in the present fRG approach are
summarised in \Fig{fig:phasediagramII} and \Fig{fig:phasediagram}.
The respective DSE results and an exhaustive discussion can be found
in the recent review \cite{Fischer:2018sdj}, for original work see
e.g.\ \cite{Qin:2010nq, Fischer:2011mz, Fischer:2012vc,
  Fischer:2013eca, Fischer:2014ata, Fischer:2014vxa, Mueller:2015fka,
  Eichmann:2015kfa, Gao:2015kea, Gao:2016qkh}.

It goes without saying that the approximations applied in the
different works differ vastly. However, all DSE works share a common
feature with the present fRG work: The most general quark-gluon
interactions are not fully taken into account and both, momentum
dependencies as well as tensor structures are missing. Typically, this
causes a lack of infrared strength of dynamical chiral symmetry
breaking \cite{Mitter:2014wpa, Cyrol:2017ewj}. This is compensated for
by an infrared enhancement similar in spirit to that described in
\app{app:IR-enhancement}. Accordingly, the systematic error inherent
to these approximations (apart from other error sources) relates to
the strength of this infrared enhancement.

As also discussed in \app{app:IR-enhancement}, such a setup guarantees
the value of the chiral condensate and the correct temperature
dependence of the order parameters at vanishing density or chemical
potential. However, it potentially lacks quantitative precision for
the $\mu_B$-dependence, as the $\mu_B$-dependence of the infrared
enhancement, that is the $\mu_B$-dependence of the missing tensor
structures in the quark-gluon vertex, is not known. On the other hand,
while the explicit $\mu_B$-dependence is only present in the
quark-gluon correlations, the chemical potential dependence of the
gluon propagator has been shown to be very small, see
\fig{fig:inZA}. Accordingly, the respective systematic error of
dropping this dependence that has been used in various works is
presumably very small.

Now we adopt these observations to the phase structure computations
with the DSE. All of them exhibit a larger curvature
$\kappa^\textrm{DSE}$ in comparison to the lattice results and the fRG
results of the present work, which are compatible. The larger
curvature at low density implies a stronger $\mu_B$-dependence of,
e.g., order parameters, which suggests -in a linear error propagation-
also a CEP at too small chemical potential and temperature. Indeed,
the DSE curvatures and positions of the CEPs can be ordered
accordingly, see \Fig{fig:phasediagramII}.

Let us now concentrate on \cite{Fischer:2014ata}. In this work, as in
the present one, the phase boundary is well described by only the
leading contribution proportional to $\kappa$ in
\eq{eq:curv}. Assuming that all physical effects influence the
$\mu_B$-dependence only linearly, we can rewrite the curvature term of
the DSE phase boundary as follows:
\begin{align}
  \kappa_\textrm{\tiny{DSE}}\0{\mu_{B,\textrm{\tiny{DSE}}}^2}{T_c^2}=
  \kappa_\textrm{\tiny{fRG}}  \0{\mu_B^2}{T_c^2}\,,\quad \textrm{where}
  \quad \mu_{B} = \sqrt{\0{\kappa_\textrm{\tiny{DSE}}}{
  \kappa_\textrm{\tiny{fRG}} } }\mu_{B,\textrm{\tiny{DSE}}}\,,
\end{align}
and we note that $T_c$ at vanishing chemical potential is practically
the same in both cases.  Using the value
$\kappa_\textrm{\tiny{DSE}}=0.0238$ and
${\mu_{B,\textrm{\tiny{DSE}}_\textrm{\,\tiny{CEP}}}} =488\,$MeV from
\cite{Fischer:2014ata, Fischer:2018sdj}, we are led to
\begin{align}
  { \mu_B}_\textrm{CEP} = 631\,\textrm{MeV} \,,
\end{align}
to be compared with $ { \mu_B}_\textrm{CEP}=635$\, MeV in the present
work, see \eq{eq:CEP2+1}. This is promisingly close. However, we rush
to add that such a `linear' error analysis can be deceptive. For
example, the analysis of the $\mu_B$-dependence of the baryonic
contributions in \cite{Eichmann:2015kfa, Fischer:2018sdj} show that a
smaller $\kappa$ induced from baryonic fluctuations comes with a
smaller $ { \mu_B}_\textrm{CEP}$ in contradistinction to the
discussion here, involving the direct $\mu_B$-dependence in the
quark-gluon vertex. This calls for a more elaborate combined error
analysis in functional approaches, and we hope to report on this in
the near future.

\begin{figure}[t]
  \includegraphics[width=1\columnwidth]{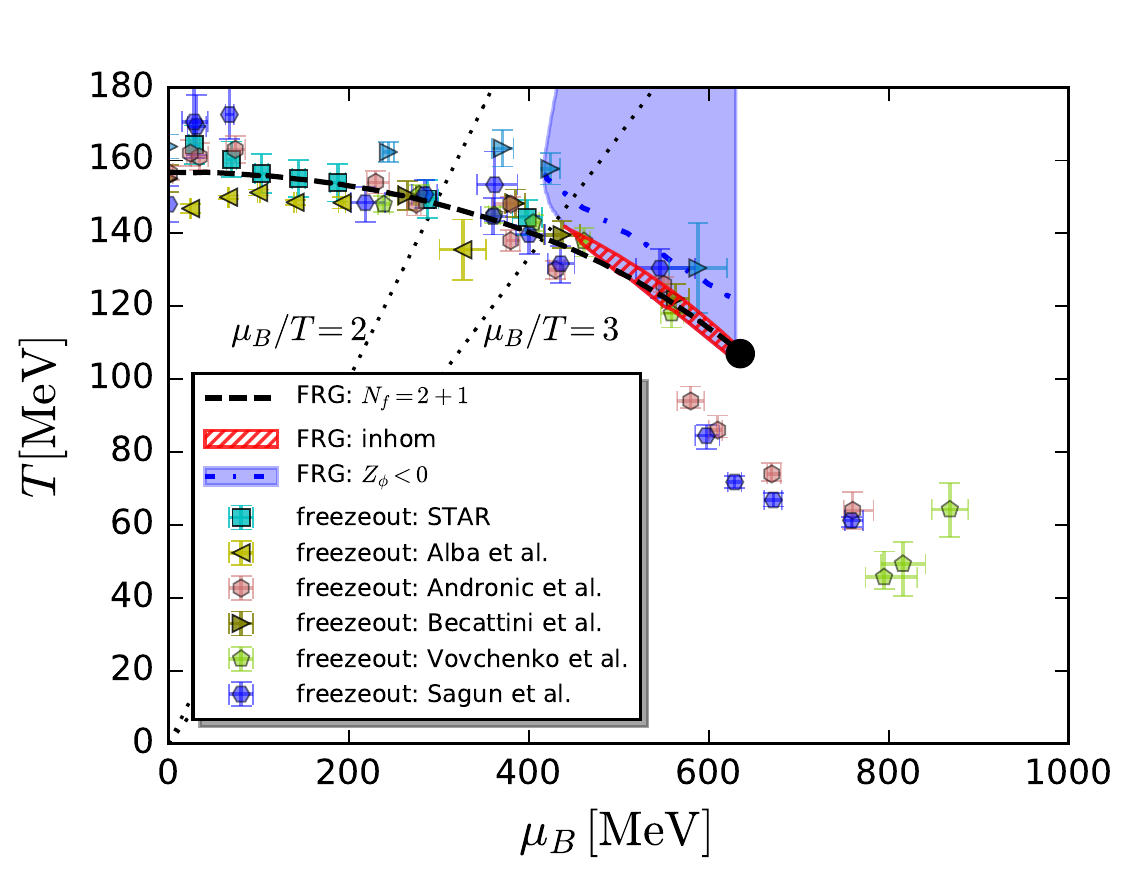}
  \caption{Phase diagram for $N_f=2+1$ flavour QCD in comparison to
    freeze-out data. The crossover temperature has been determined
    through the peak position of the thermal susceptibility of the
    renormalised light chiral condensate $\Delta_{l,R}$, see
    \eq{eq:chiralcondren}, \eq{eq:RenCondSigma}:
    $\partial_T \Delta_{l,R}$ at fixed baryon chemical potential
    $\mu_B$. For more details see \sec{sec:num}.  The phase boundary
    globally agrees well with recent lattice results, and in
    particular the curvature of the phase boundary for small chemical
    potential, \mbox{$\kappa=0.0142(2)$}, is consistent with recent
    lattice results. We find a critical end point at
    $(T_\text{\tiny CEP},{\mu_B}_{\text{\tiny CEP}})=(107, 635)\,\text{MeV}$. \\[0.5ex]
    \textit{fRG: inhom:} The blue area depicts the regime with a
    negative slope of the mesonic dispersion at vanishing spatial
    momentum, that is $Z_\phi(0)<0$, see \fig{fig:Zphip0}. There, the
    the meson dispersion has a minimum at a nonvanishing spatial
    momentum. This is a strong indication for an inhomogeneous regime.
    We also show the minimum of $Z_\phi(0)$ in $T$ at a fixed value of
    $\mu_B$ with the blue dashed line.  The red hatched area shows
    where the inhomogeneous regime overlaps with a sizable homogeneous
    chiral condensate.  The latter is defined as the size of the
    (reduced) chiral condensate where the thermal susceptibility
    $\partial \Delta_{l,R}/\partial T$ at fixed $\mu_B$ reaches 80$\%$
    of its peak height (from above), and larger.  In this region, a
    competition between homogeneous and inhomogeneous condensation is
    expected and the phase structure could be altered significantly.
    For a detailed discussion see \sec{sec:Inhom}.\\[0.5ex]
    \textit{ Freeze-out data}: \cite{Adamczyk:2017iwn} (STAR),
    \cite{Alba:2014eba} (Alba {\it et al.}), \cite{Andronic:2017pug} (
    Andronic {\it et al.}), \cite{Becattini:2016xct} (Becattini {\it
      et al.}), and \cite{Vovchenko:2015idt} (Vovchenko {\it et
      al.}). Note that freeze-out data from Becattini {\it et al.}
    with (blue) and without (dark green) afterburning corrections are
    shown in two different colors.}\label{fig:PhasediagramInhom}
\end{figure}

Still, our preliminary analysis of the systematic error in
state-of-the-art functional approaches allows for a simple estimate of
a region where the critical end point, or more precisely the onset of
a new phase of matter, is most likely. For this combined estimate we
rely on the DSE results from \cite{Fischer:2014ata}, as this
computation has the most complete $\mu_B$-dependence of the gluon
propagator and the quark-gluon vertex to date. This is also reflected
by the fact that it features the smallest curvature
$\kappa_\textrm{DSE}$ of all the DSE computations, as well as the
largest ratio
$ {\mu_B}_{_{\tiny{\text{CEP}}}} / T_{_{\tiny{\text{CEP}}}}$. Now we
correct the DSE curvature by hand with the factor
$\kappa_\textrm{fRG}/\kappa_\textrm{DSE}$. As the above systematic
error analysis for the DSE also suggests that the CEP is at too low
chemical potential, the rescaled CEP from the DSE gives us a lower
estimate on the CEP on the $T_c(\mu_B)$ curve of fRG and lattice with
$\mu_B\approx 480$\, MeV. Moreover, at $\mu_B \gtrsim 500$\,MeV the
potential inhomogeneous regime intersects the chiral transition
region, see \fig{fig:PhasediagramInhom}, and the competition between
homogeneous and inhomogeneous condensation could modify the phase
structure there. Below these values we see no indications for a
critical end point. In combination, this leads us to the estimate
\begin{align}\label{eq:EducatedGuess}
  (135\,,\, 480\, )\,\textrm{MeV}\lesssim
  (  {T}_{_{\tiny{\text{CEP}}}}\,,\,{\mu_B}_{_{\tiny{\text{CEP}}}})
  \lesssim  ( 103\,,\, 660)\,\textrm{MeV} \,, 
\end{align}
where it is indicated with $\sim$ that the upper and lowers bounds are
not strict, as they carry additional systematic errors of our
approximation that are difficult to estimate.  Below the lower bound
we see neither an indication for a CEP nor signals for the failure of
the current approximation. This is also sustained by the
investigations in \cite{Braun:2017srn, Braun:2018bik, Braun:2019a},
where the Fierz-complete basis of four-quark scattering vertex has
been taken into account: Except of a diquark channel at very large
chemical potential, $\mu_B/T\gtrsim 7$, only the scalar-pseudoscalar
channel gives large contributions.

%%%%%%%%%%%%%%%%%%%%%%%%%%%%% 
%%%%%%%%%%%%%%%%%%%%%%%%%%%%%
In turn, for chemical potentials above the regime in
\eq{eq:EducatedGuess} either a CEP has already been observed or the
standard chiral symmetry breaking pattern has broken down.  For the
upper bound on the CEP region we therefore assume that the system does
not enter the critical region in the vicinity of the CEP we find here.
A simple estimate of the phase boundary without a CEP at
$(T_\text{\tiny CEP},{\mu_B}_{\text{\tiny CEP}})=(107,
635)\,\text{MeV}$ is then given by an extrapolation of the boundaries
of the width of the chiral transition in \Fig{fig:phasediagram} from
the crossover region towards larger $\mu_B$. The intersection of these
boundaries then yields the r.h.s.\ of \eq{eq:EducatedGuess}. To be
more specific, we fit the upper and lower bound of the width of the
(2+1)-flavour chiral phase boundary in \Fig{fig:phasediagram} by the
polynomial in \eq{eq:curv} up to fourth order for chemical potentials
$\mu_B \leq 630$\,MeV. The two curves intersect at $(103, 660)$\,MeV.

We emphasise again, that these considerations are only the first step
towards a combined functional analysis of the CEP. Moreover,
\eq{eq:EducatedGuess} should be rather seen as the regime with either
a CEP or interesting new phenomena such as a phase with an
inhomogeneous condensate. As already mentioned in the beginning of
this Section, in particular the present estimate is based on a linear
error analysis in a nonlinear problem and hence has to be taken with
a grain of salt. Despite the obvious deficiencies of the present
analysis this combined functional approach at large chemical
potential, while also utilising benchmark results from lattice
simulations at smaller chemical potential will finally enable us to
provide quantitative predictions for the large density regime
including the potential CEP in the next few years.

%%%%%%%%%%%%%%%%%%%%%%%%%%%%%%%%%%%%%%%%%%%%%%%%%%%%%%%%%%% 

\section{Summary and outlook}
\label{sec:sum}

In this work we have studied the QCD phase structure at finite
temperature and baryon chemical potential as it emerges from the
fundamental dynamics of QCD. To this end, we extended the dynamical
hadronisation technique within the functional renormalisation group
approach to QCD towards the inclusion of in-medium effects. Our
results therefore constitute a new state-of-the-art regarding the
understanding of the QCD phase diagram based on the dynamics of quarks
and gluons.

We have investigated various quantities, including their temperature
and chemical potential dependence.  In particular, we have computed
order parameters for chiral symmetry breaking, e.g.\ the renormalised
chiral condensate, see \fig{fig:DeltalR}, as well as the Polyakov loop
expectation value, see \fig{fig:mf}. In addition, we studied the
constituent quark mass, the pion and $\sigma$-meson masses, the wave
function renormalisations/propagators, the running strong couplings
and the interactions of quarks and the mesonic low energy degrees of
freedom. This allowed us to gain insights into the interplay of
different degrees of freedom at different momentum-, temperature- and
density-scales.
  
Most importantly, we have obtained the phase diagram of $N_f = 2$ and
$N_f = 2 + 1$ flavour QCD at finite temperature and baryon chemical
potential, see \sec{sec:PhaseStructure}. Various aspects of our
results, including comparisons to other recent works, are shown in
Figures \ref{fig:phasediagramII}, \ref{fig:phasediagram} and
\ref{fig:PhasediagramInhom}. Our result for the curvature of the phase
boundary for $2 \!+\! 1$ quark flavours at small chemical potential is
$\kappa_{_{\tiny{N_f=2+1}}}=0.0142(2)$, which is fully compatible with
the most recent lattice results \cite{Bellwied:2015rza,
  Bazavov:2018mes}. In addition, we find a critical end point at
$(T_{_{\tiny{\text{CEP}}}},{\mu_B}_{_{\tiny{\text{CEP}}}})=(107,
635)$\,MeV.  We compare our results for $N_f = 2 + 1$ and $N_f = 2$ in
\fig{fig:phasediagram}. Owing to strange quark fluctuations, the
(2+1)-flavour phase boundary is systematically lower than the
two-flavour boundary.

An intriguing aspect of our results is that we find strong indications
for an inhomogeneous regime at large chemical potentials in the region
$\mu_B\gtrsim 420$\,MeV, shown in Figures \ref{fig:phasediagramII} and
\ref{fig:PhasediagramInhom}. Since this regime intersects the phase
boundary, it could have important implications for the QCD phase
structure. For a detailed discussion see \sec{sec:Inhom}.

A preliminary analysis of the systematic error in our computation is
done in \sec{sec:SysFun}. In combination with a similar analysis of
the DSE as well as with the benchmark results from the lattice at
small chemical potential, we put forward a first suggestion for the
region of the critical end point, see \eq{eq:EducatedGuess}. This
combination of results from functional approaches at large chemical
potential and lattice results at small chemical potential will finally
allow us to pin down the location of CEP or, given potential
inhomogeneous phases and their implications, the onset of a new state
of matter at large density.

This calls for a systematic improvement of the approximations used
here. Firstly, only quark-antiquark scattering in the $\sigma$-$\pi$
channel has been taken into account for the dynamical
hadronisation. Other channels, however, may play a sizable
r$\hat{\textrm{o}}$le in particular at large $\mu_B$, such as diquark
and vector-meson channels, see, e.g., \cite{Braun:2017srn,
  Braun:2018bik}. Therefore, the present approximation for the
four-quark couplings is currently extended to a Fierz-complete basis,
see \cite{Mitter:2014wpa, Braun:2017srn, Braun:2018bik,
  Cyrol:2017ewj}. In addition, we need to allow for inhomogeneous
quark-antiquark scattering channels in order to supplement our present
results in this direction. Secondly, we have employed a minimal
extension to include strange quark dynamics in our calculation. In the
current work we have been concentrating on specific light quark
observables which are insensitive to the details of the strange
sector. However, this is insufficient for the highly interesting
investigations of strangeness at finite chemical potential and
temperature; for an fRG point of view see \cite{Fu:2018qsk,
  Fu:2018swz}. For these aspects a full (2+1)-flavour study, including
the corresponding four-quark interactions and resonances, is
required. Finally, we relied on external input for the
momentum-dependent gluon and ghost propagators in vacuum, the
in-medium screening of gluons in Yang-Mills theory and the gauge part
of the gluon effective potential.  Only the average momentum
dependences encoded in the cutoff dependence have been
considered. Full momentum dependences have so far been considered in
vacuum QCD and finite temperature Yang-Mills theory,
\cite{Mitter:2014wpa, Braun:2014ata, Cyrol:2016tym, Cyrol:2017ewj,
  Cyrol:2017qkl}.  Its extension to QCD at finite temperature and
chemical potential is under way. This also includes a complete basis
of tensor structures for in particular the quark-gluon vertex as well
as the purely gluonic vertices. The selfconsistent determination of
the gluon effective potential along the lines of \cite{Braun:2007bx}
is also on our agenda.

Further interesting applications of the present work are manifold. In
the following, we list selected active projects within the fQCD
collaboration \cite{fQCD}. A detailed study of QCD thermodynamics and
the equation of state is the most obvious application. This will
provide input for phenomenological studies of heavy-ion collisions
that is firmly rooted in QCD even at small beam energies. Since the
chiral limit is easily accessible within the present approach, the
magnetic equation of state is also being computed. Furthermore,
fluctuations and correlations of conserved charges are evaluated. The
present approach allows for both, the spectral reconstruction and the
direct computation of real-time correlation functions in QCD along the
lines of \cite{Floerchinger:2011sc, Strauss:2012dg, Haas:2013hpa,
  Kamikado:2013sia, Silva:2013foa, Christiansen:2014ypa,
  Pawlowski:2015mia, Jung:2016yxl, Pawlowski:2017gxj, Cyrol:2018xeq,
  Tripolt:2018qvi, Dudal:2019gvn, Binosi:2019ecz, Li:2019hyv}, that
are important for QCD-assisted transport, \cite{Bluhm:2018qkf}, and
QCD-assisted hydrodynamics, \cite{Dubla:2018czx}. These applications
play important r$\hat{\textrm{o}}$les in experimental searches for the
CEP, and more generally for our understanding of the QCD phase
structure. We hope to report on these matters in the near future.
%%%%%%%%%%%%%%%%%%%%%%%%%%%%%%%%%%%%%%%%%%%%%%%%%%%%%%%%%%%

\begin{acknowledgments}
  We thank R.~Alkofer, J.~Braun, C.~Fischer, J.~Papavassiliou,
  R.~Pisarski, J.~Rodriguez-Quintero, B.~J.~Schaefer, S. Valgushev,
  L.~von Smekal, N.~Wink and S.~Zafeiropoulos for discussions. We
  thank the authors of \cite{Zafeiropoulos:2019flq, Boucaud:2018xup}
  for providing us with unpublished lattice data of the $N_f=2+1$
  gluon propagator and a collaboration on related subjects,
  \cite{BDRSZ:2019}. This work is done within the fQCD collaboration
  \cite{fQCD}, and is supported by EMMI, the BMBF grant 05P18VHFCA, by
  the National Natural Science Foundation of China under Contract No.\
  11775041, and is part of and supported by the DFG Collaborative
  Research Centre SFB 1225 (ISOQUANT) as well as by the DFG under
  Germany's Excellence Strategy EXC - 2181/1 - 390900948 (the
  Heidelberg Excellence Cluster STRUCTURES). F.R.\ is supported by the
  DFG through grant RE 4174/1-1.

\end{acknowledgments}

%%%%%%%%%%%%%%%%%%%%%%%%%%%%%%%%%%%%%%%%%%%%%%%%%%%%%%%%%%%%%
%%%%%%%%%%%%%%%%%%%%%%%%%%%%%%%%%%%%%%%%%%%%%%%%%%%%%%%%%%%%%

\appendix
%%%%%%%%%%%%%%%%%%%%%%%%%%%%%%%%%%%
% allow for equations to break between column and pages to avoid
% awkward spacing in the appendix.
%%%%%%%%%%%%%%%%%%%%%%%%%%%%%%%%%%%
\begingroup
\allowdisplaybreaks
%%%%%%%%%%%%%%%%%%%%%%%%%%%%%%%%%%%

%%%%%%%%%%%%%%%%%%%%%%%%%%%%%%%%%%%
%%%%%%%%%%%%%%%%%%%%%%%%%%%%%%%%%%%

\section{Chiral condensates}\label{app:chiralcond}
Here we explain of how to extract chiral condensates within the
present approach. Results in the present approximations for the
\textit{renormalised} chiral condensate, \eq{eq:chiralcondren}, and
the \textit{reduced} chiral condensate, \eq{eq:chiralcondred}, are
presented in \Fig{fig:DeltalR_muB0} and \Fig{fig:Deltals} respectively.

Naturally, these observables are derived from the thermodynamic
potential density or free energy density defined in \eq{eq:thermpo},
evaluated on the EoM \eq{eq:EoMJ} with $J=0$. This is a finite,
renormalisation group invariant observable, as follows from its
relation to the effective action $\Gamma[\Phi_{\textrm{EoM}}]$. The
chiral condensates are proportional to
$\int_x \langle \bar q_i q_i\rangle$ with $q_i=u,d,s$ and vanishes in
the chirally symmetric regime. Up to the normalisation it is given by
the derivative of $\Gamma$ w.r.t.\ the current quark mass $m^0_{q_i}$,
to wit
\begin{align}
  \Delta_{q_i} =
  &\,  m_{q_i}^0\frac{\partial
    \Omega[\Phi_{\textrm{EoM}};T,\mu_q]}{\partial {m^0_{q_i}}}=
    m_{q_i}^0  \frac{T}{\cal V} \int_x \langle \bar q_i(x) q_i(x)\rangle\,,
                 \label{eq:chiralcond}
\end{align} 
where no sum over $i$ is implied in \eq{eq:chiralcond}. The
normalisation in \eq{eq:chiralcond} includes the trivial volume factor
$T/{\cal V}$ with the spatial volume $\cal V$. The multiplication
factor $m_{q_i}^0$ leading to the logarithmic current quark masses
derivative removes possible ambiguities relative to the condensates
from other formulations such as DSEs and the lattice. In the present
work the masses of the light quarks $u,d$ are the same, $m_u=m_d=m_l$,
and we define the light quark condensate
$\Delta_{l}= \Delta_{u}= \Delta_{d}$. Up to renormalisation terms the
condensates $\Delta_{q_i}$ read
\begin{align}\label{eq:chiralcondG}
  \Delta_{q_i}\simeq  - m_{q_i}^0
  T\sum_{n\in\mathbb{Z}} \int \frac{d^3 q}{(2 \pi)^3}
  \tr \,G_{q_i\bar q_i} (q)\,. 
\end{align}
The necessity of the renormalisation of \eq{eq:chiralcondG} is
typically circumvented by subtracting two condensates from each other.
One convenient choice is to only consider the thermal and density part
of the condensates, similar to the definition of the pressure
$p=
-(\Omega[\Phi_\textrm{EoM};\mu_q,T]-\Omega[\Phi_\textrm{EoM};0,0])$. This
leads us to the \textit{renormalised} condensate $\Delta_{q_i,R}$ with
\begin{align}\label{eq:chiralcondren}
  \Delta_{q_i,R} = \frac{1}{{\cal N}_R}\left[\Delta_{q_i}(T,\mu_q) - 
  \Delta_{q_i}(0,0)\right]\,.
\end{align}
Indeed, $\Delta_{q_i,R} = - m^0_{q_i} \partial_{q_i} p$. In
\eq{eq:chiralcondren} the normalisation ${\cal N}_R$ renders
$\Delta_{q_i,R}$ dimensionless.  ${\cal N}_R$ is a convenient
normalisation that is typically chosen to be one of the
characteristic, well-determined, scales in the theory. A common choice
is ${\cal N}_R= m_\pi^4$ for the physical case. If also being
interested in the chiral limit, ${\cal N}_R= f_\pi^4$ is better
suited.

%
%%%%%%%%%%%%%%%%%%%%%%%%%%%%%
\begin{figure}[t]
\includegraphics[width=0.98\columnwidth]{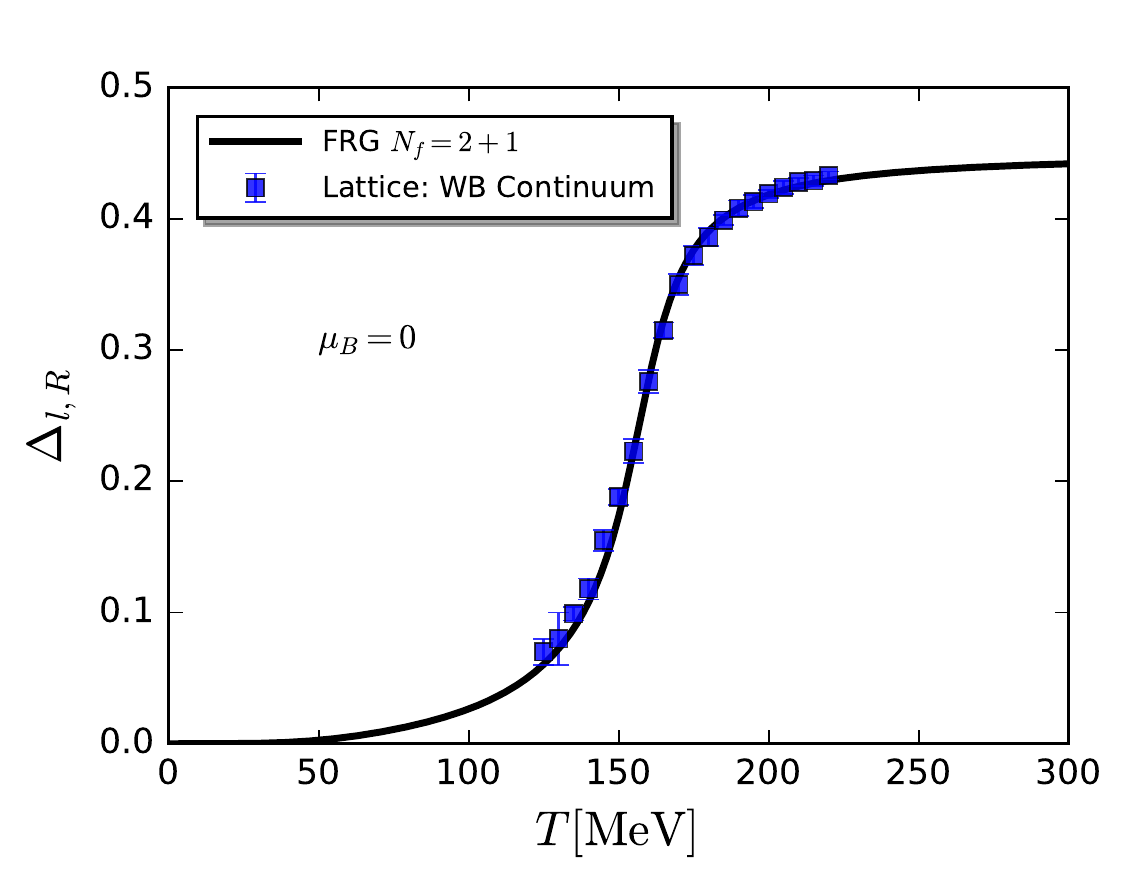}
\caption{$N_f=2+1$ renormalised light chiral condensate
  $\Delta_{l,R}$, in Eqs. (\ref{eq:chiralcondren}) and
  (\ref{eq:RenCondSigma}) as a function of the temperature with
  $\mu_B=0$, in comparison to the lattice result in
  \cite{Borsanyi:2010bp}. The normalisation constant in
  \eq{eq:chiralcondren} and \eq{eq:RenCondSigma} is chosen to match
  the scale in the lattice calculation.}\label{fig:DeltalR_muB0}
\end{figure}
%%%%%%%%%%%%%%%%%%%%%%%%%%%%%
%

Another common choice is the \textit{reduced} condensate
$\Delta_{l,s}$, which is proportional to the weighted difference
between light and strange quark condensate, $\Delta_l$ and
$\Delta_s$. It is normalised by its value in the vacuum and reads
\begin{align}\label{eq:chiralcondred}
  \Delta_{l,s}(T,\mu_q) = \frac{ \Delta_{l}(T,\mu_q) -
  \left(\frac{m_l^0}{m_s^0}
  \right)^2\Delta_s(T,\mu_q)}{
  \Delta_{l}(0,0) - \left(\frac{m_l^0}{m_s^0}\right)^2
  \Delta_s(0,0)}\,.
\end{align}
In the functional RG approach the necessity of renormalising
\eq{eq:chiralcondG} is resolved by the very definition
\eq{eq:chiralcond} to use current quark mass derivatives of the the
basic object in the approach, the {\it finite} effective action
$\Gamma[\Phi_\textrm{EoM}]$: We first use that $\Delta_l$ can be
represented as a $c_\sigma$-derivative,
\begin{align}\label{eq:mDer-cDer}
  \Delta_l =\frac12 m_l^0\frac{\partial
  \Omega[\Phi_{\textrm{EoM}};T,\mu_q]}{\partial {m_l^0}}
  = \frac12 c_\sigma
  \frac{\partial \Omega[\Phi_{\textrm{EoM}};T,\mu_q]}{ \partial {c_\sigma}}\,,
\end{align}
where we have used that $c_\sigma \propto m_l^0$ carries the only
dependence of $\Omega$ on the current quark mass $m_l^0$. This leads
us to the finite chiral condensate
\begin{align}\label{eq:chiralcondSigma}
  \Delta_{l}(T,\mu_q) = -\frac{1}{2} c_\sigma
  \,\sigma_\textrm{EoM}(T,\mu_q) \,, 
\end{align}
which is simply the term in the effective action that explicitly
breaks chiral symmetry. We emphasise that \eq{eq:chiralcondSigma} is
an exact relation. Using \eq{eq:chiralcondSigma} in
\eq{eq:chiralcondren} we arrive at
\begin{align}\label{eq:RenCondSigma}
  \Delta_{l,R}(T,\mu_q) = -\frac{c_\sigma}{{2 \,\cal N}_{R}}
  \left[ \sigma_\textrm{EoM}(T,\mu_q) -
  \sigma_\textrm{EoM}(0,0)\right]\,. 
\end{align}
This is readily computed within the present approach. In
\fig{fig:DeltalR_muB0} we show the results within the current
approximation in comparison to the continuum extrapolated lattice
result in \cite{Borsanyi:2010bp}, for results at finite baryon
chemical potentials see \fig{fig:DeltalR} in
\sec{sec:Conf+Chiral}. The normalisation constant in
\eq{eq:chiralcondren} and \eq{eq:RenCondSigma} is chosen to match the
scale in the lattice calculation.

In an approach with (dynamical) hadronisation also for the strange
quark sector also the reduced condensate can be represented in terms
of the meson field expectation values. Similarly to \eq{eq:mDer-cDer}
we get
\begin{align}\label{eq:mDer-cDer-s}
  \Delta_s =m_s^0\frac{\partial
  \Omega[\Phi_{\textrm{EoM}};T,\mu_q]}{\partial {m_s^0}}
  = c_{\sigma_s}
  \frac{\partial \Omega[\Phi_{\textrm{EoM}};T,
  \mu_q]}{ \partial {c_{\sigma_s}}}\,,
\end{align}
leading to 
\begin{align}\label{eq:CondSigma_s}
  \Delta_{s}(T,\mu_q) = -\frac{1}{\sqrt{2} }c_{\sigma_s}\,
  \sigma_{s,\textrm{EoM}}(T,\mu_q)\,. 
\end{align}
With \eq{eq:chiralcondSigma} and \eq{eq:CondSigma_s} we arrive at 
\begin{align}\label{eq:RedCondSigma}
  \Delta_{l,s}(T,\mu_q) = 
  \frac{\left(\sigma -\sqrt{2}\frac{c_\sigma}{c_{\sigma_s}}
  \sigma_s\right)_{T,\mu_q}}{
  \left(\sigma - \sqrt{2}\frac{c_\sigma}{c_{\sigma_s}}
  \sigma_s\right)_{0,0}}\,,  
\end{align}
where we have used that 
\begin{align}\label{eq:ctomApp}
  \frac{m_l^0}{m^0_s}= \frac{c_\sigma}{c_{\sigma_s}} \,, 
\end{align}
see also \eq{eq:ctom} and \sec{sec:TruncAct} for its derivation. It is
implicitly understood that $\sigma$ and $\sigma_s$ in
\eq{eq:RedCondSigma} are evaluated on the respective EoMs.

In the current work we utilise the simple approximation \eq{eq:sigmas}
for the respective strange condensate.  Using \eq{eq:sigmas} in the
reduced condensate \eq{eq:RedCondSigma} leads us to the simple
expression
\begin{align}\label{eq:RedCondSigmaApprox}
  \Delta_{l,s}(T,\mu_q) = 
  \frac{ \sigma(T,\mu_q) }{ \sigma(0,0) }
  \frac{ 1-\frac{m_l^0}{m_l(T,\mu_q)}}{
  1-\frac{m_l^0}{m_l(0,0)}}\,, 
\end{align}
Notably, the explicit dependence on the current strange quark mass
$m^0_s$ has dropped out. \Eq{eq:RedCondSigmaApprox} also makes evident
that it is even nonvanishing in the flavour-symmetric case with
$m_l^0=m_s^0$ even though both the numerator and the denominator in
\eq{eq:RedCondSigma} vanish.

\Eq{eq:RedCondSigma} still contains explicitly the parameters of
explicit chiral symmetry breaking in the initial action,
$c_\sigma, c_{\sigma_s}$. Now we rewrite \eq{eq:RedCondSigma} in terms
of the full effective action $\Gamma_{k=0}$. To that end we utilise
the equations of motion. For constant solutions
$\sigma_\textrm{EoM}, \sigma_{s,\textrm{EoM}}$ they read
\begin{align}\nonumber 
  \sigma_\textrm{EoM}=
  &\,
    \frac{c_\sigma}{V^{(1,0)}_k(\rho_\textrm{EoM},
    \rho_{s,\textrm{EoM}})}, \\[1ex] 
  \sigma_{s,\textrm{EoM}}=
  &\,\frac{1}{\sqrt{2}} \frac{c_{\sigma_s}}{
    V^{(0,1)}_k(\rho_\textrm{EoM},\rho_{s,\textrm{EoM}})}  \,, 
\label{eq:EoMsigmas}\end{align}
where $V^{(1,0)}_k(\rho,\rho_{s}) =\partial_\rho V_k(\rho,\rho_{s})$
and
$V^{(0,1)}_k(\rho,\rho_{s}) =\partial_{\rho_s} V_k(\rho,\rho_{s})$.
Inserting \eq{eq:EoMsigmas} for $k=0$ in \eq{eq:RedCondSigma} yields
\begin{align}\label{eq:Deltals-pre}
  \Delta_{l,s} = \left.\frac{\left(\frac{1}{m_\pi^2} -
  \frac{1}{ V^{(0,1)}(\rho,\rho_{s})}\right)_{T,\mu_q}}
  {\left(\frac{1}{m_\pi^2} -
  \frac{1}{ V^{(0,1)}(\rho,\rho_{s})}
  \right)_{0,0}}\right|_{\textrm{EoM}}\,,
\end{align}
where the fields $\rho,\rho_s$ are evaluated on the respective EoMs
\eq{eq:EoMsigmas} at finite $T,\mu_q$ and in the
vacuum. \Eq{eq:Deltals-pre} depends on the unrenormalised pion mass
$m_\pi^2=V^{(1,0)}(\rho_\textrm{EoM})$, and the effective potentials
in \eq{eq:Deltals-pre} are evaluated at vanishing cutoff scale,
$V(\rho,\rho_s)=V_{k=0}(\rho,\rho_s)$. \Eq{eq:Deltals-pre} only
depends on the full effective potential at vanishing cutoff scale
$k=0$: the apparent dependence on the parameters of the initial
effective action has been removed. In the chiral limit both the
numerator and the denominator of the reduced condensate vanish as they
should. A convenient reparameterisation of \eq{eq:Deltals-pre} leads
us to
\begin{align}\label{eq:Deltals-final}
  \Delta_{l,s} =\left. \frac{ \sigma(T,\mu_q)}{ \sigma(0,0)} \frac{  1
  -  \left[\frac{ V^{(1,0)}(\rho,\rho_{s}) }{
  V^{(0,1)}(\rho,\rho_{s}) }\right]_{T,\mu_q}}
  {1 - \left[  \frac{ V^{(1,0)}(\rho,\rho_{s}) }{
  V^{(0,1)}(\rho,\rho_{s}) }\right]_{0,0}}\right|_{\textrm{EoM}}\,. 
\end{align}
Its computation only requires the knowledge of the full effective
potential $V(\rho,\rho_s)$ at vanishing cutoff scales at the
$T,\mu_q$-dependent EoMs $\sigma_\textrm{EoM},\sigma_{s,\textrm{EoM}}$:
the $T,\mu_q$-dependent pion and 
masses. The flow equation for the effective potential and the
relevance analysis in \sec{sec:Props+Couplings} and
Appendices~\ref{app:dth}, \ref{app:dtlambda} allows us also to provide
a simple expression for $V(\rho,\rho_s)$. The effective potential is
generated from the $\sigma, \sigma_s$ dependences of the Yukawa terms.
These terms have the full $N_f$-symmetry which is unspoiled by the
flow as neither $c_\sigma$ nor $c_{\sigma_s}$ are present. In the
present approximation we only consider the diagonal terms in the full
potential.  This is consistent as long as the potential is mostly
driven by the quark contribution: This approximation is well justified
for larger temperatures and small baryon chemical potential, but may
fail for larger chemical potential $\mu_B$ and small temperatures.
This is one of the reasons why we do not consider the reduced
condensate as an order parameter for chiral symmetry breaking here, its
reliability at large $\mu_B$ requires a better approximation. In the
present quark-dominated approximation the potential is simply a sum of
the same potential $V_k$ for both $\rho$ and $\rho_s$ separately. This
yields
\begin{align}\label{eq:SumofV}
  V_k(\rho,\rho_s) \approx    V_k(\rho) + \frac12 V_k(2 \rho_s)\,, 
\end{align}
which is a quantitative approximations for cutoff scales
$k\gtrsim 200-300\,$MeV. With \eq{eq:SumofV} we arrive at
\begin{align}\label{eq:Deltals-finalApprox}
  \Delta_{l,s} =\left. \frac{ \sigma(T,\mu_q)}{
  \sigma(0,0)} \frac{  1
  -  \left[\frac{ V'(\rho) }{
  V'(2\rho_{s}) }\right]_{T,\mu_q}}
  {1 - \left[  \frac{ V'(\rho) }{
  V'(2\rho_{s}) }\right]_{0,0}}\right|_{\textrm{EoM}}\,,
\end{align}
\Eq{eq:Deltals-finalApprox} can be evaluated within our current
approximation, see \Fig{fig:Deltals}. Note that the reduced
condensate, $\Delta_{l,s}$, in contradistinction to the renormalised
condensate, $\Delta_{l,R}$, is sensitive to the correct relative
treatment of strange and light quark sector. In the current work we
have used a simple approximation for the strange quark sector as none
of our observables is sensitive to the subleading differences between
light quark and strange quark sector. Put differently, we may use the
reduced condensate for fixing the strange current quark mass $m_s^0$
or $c_{\sigma_s}$, alternatively to the ratio of the decay constants
$f_K/f_\pi$. Given the large value of the constituent strange quark
mass in the Landau gauge, $m_s^0\gtrsim 500$\, MeV, see e.g.\
\cite{Fischer:2018sdj}, we expect $m_s^0\gtrsim 150$\, MeV for this
adjustment, for a detailed discussion see \sec{sec:TruncAct} around
\eq{eq:ctom}. There we also referred to the $N_f=2+1$ flavour ratio
from lattice simulations, $f_K/f_\pi \approx 27$, which is obtained
here for
\begin{align}
  \Delta \bar m_{sl} = 150\, \textrm{MeV}\,,\quad
   c_{\sigma_s}= 97.5\,{\textrm{GeV}}^3\,.
\end{align}
Note also that $c_{\sigma_s}$ and hence the ratio varies rapidly with
$\Delta \bar m_{sl}$ from current quark mass ratios
$m_s^0/m_l^0=c_{\sigma_s}/c_\sigma = 14,27,30,34$ for
$\Delta \bar m_{sl}=120,150,155,160$\,MeV. While this has a minimal
impact on the reduced condensate $\Delta_{l,s}$ at small temperatures
this variation changes the temperature-dependence of $\Delta_{l,s}$ at
large temperature. This is clearly seen in \fig{fig:Deltals}. Note
that while this strange quark mass variation has no impact on the
light quark and gluon correlation functions and observables, it of
course is relevant for observables and correlations with strange quark
content. In particular we see quantitative agreement of the reduced
condensate for all temperatures studied here for current quark ratios
close to the physical one.
%
%%
%%%%%%%%%%%%%%%%%%%%%%%%%%%%%
\begin{figure}[t]
\includegraphics[width=0.95\columnwidth]{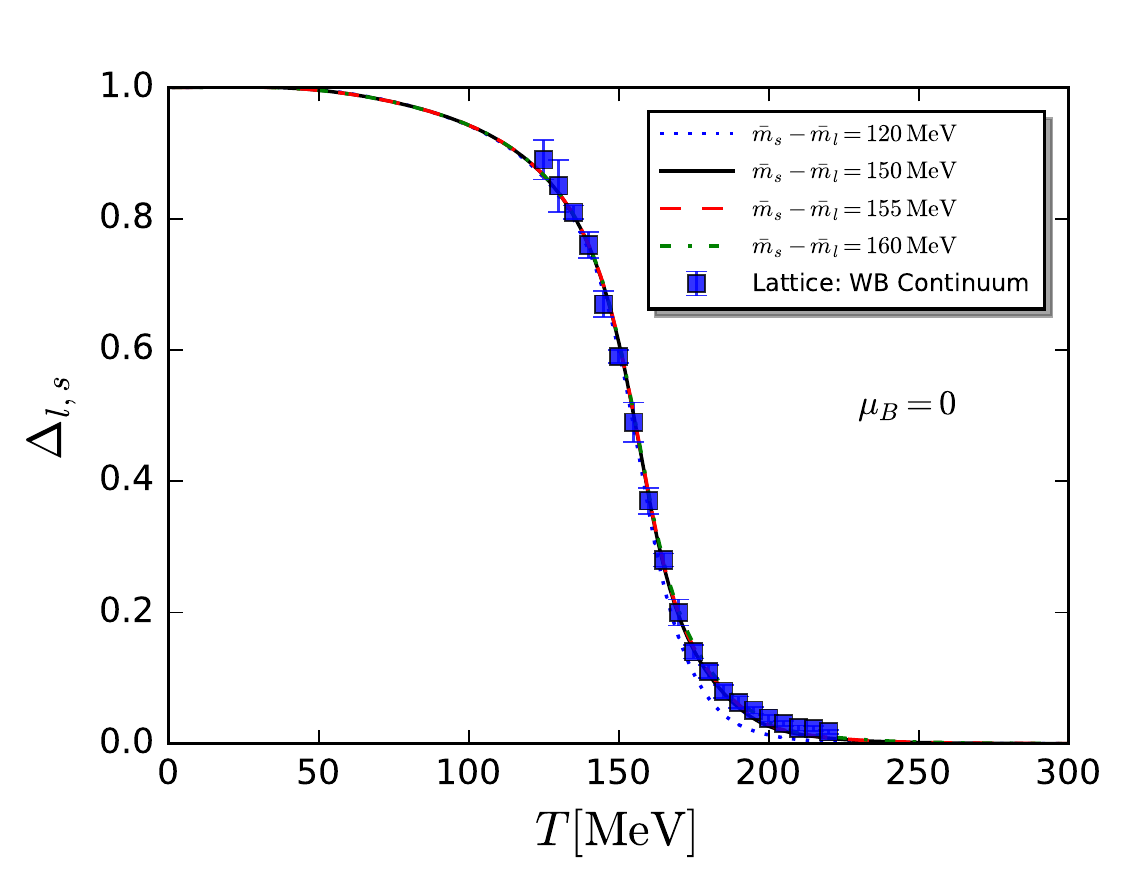}
\caption{The reduced condensate in the approximation
  \eq{eq:Deltals-finalApprox} for the constituent quark mass
  difference
  $\Delta\bar m_{sl}=\bar m_s-\bar m_l=120,150,155,160$\,MeV for
  current quark mass ratios of
  $m_s^0/m_l^0=c_{\sigma_s}/c_\sigma \approx 14,27,30,34$ in
  comparison to the lattice results in \cite{Borsanyi:2010bp}. We
  observe quantitative agreement for all temperatures for ratios close
  to the physical one from $N_f=2+1$ flavour lattice simulations,
  $m_s^0/m_l^0\approx 27$ from \cite{Aoki:2013ldr}.}
\label{fig:Deltals}
\end{figure}
%%%%%%%%%%%%%%%%%%%%%%%%%%%%%
%%
%

We close this Appendix with a discussion of possible improvements of
the computation of the condensates (or other observables) utilising
the flow equation of the condensates instead of the their
representation in terms of the effective action/potential at
$k=0$. Such an improvement is based on the fact that in a given
approximation each flow-step generates terms (information) that is
dropped when projecting on the approximation of the effective action
at hand. Accordingly, a flow representation of a specific observables
may keep this information.

Of course, without approximation both ways have to give the same
results, which can be cast in form of an integrability condition,
\begin{align}\label{eq:integrability}
  \left[ \partial_{c_\sigma}\,,\, \partial_t\right]\,
  \Gamma_k[\Phi_\textrm{EoM}] \equiv 0\,,
\end{align}
for more details see \cite{Pawlowski:2005xe, Pawlowski:2015mlf,
  Braun:2018svj}. While being trivial for the full theory,
\eq{eq:integrability} is nontrivial within a given approximation. For
the flow of the chiral condensate $\Delta_{q_i}$ we use the flow
equation for the effective action,
\begin{align}\nonumber 
  \partial_t \Delta_{q_i} =&\, m^0_{q_i}\frac{\partial}{\partial
                             m^0_{q_i}}\partial_t \Gamma[
                             \Phi_\textrm{EoM}]\\[1ex] 
  = &\,\frac12 m^0_{q_i}\frac{\partial}{\partial
      m^0_{q_i}}  \Bigl( \Tr\, G_k[
      \Phi_\textrm{EoM}]\,\partial_t R_k\Bigr)\,.
\label{eq:FlowDelta}\end{align}
Note that all the dynamical hadronisation terms in the flow vanish on
the equation of motion within our choice as
$\langle \partial_t \hat{\phi}_k\rangle_{\textrm{EoM} }=0$. Seemingly,
\eq{eq:FlowDelta} carries explicit and implicit dependences on the
current quark mass. The implicit ones are that of the
$\Phi_{\textrm{EoM}}$ and that of the couplings. The latter
dependences can be computed with the same recursive relations as has
been put forward in \cite{Fu:2015naa} for the
$\mu_q$-dependence. However, \eq{eq:FlowDelta} can be simplified with
\eq{eq:mDer-cDer}, and we write for the light quark condensate
\begin{align}\nonumber 
  \partial_t \Delta_{l} = &\,\frac12 \,c_\sigma
                            \partial_{c_\sigma} \partial_t
                            \Gamma[\Phi_{\textrm{EoM}}]
  \\[1ex]\nonumber 
  =&\,\frac14 \, c_\sigma
     \partial_{c_\sigma}  \Bigl(\Tr\,G_k[\Phi_\textrm{EoM}]
     \,\partial_t R_k\Bigr)\\[1ex]
  =&\,\frac14 \, c_\sigma
     \int_x \frac{\partial \Phi_{\textrm{EoM}}}
     {\partial {c_\sigma}}
     \frac{\delta}{\delta \Phi}
     \left. \Bigl(\Tr\,G_k[\Phi] \,\partial_t R_k
     \Bigr)\right|_{\Phi_\textrm{EoM}} \,,
\label{eq:FlowDeltalc}\end{align}
where we have used that the diagrams depends on $c_\sigma$ only via
the equations of motion $\Phi_\textrm{EoM}$. As in \eq{eq:FlowDelta}
all the dynamical hadronisation terms in the flow vanish on the
equation of motion within our choice as
$\langle \partial_t \hat{\phi}_k\rangle_{\textrm{EoM} }=0$. For the
strange quark condensate we get
\begin{align} 
  \partial_t \Delta_{s} 
  =\,\frac12 \, c_{\sigma_s}
  \int_x \frac{\partial \Phi_{\textrm{EoM}}}{
  \partial {c_{\sigma_s}}} \frac{
  \delta}{\delta \Phi}
  \left. \Bigl(\Tr\,G_k[\Phi] \,\partial_t R_k
  \Bigr)\right|_{\Phi_\textrm{EoM}} \,, 
\label{eq:FlowDeltasc}\end{align}
The only nonvanishing solutions of the EoM are $\sigma_\textrm{EoM}$,
$\sigma_{s, \textrm{EoM}}$ and
$L_\textrm{EoM}, \bar L_{\textrm{EoM}}$. In a further approximation we
only consider the direct contributions and neglect the subleading
dependencies of $\sigma_s, L, \bar L$ on $c_\sigma$ via the change of
$\sigma_\textrm{EoM}$, as well as that of $\sigma, L, \bar L$ on
$c_{\sigma_s}$ via the change of $\sigma_{s,\textrm{EoM}}$,
\begin{align}\label{eq:subleading-dc}
  \frac{ \partial (\sigma_s,L,\bar L)_\textrm{EoM} }{
  \partial c_\sigma}
  \approx 0\approx 
  \frac{ \partial (\sigma,L,\bar L)_\textrm{EoM} }{
  \partial c_{\sigma_s}}\,.
\end{align}
In the current approximation this leads us to
\begin{align}\label{eq:dtDeltalcsimple}
  \partial_t \Delta_l =\frac{c_\sigma}{2}
  \frac{\partial\sigma_{\textrm{EoM}}}{\partial {c_\sigma}}\,
  \partial_\sigma 
  \left[  \frac12 \, \Tr \,
  G_{\phi\phi}\,\partial_t R_\phi  - \Tr \, G_{q\bar q}\,
  \partial_t R_q \right]\,,
\end{align}
and
\begin{align}\label{eq:dtDeltascsimple}
  \partial_t \Delta_s = \, c_{\sigma_s}
  \frac{\partial\sigma_{s,\textrm{EoM}}}{\partial c_{\sigma_s}}\,
  \partial_{\sigma_s} 
  \left[  \frac12 \, \Tr \,
  G_{\phi\phi}\,\partial_t R_\phi - \Tr \, G_{q\bar q}\,\partial_t R_q\right]\,,
\end{align}
For the explicit result we first concentrate on the light
condensate. Within the approximation \eq{eq:subleading-dc} we get
\begin{align}\label{eq:dcsigma}
  c_{\sigma} \frac{\partial  \Gamma^{(1)}_{\sigma}[
  \Phi_\textrm{EoM}]}{\partial c_\sigma} =
  \Gamma^{(2)}_{\sigma \sigma}[\Phi_\textrm{EoM}]
  \,c_\sigma \frac{\partial
  \sigma_{\textrm{EoM}}}{\partial {c_\sigma}}  -
  c_\sigma  =0 \,, 
\end{align}
\Eq{eq:dcsigma} leads us to
\begin{align}\label{eq:dc-sigma}
  c_\sigma  \frac{\partial \sigma_{\textrm{EoM}}}{\partial {c_\sigma}} =
  \frac{ c_\sigma}{m_\sigma^2 }\,,\quad \textrm{with} \quad m_\sigma^2
  = \left. \partial_\sigma^2 V_k\right|_{\sigma_\textrm{EoM}}\,. 
\end{align}
Collecting all the ingredients this leads us to
\begin{align}\nonumber 
  \partial_t  \Delta_l  =
  &\,  \frac12 \frac{c_\sigma}{m^2_\sigma}
    \Biggl[ \frac{h_k}{2} 
    \Tr G_{q\bar q}^2 \,\partial_t R_q \\[1ex]
  &\, - \frac{1}{2}\Tr\,\left(  G_{\sigma\sigma}^2
    V^{(3)}_{\sigma\sigma\sigma } +
    G_{\pi\pi}^2 V^{(3)}_{\sigma\pi\pi} \right)
    \,\partial_t R_\phi\Biggr]\,.
\label{eq:FlowDeltalfinalc}\end{align}
Note that the expression in the square bracket is nothing but
$\partial_\sigma \dot V_k$ and $m_\sigma^2$ is the second
$\sigma$-derivative of $V_k$, see \eq{eq:dc-sigma}. Now we use that
the flow of $\sigma_{\textrm{EoM},k}$ follows straightforwardly from
the cutoff dependence EoM $\partial_\sigma V_k=c_\sigma$. The
$t$-derivative of the EoM reads
\begin{subequations}\label{eq:dtsigmaEoM}
\begin{align}
  & \partial_t \sigma_{\textrm{EoM},k} V^{(2)}_k
    + \partial_t V^{(1)}_k= 0\,,
\end{align}
which can be solved for the flow of $\sigma_{\textrm{EoM},k}$, 
\begin{align}\label{eq:miniflow}
  & \to \partial_t \sigma_{\textrm{EoM},k} = -\left.
    \frac{ \dot{V}_k^{(1)} }{{V_k^{(2)}}}\right|_{\sigma=
    \sigma_{\textrm{EoM},k} }\,. 
\end{align}
\end{subequations}
Accordingly, \eq{eq:FlowDeltalfinalc} is nothing but the flow of
$\sigma_{\textrm{EoM},k}$ up to a prefactor $-c_\sigma$. This follows
already from \eq{eq:chiralcondSigma}, to wit
\begin{align}\label{eq:FlowDeltal}
  \frac{1}{c_\sigma} \partial_t  \Delta_{l} =
  -\frac{1}{2}\partial_t \sigma_{\textrm{EoM},k}\,,  
\end{align}
which proves that \eq{eq:integrability} holds nontrivially in the
present approximation. Note also that at large cutoff scales the term
in the second line of \eq{eq:FlowDeltalfinalc} from the mesonic loop
drops quickly proportional to $\sigma_{\textrm{EoM}}$.  The prefactor
of the first term approaches the current quark mass
$\frac{c_\sigma\,h_k }{m^2_\sigma} \to 2 m_l^0$. Hence, in this limit
\eq{eq:FlowDeltalfinalc} reduces to the explicit $m_l^0$-derivatives
in \eq{eq:FlowDelta}.  In turn, for smaller cutoff scales
\eq{eq:FlowDeltalfinalc} carries also the implicit ones in
\eq{eq:FlowDelta}.
%%%%%%%%%%%%%%%%%%%%%%%%%%%%%%%%%%%
%%%%%%%%%%%%%%%%%%%%%%%%%%%%%%%%%%%

%%%%%%%%%%%%%%%%%%%%%%%%%%%%%%%%%%%
%%%%%%%%%%%%%%%%%%%%%%%%%%%%%%%%%%%
\section{Dynamical hadronisation}
\label{app:DynHad-c}

Here we carefully derive the hadronisation relations for the given
hadronisation \eq{eq:HadABC}, which is recalled here,
\begin{align}\label{eq:HadABC-app}
  \langle   \partial_t \hat \phi_k\rangle=
  &\dot{A}_k \,\bar{q}\tau q+\dot{B}_k
    \,\phi + \frac1{\sqrt{2 N_f}} \dot{C}_k \hat e_\sigma\,,
\end{align}
with $\hat e_\sigma$ is a conveniently normalised vector in the
$\sigma$-direction, $\hat e_\sigma^2=1/\sqrt{2 N_f}$. Now we use the
shift in the $\sigma$-direction, $\dot {C}_k(\Phi) \, \hat e_\sigma$,
for absorbing the current quark mass of up- and down-quarks into the
mesonic field. Here we allow a field dependence of the coefficient
$\dot {C}_k$. Then, the scalar quark two-point function
\begin{align}\label{eq:scalarG2}
  (\bar q \tau^0 q) = (\bar u u + \bar d d)/2\,,
\end{align}
is proportional to $\sigma$ and its symmetry part vanishes for all
momenta $p$ and cutoff scales $k$,
\begin{align}\label{eq:m_cur=0}
  \Gamma^{(2)}_{(\bar q\tau^0 q)}[\sigma](p)+
  \Gamma^{(2)}_{(\bar q\tau^0 q)}[-\sigma](p)\equiv 0\,. 
\end{align}
This entails, that the flow of the symmetric part vanishes, 
\begin{align}\label{eq:d_t-m_cur=0}
  \partial_t \Gamma^{(2)}_{(\bar q\tau^0 q)}[\sigma](p)+\partial_t
  \Gamma^{(2)}_{(\bar q\tau^0 q)}[-\sigma](p)\equiv 0\,. 
\end{align}
Moreover, it also leads to vanishing flow diagrams,  
\begin{align}\label{eq:diag-m_cur=0}
  \textrm{Flow}^{(2)}_{(\bar q\tau^0 q)}[\sigma](p)+
  \textrm{Flow}^{(2)}_{(\bar q\tau^0 q)}[-\sigma](p)\equiv 0\,. 
\end{align}
For the determination of $\dot{C}_k$ we evaluate the flow equation for
the scalar part of the quark two point function, derived from
\eq{eq:FlowQCD} with the choice \eq{eq:HadABC-app}. It reads
\begin{align}
  \partial_t \Gamma_{(\bar q\tau^0 q)}^{(2)}+
  \dot{A}_k\, \left( \frac{\delta \Gamma}{\delta \sigma} +
  c_\sigma\right) + \dot{C}_k\,
  \Gamma^{(3)}_{(\bar q \tau^0q) \sigma}=
  \textrm{Flow}^{(2)}_{(\bar q\tau^0 q)}
  \,,\label{eq:Gbarqq}
\end{align}
where the field-dependence of all correlation functions has been
suppressed, and the subscript ${}_{(\bar q q)}$ indicates the
projection on the scalar part of the quark two-point function
$\bar q q = 2 \bar q \tau^0 q$. Upon symmetrising \eq{eq:Gbarqq} the
flow of the quark two-point function,
$\partial_t \Gamma_{(\bar q\tau^0 q)}^{(2)}$ and the diagrams,
$ \textrm{Flow}^{(2)}_{(\bar q\tau^0 q)}$, drop out due to
\eq{eq:d_t-m_cur=0}, \eq{eq:diag-m_cur=0}. This leaves us with
\begin{align}\label{eq:dotC-det}
  \dot{C}_k= - {\dot{A}_k}\, \frac{ 1}{
  \Gamma^{(3)}_{(\bar q \tau^0 q) \sigma}[\sigma]}\,\frac{
  \Gamma^{(1)}_\sigma[\sigma]+\Gamma^{(1)}_\sigma[
  -\sigma]+2 c_\sigma}{2} 
  \,, 
\end{align}
where we have used that \eq{eq:m_cur=0} leads us to
$ \Gamma^{(3)}_{(\bar q \tau^0 q) \sigma}[-\sigma]=
\Gamma^{(3)}_{(\bar q \tau^0 q) \sigma}[\sigma]$.  In the chiral limit
the numerator in \eq{eq:dotC-det} vanishes and $\dot{C}_k\equiv 0$ as
expected.  For finite current quark masses \eq{eq:dotC-det} the
symmetric sum
$ \Gamma^{(1)}_\sigma[\sigma]+\Gamma^{(1)}_\sigma[-\sigma] = - 2
c_\sigma$. Hence, we finally conclude
\begin{align}\label{eq:C_k=0}
\dot{C}_k\equiv 0\,. 
\end{align}
This is the expected property following from the symmetry constraints
on the $\sigma$-dependence we have imposed in \eq{eq:m_cur=0}, see
also the discussion in \sec{sec:FlowQCD}. With \eq{eq:C_k=0} we are
led to
\begin{align}
  \partial_t \Gamma_{(\bar q\tau^0 q)}^{(2)}+
  \dot{A}_k\, \left( \frac{\delta \Gamma}{\delta \sigma}
  +c_\sigma\right) =
  \textrm{Flow}^{(2)}_{(\bar q\tau^0 q)}
  \,,\label{eq:Gbarqqfinal}
\end{align}
As \eq{eq:Gbarqqfinal} depends on $\Gamma^{(1)}_\sigma$, we also study
the respective flow in order to discuss the selfconsistence of the
current expansion schemes as well as the optimisation of the
approximation \eq{eq:action} used in the present work. The flow of
$\Gamma^{(1)}_\sigma$ at $(q,\bar q)=0$ reads
\begin{align}
 \partial_t \Gamma_{\sigma}^{(1)}=
  \textrm{Flow}^{(1)}_{\sigma}
  \,,\label{eq:dotV-inserted}
\end{align}
and is that in the chiral limit: it does not depend on $c_\sigma$, see
\sec{eq:GammaSigmaSym}. Accordingly the dynamical hadronisation put
forward here carries the full chiral symmetry at each cutoff scale due
to the shift in $J_\sigma$. Explicit chiral symmetry breaking is
entirely carried by the linear term in $\sigma$.  Inserting
\eq{eq:action} in \eq{eq:Gbarqqfinal}, we arrive at
\begin{align}
  \partial_t h_k(\rho)  + 
  m_\pi^2(\rho) \,\dot{A}  = \frac1\sigma
  \textrm{Flow}^{(2)}_{(\bar q \tau^0 q)}\,,
  \label{eq:Gbarqqexplicitfinal}
\end{align}
where $m_\pi^2(\rho)=\partial_\rho V(\rho)$. In terms of dimensionless
variables and after rescaling with the wave function renormalisations
we are led to \eq{eq:hflow}.

In summary we are left with the two key equations
\eq{eq:dotV-inserted} and \eq{eq:Gbarqqexplicitfinal}. In particular
the latter one, \eq{eq:Gbarqqexplicitfinal}, relates the
$\rho$-dependence of the pion mass function $m_\pi^2(\rho)$ to that of
the Yukawa coupling. This entails that the field-dependence of the
Yukawa coupling is in one-to-one correspondence to that of the
effective potential, see also \cite{Pawlowski:2014zaa}. Since we do
not take into account the $\rho$-dependence of $h_k(\rho)$, the
respective systematic error is minimised for an expansion point in the
Taylor expansion in $\rho$, that minimises the higher Taylor
coefficients of $m_\pi^2(\rho)$:
\begin{align}\label{eq:min-rho}
  \min_{\rho} |\partial_\rho^n m_\pi^2(\rho)| = \min_{\rho}|
  \partial_\rho^{n+1} V(\rho)|\,.
\end{align}
The Taylor coefficients in \eq{eq:min-rho} are minimised at the
flowing minimum, the $k$-dependent EoM for
$\kappa_k = \sigma^2_{\textrm{EoM},k}/2$. Note however that this
expansion point does not lead to best convergence for the flow of the
effective potential, \eq{eq:dotV-inserted}. The latter flow shows best
convergence for a fixed expansion point close to the final minimum
$\sigma_{\textrm{EoM}}=\sigma_{\textrm{Eom},k=0}$. Accordingly, for
small cutoff scales a fixed expansion point fares better than the
running one in terms of convergence. It is suggestive to switch of the
running of the renormalised expansion point at
\begin{align}
  \left.  \partial_t \bar \kappa_k\right|_{k_\textrm{peak}} =0\,,
  \quad \textrm{with} \quad &\bar\kappa_k =
                                                             \bar \rho_{\textrm{EoM},k} \,,
\end{align}
see \eq{eq:EoMBarRho} in \sec{sec:EffPot}. However, if the frozen
expansion point is too far away from the physical configuration
$\rho_{\textrm{EoM}}=\bar \rho_{\textrm{EoM},k=0}\leq \bar \rho_{
  \textrm{EoM}, k_{\textrm{peak}}}$, the latter may not be in the
radius of convergence of the Taylor expansion. This is checked by
stopping the flow of the minimum later and checking the stability of
the results under this procedure. This subtlety can be avoided
completely by a field-dependent Yukawa-coupling (and further
field-dependent couplings such as the wave function renormalisations),
see e.g.\ \cite{Pawlowski:2014zaa}. This will be studied in future
work.
%%%%%%%%%%%%%%%%%%%%%%%%%%%%%%%%%%%
%%%%%%%%%%%%%%%%%%%%%%%%%%%%%%%%%%%

%%%%%%%%%%%%%%%%%%%%%%%%%%%%%%%%%%%%%%%%%%%%%%%%%%%%%%%%%%%%%
%%%%%%%%%%%%%%%%%%%%%%%%%%%%%%%%%%%%%%%%%%%%%%%%%%%%%%%%%%%%%

\section{Strange quark and the minimal extension to $N_f=2+1$}
\label{app:Nf2p1}

In this work we adopt an minimal approach to extend the calculation of
$N_f=2$ to that of $N_f=2+1$. First of all, the contribution of the
strange quark loop to the gluon anomalous dimension is included, i.e.,
another term denoted by $\eta_{A,k}^{s}$ is added in \eq{eq:etaAT},
which reads
 \begin{align}\label{eq:etaATNf2p1}
   \eta_{A,k}=\eta_{A,\textrm{vac}}^{\text{QCD}} + \Delta\eta_{A}^{\text{YM}}
   +\Delta\eta_{A}^{l} + \eta_{A}^{s}\,.
\end{align}
The in-medium gluon vacuum polarisation generated by the light quarks,
$\Delta\eta_{A}^{l}$, is given by the in-medium part of
\eq{eq:DeltaAqexpl} and \eq{eq:eq:DeltaAqexplMed} with
$\bar m_q = \bar m_l$, $N_f = 2$ and the light-quark--gluon coupling,
$g_{\bar l A l}$, defined in \eq{eq:dtg} and \eq{eq:dtglexp}. For the
strange quark contribution, $\eta_{A}^{s}$, we use \eq{eq:DeltaAqexpl}
directly with $\bar m_q = \bar m_s$, defined in \eq{eq:ml+s_const},
$N_f = 1$ and the the strange-quark--gluon coupling, $g_{\bar s A s}$,
defined below. Note that $\eta_{A}^{s}$ not only contains the thermal
contribution, as does $\Delta\eta_{A}^{q}$, but also includes the
vacuum contribution, since our input for the unquenched QCD
calculation in vacuum, i.e.\ the first term on the r.h.s.\ of
\eq{eq:etaATNf2p1}, in only for two flavours \cite{Cyrol:2017ewj}.

Furthermore, the quark-gluon coupling for the strange quark is
distinguished from that for the $u$- and $d$-quarks, and its flow
equation is given by
\begin{align}
  \partial_t g_{\bar{s}As}=&\left( \frac12\eta_A+\eta_q\right)
                             g_{\bar{s}As}  +
                             \overline{\textrm{Flow}}^{(3),A}_{(\bar s A s)}\,. 
 \label{eq:dtgs}
\end{align}
where the last term is defined in \eq{eq:dtgA}. In comparison to
\eq{eq:dtg}, we note that in \eq{eq:dtgs} we have neglected the
contribution of mesonic fluctuations to the running of
$g_{\bar{s}As}$. This will be improved in our future work. All
explicit expressions for the quantities discussed here are given in
\app{app:gluon} and \ref{app:dtqg}.

%%%%%%%%%%%%%%%%%%%%%%%%%%%%%%%%%%%%%%%%%%%%%%%%%%%%%%%%%%%%%
%%%%%%%%%%%%%%%%%%%%%%%%%%%%%%%%%%%%%%%%%%%%%%%%%%%%%%%%%%%%%

%%%%%%%%%%%%%%%%%%%%%%%%%%%%%%%%%%%
%%%%%%%%%%%%%%%%%%%%%%%%%%%%%%%%%%%

\section{Direct sum versus full back coupling}\label{app:YMtoQCD}

Here, we motivate the approximation we use for the gauge field
propagator. The goal is to find a simple, yet quantitatively reliable
scheme. This discussion is largely based on \cite{Braun:2014ata,
  Rennecke:2015} and reiterated here for completeness. The most
prominent feature in the gauge sector is confinement. In linear
covariant gauges it manifests itself in the emergence of a gluon mass
gap, cf.\ \eq{eq:mA}, together with an enhanced ghost propagator in
the IR \cite{Gribov:1977wm, Kugo:1979gm, Zwanziger:2003cf}. This
implies that in order to capture the mass gap, the full momentum
dependence of the gauge field propagators has to be resolved. This is
a formidable task which requires sophisticated truncation schemes and
considerable numerical effort, e.g.\ \cite{Fischer:2008uz,
  Aguilar:2008xm, Boucaud:2008ky, Fister:2011uw, Mitter:2014wpa,
  Cyrol:2016tym, Cyrol:2017qkl}. Here, we motivate an approximation
scheme which uses the full gauge field propagators from
fist-principles computations of the pure gauge theory as input, while
matter effects, i.e.\ quark and hadron fluctuations, and their
backreaction on the gauge sector, are computed selfconsistently.

Within the truncation scheme described in \sec{sec:TruncG}, the
propagators enter the flow equations through the corresponding
anomalous dimensions and mass parameters. The gluon anomalous
dimension, for instance, is in general a function of the strong
couplings and the mass gap,
\begin{align}
  \eta_{A,k}^\text{QCD} = \eta_{A,k}^\text{QCD}(\alpha_{\bar q A q},
  \alpha_{\bar c A c},\alpha_{A^3},\alpha_{A^4}, \bar m_{A}^2, \bar m_{q}^2)\,,
\end{align}
where we omitted the $k$-dependence on the couplings and masses for
the sake of brevity. Within both, the fRG and DSE approaches this can
always be split into a part where only diagrams of the pure gauge
theory contribute,
$\eta^\textrm{glue}(\alpha_{\bar c A c},\alpha_{A^3},\alpha_{A^4},
\bar m_{A}^2)$, and the matter contribution,
$\eta^\textrm{quark}_{A}(\alpha_{\bar q A q}, \bar m_{q}^2)$. We note
that the couplings themselves are the ones of QCD, i.e.\ the couplings
in the gauge sector also depend on the ones in the matter sector and
vice versa. Hence, the splitting of the gluon anomalous dimension into
gauge and matter part is purely formal. A first simplification can be
made by using that the different gauge couplings only differ from each
other in the low-energy regime where the mass gap is generated, i.e.\
where gluons decouple in QCD. Hence, we parametrise the anomalous
dimension as
\begin{align}\label{eq:etasum}
  \eta_{A,k}^\text{QCD} = \eta^{\textrm{glue}}_{A}(\alpha_s,\bar m_{A}^2) +
  \eta^\textrm{quark}_{A}(\alpha_{\bar q A q}, \bar m_{q}^2)\,,
\end{align}
where $\alpha_s$ stands for $\alpha_{c \bar c A }$, $\alpha_{A^3}$ or
$\alpha_{A^4}$. Note that the gluon mass gap is part of the gauge
contribution. As discussed in \sec{sec:TruncG}, the glue diagrams of
the anomalous dimension in Yang-Mills theory and in QCD agree, and it
can be parameterised in terms of the couplings $\alpha_s$ and the mass
gap $m_A^2$. Accordingly, the differences between the glue parts in
QCD and Yang-Mills theory come from the differences in $\alpha_s$ and
$m_A^2$. We have
\begin{align}\nonumber 
  \left.\eta_A^{\textrm{glue}}\right|_{\textrm{YM}}=
  &\,
    \eta_{A,k}^\textrm{YM}= \eta_A^{\textrm{glue}}
    \left(\alpha^{\textrm{YM}},
    (\bar m^{\textrm{YM}}_{A})^2\right)\,, \\[1ex]
  \left.\eta_A^{\textrm{glue}}\right|_{\textrm{QCD}}=
  &\,
    \eta_A^{\textrm{glue}}\left(\alpha_s,
    \bar m_{A}^2\right)\,, 
\label{eq:eta(alpha)-App}\end{align}
where $\alpha_s$ and $m_A^2$ are the QCD couplings and mass gap
respectively. This suggest a possible approximation scheme for the
gluon propagator: One takes the Yang-Mills anomalous dimension
$\eta_{A}^\text{YM}$ in \eq{eq:eta(alpha)-App} as an external input
from a first-principles computation, computes $\alpha_s$,
$\bar m_{A}^{2}$ and
$\eta_{A}^\textrm{quark}(\alpha_{\bar q A q}, \bar m_{q}^2)$ as
functions of the RG scale $k$ and puts everything together according
to \eq{eq:etasum}.

%%
%%%%%%%%%%%%%%%%%%%%%%%%%%%%%
\begin{figure}[t]
\includegraphics[width=0.98\columnwidth]{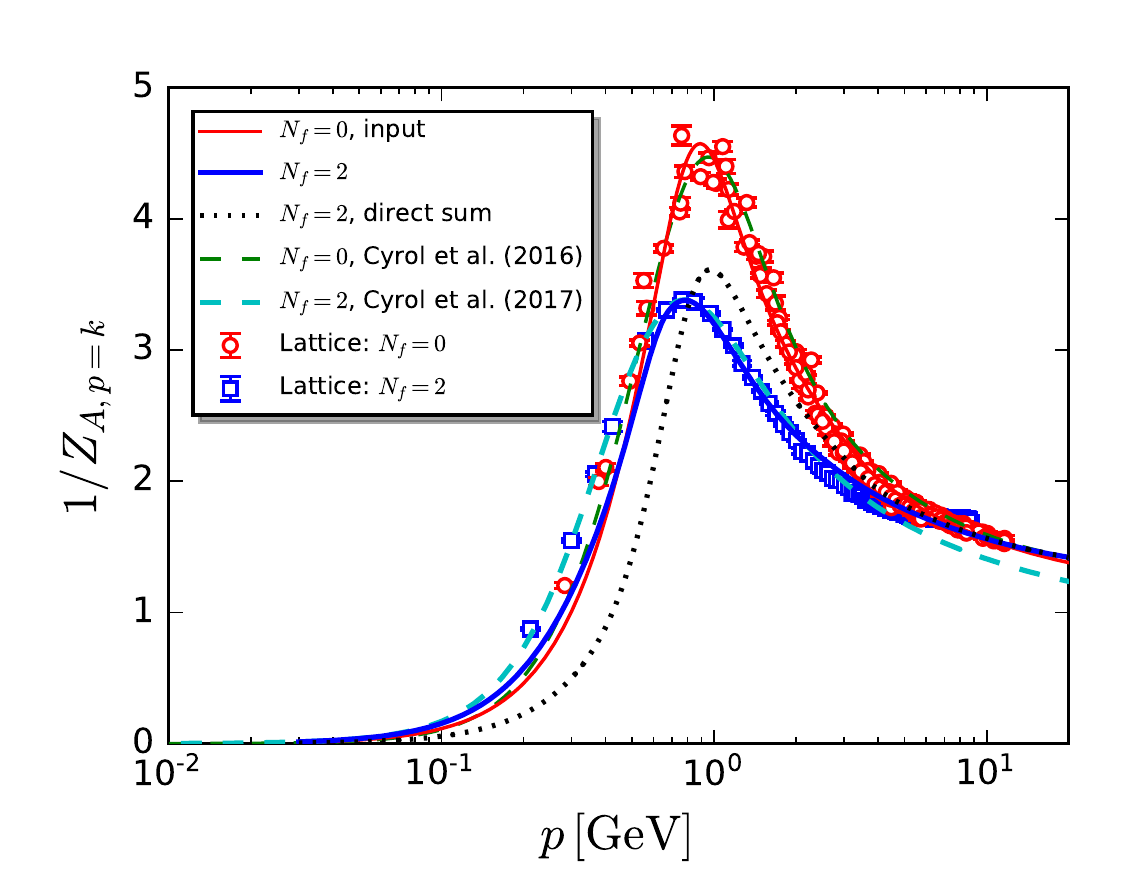}
\caption{Comparison between the $N_f=2$ propagators obtained from
  lattice gauge theory and fRG results at vanishing temperature and
  density. The solid red thin line shows the gluon propagator of the
  pure gauge theory we use as input. The open red circles are the
  corresponding lattice results \cite{Sternbeck:2006cg}. The blue open
  squares are the lattice results for two-flavour QCD
  \cite{Sternbeck:2012qs}. The solid blue thick line shows the fRG
  result using \eq{eq:etasum} and the dotted black line shows the
  result based on \eq{eq:etasumcheap}. Both results are taken from
  \cite{Braun:2014ata, Rennecke:2015}. Furthermore, results from the
  fRG calculations with the state-of-the-art truncations at vacuum for
  the pure Yang-Mills theory \cite{Cyrol:2016tym} and $N_f=2$
  unquenched QCD \cite{Cyrol:2017ewj} are shown for comparison, which
  are denoted by the green and cyan dashed lines, respectively.}
\label{fig:gluepropcomp}
\end{figure}
%%%%%%%%%%%%%%%%%%%%%%%%%%%%%
%%

The advantage of this procedure is that the information about the full
momentum dependence, and in particular the emergent mass gap, is
encoded in the quantitatively reliable external input. The corrections
from matter fluctuations as well as their feedback on the gauge
contribution are computed `locally', i.e.\ the full RG scale
dependence of the correlation functions is resolved, but only one
momentum configuration (typically vanishing momenta) is taken into
account. A simple reduction of this procedure is to ignore the
feedback from the matter sector onto the gauge sector and simply
define
\begin{align}\label{eq:etasumcheap}
  \eta_{A,k}^\text{QCD} = \eta_{A,k}^\text{YM}+
  \eta_{A,k}^\textrm{quark}(\alpha_{\bar q A q}, \bar m_{q}^2)\,.
\end{align}
This amounts to only adding the gluon vacuum polarisation to the input
from YM.

To check the quality of both approximations we compare the resulting
gluon propagators to results from lattice gauge theory
\cite{Sternbeck:2006cg, Sternbeck:2012qs} at vanishing temperature and
chemical potential in \Fig{fig:gluepropcomp}. In this figure we also
show the benchmark fRG calculations with the state-of-the-art
truncation of pure Yang-Mills theory in vacuum, \cite{Cyrol:2016tym},
(thin green dashed line), and $N_f=2$ flavour QCD \cite{Cyrol:2017ewj}
(thick cyan dashed line). The fRG results in the current approximation
have been taken from \cite{Braun:2014ata, Rennecke:2015}: The red line
shows the input propagator from the pure gauge theory. The open
circles are the corresponding lattice data in
\cite{Sternbeck:2006cg}. The open blue squares show the lattice
results for two-flavour QCD \cite{Sternbeck:2012qs}. The solid blue
line shows our results within the approximation based on
\eq{eq:etasum}. We used $\alpha_s = \alpha_{\bar c A c}$.

Indeed, this approximation leads to a perfect agreement with the
corresponding lattice results. To facilitate a comparison between $p$-
and $k$-dependent quantities, we have identified $k=p$. Note that the
momentum scale is arbitrary here. If the feedback to the gauge sector
is ignored, i.e.\ the approximation in \eq{eq:etasumcheap} is used, we
find the dotted black line in \Fig{fig:gluepropcomp}. While it still
is a good approximation to the full propagator, it is clearly
outperformed by \eq{eq:etasum}.

This results motivates the use of a local approximation, i.e.\
resolving only the RG scale dependence of the correlation functions,
for the effects of matter fluctuations, as long as the relevant
nonlocal information, i.e.\ the genuine momentum dependence, is taken
into account through the input. This is crucial for the present work,
since we compute the thermal and density corrections to the gluon
vacuum polarisation in a local approximation directly, while we use
the gluon anomalous dimension of QCD at $T = 0$ as well as the thermal
effects of YM as input, cf.\ \eq{eq:etaAT}.

%%%%%%%%%%%%%%%%%%%%%%%%%%%%%%%%%%%
%%%%%%%%%%%%%%%%%%%%%%%%%%%%%%%%%%%

%%%%%%%%%%%%%%%%%%%%%%%%%%%%%%%%%%%%%%%%%%%%%%%%%%%%%%%%%%%%%
%%%%%%%%%%%%%%%%%%%%%%%%%%%%%%%%%%%%%%%%%%%%%%%%%%%%%%%%%%%%%
\section{Initial conditions \& infrared enhancement}
\label{app:num}

\subsection{Initial conditions} \label{app:initial} The UV cutoff
scale where the flow for the effective action $\Gamma_k[\Phi]$, or
rather that of the correlation functions $\Gamma^{(n)}_k[\Phi]$, is
initiated, is chosen as
\begin{align}
\Lambda = 20\,\text{GeV}\,.
\end{align}
At this scale, only the fundamental couplings of QCD, i.e.\ the strong
coupling and the quark masses, are relevant. We initiate the flow with
a nonvanishing strong coupling $\alpha_s$,
\begin{align}\label{eq:initalalpha}
  \alpha_{\bar q Aq,\Lambda} =  \alpha_{A^3,\Lambda} = \alpha_{A^4,\Lambda}=
  \alpha_{\bar c A c,\Lambda}=\alpha_{s,\Lambda}\,,
\end{align}
such that all avatars of the strong coupling agree at the initial
scale. The initial strong coupling $\alpha_s (\Lambda)$ is constrained
by the requirement of the RG consistency in \eq{eq:RG-consistency},
and we find that \eq{eq:RG-consistency} is satisfied for
\begin{align}\label{eq:initalalpha-values}
  \left. \alpha_{s, \Lambda}\right|_{N_f=2+1}=0.235\,,\qquad \left.
  \alpha_{s, \Lambda}\right|_{N_f=2}=0.21\,.
\end{align}
The only other input at our perturbative initial scale are the current
quark masses. We identify the up and down quark masses, $m_u=m_d=m_q$,
and fix their value by the pion mass in the vacuum to be $m_\pi=138$\,
MeV. Following our discussion in \sec{sec:MatterProps}, we emphasise
that the curvature mass $\bar m_\pi$ of the pion has been shown to be
in very good agreement with the pion pole mass $m_{\pi,\textrm{pol}}$
defined by $\Gamma^{(2)}_{\pi\pi}(p^2 = -m^2_{\pi,\textrm{pol}})=0$,
see \cite{Helmboldt:2014iya, Pawlowski:2017gxj}. This can be traced
back to the mild momentum dependence of the pion correlation function,
\cite{ Helmboldt:2014iya, Mitter:2014wpa, Cyrol:2017ewj} as well as
its relatively small width. For the $\sigma$ resonance this is less
obvious and in particular the width is not expected to be
small. Accordingly the curvature mass $m_\sigma$ may not be close to
the position of the smallest scalar resonance. In the present work it
is computed to be $m_\sigma=485$\, MeV. As discussed in
\sec{sec:TruncG}, the light quark mass is absorbed in a shift in the
$\sigma$-field and simply leads to a linear term in the potential,
$c_\sigma\, \sigma$, see \sec{sec:TruncAct}. We arrive at
\begin{align}\label{eq:ExChiral}
  c_\sigma = 3.6\, \textrm{GeV}^3\,,\quad\; \textrm{with }\quad\; 
  m_{\pi,k=0}= 137\, \textrm{MeV}\,.
\end{align}
Note that $\bar c_{\sigma,k}=c_\sigma/Z_{\phi,k}^{1/2}$ with
$c_\sigma$ being independent of $k$. This leads to
\begin{align}
  \partial_t \bar c_k = \frac{\eta_{\phi,k}}{2}\, \bar c_k\,.\label{}
\end{align}
The strange current quark mass is fixed as $m_{s,\Lambda}=120$\, MeV
by the ratio $f_K/f_\pi$, see the discussion in \sec{sec:TruncAct} and
Table~\ref{tab:InitialValues}.  This amounts to
\begin{align}\label{eq:ExChirals}
  c_{\sigma_s} = 48.8\, \textrm{GeV}^3\,,\quad\; \textrm{with }\quad\; 
  \Delta \bar m_{sl}= 120\, \textrm{MeV}\,.
\end{align}
We have checked that a variation of the difference of the constituent
quark masses
$100\, \textrm{MeV}\lesssim \Delta\bar m_{sl}\lesssim 200$\,MeV does
not influence our results for light quark and gluon correlation
functions and observables.

In the perturbative regime all other couplings are irrelevant, so they
are either vanishing or have irrelevant RG flows. This entails in
particular that the full four-quark coupling
$ \bar \lambda_{q,\textrm{eff}}$ (that on the EoM for $\phi$) should
be infinitesimally small,
\begin{align}\label{eq:lambdaL}
  \bar \lambda_{q,\textrm{eff}} = \frac{\bar h_\Lambda^2}{2
  \bar m_{\phi,\Lambda}^2}\approx 0\,.
\end{align}
where we have used the trivial initial effective potential
\begin{align}\label{eq:Vinitial}
  \bar V_{\Lambda}(\bar \rho)&=\bar m^2_{\phi,\Lambda}
                               \bar \rho\,.
\end{align}
\Eq{eq:lambdaL} fixes the ratio of the meson mass parameter and the
Yukawa coupling. Dynamical hadronisation entails that results should
only depend on the combination \eq{eq:lambdaL}, and not on
$\bar m_{\phi,\Lambda}^2$ and $\bar h_\Lambda$ separately.  Moreover,
if $\bar \lambda_{q,\textrm{eff}} $ is small in comparison to its
flow,
$\bar \lambda_{q,\textrm{eff}} /\partial_t \bar
\lambda_{q,\textrm{eff}} \approx 0$, \eq{eq:lambdaL} holds, and the
results at $k=0$ do not depend on $\bar \lambda_{q,\textrm{eff}} $. We
have checked that this is indeed the case as well as the independence
on $\bar m_{\phi,\Lambda}^2$ and $\bar h_\Lambda$ separately, see also
\cite{Braun:2014ata}. In fact, this behaviour is guaranteed by an
IR-attractive fixed point in the weak-coupling regime
\cite{Gies:2002hq}.

The above scenario holds for initial cutoff scales of
$\Lambda \gtrsim 10 $\, GeV, as we are safely in the perturbative
regime of QCD. Then the above initial conditions should be applicable
with appropriately adjusted values for the RG-consistent strong
coupling. We have investigated the dependence of our results on the
value of $\Lambda$ in this regime. We found that the
$\Lambda$-dependence is negligible.

\subsection{Infrared enhancement}\label{app:IR-enhancement}

In \sec{sec:QuarkGluon} we have discussed a potential infrared
enhancement of the quark-gluon coupling in order to compensate for the
dynamics of the missing nonclassical tensor structures in the present
approximation. We follow the approach proposed in
Ref. \cite{Braun:2014ata}: The potentially missing interaction
strength is compensated for by a phenomenological infrared enhancement
of the vector tensor structure. This is done with the replacement,
\begin{align}\label{eq:IR-fct}
  \partial_{t} \bar{g}_{\bar q q A} \rightarrow
  \bar{g}_{\bar q q A}\, \partial_{t} \varsigma_{a,b}(k)
  +\varsigma_{a,b}(k)  \partial_{t} \bar{g}_{\bar q q A}\,,
\end{align}
with the infrared enhancement function given by
\begin{align}\label{eq:IR-enhance}
  \varsigma_{a,b}(k)&=1+a\frac{(k/b)^\delta}{\exp[ (k/b)^\delta]-1}\,.
\end{align}
This function behaves as $\varsigma_{a,b}(k)\rightarrow 1$ with
$k > b$ and $\varsigma_{a,b}(k)\rightarrow 1+a$ with $k < b$. In this
work $b=$ 2 GeV and $\delta=2$ are chosen. The strength of the
enhancement for $N_f=2+1$, $a=0.034$, and for $N_f=2$, $a = 0.008$, is
fixed by fitting the physical constituent quark mass. Note that this
IR enhancement is only applied for the quark-gluon vertex, and the
three gluon coupling $g_{A^3}$ is not enhanced. This takes into
account that the infrared enhancement takes into account missing
tensor structures, which is relevant for the quark-gluon vertex, but
not for the three-gluon vertex.

\subsection{Scales}\label{app:Scales}
QCD in the chiral limit, that is with vanishing current quark masses
$m_{l,s}^0=0$, has no scale on the classical level, and all observables
can only be measured in the dynamically created scales: The
confinement scale, carried e.g.\ by $\Lambda_\textrm{QCD}$ or the
string tension, and the chiral symmetry breaking scale, carried e.g.\
in the chiral condensate or $f_{\pi,\chi}$.

At the physical point with $m_{l,s}^0\neq 0$ all scales are measured
in the above dynamical scales, for more details and the possible
choice of observables see \sec{sec:TruncAct}. In particular, this
leads us to
\begin{align}\label{eq:fpifpi}
  \frac{  f_{\pi}}{f_{\pi,\chi}} = \frac{93}{88}\,, \qquad\qquad
  \frac{  f_K}{f_{\pi,\chi}} =\frac{111}{88}\,.
\end{align}
Naturally, the scales at the physical point may also be measured in
units of the reduced condensate, $\Lambda_\textrm{QCD}$ or the string
tension. However, the pion decay constant has a direct physical
meaning and hence is preferable.

These scales may also be set in comparison to lattice simulations,
which has been done in \cite{Mitter:2014wpa, Cyrol:2017ewj} for
$N_f=2$ flavour QCD. In the present work we took over the scales from
\cite{Mitter:2014wpa, Cyrol:2017ewj}, fixing the scales at
$\Lambda=20$\, GeV in a direct extension to $N_f=2+1$. RG-consistency,
see \sec{sec:RGconsistency}, then determines the value of the
$N_f=2+1$ flavour strong coupling $\alpha_{s,\Lambda}$ at the initial
scale, providing us with a prediction of the $N_f=2+1$ gluon
propagator, see \fig{fig:inZA_lattice2-2+1}. This prediction is in
quantitative agreement with the continuum-extrapolated lattice data
from \cite{Zafeiropoulos:2019flq, Boucaud:2018xup, BDRSZ:2019} with a
pion mass of $m_\pi = 139$\,MeV.
%
%%%%%%%%%%%%%%%%%%%%%%%%%%%%%
\begin{figure}[t]
\includegraphics[width=0.98\columnwidth]{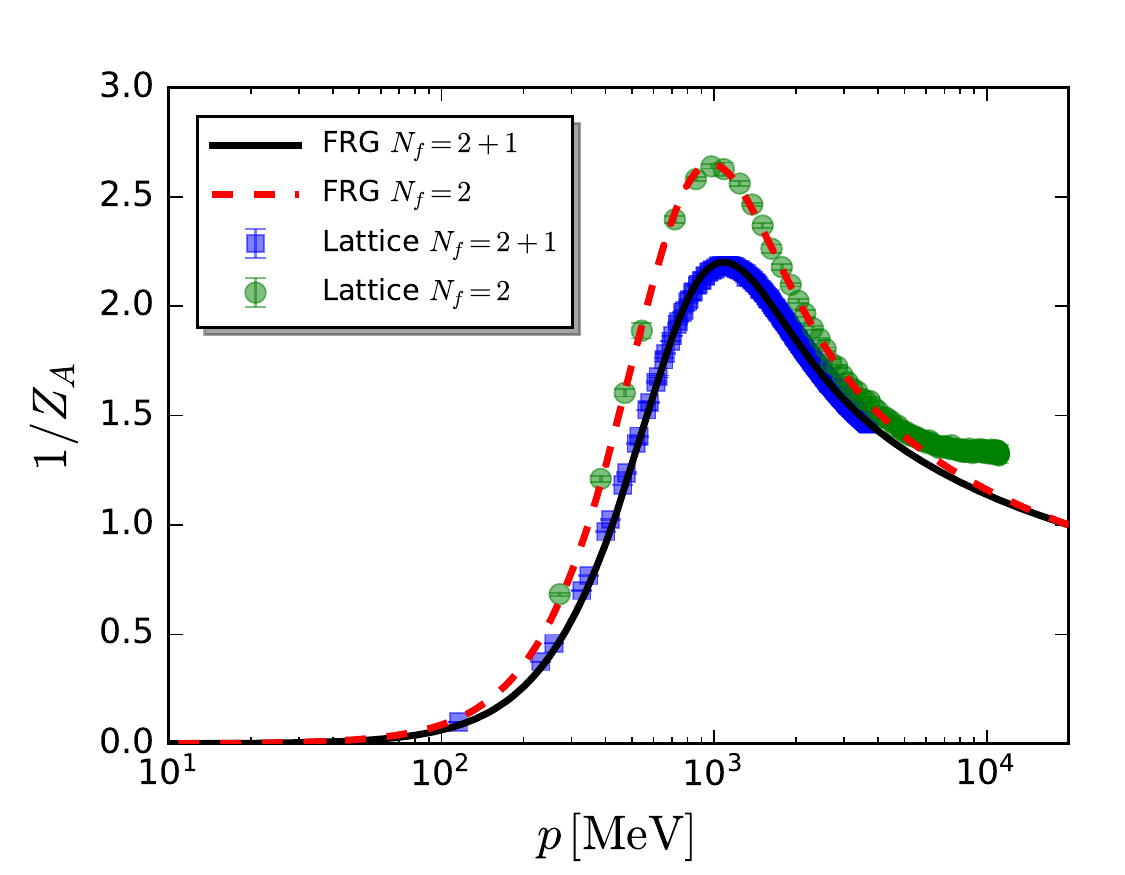}
\caption{$N_f=2$ flavour and $N_f=2+1$ gluon dressing functions
  $1/Z_{A}$ as function of momentum. The $N_f=2$ flavour gluon
  dressing computed with the fRG in \cite{Cyrol:2017ewj} is the input
  in the present work. It is in quantitative agreement with the
  respective lattice results. The lattice data here are taken from
  \cite{Sternbeck:2012qs}. The $N_f=2+1$ flavour gluon dressing shown
  here is a genuine prediction of the present computation. It is shown
  to be in quantitative agreement with the respective lattice results
  \cite{Zafeiropoulos:2019flq, Boucaud:2018xup, BDRSZ:2019}. }
\label{fig:inZA_lattice2-2+1}
\end{figure}
%%%%%%%%%%%%%%%%%%%%%%%%%%%%%
%
This completes our scale setting for the physical point. All
quantities are measured in the respective absolute units.

For the sake of completeness we also provide results for the pion
decay constant. This computation is also interesting in terms of the
predictive power as well as the limitations of the present
approximation. The pion decay constant can be determined from the
momentum dependence of the quark propagator dressings $Z_q(p)$ and
$M_q(p)$, defined in \eq{eq:G2qphi}. These dressings are linked to the
Bethe-Salpether wave function of the pion, for reviews see e.g.\
\cite{Fischer:2006ub, Bashir:2012fs, Eichmann:2016yit},
\begin{align}\label{eq:fpiMq}
  f^2_\pi = \frac{3}{2 \pi^2} \int_0^\infty d p\,  \frac{p^3\, M_q(p) }{\left[ p^2
  +M_q(p)\right]^2}
  \left[M_q(p) - p \frac{\partial M_q(p)}{\partial p}\right]\,.
\end{align}
\Eq{eq:fpiMq} is sensitive to the full momentum dependence of the
quark mass function $M_q(p)$.  This dependence is suppressed in the
flows due to the presence of the regulator term in the
propagators. The letter suppression is the argument for the
momentum-independent approximation in the present work with
$m_{l,k}=M_{q,k}(p=0)$. Note also that the results in
\cite{Cyrol:2017ewj} have been obtained for a fixed expansion point or
background with $\partial_t \kappa=0$ with a four-dimensional flat
regulator. Both properties can be taken account in the present work by
evaluating the mass function on a fixed background
$\kappa=1/2 \sigma_\textrm{EoM}$ at vanishing cutoff $k=0$ as well as
rescaling the cutoff scale $k\to 3/4 k$.  Moreover, we have to rescale
the mass function $M_{q,k}\to M_{q,k} m_{l,0}/M_{q,0}$ in order to
have the same infrared value. This leads us to \fig{fig:Mq-ml2}, which
first of all confirms impressively the validity of our approximation
scheme.
%
%%
%%%%%%%%%%%%%%%%%%%%%%%%%%%%%
\begin{figure}[t]
\includegraphics[width=0.95\columnwidth]{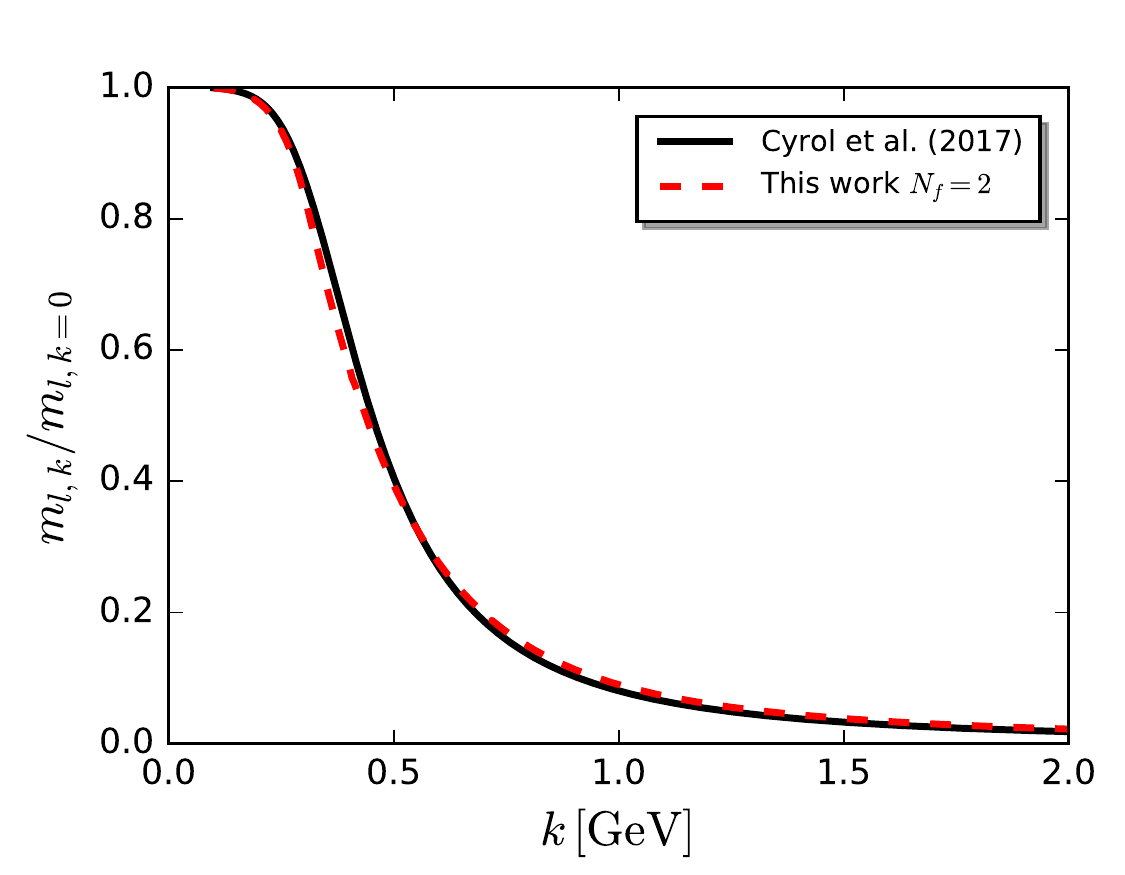}
\caption{Two-flavour QCD quark mass functions from
  \cite{Cyrol:2017ewj} and the present work for a fixed background
  $\sigma=\sigma_{\textrm{EoM}}$ at vanishing cutoff scale. In
  \cite{Cyrol:2017ewj} the full momentum-dependent mass $M_{q,k}(p)$
  function in $N_f=2$ flavour QCD has been computed, and we have
  $m_{l.k}/m_{l.k=0} = M_{q,k}(0)/M_{q,k=0}(0)$. }
\label{fig:Mq-ml2}
\end{figure}
%%%%%%%%%%%%%%%%%%%%%%%%%%%%%
%%
Moreover it allows us to use the rescaled quark mass function
$M_{q,\textrm{res}}(p)$ of \cite{Cyrol:2017ewj} for an estimate of
$f_\pi$ in the present work with \eq{eq:fpiMq}. The rescaled quark
mass function reads
\begin{align}\label{eq:MqEasy2}
  M_{q,\textrm{res}}(p) = M_{q,k=0}(p)\frac{m_{l,k=0}}{M_{q,k=0}(0)}\,, 
\end{align}
and the result for the pion decay constant at the physical point is
\begin{align}\label{eq:fpi2}
  f_\pi \approx 96.0\, \textrm{MeV}\,, 
\end{align}
see also Table~\ref{tab:InitialValues}. In the current approximation
the $N_f=2+1$ flavour mass function has the same cutoff behaviour as
the $N_f=2$ flavour one. This is related to the fact, that the strange
quark feeds into the light quark mass function only indirectly over
the gluon propagator and running coupling $\alpha_{\bar l A l}$ in the
quark-gluon vertex. Rescaling the mass functions with their value at
$k=0$ makes this very apparent, see \fig{fig:ml2-ml2+1}.

We conclude that the agreement of the rescaled $M_{q,k}(0)$ with the
$N_f=2+1$ flavour light quark mass function $m_{l,k}$ works equally
well as for the two-flavour case depicted in \fig{fig:Mq-ml2}. The
respective plot is shown in \fig{fig:Mq-ml2+1}.

Consequently we can use the same estimate as in \eq{eq:MqEasy2} for
the $N_f=2+1$ flavour case. This leads us to
\begin{align}\label{eq:fpi2+1}
  f_\pi \approx 93.0\, \textrm{MeV}\,, 
\end{align}
see also Table~\ref{tab:InitialValues}. Finally we are interested in
the pion decay constant in the chiral limit. The respective light
quark mass function $m_{l,\chi}$ at $k=0$ is given by
$m_{l,\chi} = 319$\, MeV, leading to
\begin{align}\label{eq:fpichi}
f_{\pi,\chi} \approx 88.6\,\textrm{MeV} 
\end{align}
In summary the scale setting in the present work is selfconsistent
with both the confinement scales, here determined via the gluon
dressing function, as well as the chiral symmetry breaking ones, here
determined with the pion decay constant. This is further nontrivial
evidence for the reliability of the present approximation for not too
large chemical potential.

%
%%
%%%%%%%%%%%%%%%%%%%%%%%%%%%%%
\begin{figure}[t]
\includegraphics[width=0.95\columnwidth]{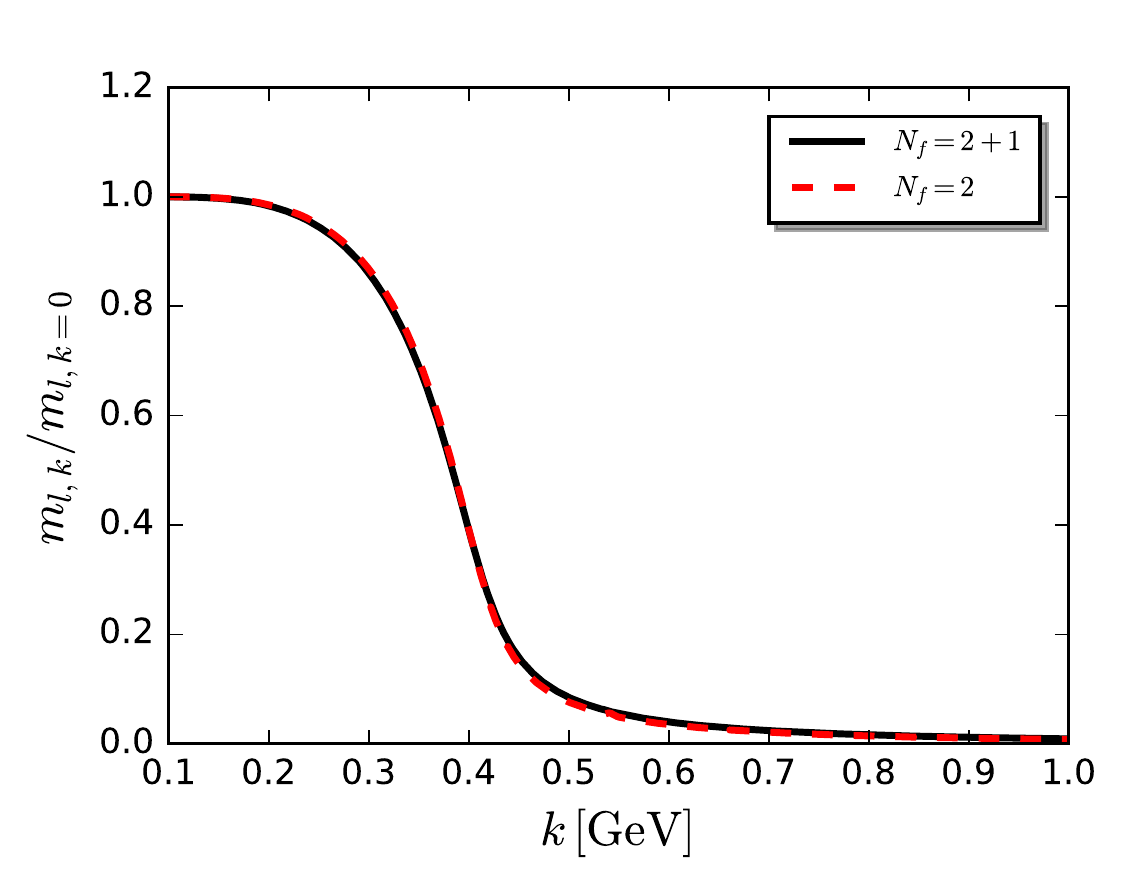}
\caption{$N_f=2$ and $N_f=2+1$ light quark mass functions as a
  function of the cutoff scale $k$, normalised by their value at
  vanishing cutoff scale $k=0$.}
\label{fig:ml2-ml2+1}
\end{figure}
%%%%%%%%%%%%%%%%%%%%%%%%%%%%%
%%
%
%%
%%%%%%%%%%%%%%%%%%%%%%%%%%%%%
\begin{figure}[b]
\includegraphics[width=0.95\columnwidth]{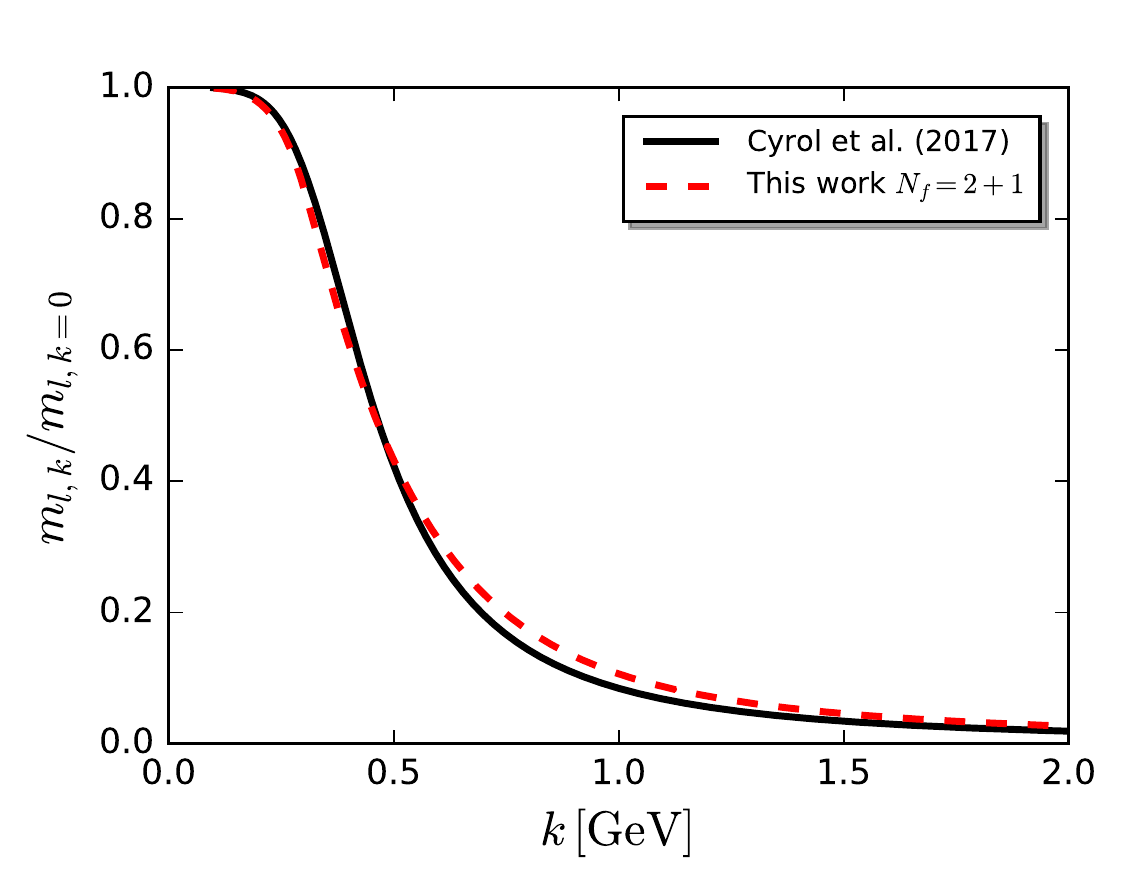}
\caption{$N_f=2$ quark mass function from \cite{Cyrol:2017ewj} and the
  $N_f=2+1$ light quark mass function from present work in the fixed
  expansion. }
\label{fig:Mq-ml2+1}
\end{figure}
%%%%%%%%%%%%%%%%%%%%%%%%%%%%%
%%

%%%%%%%%%%%%%%%%%%%%%%%%%%%%%%%%%%%%%%%%%%%%%%%

\section{Flow of the effective potential}
\label{app:flowV}

Within the approximation discussed in \sec{sec:EffPot}, the effective
potential $V_{\mathrm{mat}, k}(\rho, L,\bar L)$ in \eq{eq:EffPot}
satisfies the flow equation \eq{eq:FlowQCD}, which we recall,
\begin{align}
  \partial_{t}V_{\mathrm{mat}, k}(\rho)
  &=\frac{k^{4}}{4\pi^{2}}\big[(N_{f}^{2}-1)
    l_{0}^{(B,4)}(\tilde{m}_{\pi,k}^{2},\eta_{\phi,k};T)
    \nonumber \\[1ex]
  &\quad+l_{0}^{(B,4)}(\tilde{m}_{\sigma,k}^{2},
    \eta_{\phi,k};T)\nonumber\\[1ex]
  &\quad-4N_{c}N_{f}l_{0}^{(F,4)}(\tilde{m}_{q,k}^{2},
    \eta_{q,k};T,\mu_q)\big]\,,\label{eq:VflowApp}
\end{align}
where the threshold functions $l_{0}^{(B/F,4)}$ are given in
\eq{eq:l0Bd} and \eq{eq:l0Fd} in \app{app:threshold}, and the Polyakov
loops $L,\bar L$ are not shown explicitly for brevity. The right hand
side of \eq{eq:VflowApp} also depends on the dimensionless
renormalised quark and meson masses, i.e.,
\begin{align}\nonumber 
  \tilde{m}_{q,k}^{2}
  &=\frac{h_{k}^{2}\rho}{2k^{2} Z_{q,k}^{2}}\,,
    \quad \quad\tilde{m}_{\pi,k}^{2}=
    \frac{V_{k}^{\prime}(\rho)}{k^{2} \bar Z_{\phi,k}}\,,
  \\[1ex]
  \tilde{m}_{\sigma,k}^{2}
  &=\frac{V_{k}^{\prime}(\rho)+2\rho 
    V_{k}^{\prime\prime}(\rho)}{k^{2}\bar Z_{\phi,k}}\,.
                          \label{eq:DimlessMass} 
\end{align}
Performing a Taylor expansion of the effective potential, one arrives
at
\begin{align}
  V_{\mathrm{mat}, k}(\rho)
  &=\sum_{n=0}^{N_v}\frac{\lambda_{n,k}}{n!}(\rho
    -\kappa_k)^n\,, \label{eq:VTaylor}
\end{align}
where the expansion point $\kappa_k$ can depend on the RG scale. Note
that the expansion coefficients $\lambda_{n,k}$ should not be confused
with the four-fermion coupling $\lambda_{q}$ in \eq{eq:action}. It is
more convenient to work with RG-invariant variables, which are denoted
in this paper by quantities with a bar, except for the strong
couplings in \eq{eq:alphasGluon} and \eq{eq:alphasMatter}, for
example, $\bar\rho= \bar Z_{\phi,k} \rho$,
$\bar\kappa_k= \bar Z_{\phi,k} \kappa_k$, and
$\bar{\lambda}_{n,k}=\lambda_{n,k}/\bar Z_{\phi,k}^n$ with
$\bar Z_\phi$ defined in \eq{eq:barZphi}. The
$\eta^{\ }_{\Phi_i, k}$'s in \eq{eq:VflowApp} are the anomalous
dimensions for respective fields, which are given in \eq{eq:eta}.  We
reformulate the effective potential in terms of RG-invariant variables
through
$V_{\mathrm{mat}, k}(\rho)=\bar V_{\mathrm{mat}, k}(\bar \rho)$, i.e.,
\begin{align}
  \bar V_{\mathrm{mat}, k}(\bar \rho)
  &=\sum_{n=0}^{N_v}\frac{\bar
    \lambda_{n,k}}{n!}(\bar \rho-\bar \kappa_k)^n\,.
                         \label{eq:VbarTaylor}
\end{align}
The expansion point $\bar \kappa_k$ can be chosen freely, provided
that the physics of our concern is within the radius of convergence of
the expansion \cite{Pawlowski:2014zaa}. Expanding about a fixed
background is a convenient choice, see e.g.\ \cite{Pawlowski:2014zaa,
  Braun:2014ata, Fu:2015naa, Rennecke:2016tkm}, where the
unrenormalised expansion point is chosen to be independent of $k$. In
this work we adopt the physical point expansion. In this expansion
$\bar \kappa_k$ is chosen to be the minimum of the effective potential
at each scale $k$, i.e.\ it is given by the solution of the EoM,
\begin{align}
  \frac{\partial}{\partial \bar \rho}\Big(\bar V_{k}(\bar \rho)-\bar c_k
  \bar \sigma \Big)\bigg \vert_{\bar\rho=\bar \kappa_k}&=0\,. \label{eq:Vstat}
\end{align}
With (\ref{eq:VflowApp}) and (\ref{eq:Vstat}), see also
\eq{eq:dtsigmaEoM}, one obtains the flow of the expansion point in
\eq{eq:VbarTaylor}, to wit, 
\begin{align}
  \partial_t \bar \kappa_k=&-\frac{\bar c_k^2}{\bar \lambda_{1,k}^3+
                             \bar c_k^2 \bar \lambda_{2,k}}\bigg[\eta_{\phi,k}
                             \Big(\frac{\bar \lambda_{1,k}}{2}+\bar\kappa_k
                             \bar \lambda_{2,k}\Big)\nonumber \\[1ex] 
                           &+\left. \partial_{\bar\rho} \partial_{t}\big|_{\rho}
                             \bar V_{k}(\bar \rho)\big)\right|_{\bar\rho=\bar \kappa_k}
                             \bigg]\,.\label{}
\end{align}
The flow equations of the expansion coefficients of the effective
potential $\bar\lambda_{n,k}$, which describe $2n$-meson scattering,
are
\begin{align}
  \partial_t \bar\lambda_{n,k}=
  &\partial_{\bar\rho}^n
    \big(\partial_{t}\big|_{\rho}
    \bar V_{k}(\bar \rho)\big)
    \Big |_{\bar\rho=\bar \kappa_k}+
    n\eta_{\phi,k}\bar\lambda_{n,k}\nonumber \\[1ex] 
  &+\big(\partial_t \bar \kappa_k+\eta_{\phi,k}
    \bar\kappa_k\big)\bar\lambda_{n+1,k}\,. \label{}
\end{align}
In this work the maximal order of the Taylor expansion for the
effective potential in \eq{eq:VbarTaylor} is chosen to be $N_v=5$, and
we have checked its convergence for the $\mu_B$-range studied here,
$0\leq \mu_B \lesssim 600$\,MeV.

%%%%%%%%%%%%%%%%%%%%%%%%%%%%%%%%%%%%%%%%%%%%%%%%%%%%%%%%%%%%%

\section{Glue potential}
\label{app:gluepot}

We employ the parameterisation put forward in \cite{Lo:2013hla} for
the glue potential in \eq{eq:Vglue}. The advantage is that, besides
the expectation value of the Polyakov loop and the pressure in
Yang-Mills theory, quadratic fluctuations of the Polyakov loop are
taken into account. For a detailed discussion we refer to
\cite{Fu:2015naa}. The glue potential reads
\begin{align}
  V_\text{glue}(L,\bar{L})
  &= -\frac{a(T)}{2} \bar L L + b(T)
    \ln M_H(L,\bar{L})\nonumber \\[1ex]
  &\quad + \frac{c(T)}{2} (L^3+\bar L^3) + d(T) (\bar{L} L)^2\,,
                             \label{eq:polpot}
\end{align}
with the Haar measure 
\begin{align}
  M_H (L, \bar{L})
  &= 1 -6 \bar{L}L + 4 (L^3+\bar{L}^3) - 3  (\bar{L}L)^2\,.
\end{align}
The temperature dependence of coefficients on the r.h.s.\ of
\eq{eq:polpot} enters through
\begin{align}
  x(T)
  &= \frac{x_1 + x_2/(t_{\text{\tiny{YM}}}+1) + x_3/(t_{\text{\tiny{YM}}}+1)^2}{1 +
    x_4/(t_{\text{\tiny{YM}}}+1) + x_5/(t_{\text{\tiny{YM}}}+1)^2}\,,\label{eq:xT}
\end{align}
for $x\in \{a,c,d\}$, and 
\begin{align}
  b(T )
  &= b_1 (t_{\text{\tiny{YM}}}+1)^{-b_4}\left (1 -e^{b_2/(t_{\text{\tiny{YM}}}+1)^{b_3}} \right)\,.
          \label{eq:bT}
\end{align}
The constants in \ref{eq:xT} and \ref{eq:bT} are taken from
\cite{Lo:2013hla}, and are also collected in \Tab{tab:coeffs} for
convenience. $t_{\text{\tiny{YM}}}$ is related to the reduced
temperature of Yang-Mills theory, and it has been found in
\cite{Pawlowski:2010ht, Haas:2013qwp, Herbst:2013ufa} that unquenching
effects in QCD are well captured through a linear rescaling of the
reduced temperature of the pure gauge theory,
\begin{align}
  t_{\text{\tiny{YM}}}
  &\rightarrow \alpha\,t_{\text{\tiny{glue}}}
    \quad \text{with}\quad t_{\text{\tiny{glue}}}=
    (T-T_c^\text{\tiny{glue}})/T_c^\text{\tiny{glue}},\label{}
\end{align}
where $\alpha=0.57$, $T_c^\text{\tiny{glue}}$=250 MeV for $N_f=2$ and
225 MeV for $N_f=2+1$ are used in this work. For more relevant
details, see the foregoing references.

%
 %%%%%%%%%%%
\begin{table}[tb!]
  \centering
  \begin{tabular}{c||c|c|c|c|c}
     & 1 & 2 & 3 & 4 & 5 \rule{0pt}{2.6ex}\rule[-1.2ex]{0pt}{0pt}\\ \hline\hline
    $a_i$ &-44.14& 151.4 & -90.0677 &2.77173 &3.56403 \\\hline
    $b_i$ &-0.32665 &-82.9823 &3.0 &5.85559  &\\\hline
    $c_i$ &-50.7961 &114.038 &-89.4596 &3.08718 &6.72812\\\hline
    $d_i$ & 27.0885 &-56.0859 &71.2225 &2.9715 &6.61433\\
  \end{tabular}
  \caption{Constants in \eq{eq:xT} and \eq{eq:bT} for the glue
    potential.}
  \label{tab:coeffs}
\end{table}
%%%%%%%%%%%%
%
%%%%%%%%%%%%%%%%%%%%%%%%%%%%%%%%%%%%%%%%%%%%%%%%%%%%%%%%%%%%%
%%%%%%%%%%%%%%%%%%%%%%%%%%%%%%%%%%%%%%%%%%%%%%%%%%%%%%%%%%%%%

\section{Gluon anomalous dimension}
\label{app:gluon}

%
%%%%%%%%%%%%%%%%%%%%%%%%%%%%%
\begin{figure*}[t]
\includegraphics[width=0.98\textwidth]{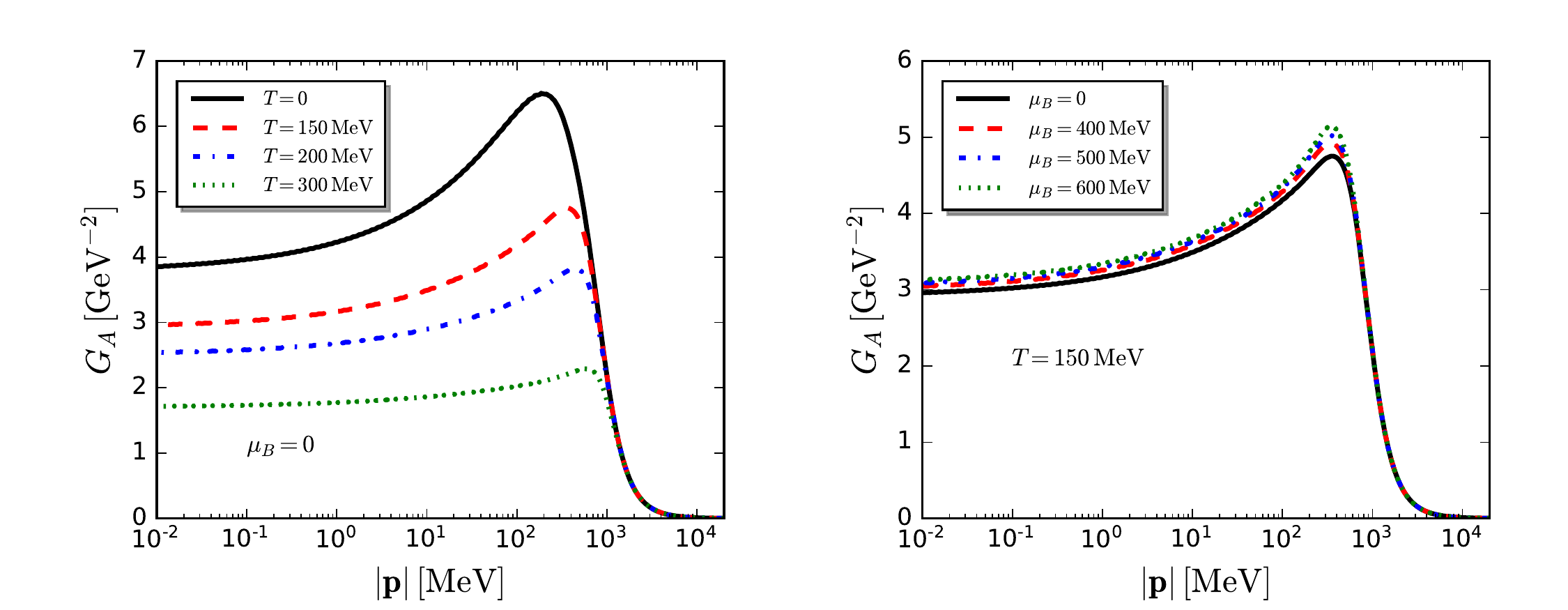}
\caption{$N_f=2+1$ gluon propagator $G_A=1/(Z_{A,k}k^2)$ at vanishing
  frequency $p_0=0$ as a function of spatial momenta $|\bm{p}|$ at
  different temperatures (left panel) and baryon chemical potentials
  (right panel).}\label{fig:GA}
\end{figure*}
%%%%%%%%%%%%%%%%%%%%%%%%%%%%%
%
Implementing the projection as shown in \eq{eq:etaAQL}, one obtains
the contribution of the quark loop to the gluon anomalous dimension,
which reads
\begin{align}
  \eta_{A}^{q}
  &=-\frac{N_f}{\pi^2} g_{\bar q A q,k}^{2}\int_0^1 d x
    \bigg[(1-\eta_{q,k}) \sqrt x+\eta_{q,k} x\bigg]\nonumber\\[1ex]
  &\times \int_{-1}^1 d \cos \theta\bigg[
    \Big(\mathcal{FF}_{(1,1)}(\tilde{m}_{q,k}^{2},\tilde{m}_{q,k}^{2})
    \nonumber\\[1ex]
  &-\mathcal{FF}_{(2,1)}(\tilde{m}_{q,k}^{2},\tilde{m}_{q,k}^{2})\Big)
    +\Big(\sqrt x \cos^2\!\theta-\cos \theta\Big)\nonumber\\[1ex]
  &\times\Big(1+r_F(x')\Big)\Big(\mathcal{FF}_{(2,1)}(\tilde{m}_{q,k}^{2},
    \tilde{m}_{q,k}^{2})\nonumber\\[1ex]
  &-\frac{1}{2}\mathcal{FF}_{(1,1)}(\tilde{m}_{q,k}^{2},\tilde{m}_{q,k}^{2})
    \Big)\bigg]\,,\label{eq:DeltaAqexpl}
\end{align}
with the fermionic regulator in \eq{eq:rF}. We have defined
$x=\bm{q}^2/k^2$, $x'=(\bm{q}-\bm{p})^2/k^2$. $\bm{q}$ and $\bm{p}$
are the loop and external momenta, respectively, and $\theta$ is the
angle between them.  In order to take into account the screening
effect of the quark on the gluon propagator more efficiently, we
choose the external momentum to be $|\bm{p}|=k$ in
\eq{eq:DeltaAqexpl}. This is necessary for consistence, since we also
evaluate the external input for the gluon anomalous dimension at
$|\bm{p}|=k$. The in-medium contribution, which we use for the
light-quark contribution, is then
\begin{align} \label{eq:eq:DeltaAqexplMed}
\Delta\eta_{A}^{q} = \eta_{A}^{q} - \eta_{A}^{q}\big|_{T,\mu = 0}\,.
\end{align}
The threshold functions in \eq{eq:DeltaAqexpl} are given in
\eq{eq:FF}. Moreover, the second term on the r.h.s.\ of \eq{eq:etaAT},
viz.\ $\Delta\eta_{A}^{\text{glue}}$, which is the thermal part of the
pure glue contribution to the gluon anomalous dimension, has to be
specified. Again, we build on the results in Yang-Mills theory in
\cite{Cyrol:2017qkl}. As discussed there,
$\Delta\eta_{A}^{\text{glue}}$ is captured quantitatively through the
inclusion of a thermal screening mass of the gluon,
$\Delta m_\textrm{scr}^2(k,T)$. In addition to the temperature, it has
to depend on the RG scale since thermal effects are rapidly suppressed
for $k \gtrsim 2\pi T$. Therefore, in the following we discuss the
modification of the gluon anomalous dimension resulting from its
thermal screening mass. Beginning with the inverse gluon propagator,
\begin{align}
  G_A^{-1}(k)&=Z_{A,k}k^2\,,\label{}
\end{align}
one obtains a modified one with the screening mass,
\begin{align}
  \bar G_A^{-1}(k)
  &=\bar Z_{A,k}k^2=Z_{A,k}k^2+
    \Delta m_\textrm{scr}^2(k,T)\,.\label{eq:ZAbar}
\end{align}
Differentiating both sides with respect to $t$ leads to
\begin{align}
  \bar Z_{A,k}k^2(2-\bar \eta_{A,k})
  &=
    Z_{A,k}k^2(2-\eta_{A,k})+\partial_t
    \big(\Delta m_\textrm{scr}^2(k,T)\big)\,.
                                      \label{eq:ZAbar2}
\end{align}
where $\bar \eta_{A}=-\partial_t \bar Z_{A}/ \bar Z_{A}$ is the
anomalous dimension with the screening mass. Replacing $Z_{A}$ in
\eq{eq:ZAbar2} by resorting to \eq{eq:ZAbar}, one is led to
\begin{align}
  \bar \eta_{A}
  &=\eta_{A}+\frac{\Delta m_\textrm{scr}^2(k,T)}{\bar
    Z_{A}k^2}(2-\eta_{A})\nonumber\\[1ex]
  &-\frac{1}{\bar Z_{A}k^2}\partial_t
    \big(\Delta m_\textrm{scr}^2(k,T)\big)\,.\label{}
\end{align}
Guided by the results of \cite{Cyrol:2017qkl}, we propose an ansatz
for the screening mass as follows:
\begin{align}
  \Delta m_\textrm{scr}^2(k,T)
  &=(c\,T)^2 \exp\Big[-\Big(\frac{k}{\pi T}
    \Big)^n\Big]\,.\label{eq:screenmass}
\end{align}
Here $c=2$ is adopted for $N_f=2+1$, which is consistent with the
result in \cite{Cyrol:2017qkl}. Furthermore, we choose $n=2$ in
\eq{eq:screenmass}. We have also investigated the dependence of our
results on the parameters in \eq{eq:screenmass}, and find that
variations of the parameters result in only a mild change of the
pseudocritical temperature of the chiral phase transition.

In \sec{sec:Props+Couplings} we show the momentum (or cutoff)
dependence of the gluon dressing function $1/Z_A(p)$ for different
temperatures and baryon chemical potentials. In particular the
dependence on $\mu_B$ is very mild and only concerns the infrared. The
in-medium behaviour is more pronounced in the gluon propagator
depicted in \fig{fig:GA}. However, we emphasise that the dressing
functions are the relevant quantities for the flow equations in the
present work.

%%%%%%%%%%%%%%%%%%%%%%%%%%%%%%%%%%%%%%%%%%%%%%%%%%%%%%%%%%%%%
%%%%%%%%%%%%%%%%%%%%%%%%%%%%%%%%%%%%%%%%%%%%%%%%%%%%%%%%%%%%%

\section{Scalar anomalous dimension}
\label{app:etaphi}

The anomalous dimensions of the mesons for vanishing external momentum
$p=0$ and nonvanishing momentum $(p_0=0,|\bm{p}|=k)$ are given by
\eq{eq:etaphi} and \eq{eq:etaphipk}, respectively. This leads us to
\begin{align}\nonumber 
  \eta_{\phi}(0) =
  &\, \frac{\bar Z_\phi}{Z_\phi(0)} \\[1ex]
  &\,\times \frac{1}{6\pi^2}\bigg
    \{\frac{4}{k^{2}}\bar{\kappa}_{k}(\bar{V}^{''}_{k}(
    \bar{\kappa}_{k}))^{2}\mathcal{BB}_{(2,2)}(\tilde{m}_{\pi,k}^{2},
    \tilde{m}_{\sigma,k}^{2};T)\nonumber \\[1ex]
  &\,+N_{c}\bar{h}_k^{2}\Big[(2\eta_{q,k}-3)\mathcal{F}_{(2)}(
    \tilde{m}_{q,k}^{2};T,\mu_q)\nonumber\\[1ex] 
  &\,-4(\eta_{q,k}-2)\mathcal{F}_{(3)}(\tilde{m}_{q,k}^{2};T,\mu_q)\Big]\bigg\}\,,
\label{eq:etaphiexp}
\end{align}
where the ratio $\bar Z_\phi/Z_\phi(0)$ takes into account that the
flow on the right hand side of \eq{eq:etaphiexp} has been defined with
derivatives w.r.t.\ the renormalised field $\bar\phi=\bar
Z_\phi\phi$. The term with the threshold function
$\mathcal{BB}_{(2,2)}$ arises from the mesonic loop of the flow
equation for the meson propagator in \Fig{fig:propas}. The terms
involving $\mathcal{F}_{(n)}$ are related to the quark loop. Explicit
expressions for all threshold functions are collected in
\app{app:threshold}.  For \eq{eq:etaphipk} we have with
$(0,k)= (p_0=0,|\mathbf{p}|=k)$,
\begin{align}
  &\eta_{\phi}(0,k)\nonumber \\[1ex]
  =&\frac{2}{3\pi^2}\frac{1}{k^{2}}\bar{\kappa}_{k}(\bar{V}^{''}_{k}(
     \bar{\kappa}_{k}))^{2}\mathcal{BB}_{(2,2)}(\tilde{m}_{\pi,k}^{2},
     \tilde{m}_{\sigma,k}^{2};T)\nonumber \\[1ex]
  &-\frac{N_{c}}{\pi^2}\bar h_k^2\int_0^1 d x\bigg[(1-\eta_{q,k}) \sqrt x
    +\eta_{q,k} x\bigg]\nonumber\\[1ex]
  &\times \int_{-1}^1 d \cos \theta\Bigg\{\bigg[
    \Big(\mathcal{FF}_{(1,1)}(\tilde{m}_{q,k}^{2},
    \tilde{m}_{q,k}^{2})-\mathcal{F}_{(2)}(
    \tilde{m}_{q,k}^{2}) \Big)\nonumber \\[1ex]
  &-\Big(\mathcal{FF}_{(2,1)}(\tilde{m}_{q,k}^{2},
    \tilde{m}_{q,k}^{2})-\mathcal{F}_{(3)}(\tilde{m}_{q,k}^{2})
    \Big)\bigg] \nonumber \\[1ex]
  &+\bigg[\Big(\sqrt x -\cos \theta\Big)
    \Big(1+r_F(x')\Big)\mathcal{FF}_{(2,1)}(
    \tilde{m}_{q,k}^{2},\tilde{m}_{q,k}^{2})\nonumber \\[1ex]
  &-\mathcal{F}_{(3)}(\tilde{m}_{q,k}^{2})\bigg]
    -\frac{1}{2}\bigg[\Big(\sqrt x -\cos \theta\Big)
    \Big(1+r_F(x')\Big)\nonumber \\[1ex]
  &\times\mathcal{FF}_{(1,1)}(\tilde{m}_{q,k}^{2},
    \tilde{m}_{q,k}^{2})-\mathcal{F}_{(2)}(
    \tilde{m}_{q,k}^{2})\bigg]\Bigg\}\,,
\label{eq:etaphipkexp}
\end{align}
with $x=\bm{q}^2/k^2$ and $x'=(\bm{q}-\bm{p})^2/k^2$ as same as in
\eq{eq:DeltaAqexpl}. Note that in \eq{eq:etaphipkexp} we have
neglected the external momentum dependence of the meson loop for
simplicity in the numerical calculations. This is a good approximation
for both, small and large $k$, since meson loops are suppressed by the
meson masses in the IR and decouple in the UV. Only in the phase
transition region and in particular at the CEP, corrections to the
mesonic part of \eq{eq:etaphipkexp} could become relevant.

\section{Quark anomalous dimension}
\label{app:etaq}

The quark anomalous dimension $\eta_{q,k}$ in \eq{eq:etapsi} reads
\begin{align}
  \eta_{q,k}=
  &\frac{1}{24\pi^2N_{f}}(4-\eta_{\phi,k})\bar{h}_{k}^{2}\nonumber \\[1ex] 
  &\times\bigg\{(N_{f}^{2}-1)
    \mathcal{FB}_{(1,2)}(\tilde{m}_{q,k}^{2},\tilde{m}_{\pi,k}^{2};
    T,\mu_q,p_{0,\text{\tiny{ex}}})\nonumber \\[1ex] 
  &+\mathcal{FB}_{(1,2)}(\tilde{m}_{q,k}^{2},\tilde{m}_{\sigma,k}^{2};
    T,\mu_q,p_{0,\text{\tiny{ex}}})
    \bigg\}\nonumber \\[1ex] 
  &+\frac{1}{24\pi^2}\frac{N_c^2-1}{2N_c} g_{\bar q A q, k}^{2}\nonumber \\[1ex] 
  &\times\bigg\{2(4-\eta_{A,k}) \mathcal{FB}_{(1,2)}(\tilde{m}_{q,k}^{2},0;
    T,\mu_q,p_{0,\text{\tiny{ex}}})\nonumber \\[1ex] 
  &+3(3-\eta_{q,k})\Big( \mathcal{FB}_{(1,1)}(\tilde{m}_{q,k}^{2},0;
    T,\mu_q,p_{0,\text{\tiny{ex}}})\nonumber \\[1ex] 
  &-2 \mathcal{FB}_{(2,1)}(\tilde{m}_{q,k}^{2},0;T,\mu_q,p_{0,\text{\tiny{ex}}})
    \Big)\bigg\},\label{eq:etapsiexp}
\end{align}
where the fermion-boson mixed threshold function $\mathcal{FB}$'s are
given in \app{app:threshold}. It is obvious that the contributions in
\eq{eq:etapsiexp} can be divided into two sectors, which results from
the quark-meson and quark-gluon interactions, respectively.

%
%%%%%%%%%%%%%%%%%%%%%%%%%%%%%
\begin{figure*}[t]
\includegraphics[width=0.98\textwidth]{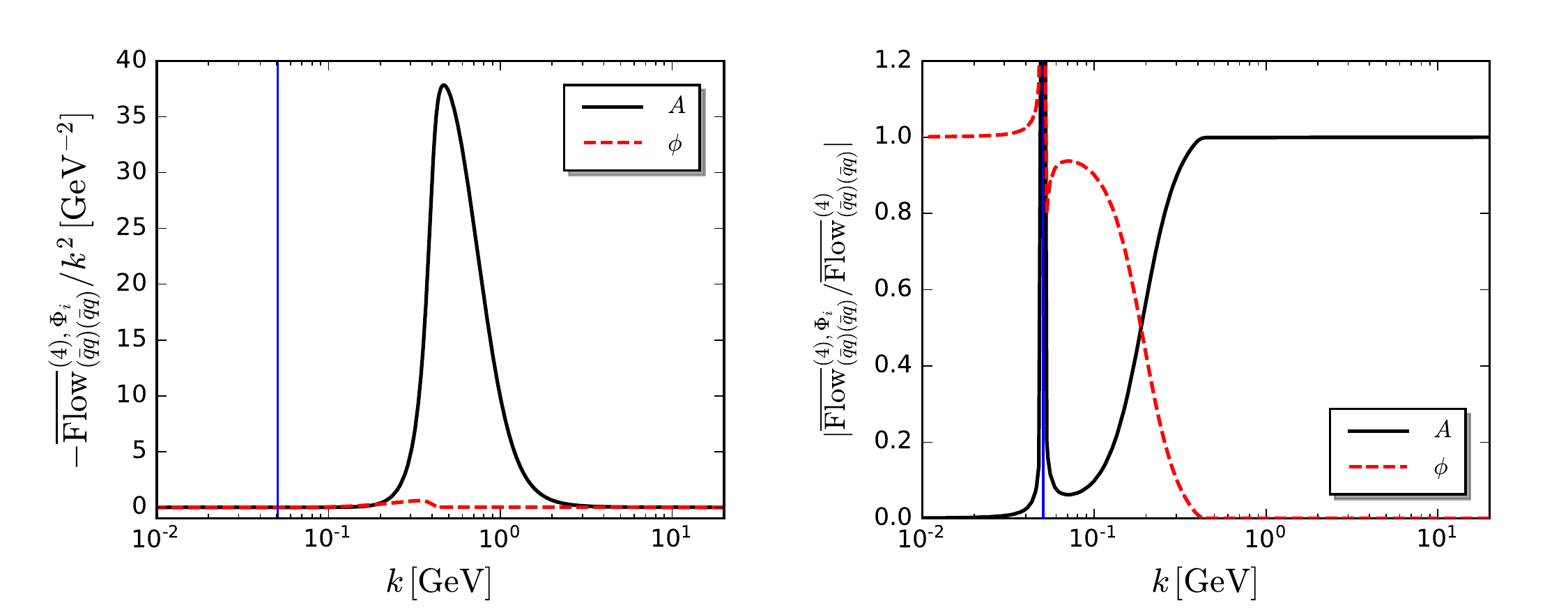}
\caption{Left panel: comparison between the flow of the $N_f=2+1$
  four--light-quark coupling arising from the gluon exchange in
  \eq{eq:dtlambdaA}, $\Phi_i=A$, and that from the meson exchange in
  \eq{eq:dtlambdaAphi}, $\Phi_i=\phi$, as a function of $k$ at
  $T=\mu_B=0$.\\[1ex] Right panel: the ratio of their respective flow
  over the total flow, i.e., the summation of \eq{eq:dtlambdaA} and
  \eq{eq:dtlambdaAphi}, where the absolute value for the ratio is
  chosen. The blue vertical line in both plots shows the position
  where the flow resulting from the meson exchange changes
  sign.}\label{fig:flow4fermi}
\end{figure*}
%%%%%%%%%%%%%%%%%%%%%%%%%%%%%
%
The external 3-momentum in \eq{eq:etapsiexp} is chosen to be
$\bm{p}=0$. However, since the fermionic Matsubara frequency,
$p_0=(2n+1)\pi T$ with $n \in \mathbb{Z}$, does not have a zero-mode,
$p_0$ cannot be chosen to be zero. In fact, a truncation that takes
into account the full frequency dependence of the quark anomalous
dimension plays a significant r$\hat{\textrm{o}}$le in obtaining
quantitative accuracy in the computation of particle number
correlations with the fRG, and is also indispensable to guarantee the
Silver Blaze properties at finite chemical potential, see
\cite{Fu:2016tey} for more relevant details. However, a frequency
dependent truncation is computationally very demanding, in particular
in a temporal gluon background field. Hence, in the present work we
use a modification of the lowest Matsubara mode. As has been discussed
in detail in \cite{Fu:2015naa}, an evaluation consistent with the
silver blaze property is done with
\begin{align}\label{eq:p_ex} 
  p_{0,\text{\tiny{ex}}} = (\pi T) \exp\{-k/(\pi T)\} - i\,\mu_q\,.
\end{align}
We use the exponential factor to suppress the artificial
$T$-dependence in the vacuum part of the flow that is introduced by
the fermionic Matsubara frequency. As discussed in \cite{Fu:2015naa,
  Fu:2016tey}, this procedure is necessary to ensure the correct
thermal behaviour without taking into account the full frequency
dependence.  The chemical potential part eliminates the $\mu$-part of
the argument $p_0+i\,\mu_q$ of the correlation functions. This keeps
the expansion point $p_{0,\text{\tiny{ex}}}+i\,\mu_q$ fixed and does
not lead to an unphysical $\mu_q$-dependence. The even quantitative
reliability of this treatment has been verified in \cite{Fu:2016tey}
through a comparison with the fully frequency-dependent calculations.

In comparison to \cite{Fu:2015naa} we use the limit
$p_{0,\text{\tiny{ex}}}+i\,\mu_q\to 0$ instead of
$p_{0,\text{\tiny{ex}}}+i\,\mu_q \to k$ for convenience. This
modification allows us to compare our vacuum results more directly to
that in \cite{Braun:2014ata}, as there the flows are evaluated at
$p_{0,\text{\tiny{ex}}}=0$. We drop the suppression factor in the
thermal part of the diagrams:
$p_{0,\text{\tiny{ex}}} \to (\pi T) - i\,\mu_q$. This leads to simpler
expressions, and the necessary suppression of thermal effects at small
$T/k$ is guaranteed by the thermal distribution functions. Both
modifications have no sizable impact on the numerical results.

\section{Flow of the Yukawa coupling}
\label{app:dth}

We have discussed the Yukawa coupling in \sec{sec:Yukawa}, and its
flow is given in \eq{eq:hflow}.  Adopting the same external frequency
and momentum as the quark anomalous dimension, the last term on the
r.h.s.\ of \eq{eq:hflow} reads
\begin{align}
  &\mathrm{Re}\big(\overline{\textrm{Flow}}^{(2)}_{(\bar q \tau^0 q)}\big)/
    \bar \sigma\nonumber\\[1ex]
  &\quad=\frac{1}{4\pi^2 N_f}\bar{h}_k^3
    \bigg[-(N_{f}^{2}-1)\nonumber \\[1ex] 
  &\qquad\times L_{(1,1)}^{(4)}(
    \tilde{m}_{q,k}^{2},\tilde{m}_{\pi,k}^{2},
    \eta_{q,k},\eta_{\phi,k};T,\mu_q,
    p_{0,\text{\tiny{ex}}})\nonumber\\[1ex]
  &\qquad+L_{(1,1)}^{(4)}(\tilde{m}_{q,k}^{2},\tilde{m}_{\sigma,k}^{2},
    \eta_{q,k},\eta_{\phi,k};T,\mu_q,p_{0,\text{\tiny{ex}}})
    \bigg]\nonumber\\[1ex]
  &\qquad-\frac{3}{2\pi^2}\frac{N_c^2-1}{2N_c}
    g_{\bar q A q, k}^{2}\bar{h}_k
    \nonumber \\[1ex]  
  &\qquad\times L_{(1,1)}^{(4)}(\tilde{m}_{q,k}^{2},0,
    \eta_{q,k},\eta_{A,k};T,
    \mu_q,p_{0,\text{\tiny{ex}}})\,.\label{eq:hexp}
\end{align}
\section{Flow of the four-quark coupling}
\label{app:dtlambda}

The flow diagrams of the four-quark coupling arising from the gluon
exchange in the first line in \Fig{fig:flow4q}, after projected onto
the $\sigma$-$\bm\pi$ channel, are denoted with
$ \overline{\textrm{Flow}}^{(4),A}_{(\bar q q)(\bar q q)}$. The flow
diagrams arising from the meson exchange or denoted with
$\overline{\textrm{Flow}}^{(4),\phi}_{(\bar q q)(\bar q q)}$. The
gluon exchange diagrams read
\begin{align}
  \overline{\textrm{Flow}}^{(4),A}_{(\bar q q)(\bar q q)}
  &=-\frac{3}{2\pi^2}\frac{N_c^2-1}{2N_c}
    \Big(\frac{3}{4}-\frac{1}{N_c^2}\Big) g_{\bar q A q,k}^{4}
    \nonumber \\[1ex] 
  &\quad\times\bigg\{\frac{2}{15}(5-\eta_{A,k})\Big[
    \mathcal{FB}_{(1,3)}(\tilde{m}_{q,k}^{2},0)
    \nonumber \\[1ex] 
  &\quad-\bar{m}_{q,k}^{2}
    \mathcal{FB}_{(2,3)}(\tilde{m}_{q,k}^{2},0)
    \Big]\nonumber \\[1ex] 
  &\quad+\frac{1}{12}(4-\eta_{q,k})\Big[\mathcal{FB}_{(2,2)}(
    \tilde{m}_{q,k}^{2},0)\nonumber \\[1ex] 
  &\quad-2\bar{m}_{q,k}^{2}\mathcal{FB}_{(3,2)}(
    \tilde{m}_{q,k}^{2},0)\Big]\,,
 \label{eq:dtlambdaA}
\end{align}
Here, the arguments $T$, $\mu_q$ and $p_{0,\text{\tiny{ex}}}$ for
threshold functions are not shown explicitly. The right hand side of
\eq{eq:dtlambdaA} is obtained as follows: It contains expressions such
as $(q\cdot q^F)(\gamma \cdot q)(\gamma \cdot q^F)/q^2$ with loop
momentum $q$. Here, $q_{\mu}^F\equiv(q_0, (1+r_F) \bm{q})$ with $r_F$
in \eq{eq:Rq}. This expression arises from the quark propagator with
the 3$d$ regulator as shown in Eq. (\ref{eq:Rq}), contracted with the
transverse magnetic tensor of the gluon in Eq. (\ref{eq:magproj}). In
order to simplify the flow in Eq. (\ref{eq:dtlambdaA}) as well as
numerical calculations, we approximate the expression above as
$q^F\cdot q^F$, which is exact for the 4$d$ regulator. We have also
investigated the reliability of this approximation through the
comparison to the relevant four-quark flow in
\cite{Braun:2014ata}. There a 4$d$ regulator has been used, and we
find that the difference is negligible.

The meson exchange diagrams have the form
\begin{align}
  \overline{\textrm{Flow}}^{(4),\phi}_{(\bar q q)(\bar q q)}
  &=\frac{1}{32\pi^2}\frac{N_f^2-2}{N_fN_c}
    \bar{h}_{k}^{4}\bigg\{\frac{2}{15}(5-\eta_{\phi,k})
    \nonumber \\[1ex] 
  &\quad\times\Big[\Big(\mathcal{FBB}_{(1,1,2)}(
    \tilde{m}_{q,k}^{2},\tilde{m}_{\pi,k}^{2},
    \tilde{m}_{\sigma,k}^{2})\nonumber \\[1ex] 
  &\quad+\mathcal{FBB}_{(1,1,2)}(\tilde{m}_{q,k}^{2},
    \tilde{m}_{\sigma,k}^{2},
    \tilde{m}_{\pi,k}^{2})\nonumber \\[1ex] 
  &\quad-2\mathcal{FB}_{(1,3)}(\tilde{m}_{q,k}^{2},
    \tilde{m}_{\pi,k}^{2})
    \Big)\nonumber \\[1ex] 
  &\quad-\tilde{m}_{q,k}^{2}\Big(
    \mathcal{FBB}_{(2,1,2)}(\tilde{m}_{q,k}^{2},
    \tilde{m}_{\pi,k}^{2},\tilde{m}_{\sigma,k}^{2})
    \nonumber \\[1ex] 
  &\quad+\mathcal{FBB}_{(2,2,1)}(\tilde{m}_{q,k}^{2},\tilde{m}_{\pi,k}^{2},
    \tilde{m}_{\sigma,k}^{2})\nonumber \\[1ex] 
  &\quad-2\mathcal{FB}_{(2,3)}(\tilde{m}_{q,k}^{2},\tilde{m}_{\pi,k}^{2})\Big)\Big]
    +\frac{1}{6}(4-\eta_{q,k})\nonumber \\[1ex] 
  &\quad\times\Big[\Big(\mathcal{FBB}_{(2,1,1)}(\tilde{m}_{q,k}^{2},
    \tilde{m}_{\pi,k}^{2},\tilde{m}_{\sigma,k}^{2})\nonumber \\[1ex] 
  &\quad-\mathcal{FB}_{(2,2)}(\tilde{m}_{q,k}^{2},\tilde{m}_{\pi,k}^{2})
    \Big)\nonumber \\[1ex] 
  &\quad-2\tilde{m}_{q,k}^{2}\Big(\mathcal{FBB}_{(3,1,1)}(\tilde{m}_{q,k}^{2},
    \tilde{m}_{\pi,k}^{2},\tilde{m}_{\sigma,k}^{2})\nonumber \\[1ex] 
  &\quad-\mathcal{FB}_{(3,2)}(\tilde{m}_{q,k}^{2},\tilde{m}_{\pi,k}^{2})
    \Big)\Big]\bigg\}\,.
 \label{eq:dtlambdaAphi}
\end{align}
In \Fig{fig:flow4fermi} we compare \eq{eq:dtlambdaA} and
\eq{eq:dtlambdaAphi} at $T=0$ and $\mu_B=0$ for the purpose of
illustration. The flow divided by $k^2$ is shown in the left plot of
\Fig{fig:flow4fermi}, and the ratio between the respective flow and
their sum in the right plot. Note that the peak in the right plot
corresponds to the position where the sum of \eq{eq:dtlambdaA} and
\eq{eq:dtlambdaAphi} is vanishing, thus resulting in a divergent ratio
there. This is because in the low $k$ region meson exchange dominates,
and this contribution changes sign at about $k \approx 75$\,MeV, as
shown by the blue vertical line in both plots. One observes that gluon
exchange, i.e., the diagrams in the first line in \Fig{fig:flow4q}, is
the dominant process for $k \gtrsim 300$\,MeV. Therefore, it is
reasonable to expect that the mixed diagrams in the third line of
\Fig{fig:flow4q} play a subleading r$\hat{\textrm{o}}$le for dynamical
hadronisation. It is interesting to observe that gluon exchange
becomes dominant already in (and below) the phase transition region at
$k \approx 400$\,MeV, which can be read-off, e.g., from
\Fig{fig:lambda}. This gives an implicit upper bound on the validity
of low-energy models, namely $k \lesssim 300$\,MeV. For first
discussions in this direction, see \cite{Rennecke:2015,
  Alkofer:2018guy}. Note that this bound is in quantitative agreement
with the results in these references.

\section{Flow of the quark-gluon coupling}
\label{app:dtqg}
The running of the strong quark-gluon coupling is governed by
\eq{eq:dtg}, where the term in the second line is divided into two
parts, i.e., $ \overline{\textrm{Flow}}^{(3),A}_{(\bar q A q)}$ and
$ \overline{\textrm{Flow}}^{(3),\phi}_{{(\bar q A q)}}$, which
correspond to the relevant contributions from the quark-gluon and
quark-meson couplings, respectively. The subscript
${}_{{(\bar q A q)}}$ indicates that these expressions are the
projections of the quark-gluon flows on the running coupling, see
\eq{eq:dtg}. With this notation, the flow of the light-quark--gluon
coupling is:
  \begin{align} \label{eq:dtglexp} \nonumber \partial_t g_{\bar{l}Al}
    &= \left( \frac12\eta_A+\eta_q\right)
      g_{\bar{l}A l}\\
    &\quad +\bigg( N_f\, \overline{\textrm{Flow}}^{(3),A}_{(\bar l A
      l)} + \overline{\textrm{Flow}}^{(3),\phi}_{{(\bar q A
      q)}}\bigg)\bigg|_{N_f=2}\,.
\end{align}
The explicit form of the flow of the strange-quark--gluon coupling is
given in \eq{eq:dtgs}. The gluonic contribution reads
\begin{align}
  \overline{\textrm{Flow}}^{(3),A}_{(\bar q A q)}
  =&\frac{3}{8\pi^2N_c}
     g_{\bar{q}Aq}^{3}\tilde{m}_{q,k}^{2}\bigg
     \{\frac{2}{15}(5-\eta_{A,k})\mathcal{FB}_{(2,2)}(
     \tilde{m}_{q,k}^{2},0)\nonumber \\[1ex] 
   &+\frac{1}{3}(4-\eta_{q,k})\mathcal{FB}_{(3,1)}(
     \tilde{m}_{q,k}^{2},0)\bigg\}\nonumber \\[1ex] 
   &+\frac{3N_c}{8\pi^2}g_{\bar{q}Aq}^{2}g_{A^3}
     \bigg\{\frac{1}{20} (5-\eta_{q,k})
     \mathcal{FB}_{(1,2)}(\tilde{m}_{q,k}^{2},0)\nonumber \\[1ex] 
   &-\frac{1}{6} (4-\eta_{q,k})\mathcal{FB}_{(2,1)}(\tilde{m}_{q,k}^{2},0)+
     \frac{1}{30} (5-2\eta_{q,k})\nonumber \\[1ex] 
   &\times\mathcal{FB}_{(2,2)}(\tilde{m}_{q,k}^{2},0)
     -\frac{4}{15}(5-\eta_{A,k})\nonumber \\[1ex] 
   &\times\mathcal{FB}_{(1,2)}(\tilde{m}_{q,k}^{2},0)
     +\frac{1}{30}(10-3\eta_{A,k})\nonumber \\[1ex] 
   &\times\mathcal{FB}_{(1,3)}(\tilde{m}_{q,k}^{2},0)\bigg\}\,. 
 \label{eq:dtgA}
\end{align}
The contribution of the mesons reads
\begin{align}
  \overline{\textrm{Flow}}^{(3),\phi}_{{(\bar q A q)}}
  =&-\frac{1}{8\pi^2N_f}g_{\bar{q}Aq}
     \bar{h}_{k}^{2}\bigg\{\frac{1}{6}(4-\eta_{q,k})\nonumber
  \\[1ex] 
   &\times\Big[\mathcal{FB}_{(2,1)}(\tilde{m}_{q,k}^{2},
     \tilde{m}_{\sigma,k}^{2})\nonumber \\[1ex] 
   &+2\tilde{m}_{q,k}^{2}\mathcal{FB}_{(3,1)}(\tilde{m}_{q,k}^{2},
     \tilde{m}_{\sigma,k}^{2})\Big]+\frac{2}{15}(5-\eta_{\phi,k})
     \nonumber \\[1ex] 
   &\times\Big[\mathcal{FB}_{(1,2)}(\tilde{m}_{q,k}^{2},\tilde{m}_{\sigma,k}^{2})
     \nonumber \\[1ex] 
   &+\tilde{m}_{q,k}^{2}\mathcal{FB}_{(2,2)}(\tilde{m}_{q,k}^{2},
     \tilde{m}_{\sigma,k}^{2})\Big]\bigg\}\nonumber \\[1ex] 
   &-\frac{N_f^2-1}{8\pi^2N_f}g_{\bar{q}Aq}\bar{h}_{k}^{2}
     \bigg\{\frac{1}{6}(4-\eta_{q,k})\nonumber \\[1ex] 
   &\times\Big[\mathcal{FB}_{(2,1)}(\tilde{m}_{q,k}^{2},
     \tilde{m}_{\pi,k}^{2})\nonumber \\[1ex] 
   &+2\tilde{m}_{q,k}^{2}\mathcal{FB}_{(3,1)}(\tilde{m}_{q,k}^{2},
     \tilde{m}_{\pi,k}^{2})\Big]+\frac{2}{15}(5-\eta_{\phi,k})
     \nonumber \\[1ex] 
   &\times\Big[\mathcal{FB}_{(1,2)}(\tilde{m}_{q,k}^{2},
     \tilde{m}_{\pi,k}^{2})\nonumber \\[1ex] 
   &+\tilde{m}_{q,k}^{2}\mathcal{FB}_{(2,2)}(\tilde{m}_{q,k}^{2},
     \tilde{m}_{\pi,k}^{2})\Big]\bigg\}\,.
 \label{eq:dtgphi}
\end{align}
For the light and strange quarks we use the corresponding masses
defined in \eq{eq:ml+s_const} and set $N_f = 2$ and $N_f = 1$,
respectively, above. Note that, as discussed in \app{app:Nf2p1}, we
only take \eq{eq:dtgA} into account for the strange-quark--gluon
coupling.

%%%%%%%%%%%%%%%%%%%%%%%%%%%%%%%%%%%%%%%%%%%%%%%%%%%%%%%%%%%%%
%%%%%%%%%%%%%%%%%%%%%%%%%%%%%%%%%%%%%%%%%%%%%%%%%%%%%%%%%%%%%

\section{Regulators and threshold functions}
\label{app:threshold}

We employ $3d$ optimised regulators, \cite{Litim:2000ci,
  Litim:2001up}, throughout this work, which read
\begin{subequations}\label{eq:Regs}
\begin{align}
  R^{q}_{k}(q_{0},
  \bm{q})
  &=Z_{q,k}i\bm{
    \gamma}\cdot\bm{q}\,r_{F}(\bm{q}^{2}/k^2)\,,
    \label{eq:Rq}\\[1ex] 
  R^{\phi}_{k}(q_{0},\bm{q})
  &=Z_{\phi,k}\bm{q}^{2}r_{B}(
    \bm{q}^{2}/k^2)\,, \label{eq:Rphi}
\end{align}
for the quarks and mesons, and we choose
\begin{align}
  (R^{A}_{k})_{\mu\nu}(q_{0},
  \bm{q})&=Z_{A,k}\bm{q}^2 r_{B}(\bm{q}^{2}/k^2)
           \Big(\delta_{\mu\nu}-
           \frac{q_{\mu}q_{\nu}}{q^2}\Big)\nonumber \\[1ex] 
         &+\frac{\bm{q}^2}{\xi}r_{B}(\bm{q}^{2}/k^2)
           \Big(\frac{q_{\mu}q_{\nu}}{q^2}\Big)
           \,, \label{eq:RA}
\end{align}
for the gluons, where the Landau gauge $\xi=0$ is adopted, and
$Z_{q,k}$, $Z_{\phi,k}$, $Z_{A,k}$ are the wave function
renormalisations for the three different fields, respectively. The
shapes of the regulators are determined by the functions
\begin{align}
  r_{F}(x)&=\Big(\frac{1}{\sqrt{x}}-1\Big)\Theta(1-x)\,,\label{eq:rF}\\[1ex] 
  r_{B}(x)&=\Big(\frac{1}{x}-1\Big)\Theta(1-x)\,,  \label{eq:rB}
\end{align}
\end{subequations}
where $\Theta(x)$ is the Heaviside step function. We note that, for
the regulator of the gluon propagator, we do not distinguish the $3d$
transverse part, i.e., $\Pi_{\mu\nu}^{\text{M}}$ in \eq{eq:magproj},
from its complementary part longitudinal to the heat bath, viz. the
electric component
$\Pi_{\mu\nu}^{\text{E}}=\Pi_{\mu\nu}^{\perp}-\Pi_{\mu\nu}^{\text{M}}$,
with the $4d$ transverse tensor:
\begin{align}
  \Pi_{\mu\nu}^{\perp}(q)
  =\delta_{\mu\nu}-\frac{q_{\mu}q_{\nu}}{q^2}\,.
                           \label{eq:Piperp}
\end{align}
In the threshold functions in this work, the scalar parts of the
propagators typically enter in the form
\begin{align}
  G_{b}(q,m^{2})
  &=\frac{1}{\tilde{q}_0^2+1+m^{2}}\,,\label{eq:Gb}\\[1ex] 
  G_{f}(q,m^{2})
  &=\frac{1}{(\tilde{q}_0+i\tilde{\mu}_q)^2+1+m^{2}}\,,  \label{eq:Gf}
\end{align}
for bosons and fermions, whose Matsubara frequencies are $q_0=2n\pi T$
($n\in \mathbb{Z}$) and $(2n+1)\pi T$, respectively. Here $q_0$ and
the chemical potential are rescaled by the RG scale $k$, i.e.,
$\tilde{q}_0=q_0/k$ and $\tilde{\mu}_q=\mu_q/k$.  The gluon propagator
is just \eq{eq:Gb} with vanishing mass $m^{2}=0$, since the gluon mass
gap is encoded in the momentum dependence of its wave function
renormalisation.

To proceed, we define
\begin{subequations}\label{eq:thresholds}
\begin{align}\nonumber 
  \mathcal{F}_{(n)}(m^{2};T,\mu_q)
  &=\frac{T}{k}\sum_{n_q}
    \Big(G_{f}(q,m^{2})\Big)^n\,\\[1ex]
  \mathcal{B}_{(n)}(m^{2};T)
  &=\frac{T}{k}\sum_{n_q}
    \Big(G_{b}(q,m^{2})\Big)^n\,.
\label{eq:threshold1}\end{align}
Upon inserting Eqs. (\ref{eq:Gb}) and (\ref{eq:Gf}), one observes
\begin{align}\nonumber 
  \mathcal{F}_{(n+1)}(m^{2};T,\mu_q)
  &=-\frac{1}{n}
    \frac{\partial}{\partial m^{2}}
    \mathcal{F}_{(n)}(m^{2};T,\mu_q)\,,
  \\[1ex]
  \mathcal{B}_{(n+1)}(m^{2};T)
  &=-\frac{1}{n}\frac{
    \partial}{\partial m^{2}}\mathcal{B}_{(n)}(m^{2};T)\,. 
    \Big(G_{b}(q,m^{2})\Big)^n\,.
\label{eq:threshold2}\end{align}
Thus, one only needs to sum up the Matsubara frequencies for the
lowest-order threshold functions $\mathcal{F}_{(1)}$ and
$\mathcal{B}_{(1)}$, which yields
\begin{align}\nonumber 
  &\mathcal{F}_{(1)}(m^{2};T,\mu_q)=\frac{1}{2\sqrt{1
    +m^{2}}}\nonumber\\[1ex]
  &\hspace{.3cm}\times\Big(1-n_{F}(m^{2};T,\mu_q)
    -n_{F}(m^{2};T,-\mu_q)\Big)\,,
\label{eq:threshold3}\end{align}
and
\begin{align}
  \mathcal{B}_{(1)}(m^{2};T)=
  &\frac{1}{\sqrt{1+
    m^{2}}}\Big(\frac{1}{2}+n_{B}(m^{2};T)\Big)\,,
\label{eq:threshold4}\end{align}
\end{subequations}
where the distribution functions read
\begin{align}
  &n_{F}(m^{2};T,\mu_q)\nonumber\\[1ex]
  =&\frac{1}{\exp\bigg\{\frac{1}{T}\Big[k(1+
     m^{2})^{1/2}-\mu_q\Big]\bigg\}+1}\,, \label{eq:nF}
\end{align}
and
\begin{align}\label{eq:nB}
  n_{B}(m^{2};T)=
  &\frac{1}{\exp\bigg\{\frac{k}{T}\Big(1
    +m^{2}\Big)^{1/2}\bigg\}-1}\,.
\end{align}
So far we have discussed the flows in the absense of a nontrivial
Polyakov loop expectation value, that is $L,\bar L=1$. For general
Polyakov loop expectation values the quark distribution function in
\eq{eq:nF} in the threshold functions \eq{eq:thresholds} turns into,
\begin{align}\label{eq:nFL}
  n_F(m^2;T,\mu_q,L,\bar{L})
  &=\frac{1+2\bar{L}e^{x/T}+
    Le^{2x/T}}{1+3\bar{L}e^{x/T}+3Le^{2x/T}+e^{3x/T}}\,,
\end{align} 
with $x=k(1+m^2)^{1/2}-\mu_q$. For antiquarks, the distribution
$n_{F}(m^2;T,-\mu)$ turns into $n_F(m^2;T,-\mu,\bar L, L)$. We remark
that the simple substitution
$n_{F}(m^{2};T,\mu_q)\to n_F(m^2;T,\mu_q,L,\bar{L})$ in the flows is
part of the current approximation. There are additional contributions
from the nontrivial color trace in threshold functions that mix
bosonic and fermionic contributions, see \eq{eq:f1b1} and
\eq{eq:f1b1b1}, which we neglect here.

For $L,\bar L=1$ \eq{eq:nFL} reduces to the Fermi-Dirac distribution
of quarks, \eq{eq:nF}. In turn, for $L,\bar L=0$, \eq{eq:nFL} is the
Fermi-Dirac distribution of baryons, that is \eq{eq:nF} with
$x\to 3 x$. Note however, that this limit is not achieved for low
temperatures as the Polyakov loop expectation value decays with
$L \exp(2 m_q/T)\to\infty$ for $T\to 0$.  Nonetheless, baryon number
fluctuations show a baryonic regime for low temperatures as they
should, for more details, see \cite{Fu:2015naa}. For the sake of
brevity we drop the variables $L$ and $\bar{L}$ in the quark
distribution function.

In the flow equation \eq{eq:VflowApp} of the effective potential for
the mesons, relevant threshold functions read
\begin{align}
  l_{0}^{(B,d)}(m^2,\eta;T)=\frac{2}{d-1}\Big(1-\frac{\eta}{d
     +1}\Big)\mathcal{B}_{(1)}(m^2;T)\,,\label{eq:l0Bd}
\end{align}
and
\begin{align}
  l_{0}^{(F,d)}(m^2,\eta;T,\mu_q)
  =\frac{2}{d-1}\Big(1-\frac{\eta}{d}
  \Big)\mathcal{F}_{(1)}(m^2;T,\mu_q)\,.\label{eq:l0Fd}
\end{align}
Both $\mathcal{F}_{(n)}$ and $\mathcal{B}_{(n)}$ only contain either
fermionic or bosonic propagators of a single species. Other threshold
functions, which describe a mixture of species, are also needed.  For
instance $\mathcal{BB}_{(2,2)}$ appearing in \eq{eq:etaphiexp} for the
mesonic anomalous dimension, which consists of two kinds of bosonic
propagators of different masses. To this end, we define
\begin{align}
  &\mathcal{BB}_{(n_1,n_2)}(m_1^2,m_2^2;T)\nonumber\\[1ex]
  =&\frac{T}{k}\sum_{n_q}\Big(G_{b}(q,m_1^2)\Big)^{n_1}
     \Big(G_{b}(q,m_2^2)\Big)^{n_2}\,.
\end{align}
It is easy to obtain the analytic expression for the lowest-order
function, i.e.,
\begin{align} 
  &\mathcal{BB}_{(1,1)}(m_1^2,m_2^2;T)\nonumber\\[1ex]
  =&-\bigg\{\Big(\frac{1}{2}+n_{B}(m_1^2;T)\Big)\frac{1}{
     \big(1+m_1^2\big)^{1/2}}\frac{1}{\big(m_1^2-m_2^2
     \big)}\nonumber \\[1ex]
  &+\Big(\frac{1}{2}+n_{B}(m_2^2;T)\Big)\frac{1}{
    \big(1+m_2^2\big)^{1/2}}\frac{1}{\big(m_2^2-m_1^2
    \big)}\bigg\}\,.
\end{align}
Higher-order functions are obtained as follows:
\begin{align}
  &\mathcal{BB}_{(n_1+1,n_2)}(m_1^2,m_2^2;T)\nonumber\\[1ex]
  =&-\frac{1}{n_1}\frac{\partial}{\partial m_1^2}
     \mathcal{BB}_{(n_1,n_2)}(m_1^2,m_2^2;T)\,,
\end{align}
and analogously for derivatives with respect to $m_{b}^{2}$.

We proceed with threshold functions mixing bosonic and fermionic
propagators,
\begin{align}
  &\mathcal{FB}_{(n_f,n_b)}(m_f^2,m_b^2;T,\mu_q,p_0)
    \nonumber \\[1ex]\nonumber 
  =&\frac{T}{k}\sum_{n_q}\Big(G_{f}(q,m_f^2)\Big)^{n_f}
     \Big(G_{b}(q-p,m_b^2)\Big)^{n_b}\,.
\end{align}
In the same way, higher-order threshold functions can be obtained from
lower-order ones, i.e.,
\begin{align}
  &\mathcal{FB}_{(n_f+1,n_b)}(m_f^2,m_b^2;T,\mu_q,p_0)
    \nonumber\\[1ex]
  =&-\frac{1}{n_f}\frac{\partial}{\partial m_f^2}
     \mathcal{FB}_{(n_f,n_b)}(m_f^2,m_b^2;T,\mu_q,p_0)\,,
\end{align}
and 
\begin{align}
  &\mathcal{FB}_{(n_f,n_b+1)}(m_f^2,m_b^2;T,\mu_q,p_0)
    \nonumber\\[1ex]
  =&-\frac{1}{n_b}\frac{\partial}{\partial m_b^2}
     \mathcal{FB}_{(n_f,n_b)}(m_f^2,m_b^2;T,\mu_q,p_0)\,.
\end{align}
Note that threshold functions $\mathcal{FB}$'s are complex valued,
when the chemical potential is nonvanishing, see e.g.\
\cite{Fu:2015naa, Fu:2016tey} for relevant discussions. Here, only the
real parts of these complex functions are kept. The validity of this
procedure has been discussed in \cite{Pawlowski:2014zaa,
  Fu:2016tey}. The external momentum $p_0+i\mu_q $ has been chosen to
be $p_{0,\text{\tiny{ex}}}$, as discussed in \app{app:etaq}. It is
left to specify the relevant threshold function of lowest-order, which
reads
\begin{widetext}
\begin{align}
  &\mathcal{FB}_{(1,1)}(m_f^2,m_b^2;T,\mu_q,p_0)\nonumber \\[1ex]
  =&\frac{k^2}{2}\bigg\{-n_{B}(m_b^2;T)\frac{1}{\big(1+m_b^2\big)^{1/2}}
     \frac{1}{\Big(ip_0-\mu_q+k\big(1+m_b^2\big)^{1/2}\Big)^2-k^2
     \big(1+m_f^2\big)}\nonumber \\[1ex]
  &-\big(n_{B}(m_b^2;T)+1\big)\frac{1}{\big(1+m_b^2\big)^{1/2}}
    \frac{1}{\Big(ip_0-\mu_q-k\big(1+m_b^2\big)^{1/2}\Big)^2-k^2
    \big(1+m_f^2\big)}\nonumber\\[1ex]
  &+n_{F}(m_f^2;T,-\mu_q)\frac{1}{\big(1+m_f^2\big)^{1/2}}
    \frac{1}{\Big(ip_0-\mu_q-k\big(1+m_f^2\big)^{1/2}\Big)^2
    -k^2\big(1+m_b^2\big)}\nonumber\\[1ex]
  &+\big(n_{F}(m_f^2;T,\mu_q)-1\big)\frac{1}{\big(1+m_f^2
    \big)^{1/2}}\frac{1}{\Big(ip_0-\mu_q+k\big(1+m_f^2\big)^{1/2}
    \Big)^2-k^2\big(1+m_b^2\big)}\bigg\}\,.
\label{eq:f1b1}
\end{align}
\end{widetext}
The threshold function $L$ in the flow equation of the Yukawa coupling
in \eq{eq:hexp} is given by
\begin{align}
  &L_{(1,1)}^{(d)}(m_f^2,m_b^2,\eta_{f},\eta_{b};
    T,\mu_q,p_0)\nonumber\\[1ex]
  =&\frac{2}{d-1}\bigg[\Big(1-\frac{\eta_{b}}{
     d+1}\Big)\mathcal{FB}_{(1,2)}(
     m_f^2,m_b^2;T,\mu_q,p_0)\nonumber\\[1ex]
  &+\Big(1-\frac{\eta_{f}}{d}\Big)\mathcal{FB}_{(2,1)}(
    m_f^2,m_b^2;T,\mu_q,p_0)\bigg]\,,
\end{align}
Furthermore, we also need another class of threshold functions,
appearing in the flow equation of the four-fermion coupling induced by
the meson exchange in \eq{eq:dtlambdaAphi}. This leads us to define
\begin{align}
  &\mathcal{FBB}_{(n_f,n_{b1},n_{b2})}(m_f^2,m_{b1}^2,m_{b2}^2)\nonumber
  \\[1ex]\nonumber 
  =&\frac{T}{k}\sum_{n_q}\Big(G_{f}(q,m_f^2)\Big)^{n_f}\Big(G_{b}(
     q-p,m_{b1}^2)\Big)^{n_{b1}}\nonumber\\[1ex]
  &\times
    \Big(G_{b}(q-p,m_{b2}^2)\Big)^{n_{b2}}\,.
\end{align}
Again, all the functions $\mathcal{FBB}$'s in
\eq{eq:dtlambdaAphi} can be easily obtained from
$\mathcal{FBB}_{(1,1,1)}$, whose analytic expression reads
\begin{widetext}
\begin{align}
  &\mathcal{FBB}_{(1,1,1)}(m_f^2,m_{b1}^2,m_{b2}^2)\nonumber \\[1ex]
  =
  &\frac{k^2}{2}\bigg\{n_{B}(m_{b1}^2;T)\frac{1}{\big(1+m_{b1}^2\big)^{1/2}}
    \frac{1}{\big(m_{b1}^2-m_{b2}^2\big)}\frac{1}{\Big(\mu_q-ip_0+k\big(1+
    m_{b1}^2\big)^{1/2}\Big)^2-k^2\big(1+m_f^2\big)}\nonumber \\[1ex]
  &+\big(n_{B}(m_{b1}^2;T)+1\big)\frac{1}{\big(1+m_{b1}^2\big)^{1/2}}
    \frac{1}{\big(m_{b1}^2-m_{b2}^2\big)}\frac{1}{\Big(\mu_q-ip_0-k\big(1
    +m_{b1}^2\big)^{1/2}\Big)^2-k^2\big(1+m_f^2\big)}\nonumber\\[1ex]
  &+n_{B}(m_{b2}^2;T)\frac{1}{\big(1+m_{b2}^2\big)^{1/2}}\frac{1}{
    \big(m_{b2}^2-m_{b1}^2\big)}\frac{1}{\Big(\mu_q-ip_0+k\big(1+
    m_{b2}^2\big)^{1/2}\Big)^2-k^2\big(1+m_f^2\big)}\nonumber \\[1ex]
  &+\big(n_{B}(m_{b2}^2;T)+1\big)\frac{1}{\big(1+m_{b2}^2\big)^{1/2}}
    \frac{1}{\big(m_{b2}^2-m_{b1}^2\big)}\frac{1}{\Big(\mu_q-ip_0
    -k\big(1+m_{b2}^2\big)^{1/2}\Big)^2-k^2\big(1+m_f^2\big)}
    \nonumber\\[1ex]
  &-n_{F}(m_f^2;T,\mu_q)\frac{k^2}{\big(1+m_f^2\big)^{1/2}}
    \frac{1}{\Big(ip_0-\mu_q+k\big(1+m_f^2\big)^{1/2}\Big)^2
    -k^2\big(1+m_{b1}^2\big)}\nonumber\\[1ex]
  &\times\frac{1}{\Big(ip_0-\mu_q+k\big(1+m_f^2\big)^{1/2}
    \Big)^2-k^2\big(1+m_{b2}^2\big)}-\big(n_{F}(m_f^2;T,
    -\mu_q)-1\big)\frac{k^2}{\big(1+m_f^2\big)^{1/2}}
    \nonumber\\[1ex]
  &\times\frac{1}{\Big(ip_0-\mu_q-k\big(1+m_f^2
    \big)^{1/2}\Big)^2-k^2\big(1+m_{b1}^2\big)}\frac{1}{
    \Big(ip_0
    -\mu_q-k\big(1+m_f^2\big)^{1/2}\Big)^2-k^2
    \big(1+m_{b2}^2\big)}\bigg\}\,.
\label{eq:f1b1b1}
\end{align}
\end{widetext}
Finally, we would like to present the threshold functions in
\eq{eq:DeltaAqexpl}, which is somewhat different from those mentioned
above, since the external momentum $\bm{p}$ in the following is
nonvanishing, and the definition reads
\begin{align}
  &\mathcal{FF}_{(n_{f_1},n_{f_2})}(m_{f_1}^2,m_{f_2}^2;T,\mu_q,p_0,
    \bm{p})\nonumber \\[1ex]  
  =&\frac{T}{k}\sum_{n_q}\Big(G_{f}(q,m_{f_1}^2)\Big)^{n_f}
     \Big(G'_{f}(q-p,m_{f_2}^2)\Big)^{n_b}\,,\label{eq:FF}
\end{align}
with $G_{f}$ given in \eq{eq:Gf} and $G'_{f}$ by
\begin{align}
 &G'_{f}(q-p,m_{f_2}^2)\nonumber \\[1ex]
  =&\frac{1}{(\tilde{q}_0-\tilde{p}_0+i\tilde{\mu}_q)^2+(\tilde{
     \bm{q}}-\tilde{\bm{p}})^2\Big(1+r_F\big((\tilde{\bm{q}}
     -\tilde{\bm{p}})^2\big)\Big)^2+m_{f_2}^2}\,,
\end{align}
and here $\tilde{\bm{q}}=\bm{q}/k$ and $\tilde{\bm{p}}=\bm{p}/k$. In
the same way, one has
\begin{align}
  \mathcal{FF}_{(2,1)}(m_{f_1}^2,m_{f_2}^2)=
  &-\frac{\partial}{\partial m_{f_1}^2}
    \mathcal{FF}_{(1,1)}(m_{f_1}^2,m_{f_2}^2)\,,
\end{align}
and the explicit expression for $\mathcal{FF}_{(1,1)}$ is given as
follows
\begin{align}
  &\mathcal{FF}_{(1,1)}(m_{f_1}^2,m_{f_2}^2)\nonumber \\[1ex]
  =&\frac{k^3}{4 E_q E_{q-p}}\bigg\{\Big(-1+n_{F}(E_{q-p};T,\mu_q)+
     n_{F}(E_{q};T,-\mu_q)\Big)\nonumber\\[1ex]
  \times& \frac{1}{i p_0-E_q-E_{q-p}}+\Big(n_{F}(E_{q-p};T,-\mu_q)-
          n_{F}(E_{q};T,-\mu_q)\Big) \nonumber\\[1ex]
  \times& \frac{1}{i p_0-E_q+E_{q-p}}+\Big(n_{F}(E_{q};T,\mu_q)-
          n_{F}(E_{q-p};T,\mu_q)\Big)\nonumber\\[1ex]
  \times& \frac{1}{i p_0+E_q-E_{q-p}}+\Big(1-
          n_{F}(E_{q};T,\mu_q)\nonumber\\[1ex]
  -&n_{F}(E_{q-p};T,-\mu_q)\Big) \frac{1}{i p_0+E_q+E_{q-p}}\bigg\}\,,
\end{align}
with 
\begin{align}
  E_q=&k(1+m_{f_1}^2)^{1/2}\,,\\[1ex]
  E_{q-p}=&k\Big[(\tilde{\bm{q}}-\tilde{\bm{p}})^2\Big(1+r_F
            \big((\tilde{\bm{q}}-\tilde{\bm{p}})^2\big)\Big)^2
            +m_{f_2}^2\Big]^{1/2}\,,
\end{align}
and the modified quark distribution function $n_{F}$ defined in
\eq{eq:nFL} with $x = E_q$.

%%%%%%%%%%%%%%%%%%%%%%%%%%%%%%%%%%%
% allow for equations to break between column and pages to avoid
% awkward spacing in the appendix.
%%%%%%%%%%%%%%%%%%%%%%%%%%%%%%%%%%%
\endgroup
%%%%%%%%%%%%%%%%%%%%%%%%%%%%%%%%%%%

% The \nocite command causes all entries in a bibliography to be
% printed out whether or not they are actually referenced in the
% text. This is appropriate for the sample file to show the different
% styles of references, but authors most likely will not want to use
% it.  \nocite{*}

%\bibliography{refspec}% Produces the bibliography viaBibTeX.
\bibliography{ref-lib}% Produces the bibliography via BibTeX.

\end{document}